\documentclass{jfm}

\usepackage{graphicx}
\usepackage{natbib}
\usepackage{hyperref}
\usepackage{amsmath}
\usepackage{ar}
\usepackage{tikz}
\usepackage{pgfplots}
\usepackage{tikz-3dplot}
\usepackage{bm}
\usepackage{multirow}
\usepackage{subfig}
\usepackage{overpic}
\usepackage{ulem}

\usepackage{tikz}

\newcommand{\aver}[1]{ \! \left\langle {#1} \right \rangle \!}

\title[Stability and dynamics of the flow past 3D prisms]{Stability and dynamics of the laminar flow past rectangular prisms}

\author[A. Chiarini and E. Boujo]
{Alessandro Chiarini\aff{1}\corresp{Present address: Dipartimento di Scienze e Tecnologie Aerospaziali, Politecnico di Milano, via La Masa 34, 20156 Milano, Italy.}\corresp{\email{alessandro.chiarini@polimi.it}} and Edouard Boujo\aff{2}\corresp{\email{edouard.boujo@epfl.ch}}
}
\affiliation{
}

\affiliation{
\aff{1} Complex Fluids and Flows Unit, Okinawa Institute of Science and Technology Graduate University, 1919-1 Tancha, Onna-son, Okinawa 904-0495, Japan
\aff{2}Laboratory of Fluid Mechanics and Instabilities, \'Ecole Polytechnique F\'ed\'erale de Lausanne, CH-1015 Lausanne, Switzerland
}

\begin{document}
\maketitle

\begin{abstract}
The laminar flow past rectangular prisms is studied in the space
of length-to-height ratio ($1 \le L/H \le 5$), width-to-height ratio ($1.2 \le W/H \le 5$) and Reynolds number ($Re \lessapprox 700$); $L$ and $W$ are the streamwise and cross-flow dimensions of the prisms. 
The primary bifurcation is investigated with linear stability analysis. 
For large $W/L$, an oscillating mode breaks the top/bottom planar symmetry. 
For smaller $W/L$ the flow becomes unstable to stationary perturbations, and the wake experiences a static deflection, vertical for intermediate $W/L$ and horizontal for small $W/L$. 
Weakly nonlinear analysis and nonlinear direct numerical simulations are used for $L/H = 5$ and larger $Re$. 
For $W/H = 1.2$ and 2.25, 
the flow recovers the top/bottom planar symmetry but loses the left/right one, via supercritical and subcritical pitchfork bifurcations, respectively. 
For even larger $Re$, the flow becomes unsteady and oscillates around either the deflected (small $W/H$) or the non-deflected (intermediate $W/H$) wake. 
For intermediate $W/H$ and $Re$, a fully symmetric periodic regime is detected, with hairpin vortices shed from the top and bottom leading-edge (LE) shear layers; its triggering mechanism is discussed.
At large $Re$ and for all $W/H$, the flow approaches a chaotic state characterised by the superposition of different modes: shedding of hairpin vortices from the LE shear layers, and wake oscillations in the horizontal and vertical directions.
In some portions of the parameter space the different modes synchronise, giving rise to periodic regimes also at relatively large $Re$.
\end{abstract}


\section{Introduction}
\label{sec:introduction}

Flows past bluff bodies and their 
transitions with increasing Reynolds numbers from steady two-dimensional (2D) wake flows, through unsteady and three-dimensional (3D) flows, to fully turbulent wakes have attracted much attention
over the years \citep{oertel-1990,williamson-1996,choi-etal-2008,thompson-etal-2021}, as their relevance goes beyond the fundamental interest and encompasses several industrial applications, for example
in the field of vortex-induced vibrations
\citep{williamson-govardhan-2008}.

\subsection{2D cylinders}
Most studies 
on 2D bluff bodies have focused on
circular and square cylinders as prototypes to characterise the flow bifurcations. At the critical Reynolds number 
(based on the free stream velocity $U_\infty$ and cylinder diameter $D$) 
$Re_c \approx 45-47$ the flow undergoes a Hopf bifurcation from a symmetric steady state towards a time-periodic asymmetric state \citep{noack.eckelmann-1994-globalstability} that gives origin to the von K\'{a}rm\'{a}n vortex shedding. 
The triggering mechanism is known to result from a global instability \citep{jackson-1987-finiteelementstudy}, which arises when the region of local absolute instability is large enough \citep{chomaz-2005}, and
whose onset therefore depends on the size of the wake recirculation region and on the maximum reverse flow velocity \citep{chiarini-quadrio-auteri-2022b}.
At larger Reynolds numbers the flow undergoes a secondary instability and becomes three-dimensional. For the circular cylinder, 
mode A with spanwise wavelength $\lambda \approx 3.9 D$ becomes unstable at $Re \approx 190$, while 
mode B with 
$\lambda \approx 1.2 D$ becomes unstable at slightly larger $Re$ \citep{barkley-henderson-1996,williamson-1996,williamson-1996b}. 
A further quasi-periodic mode with $\lambda \approx 2.5 D$ has been found to become unstable at larger Reynolds numbers \citep{blackburn-lopez-2003,blackburn-etal-2005,blackburn-sheard-2010}.

For 2D rectangular cylinders, the flow dynamics and the physical mechanism of the bifurcations vary with the length-to-height ratio $L/H$.
For $Re \ge 300$, in fact, the Strouhal number of the flow increases in an almost stepwise manner with $L/H$, due to a pressure feedback-loop that locks the shedding of vortices from the shear layers separating from the leading (LE) and trailing (TE) edges \citep{okajima-1982, nakamura-etal-1991,mills-etal-1995, hourigan-thompson-tan-2001}. This stepwise increase  is closely related to the number of large-scale vortices 
that fit along the sides of the rectangular cylinders. Two different regimes are possible depending on the relative phase of the leading- and trailing-edge vortices \citep{chiarini-quadrio-auteri-2022}.  
Using  dynamic mode decomposition, 
\cite{zhang-etal-2023} observed that the 
feedback mechanism  at $Re=1000$ changes with the length-to-height ratio. For small $L/H$, the pressure feedback loop encompasses the whole separation region and the flow is controlled by the impinging shear-layer instability; for larger $L/H$, the feedback loop covers the entire chord of the cylinder and the flow is controlled by the leading-edge vortex shedding instability. 
%
The first onset of three-dimensionality also depends on $L/H$: unlike short cylinders, where the flow becomes 3D via the  modes A, B and C of the wake \citep{robichaux-balachandar-vanka-1999,blackburn-lopez-2003}, for long enough rectangular cylinders \cite{chiarini-quadrio-auteri-2022d} found that the first 3D unstable mode is of quasi-subharmonic nature (mode QS), 
and is due to an inviscid mechanism triggered by the interaction of the vortices placed over the cylinder sides.

\subsection{Axisymmetric 3D bluff bodies}
Less studies have considered  flows past 3D  bluff bodies, although they are ubiquitous in human life and engineering applications, e.g. tall buildings, chimneys, pylons, cars, trains and other ground vehicles. 
These flows  exhibit a richer physics and, already at low $Re$, the first bifurcations and their sequence depend on the body geometry. 

Unlike 2D cylinder wakes, the flow past a sphere becomes asymmetric prior to a transition to unsteady flow \citep{magarvey-etal-1961,magarvey-etal-1961b,magarvey-etal-1965}. 
The wake remains steady and axisymmetric up to $Re \approx 211$ \citep{johnson-patel-1999}, and then 
transitions to a steady asymmetric state through the regular bifurcation of an eigenmode of  azimuthal wavenumber $m=1$ \citep{tomboulides-orszag-2000}. 
The resulting wake is characterised by a pair of steady streamwise vortices and exhibits a reflectional symmetry about a longitudinal plane of arbitrary azimuthal orientation. 
At higher $Re$, a Hopf bifurcation renders the flow oscillatory. \cite{natarajan-acrivos-1993} found with stability analysis that an unsteady $m=1$ mode of the steady axisymmetric base flow becomes unstable  at $Re \approx 277$, as  confirmed by the experiments and simulations of \cite{tomboulides-etal-1993,johnson-patel-1999,tomboulides-orszag-2000}. This unsteady regime consists of hairpin vortices  shed downstream of the sphere in the same plane as that of the steady asymmetric regime. \cite{citro-etal-2017b} performed a 3D global stability analysis of the steady asymmetric  flow to characterise the eigenmode responsible for this second bifurcation. 
They found a critical Reynolds number of $Re = 272$, and deduced 
from a structural sensitivity analysis \citep{giannetti-camarri-luchini-2010} 
that the instability is driven by the region outside the asymmetric wake. 
At larger Reynolds numbers, $Re>600$, the wake loses its periodicity and the flow becomes more chaotic \citep{magarvey-etal-1961b}.

A similar bifurcation scenario has been observed for the flow past a disk 
perpendicular to the incoming flow.
At $Re  \approx 115 $, 
a regular  bifurcation leads to a 3D steady asymmetric state, with a reflectional symmetry \cite{natarajan-acrivos-1993,fabre-etal-2008,meliga-etal-2009}. 
At slightly larger $Re$, i.e. $Re \approx 121$, a Hopf bifurcation breaks the remaining reflectional symmetry and time invariance. 
At $Re \approx 140$, a third bifurcation occurs  that preserves the flow unsteadiness but restores a reflectional symmetry orthogonal to that preserved by the first bifurcation. 

Similarly, \cite{bohorquez-etal-2011} investigated the stability of the  flow past bullet-shaped bodies (slender cylindrical bodies with an elliptical leading edge and a blunt trailing edge). They found the same sequence of bifurcations, and observed that the critical $Re$ increases with the aspect ratio. Compared to the sphere and disk wakes,  the symmetry plane selected by the first regular bifurcation is preserved over a larger range of Reynolds numbers, up to $Re \approx 500$.

\subsection{Non-axisymmetric 3D bluff bodies}

Many studies on non-axisymmetric 3D bluff bodies have focused on two simple types of geometry: finite circular cylinders and rectangular prisms.

\cite{inoue-sakuragi-2008} investigated the vortex shedding past finite circular cylinders of span-to-diameter ratio  $0.5 \le W/D \le 100$  with direct numerical simulations (DNS) for $Re \le 300$. 
They found that the flow 
changes drastically depending on $W/D$ and $Re$, and identified five basic wake patterns: 
(i)~periodic oblique vortex shedding,
(ii)~quasi-periodic oblique vortex shedding,
(iii)~periodic hairpin vortex falloff,
(iv)~two stable counter-rotating vortex pairs,
and 
(v)~alternating shedding of counter-rotating vortex pairs from the free ends.
%
For the same geometry with $W/D \in [0.5,2]$,
\cite{yang-etal-2022} obtained consistent results  
by DNS and linear stability analysis for $Re \le 1000$. 

Rectangular prisms 
differ from  finite circular cylinders, as the  sharp corners 
set the location of flow separation. 
The bifurcation sequence and critical $Re$  are expected to change with the length-to-height ratio $L/H$  (as already observed for 2D rectangular cylinders) and  width-to-height ratio $W/H$. 
\cite{MarquetLarsson2015}  investigated the linear stability of the steady flow past  thin rectangular plates with $L/H=1/6$ and various widths. 
For $ 1 \le W/H \le 2$,  a pitchfork bifurcation  breaks the top/bottom planar symmetry, similar to the flow past axisymmetric bodies like spheres and disks. 
For $W/H > 2.5$, a Hopf bifurcation breaks the top/bottom symmetry and leads to a time-periodic regime, similar to 2D cylinders. 
Surprisingly, for intermediate widths $ 2 \leq W/H \leq 2.5$, a Hopf bifurcation breaks the left/right planar symmetry. 
\cite{zampogna_boujo_2023} set the width to $W/H=1.2$ and investigated the flow bifurcations 
for $1/6 \le L/H \le 3$. 
Their analysis yields a series of pitchfork and Hopf bifurcations. Two stationary modes that break respectively the top/bottom and left/right planar symmetries become first unstable. At larger $Re$, two oscillatory modes become unstable, and again break either the top/bottom or left/right symmetry. 
The critical $Re$ of the first bifurcation increases monotonically  with $L/H$, similar to  flows past 2D rectangular cylinders \citep{chiarini-quadrio-auteri-2021a} and axisymmetric bodies (disk, sphere and bullet-shaped bodies).
For the flow past a cube, this bifurcation scenario is confirmed by the low-$Re$ experiments of \cite{klotz-etal-2014} and DNS of \cite{saha-2004} and \cite{meng_an_cheng_kimiaei_2021}: 
at $Re \approx 216$ the steady symmetric flow bifurcates towards a steady asymmetric regime with only one planar symmetry; at $Re=265$ the flow undergoes a Hopf bifurcation and oscillates about the bifurcated asymmetric regime,
with a shedding of hairpin vortices similar to that past a sphere and a disk.

\subsection{Present study}
In this work we take a step forward from the works of \cite{MarquetLarsson2015} and \cite{zampogna_boujo_2023}, and characterise the sequence of bifurcations for the flow past rectangular prisms in a three-dimensional space of parameters, varying simultaneously the length-to-height and width-to-height ratios and the Reynolds number. 
We use linear and weakly nonlinear stability analyses to characterise the first bifurcations, and fully nonlinear DNS to describe the flow regimes at larger $Re$. 
We consider $1 \le L/H \le 5$, $1.2 \le W/H \le 5$, and $Re \lessapprox 700$.
The DNS focus on $L/H=5$, similar to the BARC benchmark (``Benchmark on the Aerodynamics of a Rectangular 5:1 Cylinder'', see \url{https://www.aniv-iawe.org/barc-docs}), which aims to characterise the flow past elongated bluff bodies and set standards for simulations and experiments in the turbulent regime.

The structure of the paper is as follows. 
Section \ref{sec:methods} briefly describes  the mathematical formulation and numerical methods. 
The low-$Re$ steady flow is characterised in \S\ref{sec:baseflow}, while \S\S \ref{sec:stability}-\ref{sec:structsensit} deal with the linear stability analysis
and \S\ref{sec:WNL} with the weakly nonlinear stability analysis.
Section \ref{sec:simulations} presents the DNS results and details the flow regimes after the first bifurcations for $L/H=5$. 
A concluding discussion is provided in \S\ref{sec:conclusions}.

\section{Methods}
\label{sec:methods}
\subsection{Flow configuration}


%

\begin{figure}
\centerline{  
   \begin{overpic}[width=11cm, trim=0mm 0mm 0mm 0mm, clip=true]{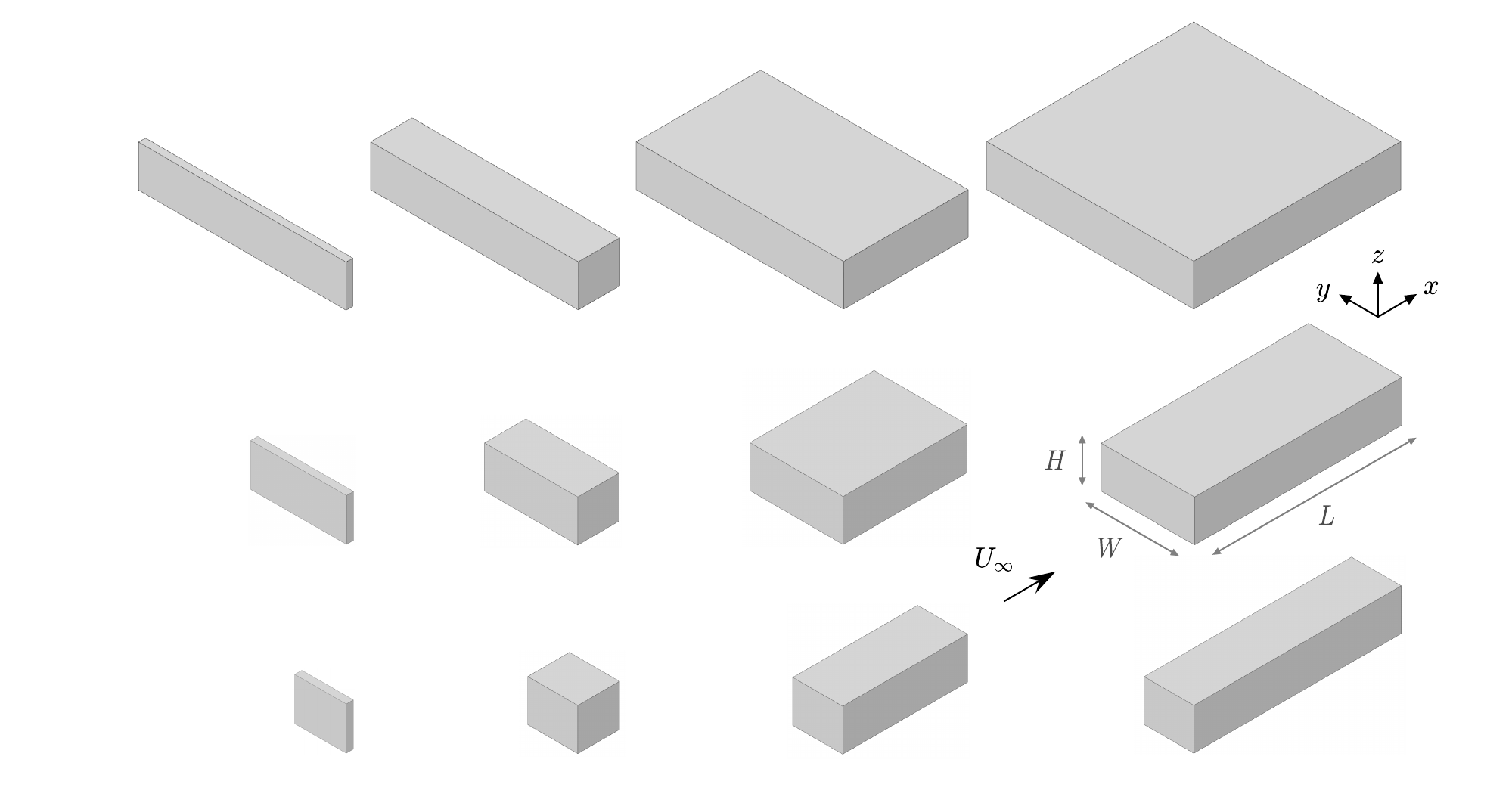}  
      \put( -3,  34){$W=5$}
      \put( -2,  19){$W=2.25$}
      \put( -1,   5){$W=1.2$}
      \put(15,   0){$L=1/6$}
      \put(33.5, 0){$L=1$}
      \put(51,   0){$L=3$}
      \put(74,   0){$L=5$}
   \end{overpic}
   \vspace{0.1cm}
}
\caption{
Overview of the prism geometry for various lengths $L$ and widths $W$.}
\label{fig:geom}
\end{figure}

The incompressible flow past rectangular prisms with aspect ratio $1 \le L/H \le 5$ and $1.2 \le W/H \le 5$ is considered; here $L$, $W$ and $H$ are the length, width and height of the prism, i.e. the sizes in the streamwise, spanwise and vertical directions; see figure~\ref{fig:geom}.
A Cartesian reference frame is placed at the leading edge of the prisms, with the $x$ axis aligned with the flow direction, and the $y$ and $z$ axes denoting the spanwise and vertical directions. The Reynolds number is based on $H$ and on the free stream velocity $U_\infty$. 
Unless otherwise stated,  all quantities are made dimensionless with $H$ and $U_\infty$.
The flow is governed by the incompressible Navier--Stokes (NS) equations for the velocity $\bm{u} = (u,v,w)$ and pressure $p$,
\begin{equation}
\frac{\partial \bm{u}}{\partial t} + \bm{u} \cdot \bm{\nabla} \bm{u} = - \bm{\nabla} p  + \frac{1}{Re} \bm{\nabla}^2 \bm{u},
\quad
\bm{\nabla} \cdot \bm{u} = 0.
\label{eq:nseq}
\end{equation}
%

\subsection{Mathematical formulation}

The onset of  instability is studied by linear stability analysis (LSA).
The field $\{\bm{u},p\}$ is divided into a time-independent base flow $\{\bm{u}_0,p_0\}$ and an unsteady contribution $\{\bm{u}_1,p_1\}$ of small amplitude $\epsilon$, i.e.
\begin{equation}
  \bm{u}(\bm{x},t) = \bm{u}_0(\bm{x}) + \epsilon \bm{u}_1(\bm{x},t), \ \text{and} \ p(\bm{x},t) = p_0(\bm{x}) + \epsilon p_1(\bm{x},t).
\end{equation} 
Using this decomposition in the NS equations \eqref{eq:nseq}, the steady base flow equations for $\{\bm{u}_0,p_0\}$ are obtained at order $\epsilon^0$, while the the linearised Navier--Stokes  equations  governing the small perturbations $\{\bm{u}_1,p_1\}$ are obtained at order $\epsilon^1$.
%
%
Using the normal-mode ansatz $\{\bm{u}_1,p_1\}(\bm{x},t)=\{\hat{\bm{u}}_1,\hat{p}_1\}(\bm{x})e^{\lambda t} + c.c.$ (where $c.c$ stands for complex conjugate)  yields an eigenvalue problem for the complex eigenvalue $\lambda = \sigma + i \omega$ and eigenvector $\{\hat{\bm{u}}_1,\hat{p}_1\}$:
\begin{equation}
\lambda \hat{\bm{u}_1} + \bm{u}_0 \cdot \bm{\nabla} \hat{\bm{u}}_1 + \hat{\bm{u}}_1 \cdot \bm{\nabla} \bm{u}_0 = - \bm{\nabla} \hat{p}_1 + \frac{1}{Re} \bm{\nabla}^2 \hat{\bm{u}}_1,
\quad
\bm{\nabla} \cdot \hat{\bm{u}}_1 = 0. 
\label{eq:eigenvalue}
\end{equation}
The linear stability of the system is determined by the sign of the real part of $\lambda$, i.e $\sigma$. 
If all $\sigma<0$, perturbations decay and the flow is stable. If $\sigma>0$ for at least one mode, perturbations grow exponentially. 
A mode is stationary or oscillatory if $\omega=0$ or $\omega>0$, respectively.
For each geometry $(L,W)$, several modes may become unstable as $Re$ increases; the critical Reynolds number $Re_c$ of each mode is computed by cubic interpolation of $\sigma(Re)$ with at least three values of $Re$ that bracket $\sigma = 0$.

\subsection{Numerical method}

Two different numerical methods are used. 
The LSA is performed with the non-commercial, finite elements  software FreeFem++ \citep{hecht-2012}, while the DNS are performed with an in-house code based on finite differences.

For the LSA (\S\ref{sec:stability}), the  steady base flow solution of  \eqref{eq:nseq} and the eigenvalue and eigenmodes solution of \eqref{eq:eigenvalue} are calculated with the same method as in \cite{zampogna_boujo_2023}.
Given the two symmetry planes $y=0$ and $z=0$ of the prism, we consider only the quarter-space $y,z\geq 0$ and build in the numerical domain  $\{x,y,z \, | \, -10\leq x \leq 20; 0\leq y,z \leq 10\}$ an unstructured tetrahedral mesh with nodes strongly clustered near the prism. See Appendix \ref{sec:app-conv} for convergence studies with respect to  mesh and domain sizes. The weak form of the equations is discretised with Arnold–-Brezzi–-Fortin MINI-elements (P1 for pressure and P1b for each velocity component). 
For $W/H > 2.5$, dimensions in the $y$ direction are normalised by $W$ and $y$ derivatives are scaled by a factor $1/W$.
The nonlinear base flow equations are solved with a Newton method, and the eigenvalue problem with a Krylov–Schur method. 
Calculations are run in parallel with a domain-decomposition method, typically on 48 processes.
Boundary conditions are as follows:
free-stream velocity ${\bm{u}}_0=(1,0,0)^T$ and zero perturbation ${\bm{u}}={\bm{0}}$ at the inlet; then, for both the base flow and the perturbations,
no-slip condition  ${\bm{u}}={\bm{0}}$ on the body,
stress-free condition  $-p \bm{n} + Re^{-1} \bnabla \bm{u} \cdot \bm{n}=\bm{0}$  at the outlet, and  
symmetric ($S$) conditions $u_{n}=0, \partial_n \bm{u}_{t}=\bm{0}$ or antisymmetric ($A$) conditions $\bm{u}_{t}=\bm{0}, \partial_n u_{n}=0$ on the symmetry planes $y=0$ and $z=0$ (where subscripts $n$ and $t$ denote normal and tangential directions, respectively), the specific choice depending on the considered field, i.e.  doubly symmetric base flow $S_y S_z$ or eigenmode belonging to one of the four possible families $S_y S_z$, $S_y A_z$, $A_y S_z$ and $A_y A_z$.
Adjoint modes, used to compute structural sensitivities, are obtained in a similar way \citep[see][for more details]{zampogna_boujo_2023}.

The 3D nonlinear simulations (\S\ref{sec:simulations}) consider $L/H=5$ and $W/H=1.2,2.25$ and $5$. 
The NS equations are solved using a DNS code introduced by \cite{luchini-2016} and written in CPL \citep{luchini-2021}, which employs second-order finite differences on a staggered grid in the three directions. 
This DNS code has been previously validated and used to compute the flow past rectangular cylinders in  the laminar \citep{chiarini-quadrio-auteri-2022d} and turbulent \citep{chiarini-quadrio-2021,chiarini-quadrio-2021b,chiarini-etal-2022} regimes. 
The momentum equation is advanced in time by a fractional step method using a third-order Runge-Kutta scheme. The Poisson equation for the pressure is solved using an iterative SOR algorithm. The presence of the prism is dealt with a second-order implicit immersed boundary method, implemented in staggered variables \citep{luchini-2013}.
%
The computational domain extends for $-25H \le x \le 75H$, $-40 H \le y \le 40 H$ and $-25H \le z \le 25H$ in the streamwise, spanwise and vertical directions, corresponding to domain sizes $L_x=100H$, $L_y=80H$ and $L_z=50H$. 
The computational domain is discretised with $N_x=1072$ and $N_z=590$ points in the streamwise and vertical directions. 
The number of points in the spanwise direction is $N_y=666$, $720$ and $804$ for $W/H=1.2, 2.25$ and $5$, respectively. The points are distributed  with a geometrical progression in the three directions, to obtain a higher resolution close to the prisms, especially close to the corners and in the wake. 
Close to the corners the grid spacing is $\Delta x = \Delta y = \Delta z \approx 0.008H$. 
The adaptive time step is  computed at every iteration to enforce a 
CFL number  below unity, yielding an average value of $\Delta t \approx 0.0066 H/U_\infty$. 
After the transient regime, all simulations are advanced for more than $1000$ convective time units $H/U_\infty$ to ensure convergence of all quantities of interest.
See appendix \ref{sec:app-conv} for a convergence study with respect to the grid resolution and the time step.

\section{Global linear stability analysis}
\label{sec:stability}

\subsection{Base flow}
\label{sec:baseflow}

\begin{figure}
\vspace{0.3cm}
\begin{centering}
\hspace{0.2cm}
\begin{tabular}{c c|c} 
 & $L=1$ & $L=5$  
\\ 
\raisebox{1.5\normalbaselineskip}[0pt][0pt]{\rotatebox[origin=c]{0}{$W=5$}}
&    
    \begin{overpic}[width=2.65cm, trim=70mm 55mm 150mm 72mm, clip=true]{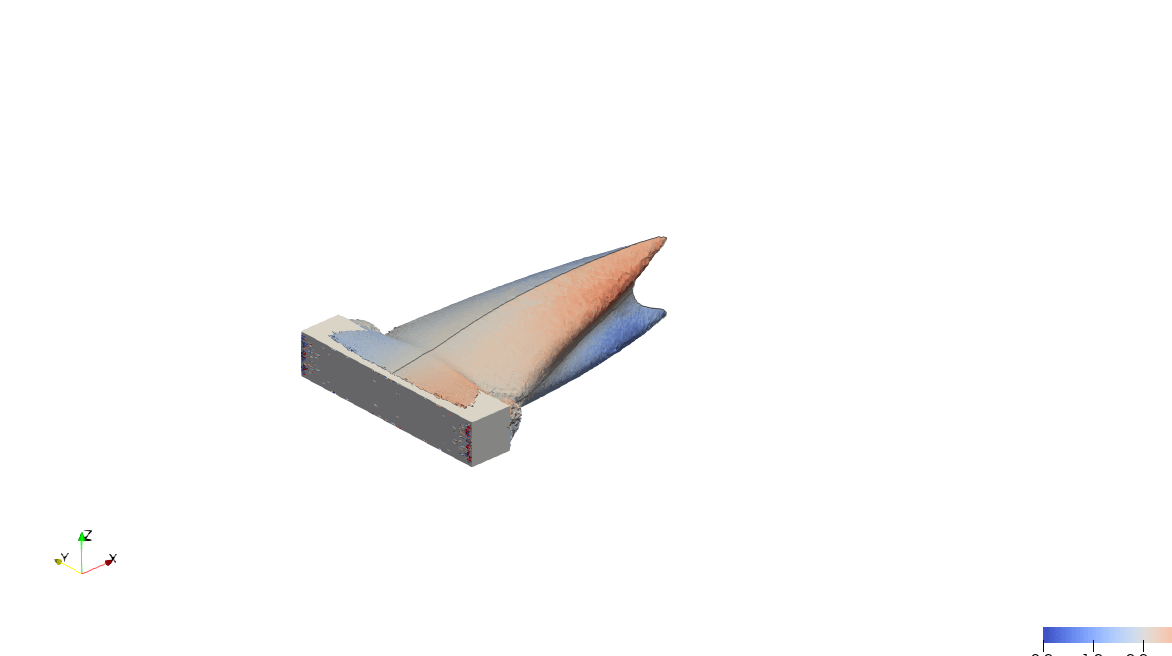}    
    \end{overpic} 
&
\hspace{0.2cm}
     \begin{overpic}[width=2.65cm, trim=70mm 55mm 150mm 72mm, clip=true]{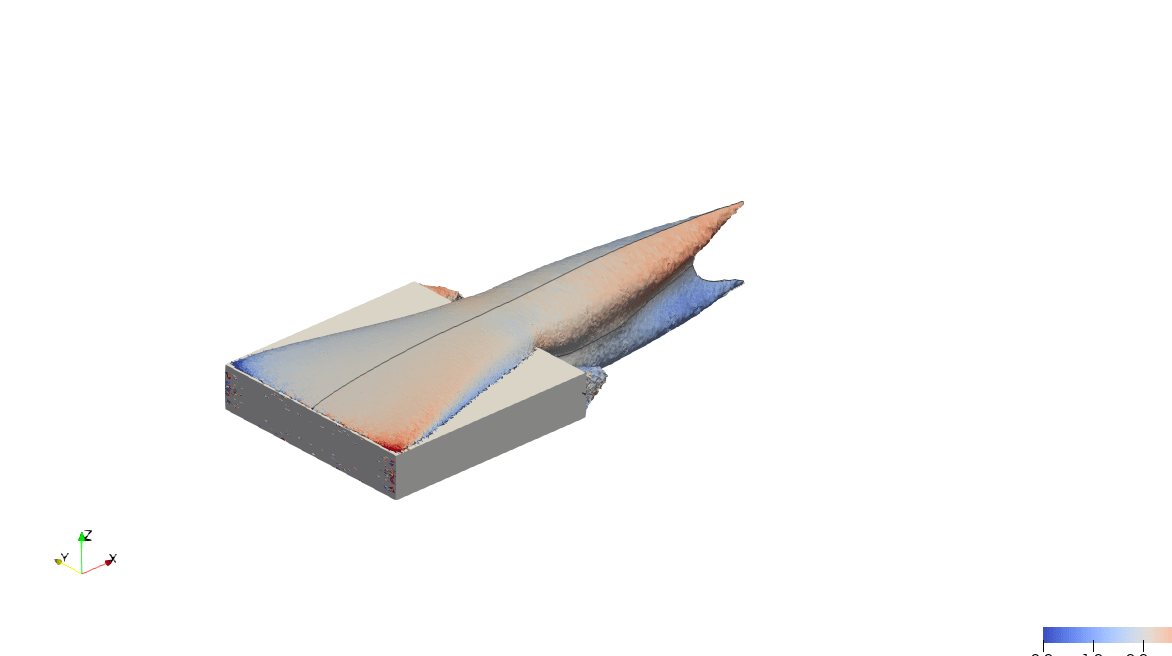}   
 	\end{overpic} 
\\
\raisebox{1.3\normalbaselineskip}[0pt][0pt]{\rotatebox[origin=c]{0}{$W=2.25$}}
&
    \begin{overpic}[width=2.65cm, trim=70mm 55mm 150mm 90mm, clip=true]{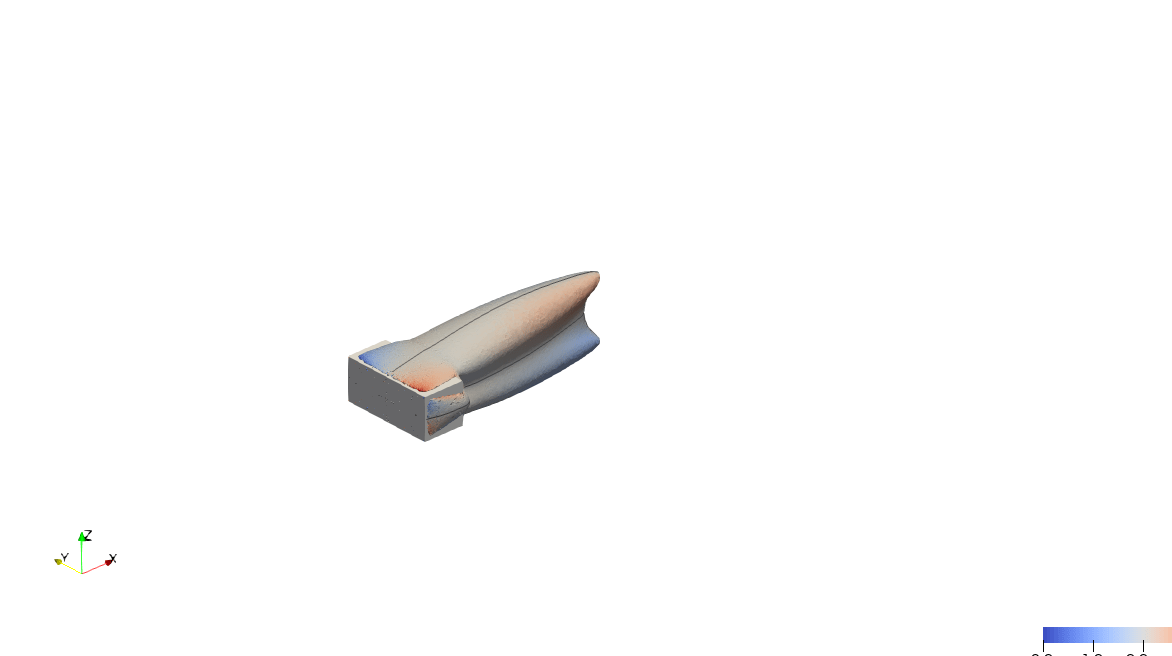}   
 	\end{overpic} 
& 
\hspace{0.2cm}
    \begin{overpic}[width=2.65cm, trim=70mm 55mm 150mm 90mm, clip=true]{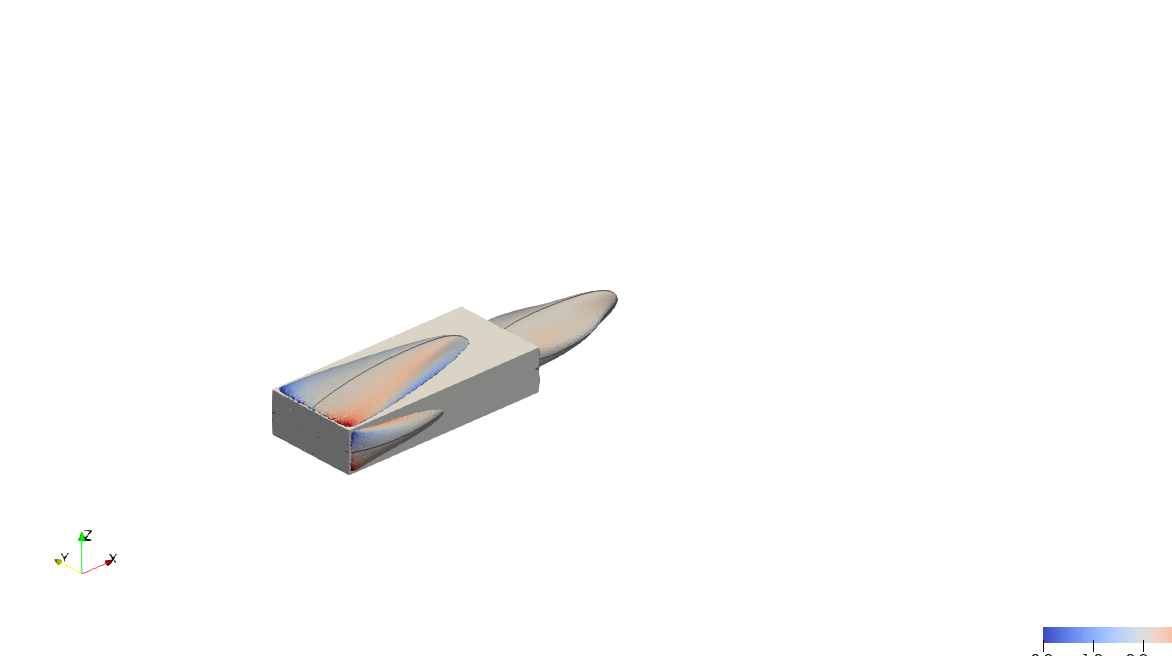}  
 	\end{overpic} 
\\
\raisebox{1.\normalbaselineskip}[0pt][0pt]{\rotatebox[origin=c]{0}{$W=1.2$}}
& 
   \begin{overpic}[width=2.65cm, trim=70mm 60mm 150mm 95mm, clip=true]{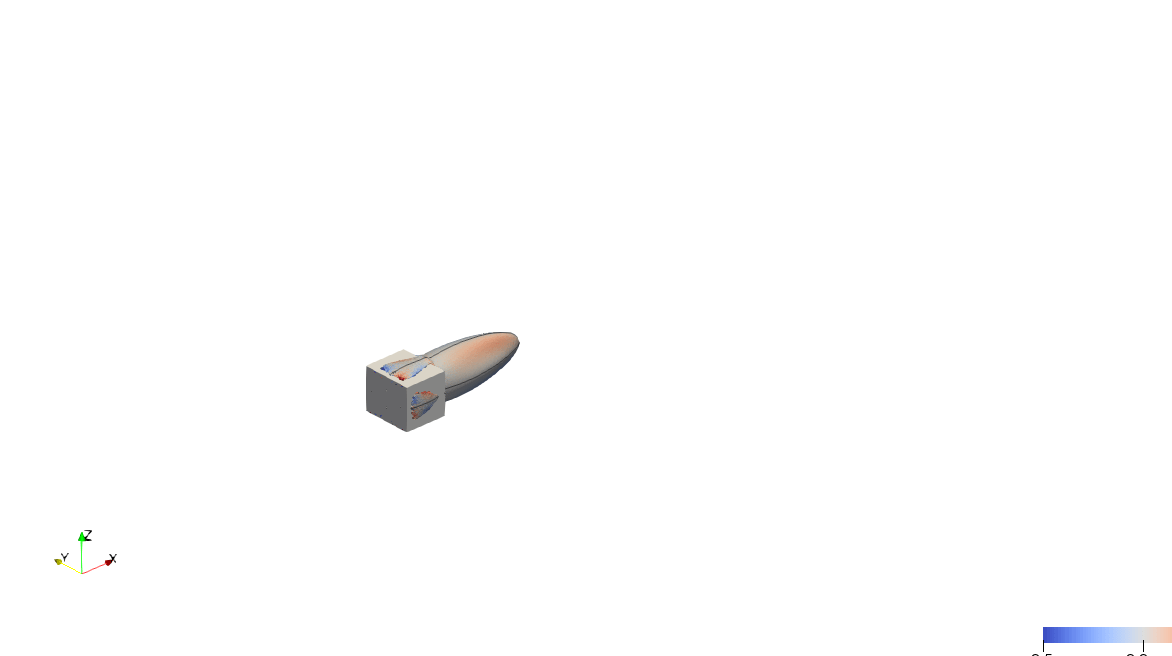} 
 	\end{overpic}
&
\hspace{0.2cm}
    \begin{overpic}[width=2.65cm, trim=70mm 60mm 150mm 95mm, clip=true]{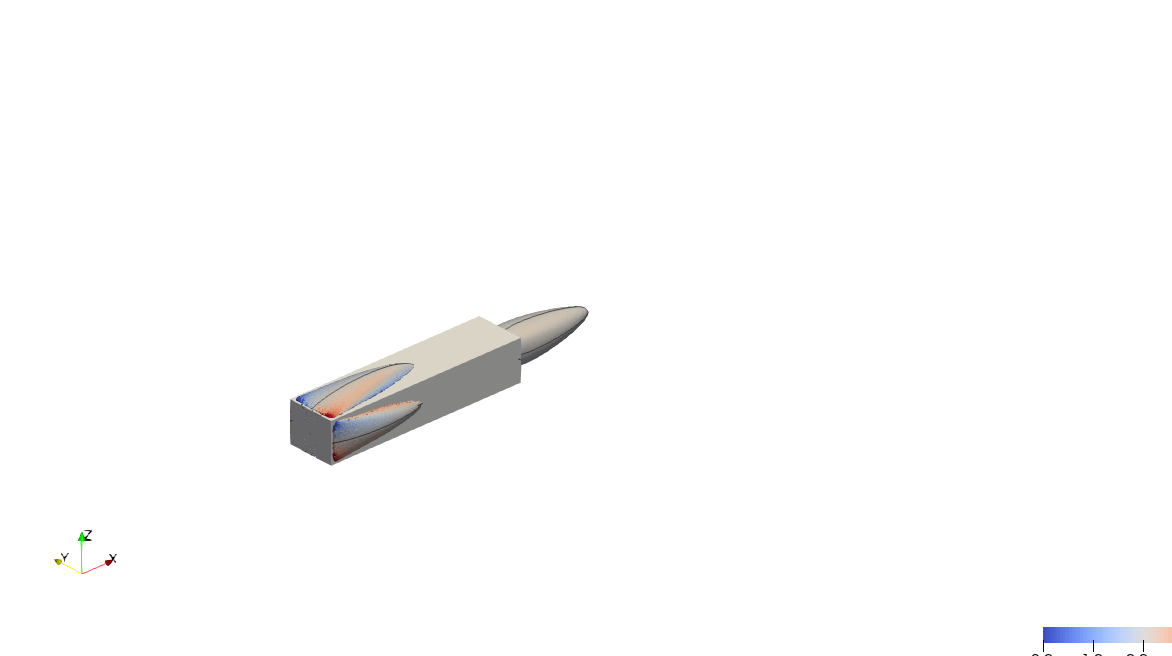} 
 	\end{overpic} 	
\\
\end{tabular} 
\vspace{-0.3cm}
\caption{
Base flow near the first bifurcation,
visualised with isosurfaces of zero streamwise velocity ($u_0 = 0$)
coloured by streamwise vorticity $\omega_x$.
\label{tab:BF}
}
\end{centering}
\end{figure}

The  steady symmetric base flow is shown in figure~\ref{tab:BF} by means of isosurfaces of  zero streamwise velocity ($u_0=0$) coloured by streamwise vorticity $\omega_x$.
The wake has a different shape depending on the prism geometry. 
Long and narrow bodies (e.g. $L=5$, $W=1.2$) tend to exhibit a convex backflow region centred about the vertical and horizontal symmetry planes $y=0$ and $z=0$, somewhat reminiscent of axisymmetric wakes. 
For these bodies the top/bottom and left/right recirculation regions originating from the leading edges reattach on the body upstream of the trailing edges.
By contrast, short and wide bodies (e.g. $L=1$, $W=5$) tend to exhibit a non-convex backflow region, with two lobes above and below the horizontal symmetry plane $z = 0$. Interestingly, moving downstream the backflow region enlarges in the vertical direction $z$ and shrinks in the spanwise one $y$, resulting in ``peanut''-shaped $y-z$ cross-sections. 
In addition, the upper/lower recirculation regions originating from the leading edges extend all the way down to the trailing edges without reattaching on the body, and are therefore connected with the main wake recirculation.

\begin{figure}
\vspace{0.3cm}
\centerline{   
    \begin{overpic}[width=4.5cm, trim=21mm 7mm 24mm 0mm, clip=true]{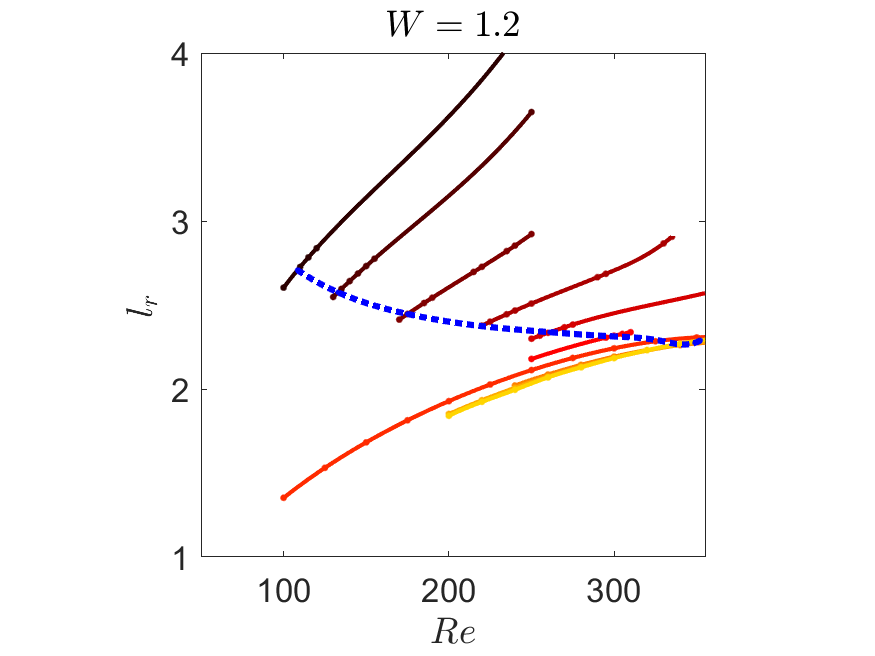}
       \put(-3,91){$(a)$}
       \put(35,86){\scriptsize $L=1/6$}  
       \put(60,71){\scriptsize $0.5$}  
       \put(68,62){\scriptsize $1$}
       \put(75,58){\scriptsize $1.5$}
       \put(89,53){\scriptsize $2$}
       \put(76,36){\scriptsize $L=5$}
 	\end{overpic}
    \begin{overpic}[width=4.5cm, trim=21mm 7mm 24mm 0mm, clip=true]{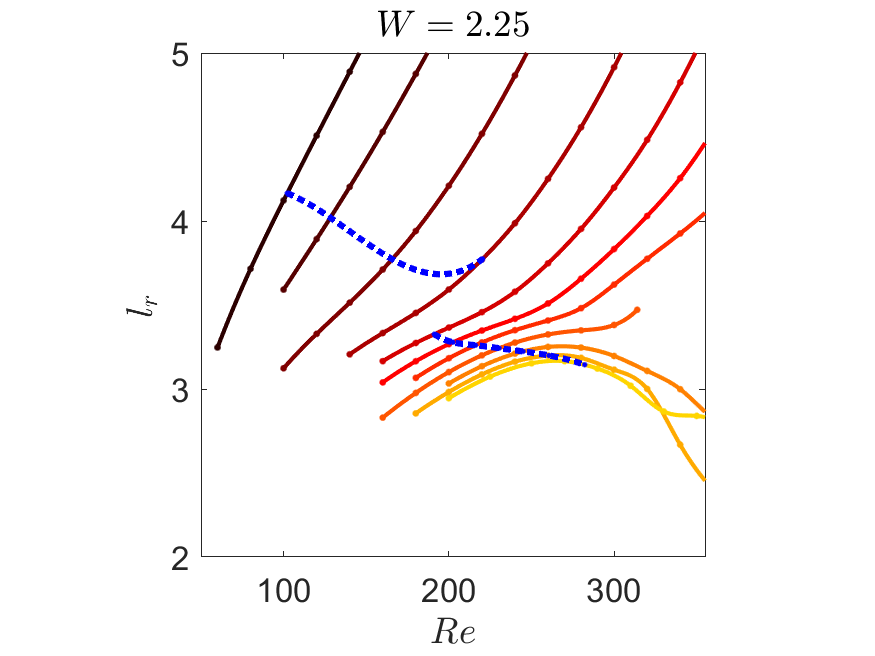}
       \put(14,86){\scriptsize $L=1/6$} 
       \put(58,86){\scriptsize $1$}
       \put(68,86){\scriptsize $1.5$}
       \put(85,86){\scriptsize $2$}
       \put(85,49){\scriptsize $3.5$}
       \put(57,34){\scriptsize $L=5$}
 	\end{overpic}
    \begin{overpic}[width=4.5cm, trim=21mm 7mm 24mm 0mm, clip=true]{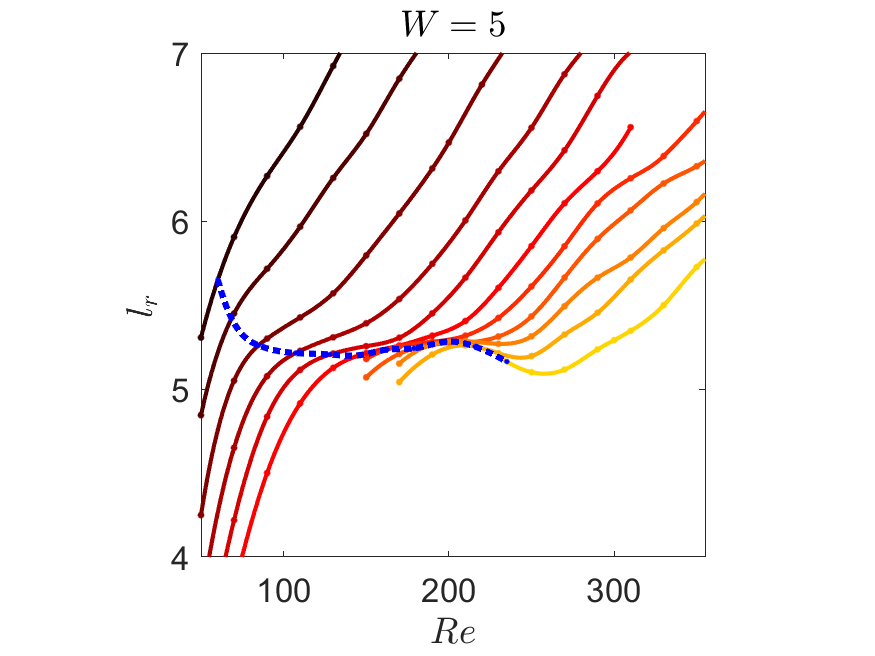}
       \put(12,86){\scriptsize $L=1/6$} 
       \put(54,86){\scriptsize $1$}
       \put(62,86){\scriptsize $1.5$}
       \put(81,81){\scriptsize $2.5$}
       \put(75,38){\scriptsize $L=5$}
 	\end{overpic}
}
\centerline{   
    \begin{overpic}[width=4.5cm, trim=16mm 0mm 24mm 7mm, clip=true]{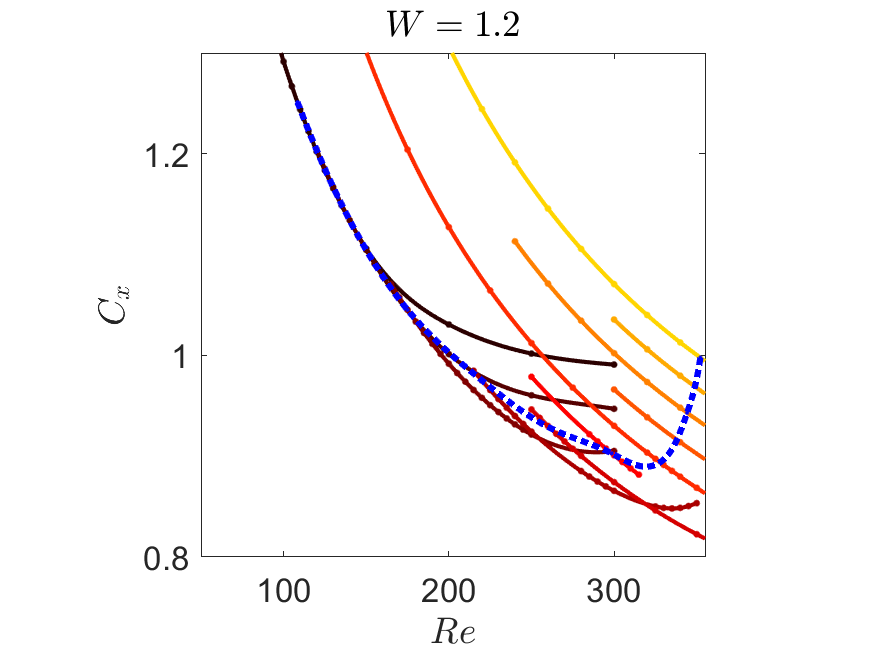}
       \put(-3,91){$(b)$}
 	\end{overpic}
    \begin{overpic}[width=4.5cm, trim=16mm 0mm 24mm 7mm, clip=true]{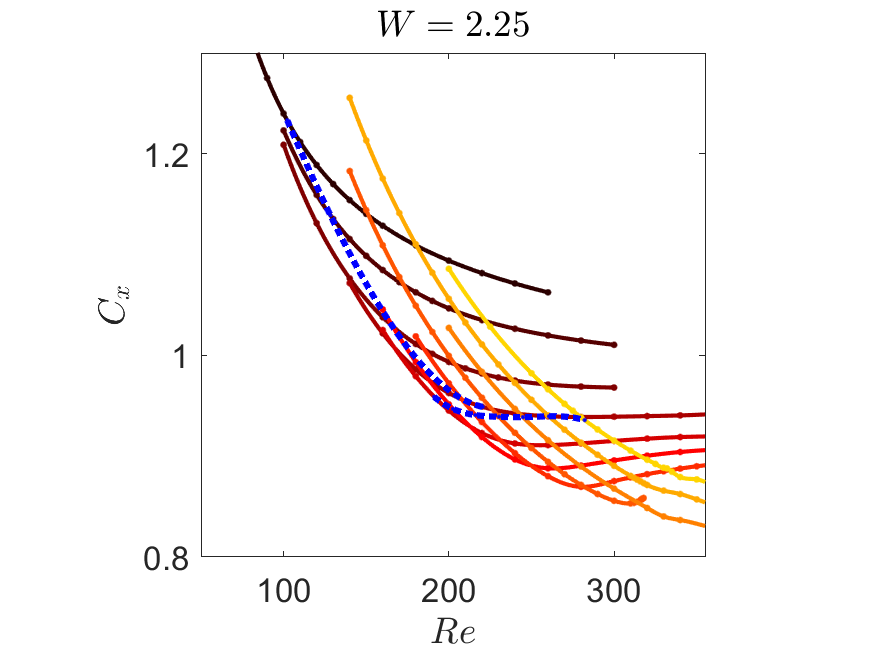}
 	\end{overpic}
    \begin{overpic}[width=4.5cm, trim=16mm 0mm 24mm 7mm, clip=true]{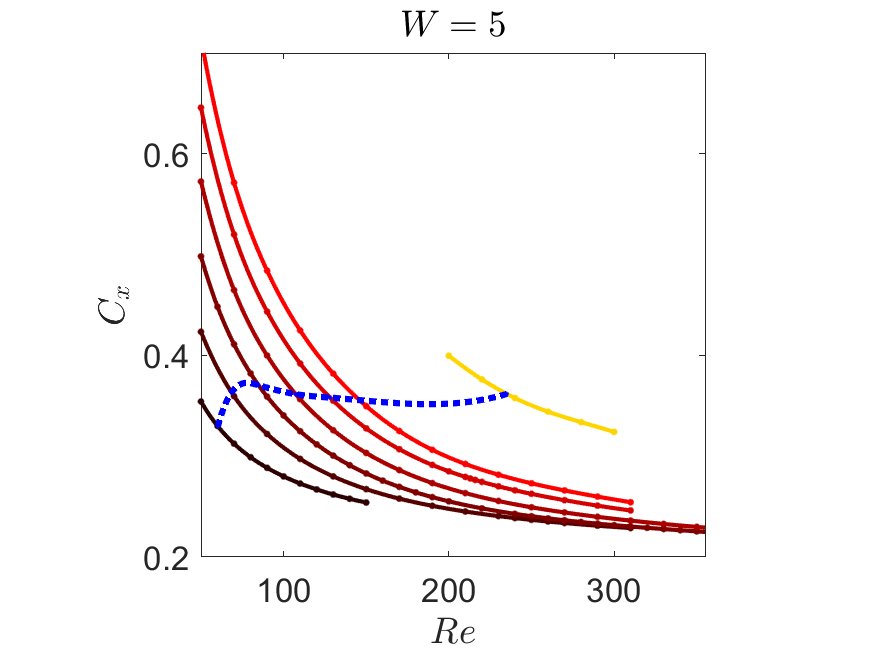}
 	\end{overpic}
}
\caption{
Base flow properties: 
$(a)$~recirculation length,
$(b)$~drag coefficient.
Solid lines: $l_r(Re)$ and $C_x(Re)$ for different body lengths $L=1/6, 0.5, 1 \ldots, 4.5, 5.$
Dotted line: $l_r$ and $C_x$ at the onset of the first bifurcation according to the LSA (\S\ref{sec:stability}).
Body width from left to right: $W=1.2, 2.25, 5$.
}
\label{fig:recirclength_and_drag}
\end{figure}

Some quantitative properties of the base flow are shown in figure~\ref{fig:recirclength_and_drag}. 
For all the body widths considered, the recirculation length $l_r$ (length of the downstream recirculation region along the symmetry axis $y=z=0$ measured from the trailing edge, figure~\ref{fig:recirclength_and_drag}$a$) generally increases with $Re$, like most 2D and 3D bluff body wakes \citep{giannetti-luchini-2007,MarquetLarsson2015}. Also, $l_r$ decreases with $L$ due to decreasing leading-edge separation angle, in agreement with the observations of \cite{zampogna_boujo_2023} for prisms with $W = 1.2$ and of \cite{chiarini-quadrio-auteri-2022b} for 2D bluff bodies of different shapes. 
The recirculation length is also seen to increase with $W$. 
Anticipating on the linear stability results (\S\ref{sec:stability}), we  follow the evolution of the recirculation length  at the first bifurcation (lowest critical $Re$; blue dashed line): $l_r$ increases with $W$, while it remains fairly independent on $L$ with an overall slightly decreasing trend. 
For $W=2.25$, we note that the value of $l_r$ at the onset of the primary bifurcation is not continuous with $L$, because the mode that drives the primary instability changes with the length of the prism and the lowest $Re_c$ jumps at $L\simeq 1.5$ (see figure~\ref{fig:neutral_curves_and_omc_vs_L}).

The base flow drag coefficient $C_x =2F_x/(\rho U_\infty H W)$,
where $F_x$ is the drag force,
decreases or stays constant with the Reynolds number in almost the complete space of parameters; see figure ~\ref{fig:recirclength_and_drag}$(b)$. 
For narrower bodies, $W=1.2$ and  $2.25$, the drag coefficient has a non-monotonic trend with $L$, first decreasing and then increasing. 
For longer bodies, $W = 5$, it monotonically increases with $L$ for all $Re$. 
At the first bifurcation (blue dashed line), $C_x$ varies with $L$ following different trends depending on the body width: it decreases and then increases for $W = 1.2$, monotonically decreases for $W = 2.25$, and remains essentially  constant for $W = 5$.

\subsection{Linear stability}
\label{sec:stability}

We now turn to the LSA. 
For each geometry, we consider the steady symmetric base flow and look at the eigenmodes that become linearly unstable as $Re$ increases. In particular, we are interested in the symmetries of the modes and  
their stationary/oscillatory nature (pitchfork/Hopf bifurcation).

\subsubsection{First bifurcation: stability diagram}

\begin{figure}
\centerline{   
    \begin{overpic}[width=8cm, trim=0mm 0mm 0mm 0mm, clip=true]{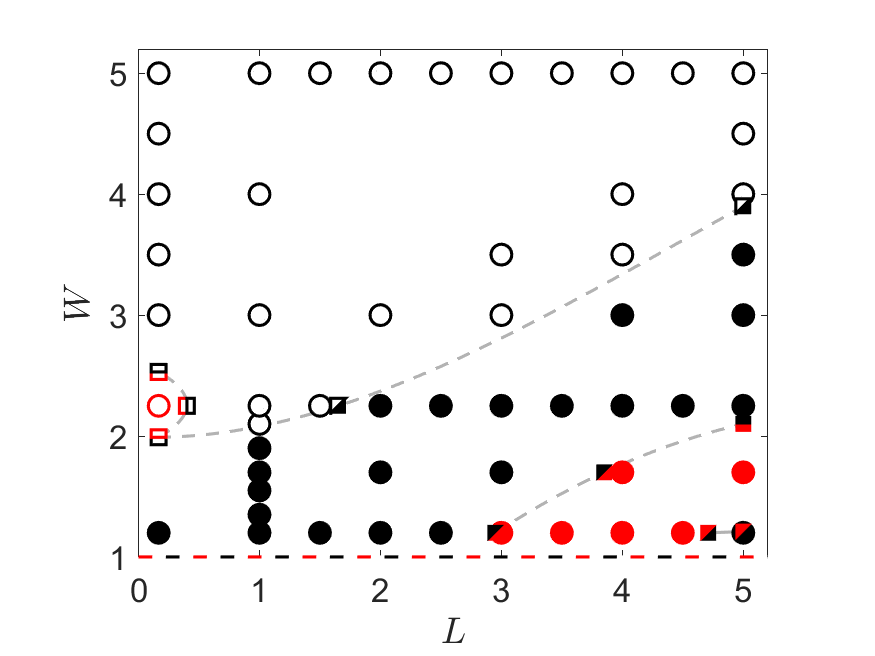} 
       \put(36,52){\small \textcolor{black}{oscillatory  $S_y A_z$} }
       \put(58,32.5){\small \textcolor{black}{stationary $S_y A_z$} }
       \put(74.5,21){\small \textcolor{red}  {stat.}} 
       \put(71.5,17){\small \textcolor{red}  {$\,\,\,A_y S_z$}} 
 	\end{overpic} 	
}
\caption{
Stability diagram for the first bifurcation.
Filled symbols: pitchfork (stationary) bifurcation; open symbols: Hopf (oscillatory) bifurcation.
Red: symmetry breaking in the horizontal $y$ direction ($A_yS_z$); black: symmetry breaking in the vertical $z$ direction ($S_yA_z$).
Circles: actual calculations; squares: interpolations along $L$ or $W$.
By symmetry, for $W=1$ the $S_y A_z$ and $A_y S_z$ eigenmodes bifurcate simultaneously.
}
\label{fig:stab_diag}
\end{figure}

We start considering the first bifurcation.
Figure~\ref{fig:stab_diag} summarises the effect of the geometry. 
Three main regions appear in the $L$-$W$ plane. 
The flow past wide bodies (large $W/L$) first become unstable to oscillatory $S_y A_z$ perturbations that break the temporal and top/bottom planar symmetries and preserve the left/right planar symmetry, leading to a periodic vortex shedding across the shorter body dimension. This is consistent with the 2D limit $W\rightarrow\infty$, and resembles the results for the flow past finite-length circular cylinders \citep{yang-etal-2022}.
Increasing $L/W$, the flow first becomes unstable to stationary $S_y A_z$ perturbations that break the top/bottom spatial symmetry and maintain the left/right one. This bifurcation leads to a static vertical deflection of the wake.
Finally, further increasing $L/W$, the flow first becomes unstable to stationary $A_y S_z$ perturbations that break the left/right symmetry while preserving the top/bottom one. This corresponds to a static horizontal deflection of the wake.
By symmetry, the $S_y A_z$ and $A_y S_z$ modes become unstable simultaneously in the special case of bodies of square cross section $W/H=1$, as observed by \cite{MarquetLarsson2015,meng_an_cheng_kimiaei_2021,zampogna_boujo_2023}.
Interestingly, the oscillatory $A_y S_z$ mode observed by \cite{MarquetLarsson2015} for thin plates $L=1/6$ of intermediate width $2 \leq W \leq 2.5$ seems restricted to a narrow region of the $L$-$W$ plane (see also the neutral curves for $W=2.25$ in figure \ref{fig:neutral_curves_and_omc_vs_L} and for $L=1/6$ and 1 in figure \ref{fig:neutral_curves_and_omc_vs_W}).

\subsubsection{Eigenmodes and neutral curves}
\label{sec:neutral-curves}

\begin{figure}
\vspace{0.3cm}
\centerline{  
 	\hspace{-0.65cm}
    \begin{overpic}[height=3.857cm, trim=6mm 9.mm 13.5mm 0mm, clip=true]{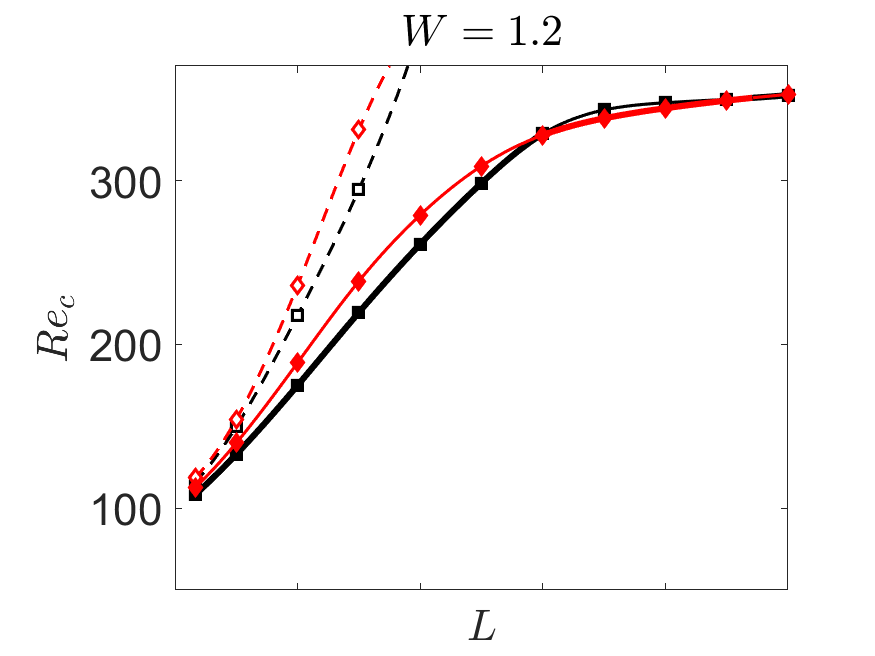} 
       \put(-2,67){$(a)$}
    \end{overpic}
 	\hspace{0.01cm}
    \begin{overpic}[height=3.857cm, trim=15mm 9.mm 14mm 0mm, clip=true]{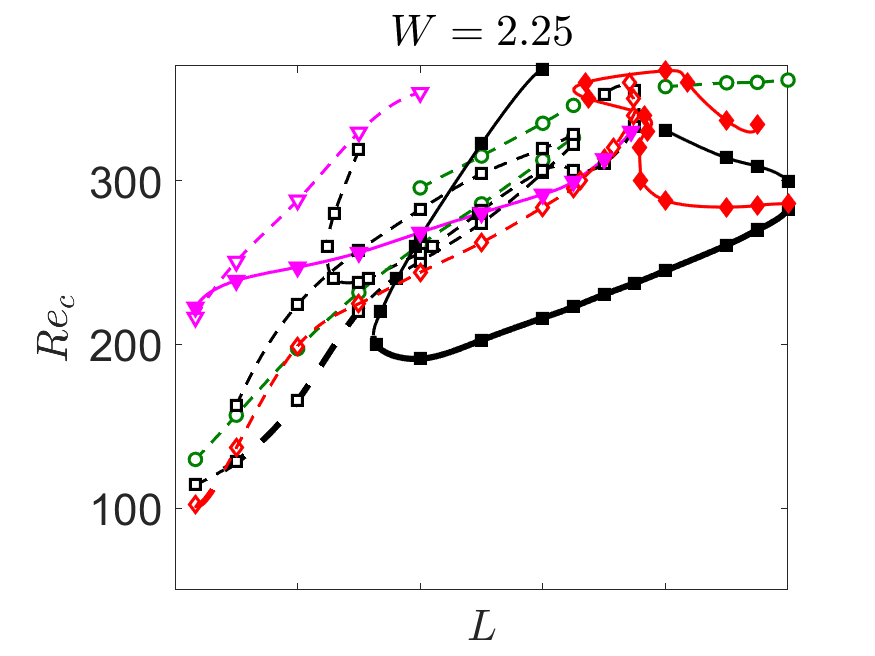}  
 	\end{overpic}
 	\hspace{0.01cm}
    \begin{overpic}[height=3.857cm, trim=15mm 9.mm 14mm 0mm, clip=true]{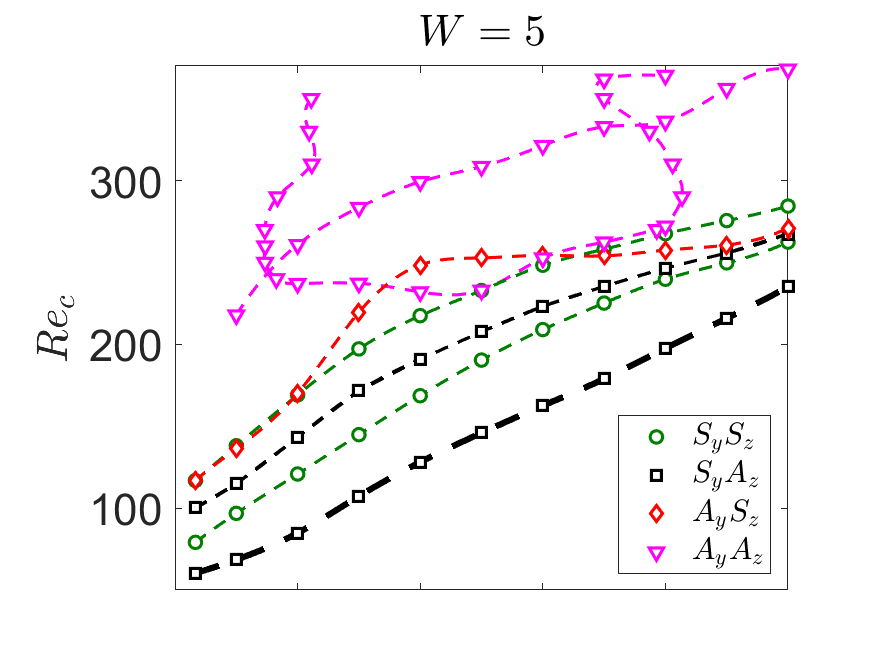}  
 	\end{overpic}
}
\vspace{0.2cm}
\centerline{  
    \begin{overpic}[height=2.925cm, trim=0mm 21mm 15.5mm 13.mm, clip=true]{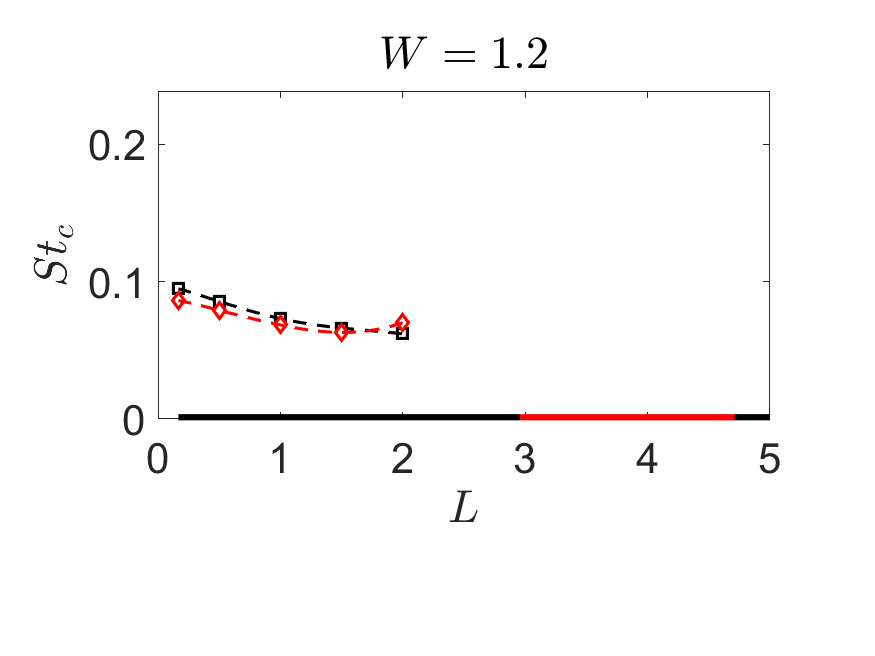}  
       \put(2,52){$(b)$}
 	\end{overpic}
 	\hspace{0.055cm}
    \begin{overpic}[height=2.925cm, trim=15mm 21mm 15.5mm 13.mm, clip=true]{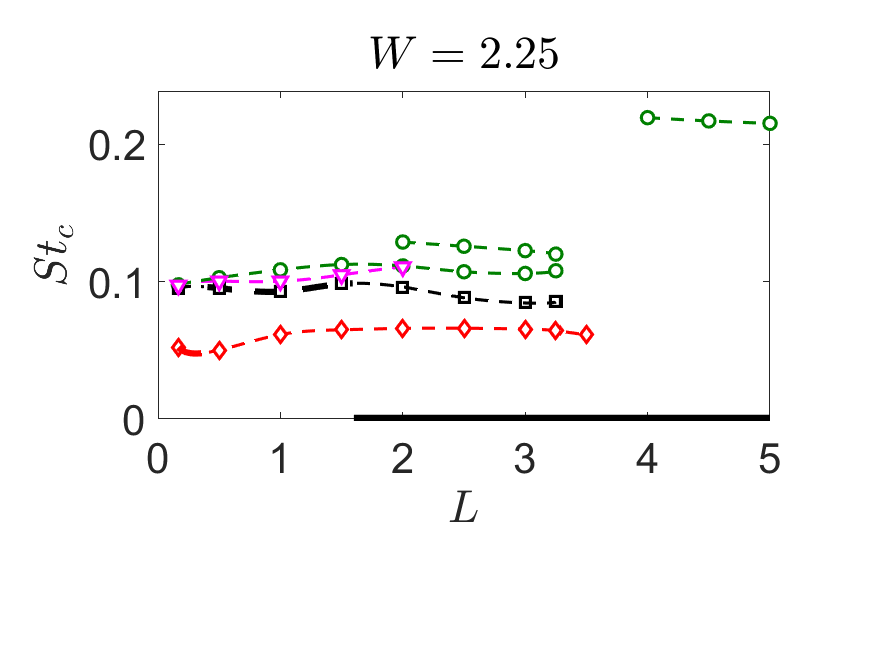}  
 	\end{overpic}
 	\hspace{0.055cm}
    \begin{overpic}[height=2.925cm, trim=15mm 21mm 0mm 13.mm, clip=true]{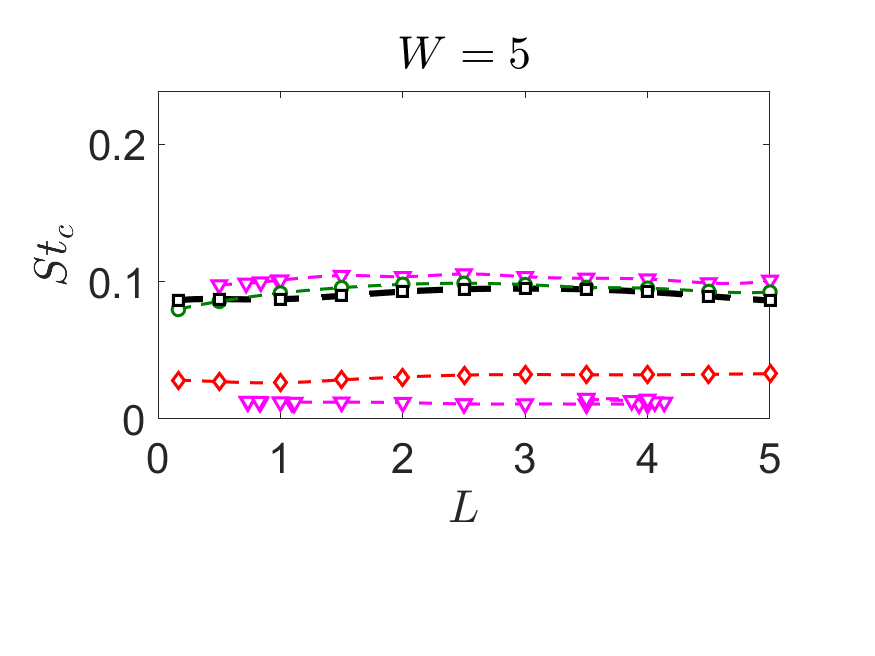}  
 	\end{overpic}
}
\caption{
$(a)$~Neutral curves: critical Reynolds number $Re_c$ as a function of body length $L$, for different body widths $W=1.2$, 2.25 and 5.
Filled symbols and solid lines indicate stationary (pitchfork) bifurcations; open symbols and dashed lines indicate oscillatory (Hopf) bifurcations.
$(b)$~Critical frequency as a function of body length $L$ for the first few bifurcations:
Strouhal number $St_c=\omega(Re_c)/(2\pi)$. 
Different lines correspond to different modes.
In both $(a)$ and $(b)$, thicker lines show the eigenmode that becomes unstable at the lowest Reynolds number.
}
\label{fig:neutral_curves_and_omc_vs_L}
\end{figure}

We now proceed to characterise in more detail the linear stability of the steady symmetric  base flow for increasing values of $Re$ and a few selected geometries. Figure~\ref{fig:neutral_curves_and_omc_vs_L} shows the critical Reynolds number and  frequency of the first bifurcations as a function of the body length $L$ for $W=1.2$, $2.25$ and $5$. 
The eigenmodes of the first two bifurcations for these widths and the two body lengths $L=1$ and $5$ are shown in figures~\ref{tab:first_two_modes}; see also 
the supplementary material.
We focus on the first few bifurcations, as the eigenmodes that become unstable at significantly larger Reynolds numbers become less relevant.

\def\hh{2.4}
\begin{figure}
\hspace{-0.2cm}
\begin{tabular}{c c | c} 
& $L=1$ & $L=5$  
\\
\raisebox{2.\normalbaselineskip}[0pt][0pt]{\rotatebox[origin=c]{90}{$W=5$}}
& 
    \begin{overpic}[height=\hh cm, trim=100mm 50mm 90mm 10mm, clip=true]{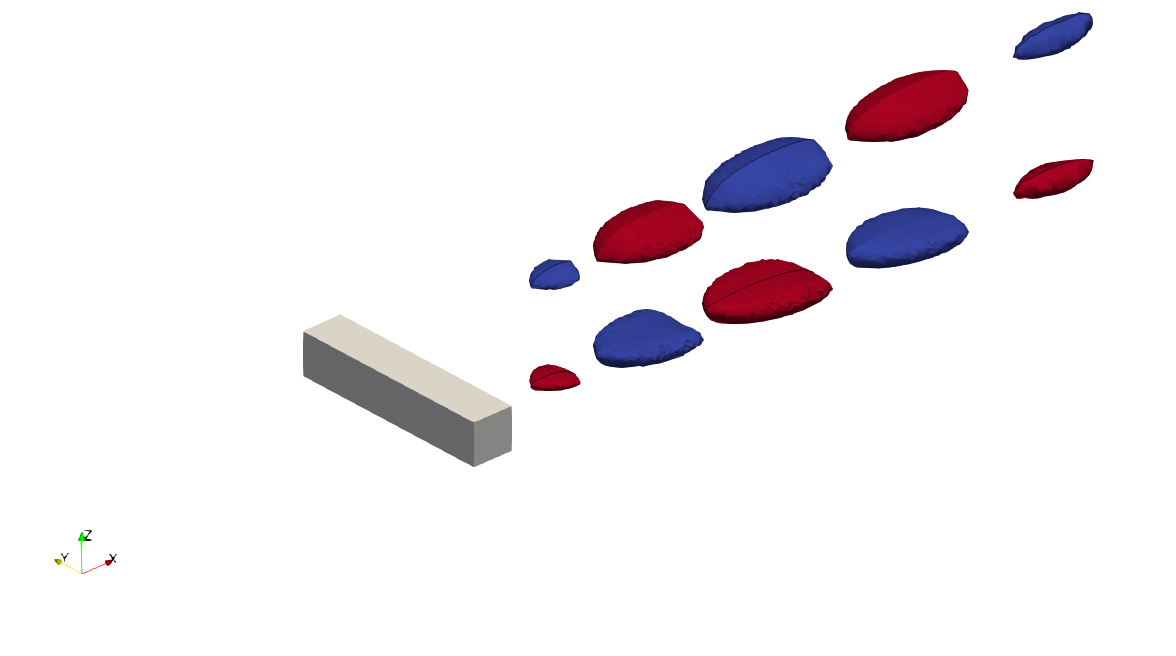}   
       \put(8,50){$S_yA_z$}
 	\end{overpic} 
 	\hspace{-0.4cm}
    \begin{overpic}[height=\hh cm, trim=100mm 50mm 90mm 10mm, clip=true]{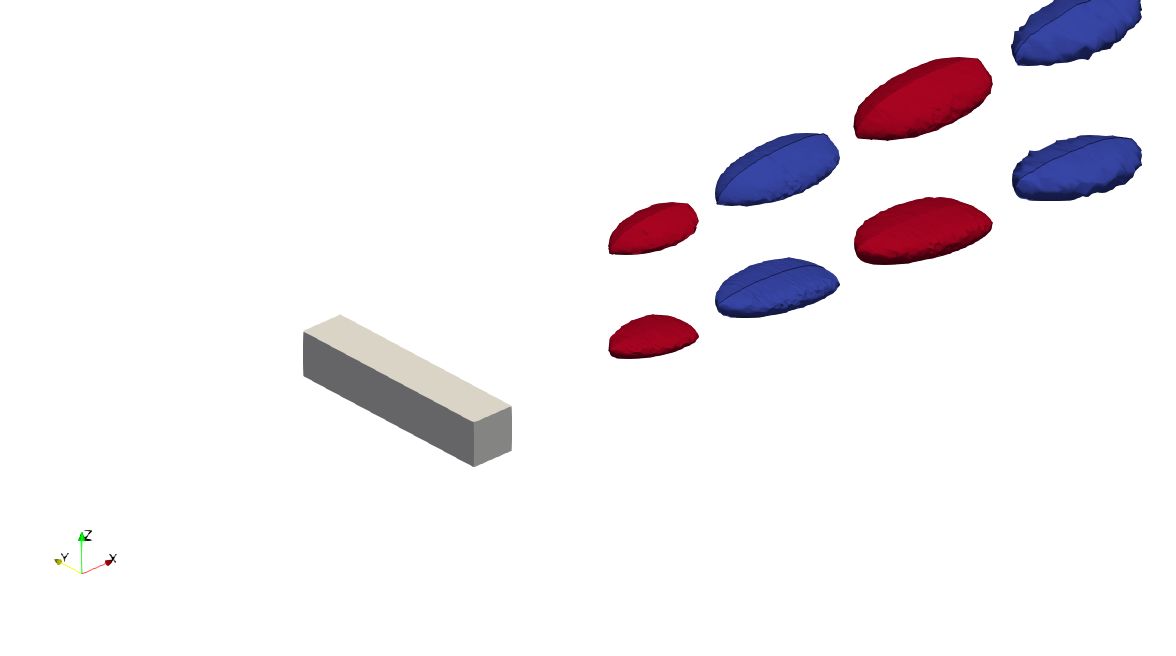}   
       \put(8,50){\textcolor[rgb]{0,0.5,0}{$S_yS_z$}}  
 	\end{overpic} 
&
 	\begin{overpic}[height=\hh cm, trim=70mm 50mm 90mm 10mm, clip=true]{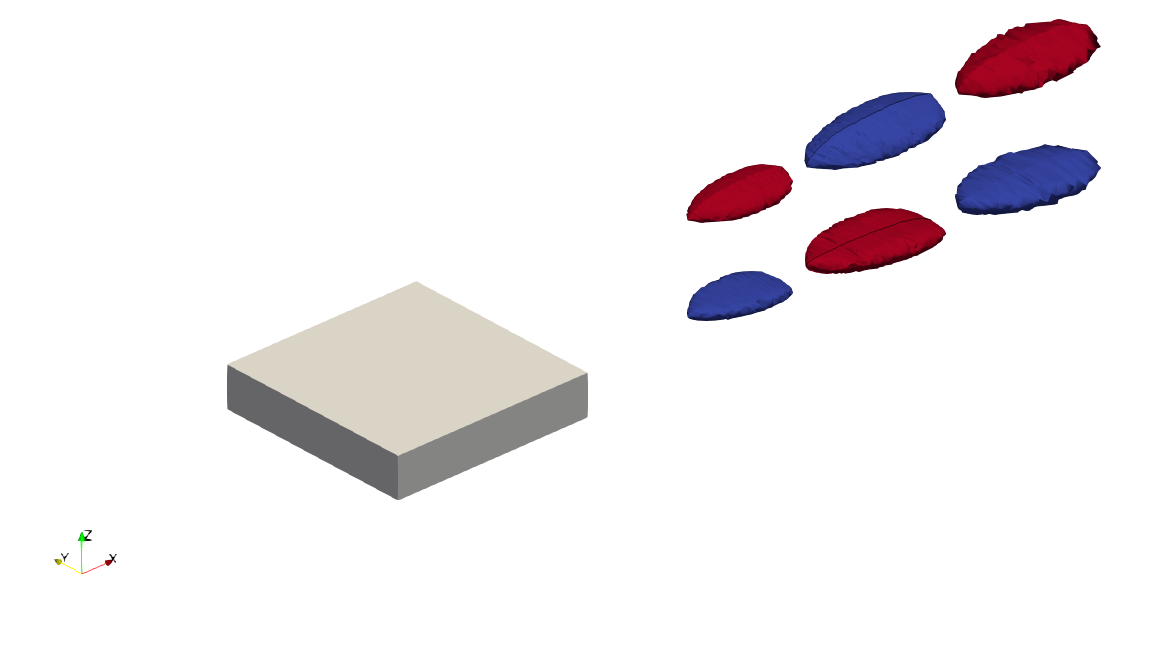}   
       \put(15,45){$S_yA_z$}
 	\end{overpic} 
 	\hspace{-0.4cm}
    \begin{overpic}[height=\hh cm, trim=70mm 50mm 90mm 10mm, clip=true]{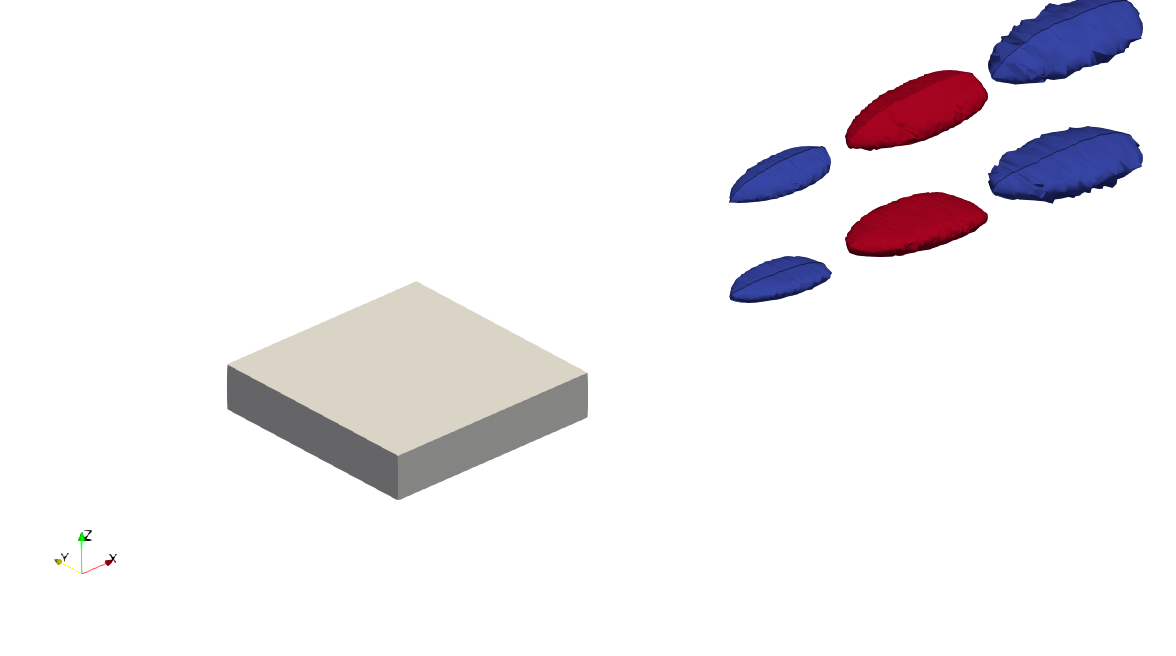}   
       \put(15,45){\textcolor[rgb]{0,0.5,0}{$S_yS_z$}}
 	\end{overpic}  	
\\
\raisebox{2.5\normalbaselineskip}[0pt][0pt]{\rotatebox[origin=c]{90}{$W=2.25$}}
& 
    \begin{overpic}[height=\hh cm, trim=100mm 50mm 90mm 10mm, clip=true]{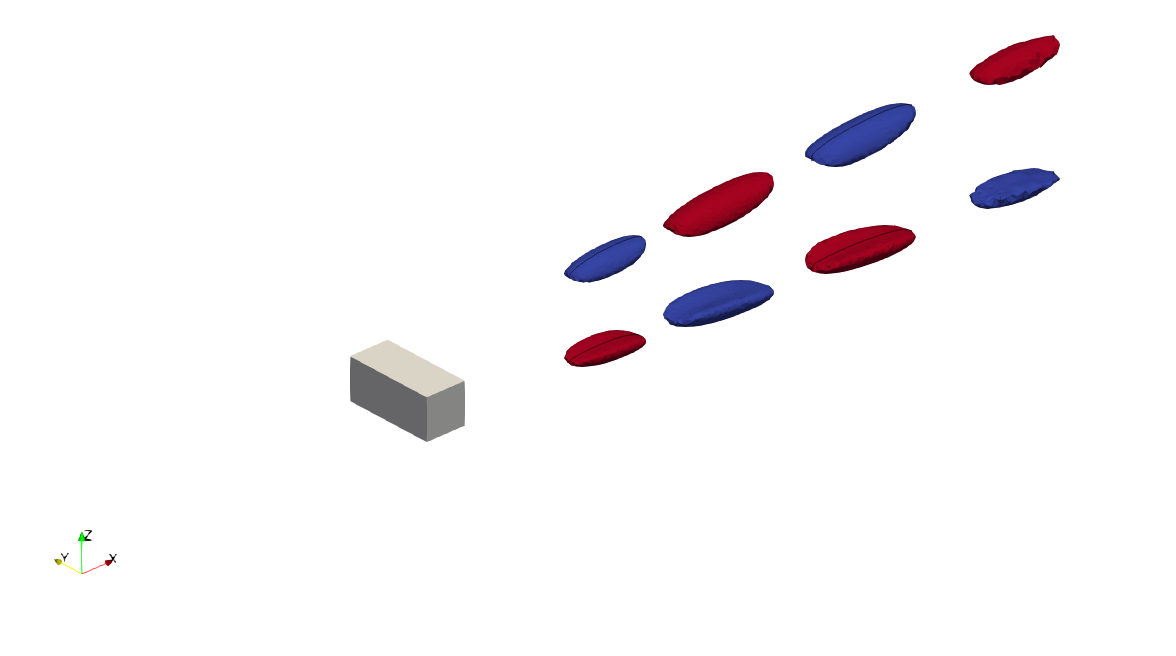}   
       \put(10,50){$S_yA_z$}
 	\end{overpic} 
 	\hspace{-0.4cm}
    \begin{overpic}[height=\hh cm, trim=100mm 50mm 90mm 10mm, clip=true]{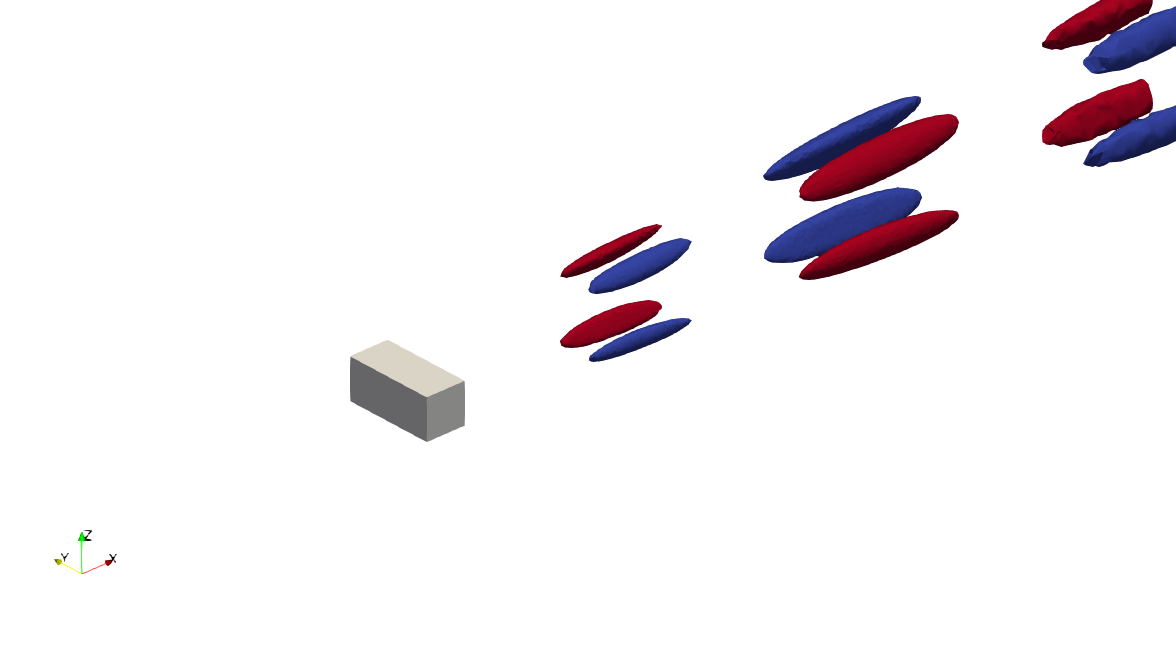}   
       \put(10,50){\textcolor{red}{$A_yS_z$}} 
 	\end{overpic} 
& 
    \begin{overpic}[height=\hh cm, trim=70mm 50mm 90mm 10mm, clip=true]{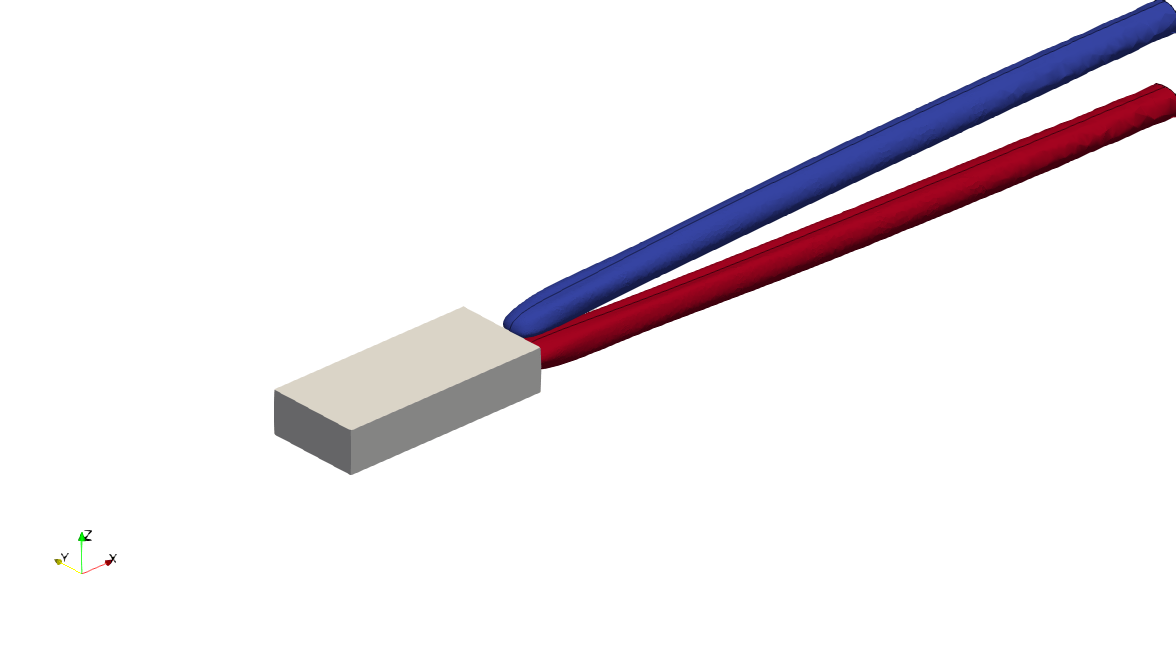} 
       \put(12,45){$S_yA_z$}  
 	\end{overpic} 	
 	\hspace{-0.4cm}
    \begin{overpic}[height=\hh cm, trim=70mm 50mm 90mm 10mm, clip=true]{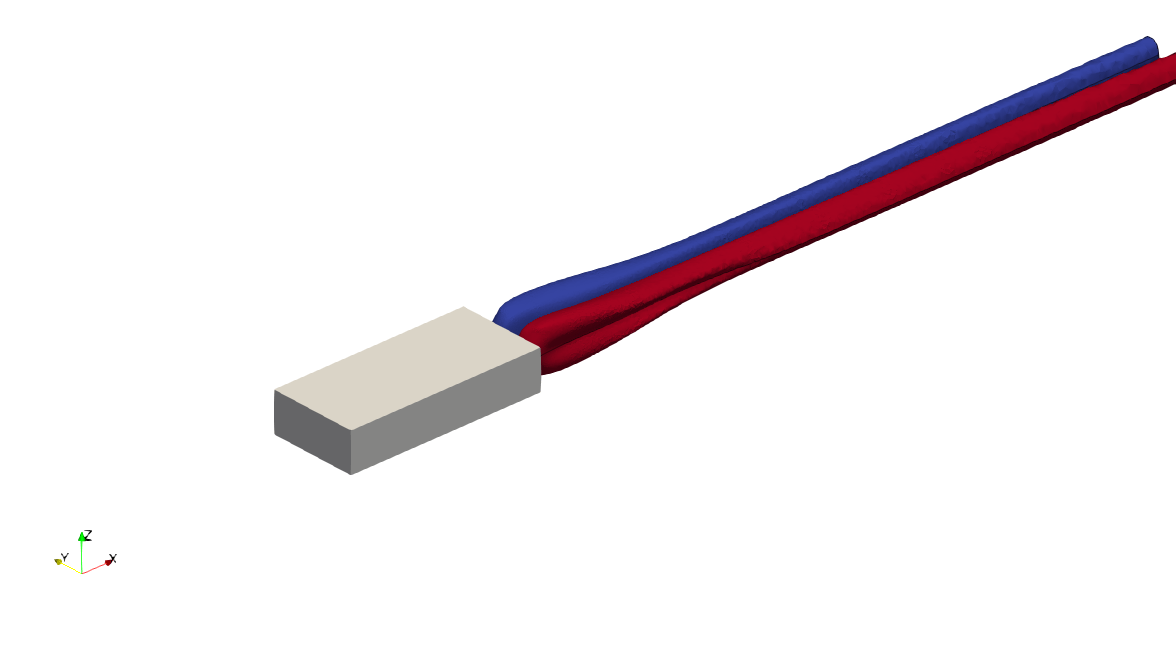}  
       \put(12,45){\textcolor{red}{$A_yS_z$}}  
 	\end{overpic} 
\\
\raisebox{3.5\normalbaselineskip}[0pt][0pt]{\rotatebox[origin=c]{90}{$W=1.2$}}
& 
    \begin{overpic}[height=2.6cm, trim=100mm 30mm 90mm 10mm, clip=true]{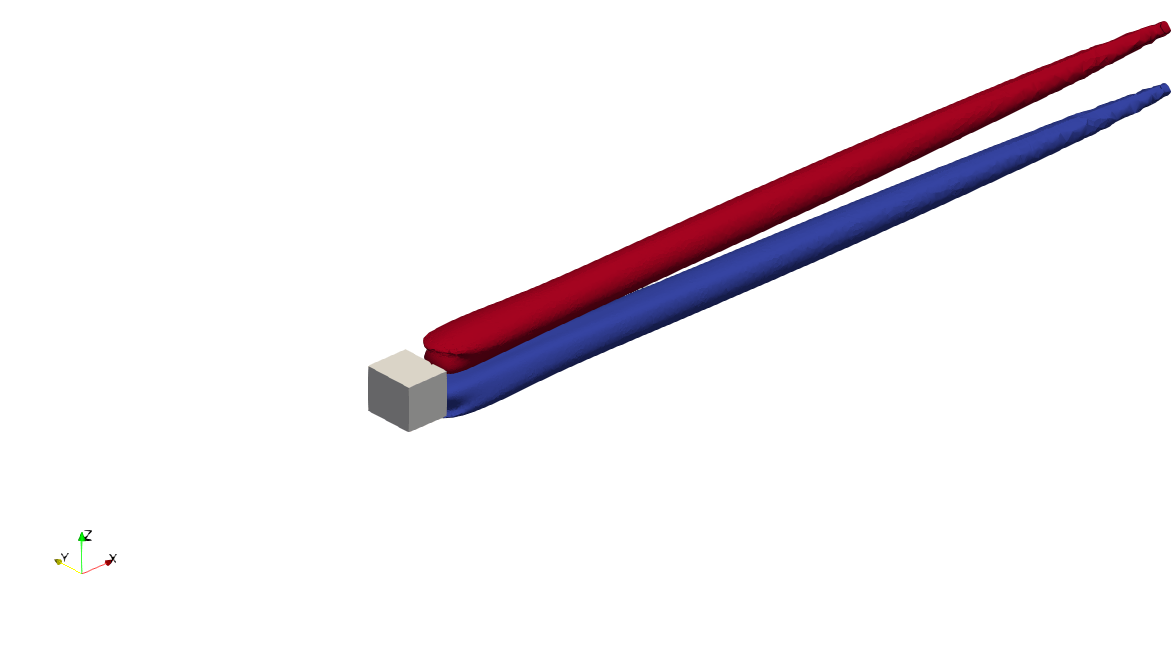}   
       \put(10,50){$S_yA_z$}
 	\end{overpic} 
 	\hspace{-0.4cm}
    \begin{overpic}[height=2.6cm, trim=100mm 30mm 90mm 10mm, clip=true]{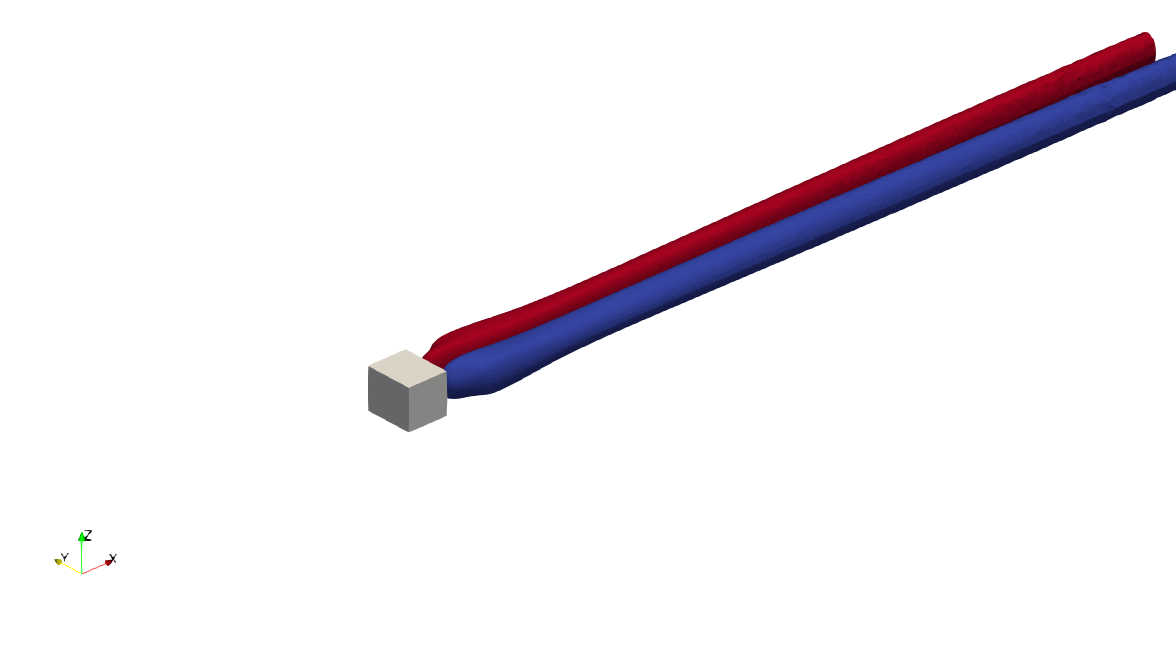}  
       \put(10,50){\textcolor{red}{$A_yS_z$}} 
 	\end{overpic}  
&
    \begin{overpic}[height=2.6cm, trim=70mm 30mm 90mm 10mm, clip=true]{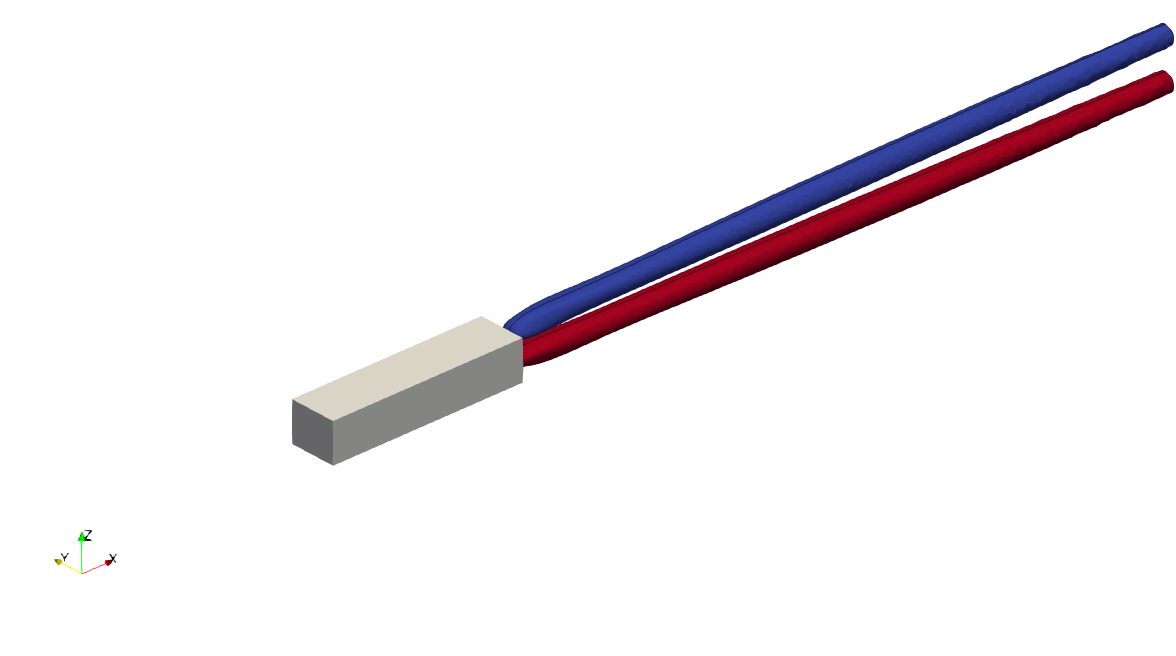}   
       \put(12,45){$S_yA_z$}
 	\end{overpic} 	  
 	\hspace{-0.4cm}
    \begin{overpic}[height=2.6cm, trim=70mm 30mm 90mm 10mm, clip=true]{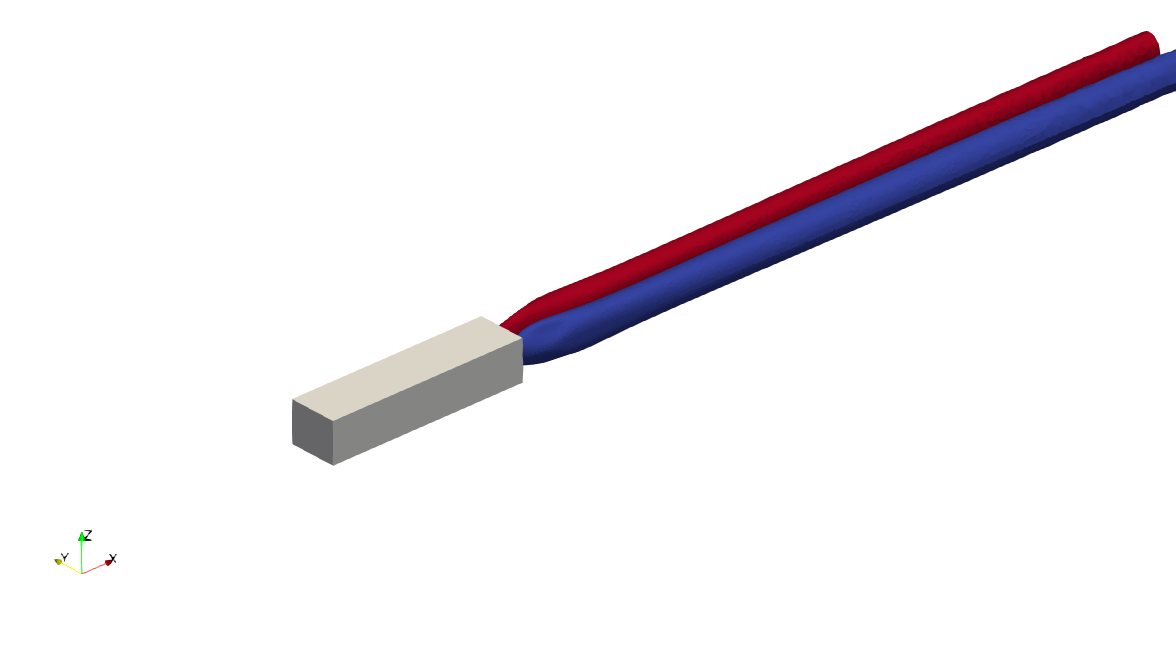}  
       \put(12,45){\textcolor{red}{$A_yS_z$}}  
 	\end{overpic}  
\\
\end{tabular} 
\vspace{-0.2cm}
\caption{
First and second bifurcating eigenmodes (left and right, respectively), visualised with isosurfaces of streamwise velocity. 
\label{tab:first_two_modes}
}
\end{figure}

For narrow bodies, $W=1.2$, the first two bifurcations of the steady symmetric base flow occur in short succession (as $W$ is close to 1), and correspond to stationary $S_y A_z$ and $A_y S_z$ modes that break either of the two planar symmetries \citep{zampogna_boujo_2023}. A pair of streamwise streaks of positive and negative streamwise perturbation velocity arise in the wake  (figure \ref{tab:first_two_modes}), leading to a vertical or horizontal displacement of the wake.
Other bifurcations occur at larger $Re$ and correspond to oscillatory $S_y A_z$ and $A_y S_z$ modes. For small $L$ these modes are almost as unstable as their stationary counterparts, but they quickly become much more stable as $L$ increases, i.e. the corresponding $Re_c$ increases faster; accordingly, for long bodies these oscillatory modes of the symmetric base flow 
are not directly relevant to the secondary stability of the deflected flow (\S\ref{sec:simW12-per-aper}). 
The associated angular frequency is of the order $\omega \simeq 0.4-0.6$ ($St = f H / U_\infty \simeq 0.06-0.09$, where $f = \omega/(2 \pi)$) for $L \leq 2$. 
The increase of $Re_c$ with $L$ is consistent with the stationary bifurcation of the flow past axisymmetric bodies: disk much thinner than its diameter $L/D \ll 1$, sphere $L/D=1$, bullet-shaped bodies $1 \leq L/D \leq 3.5$ \citep{natarajan-acrivos-1993,fabre-etal-2008,meliga-etal-2009,bohorquez-etal-2011} and with the primary oscillatory bifurcation of the flow past 2D bodies of different shape \citep{jackson-1987-finiteelementstudy,chiarini-quadrio-auteri-2022b}.

For wide bodies, $W=5$, the first bifurcation corresponds to an oscillatory $S_y A_z$ mode for all $L$ (as already observed in figure~\ref{fig:stab_diag}). The critical Reynolds number increases with $L$, and is rather well separated from the following bifurcations.
The angular frequency is $\omega \simeq 0.5-0.6$ ($St \simeq 0.08-0.10$). The unstable eigenmode is stronger in the central region (small $|y|$); see figure \ref{tab:first_two_modes}.
The second bifurcation corresponds to an oscillatory $S_y S_z$ modes for all $L$. Like the leading $S_y A_z$ mode, it is associated with vortex shedding in the upper and lower shear layers, with a similar frequency but different inter-layer phasing. Unlike the $S_y A_z$ mode that breaks the top/bottom symmetry, the $S_y S_z$ mode preserves both planar symmetries. 
In other words, in the vertical plane $y=0$ the two leading modes can be categorised as ``sinuous'' and ``varicose'', respectively. To the best of our knowledge, unsteady doubly symmetric $S_y S_z$ modes have not been reported in 3D bluff body wakes in previous works except for thin plates ($L=1/6$, $W>3.1$, \cite{MarquetLarsson2015}); see also \S \ref{sec:sim_L5_W225}.
At larger $Re$ further bifurcations are observed, including some involving doubly antisymmetric modes ($A_y A_z$). All these bifurcations are oscillatory and the corresponding frequencies are almost independent on $L$, and differ by one order of magnitude depending on the bifurcation; see figure \ref{fig:neutral_curves_and_omc_vs_L}. 

For bodies of intermediate width, $W=2.25$, the increasing trend of $Re_c$  with $L$ is similar, but  the leading mode changes with $L$: 
stationary for $L > 1.5$ (similar to narrower bodies), and oscillatory with $\omega \simeq 0.3-0.6$ ($St \simeq 0.05-0.10$) for $L < 1.5$ (similar to wider bodies). 
As mentioned previously (figure~\ref{fig:stab_diag}), for this specific width the first bifurcation breaks the left/right symmetry for very short bodies ($L<0.5$), which leads to the unusual vortex shedding across the larger body dimension observed by \cite{MarquetLarsson2015}, and breaks the top/bottom symmetry otherwise, leading to the more usual vortex shedding across the smaller dimension ($0.5 < L < 1.75$) or to a static vertical wake deflection ($L>1.75$).
Interestingly, the stationary $S_y A_z$ mode remains fully stable in the range of $Re$ investigated when $L<1.5$, while for $L>1.5$ it is restabilised for large enough $Re$.
Other bifurcations at larger $Re$ draw a more complicated pattern than for $W=1.2$ and $5$. Their order of appearance depends on $L$, and some modes are unstable only in some interval of $L$.
For almost all values of $L$, however, doubly symmetric $S_y S_z$ doubly antisymmetric $A_y A_z$ modes only correspond to the third or higher bifurcations. 

\begin{figure}
\vspace{0.3cm}
\centerline{ 
    \hspace{-0.4cm} 
    \begin{overpic}[height=3.857cm, trim=6mm 8.mm 13.5mm 1mm, clip=true]{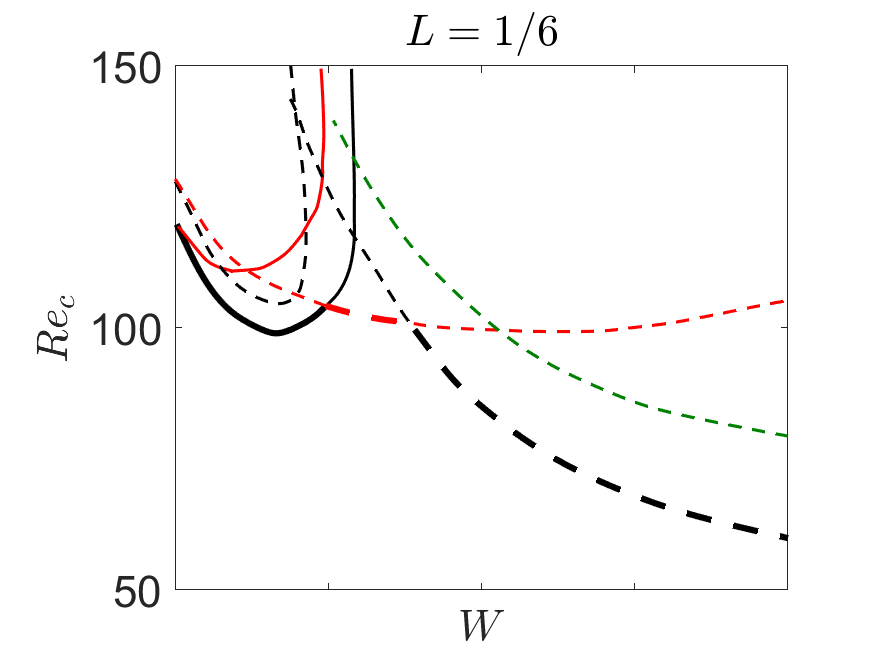}   	
       \put(-2,68){$(a)$}
    \end{overpic} 
    \hspace{0.01cm}
    \begin{overpic}[height=3.857cm, trim=15mm 8.mm 14mm 1mm, clip=true]{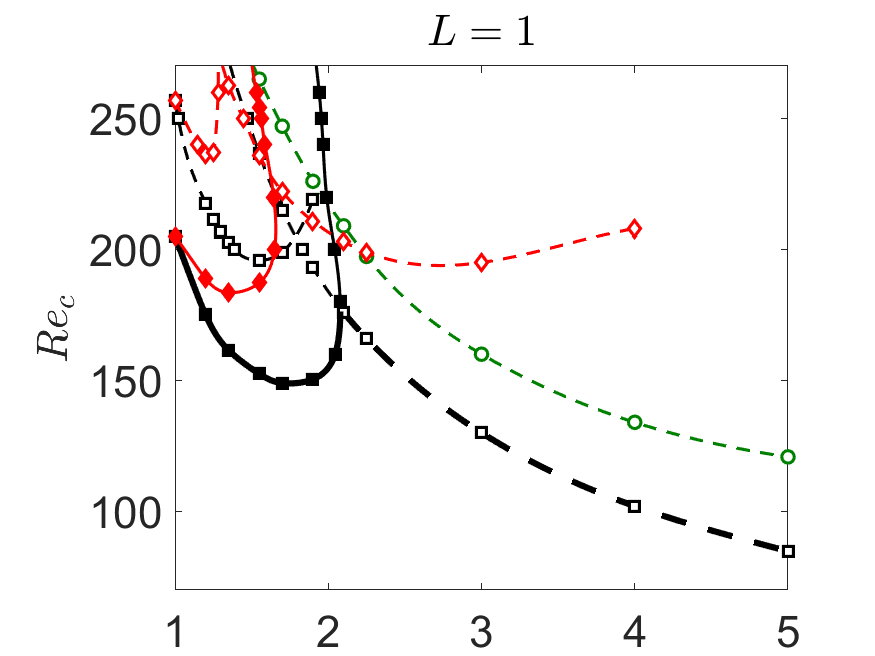}  
 	\end{overpic}
    \hspace{0.01cm}
    \begin{overpic}[height=3.857cm, trim=15mm 8.mm 14mm 1mm, clip=true]{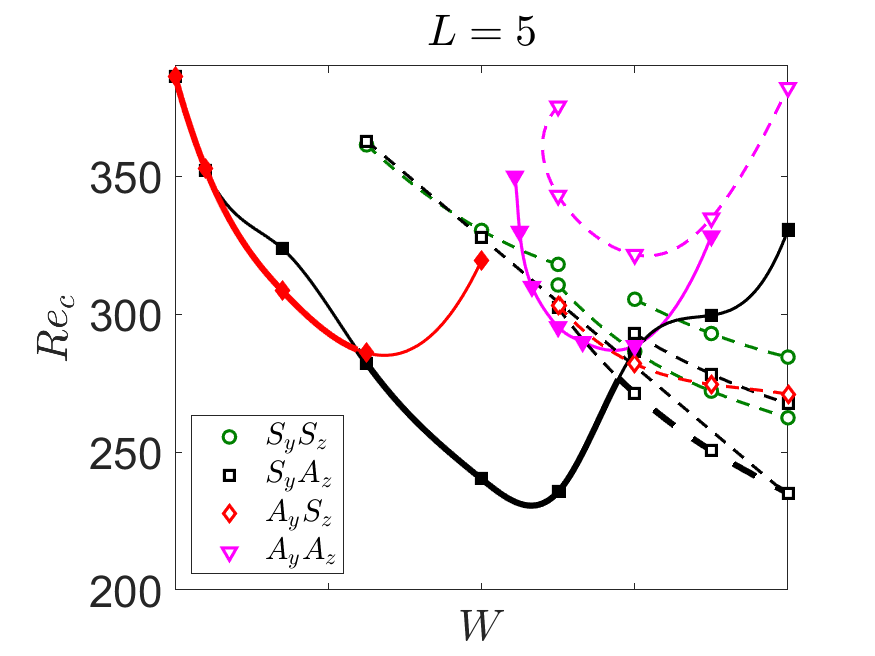}  
 	\end{overpic}
}
\vspace{0.2cm}
\centerline{ 
    \hspace{0.1cm} 
    \begin{overpic}[height=2.925cm, trim=0mm 21mm 15.5mm 13.mm, clip=true]{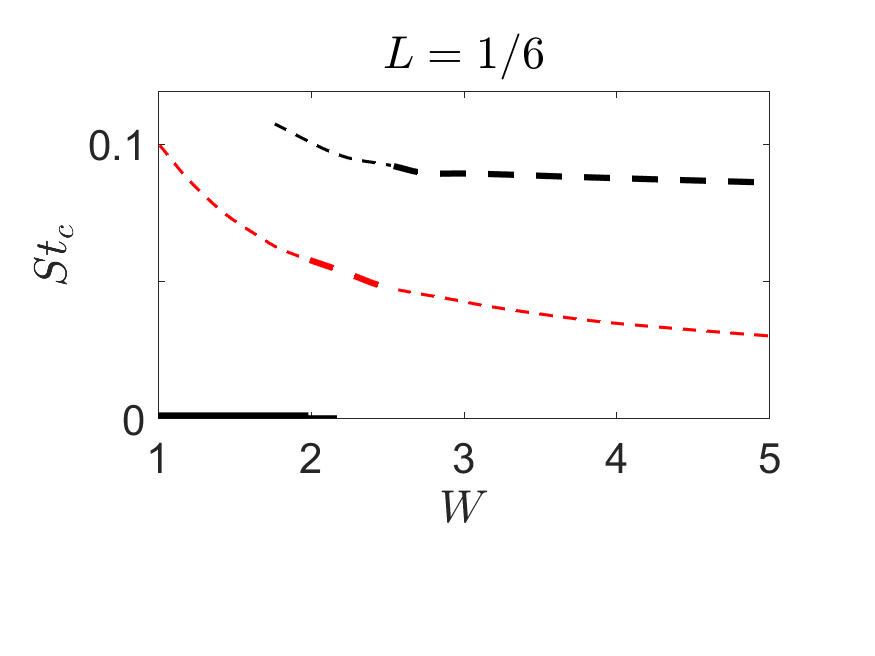}  
       \put(2,52){$(b)$} 
 	\end{overpic} 
    \hspace{0.055cm}
    \begin{overpic}[height=2.925cm, trim=15mm 21mm 15.5mm 13.mm, clip=true]{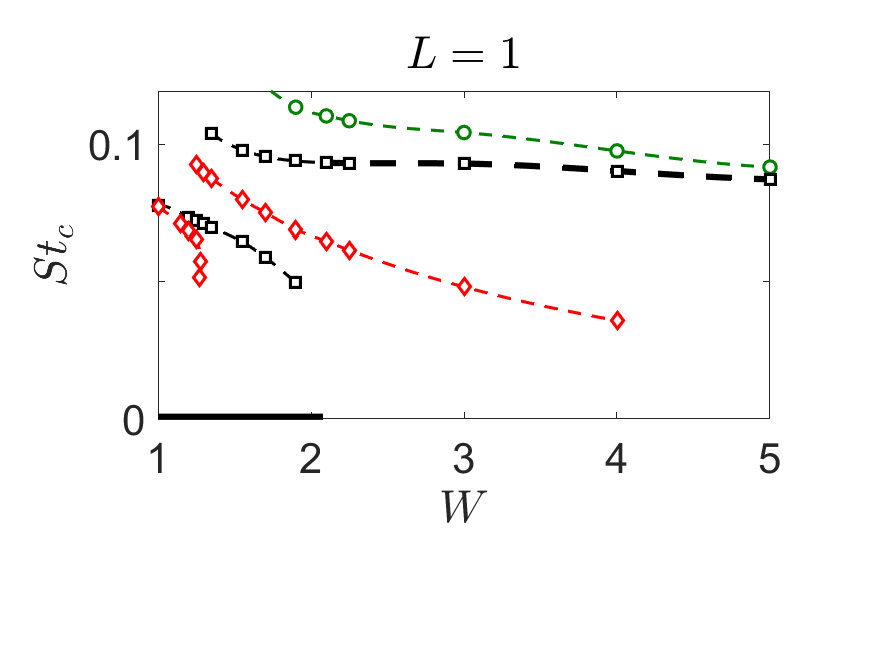}  
 	\end{overpic}
    \hspace{0.055cm}
    \begin{overpic}[height=2.925cm, trim=15mm 21mm 0mm 13.mm, clip=true]{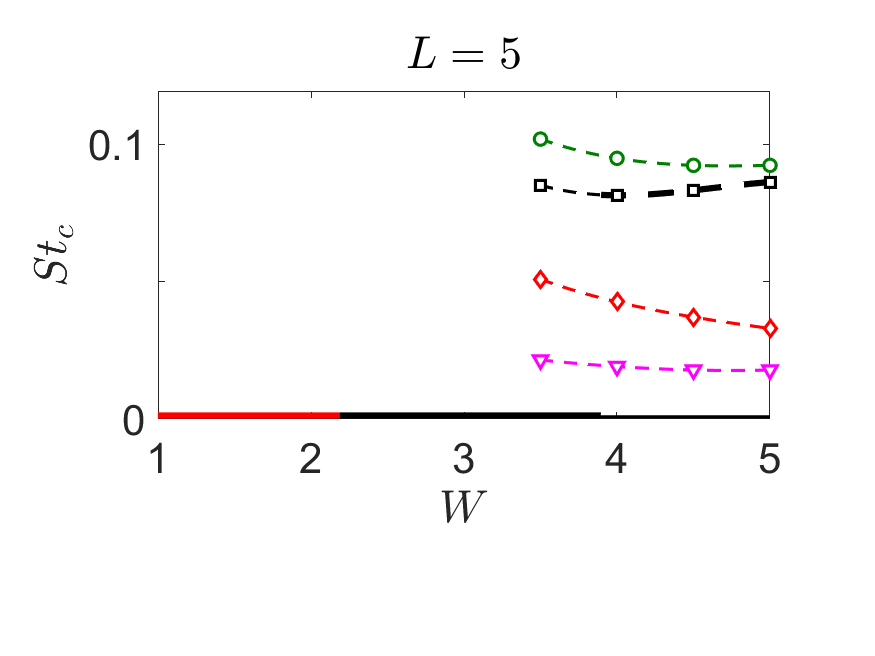}  
 	\end{overpic}
}
\caption{Same as figure \ref{fig:neutral_curves_and_omc_vs_L}, as a function of body width $W$, for different  body lengths  $L=1/6$ \citep[data from][]{MarquetLarsson2015}, $1$ and $5$. 
}
\label{fig:neutral_curves_and_omc_vs_W}
\end{figure}

Figure \ref{fig:neutral_curves_and_omc_vs_W} offers a complementary picture, with neutral curves shown as a function of the body width $W$ for the three body lengths $L=1/6$, $L=1$ and $L=5$.
The first critical $Re$ tends to decrease with $W$, albeit not monotonically: the first bifurcation lies on the envelope of different neutral curves corresponding to different modes, each mode being more unstable in a preferred interval of $W$.

\subsection{Structural sensitivity}
\label{sec:structsensit}

We now look at the structural sensitivity of the two leading modes. Bodies with $L=1$ and $5$ and $W=1.2$, $2.25$ and $5$ are considered. 
Figure~\ref{tab:first_two_modes_sensitivity} shows isosurfaces of $||\hat{\boldsymbol{u}}_1||\,||\hat{\boldsymbol{u}}_1^{\dag}||$, the product of the Euclidean norms of the eigenmode (direct mode) and the associated adjoint mode. 
This quantity was introduced by \cite{giannetti-luchini-2007} as an upper bound for the eigenvalue variation $|\delta \lambda|$ induced by a specific perturbation of the linearised Navier--Stokes  operator, namely a ``force-velocity coupling'' representing feedback from a localised velocity sensor to a localised force actuator at the same location. In this sense, the structural sensitivity is an indicator of the eigenvalue sensitivity and identifies the wavemaker \citep{monkewitz-huerre-chomaz-1993}. 
The most sensitive regions, where the direct and adjoint modes overlap, are found in the near wake of the body for all modes, which is typical of bluff body wakes. As expected, these regions with large sensitivity are symmetric with respect to both $y$ and $z$.

All the $S_y$ modes, which preserve the left/right symmetry, are most sensitive in two regions located symmetrically with respect to the horizontal symmetry plane $z=0$.
For oscillatory $S_y$ modes, observed for smaller $L/W$, the structural sensitivity is maximum exactly on the $u_0=0$ isosurface, in two flat and elongated regions separated by more than one body height; see also the cross-sections in the supplementary material. 
By contrast, for stationary $S_y$ modes observed for larger $L/W$, the structural sensitivity is maximum inside the backflow region, in more compact regions located closer to the horizontal plane.

Stationary $A_yS_z$ modes are most sensitive in two regions located symmetrically with respect to the vertical symmetry plane $y=0$, and the structural sensitivity is maximum inside the backflow region.

The peculiar oscillatory $A_y S_z$ mode observed for short lengths and intermediate widths (e.g. $L=1$, $W=2.25$) has a more convoluted structural sensitivity,  maximum outside and near the downstream end of the backflow region.

Overall, the structural sensitivity shows that the wavemaker of all the considered modes is located downstream, in relation with the wake recirculation region. For elongated bodies with $L=5$ (\S\S\ref{sec:WNL}-\ref{sec:simulations}), the structural sensitivity is null in the recirculating regions along the lateral sides of the prism (see the $u_0=0$ isosurface in figure~\ref{tab:first_two_modes_sensitivity}), meaning that these modes do not originate from the LE shear layer.

\def\hh{1.8}
\begin{figure}
\vspace{0.3cm}
\begin{centering}
\begin{tabular}{c c|c} 
& $L=1$ & $L=5$  
\\ 
\raisebox{1.\normalbaselineskip}[0pt][0pt]{\rotatebox[origin=c]{90}{$W=5$}}
& 
    \begin{overpic}[height=\hh cm, trim=120mm 68mm 110mm 32mm, clip=true]{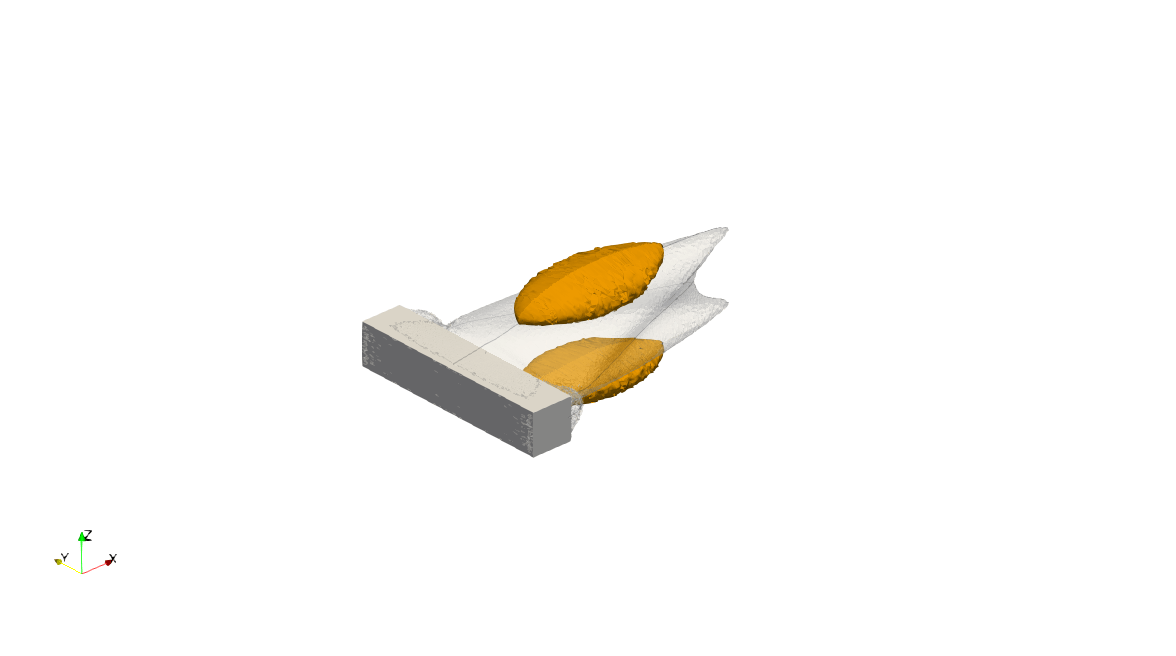}   
       \put(60,0){$S_yA_z$}
 	\end{overpic} 
 	\hspace{-0.2cm}
    \begin{overpic}[height=\hh cm, trim=100mm 65mm 130mm 35mm, clip=true]{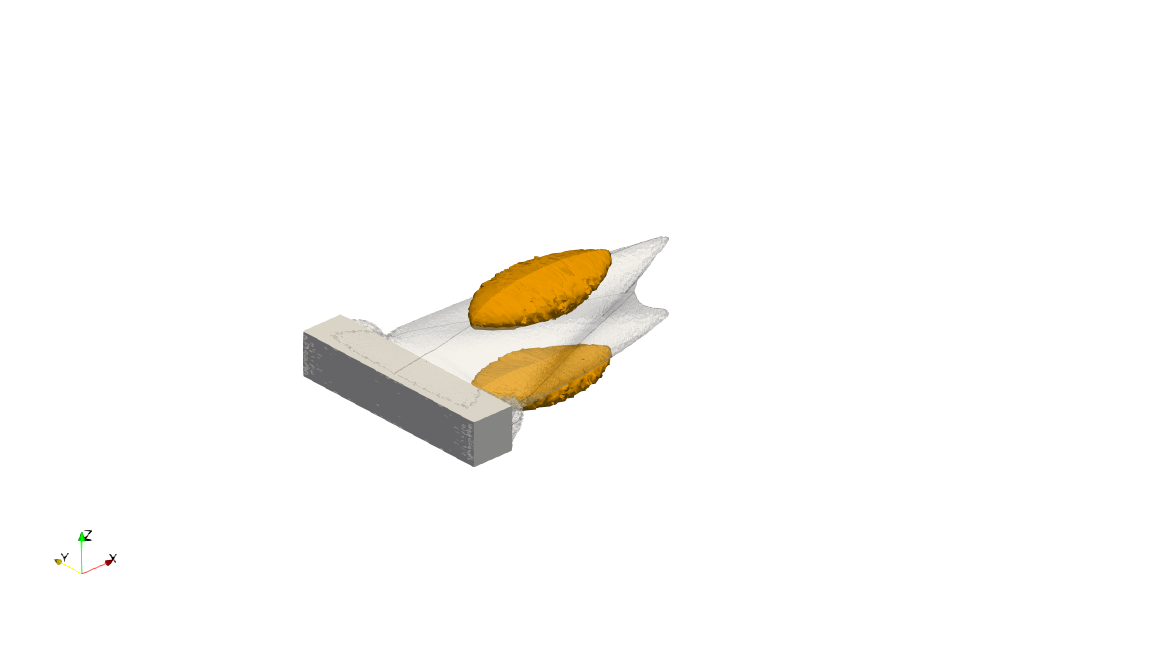}   
       \put(60,0){\textcolor[rgb]{0,0.5,0}{$S_yS_z$}}  
 	\end{overpic} 
&
    \begin{overpic}[height=\hh cm, trim=95mm 57mm 135mm 43mm, clip=true]{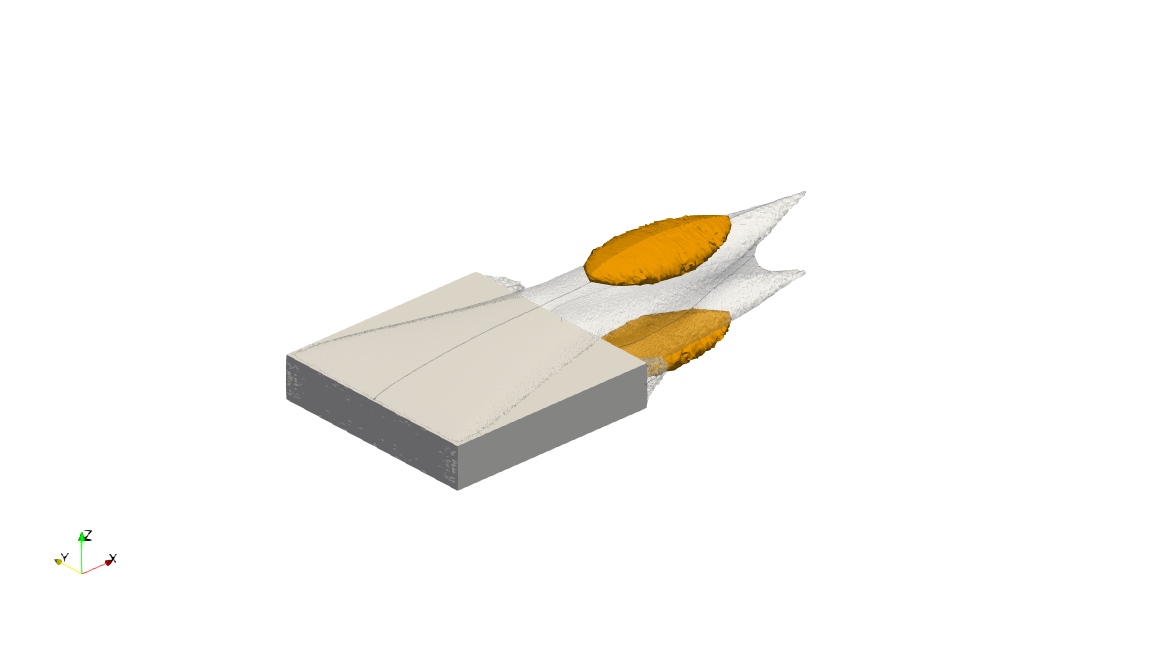}   
       \put(65,0){$S_yA_z$}
 	\end{overpic} 
 	\hspace{-0.2cm}
    \begin{overpic}[height=\hh cm, trim=75mm 53mm 155mm 47mm, clip=true]{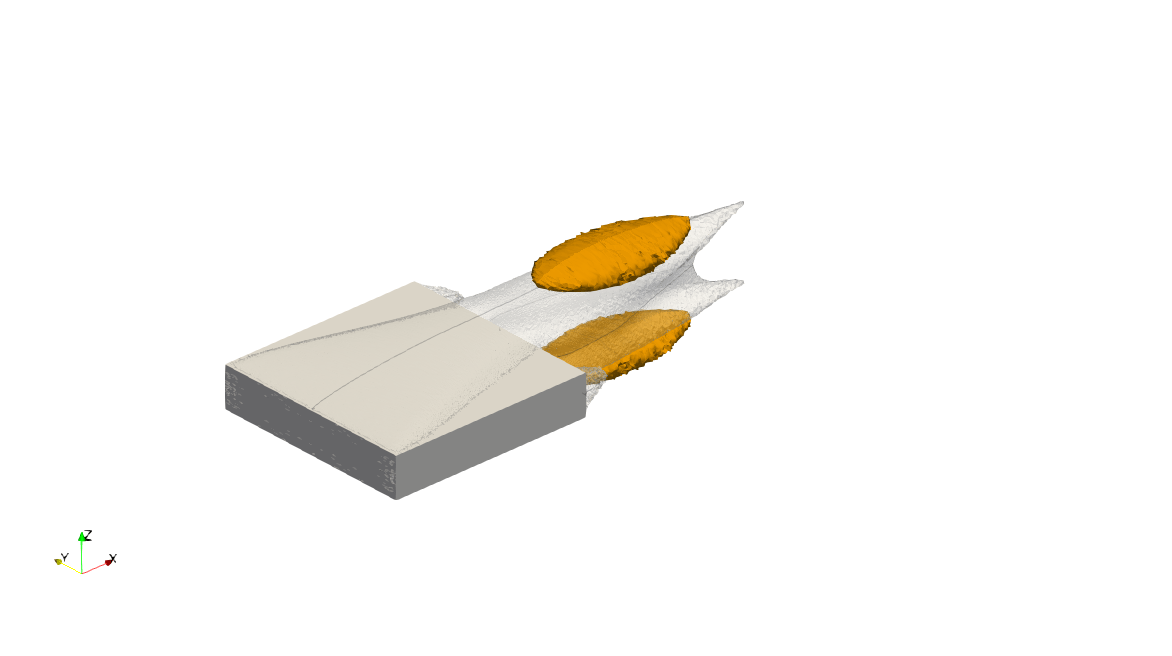}   
       \put(65,0){\textcolor[rgb]{0,0.5,0}{$S_yS_z$}}
 	\end{overpic}  
\\
\raisebox{1.5\normalbaselineskip}[0pt][0pt]{\rotatebox[origin=c]{90}{$W=2.25$}}
& 
    \begin{overpic}[height=\hh cm, trim=120mm 62mm 110mm 38mm, clip=true]{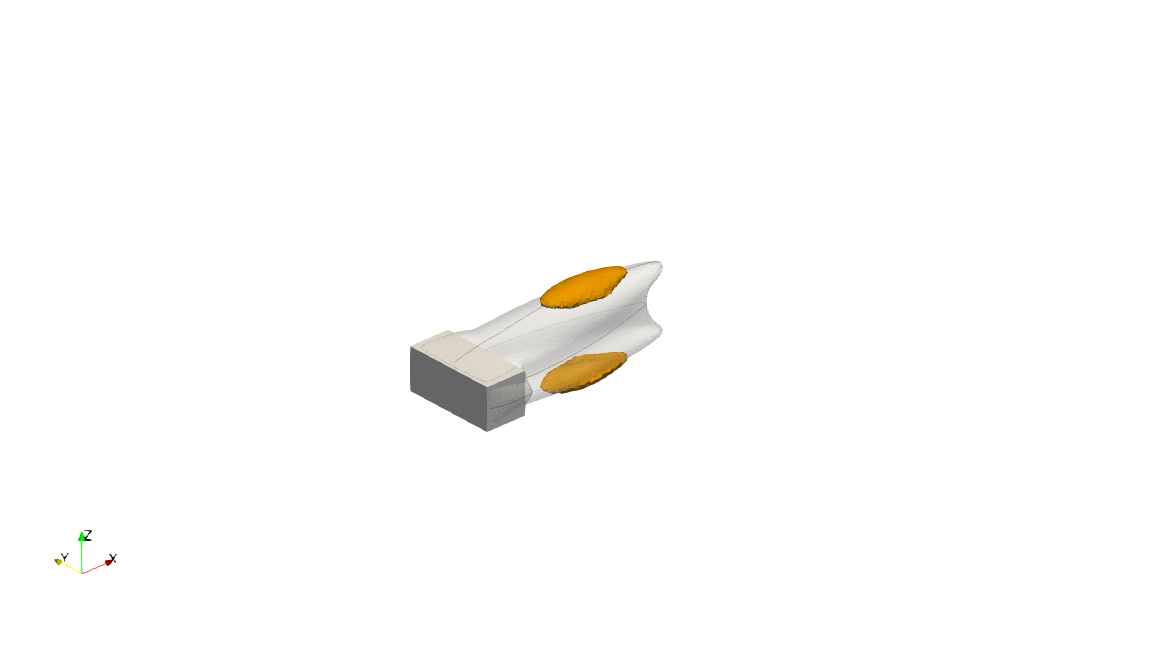}   
       \put(55,5){$S_yA_z$}
 	\end{overpic} 
 	\hspace{-0.2cm}
    \begin{overpic}[height=\hh cm, trim=100mm 60mm 130mm 40mm, clip=true]{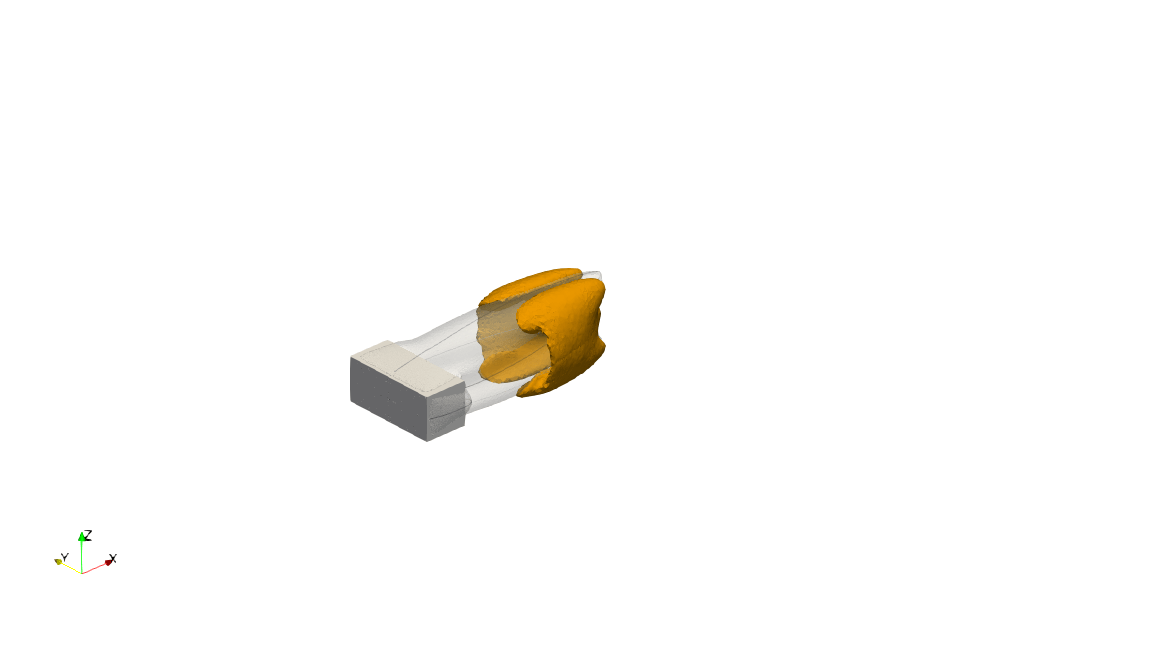}   
       \put(55,5){\textcolor{red}{$A_yS_z$}} 
 	\end{overpic}  
& 
    \begin{overpic}[height=\hh cm, trim=70mm 50mm 160mm 50mm, clip=true]{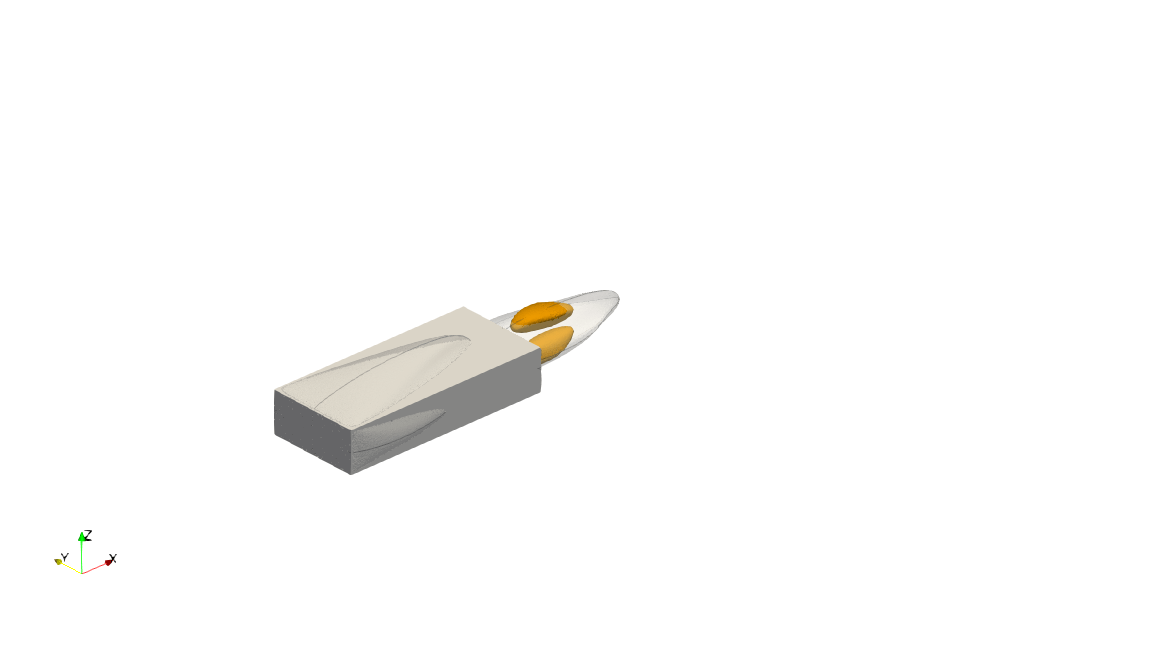} 
       \put(60,5){$S_yA_z$}  
 	\end{overpic} 	
 	\hspace{-0.2cm}
    \begin{overpic}[height=\hh cm, trim=120mm 65mm 110mm 35mm, clip=true]{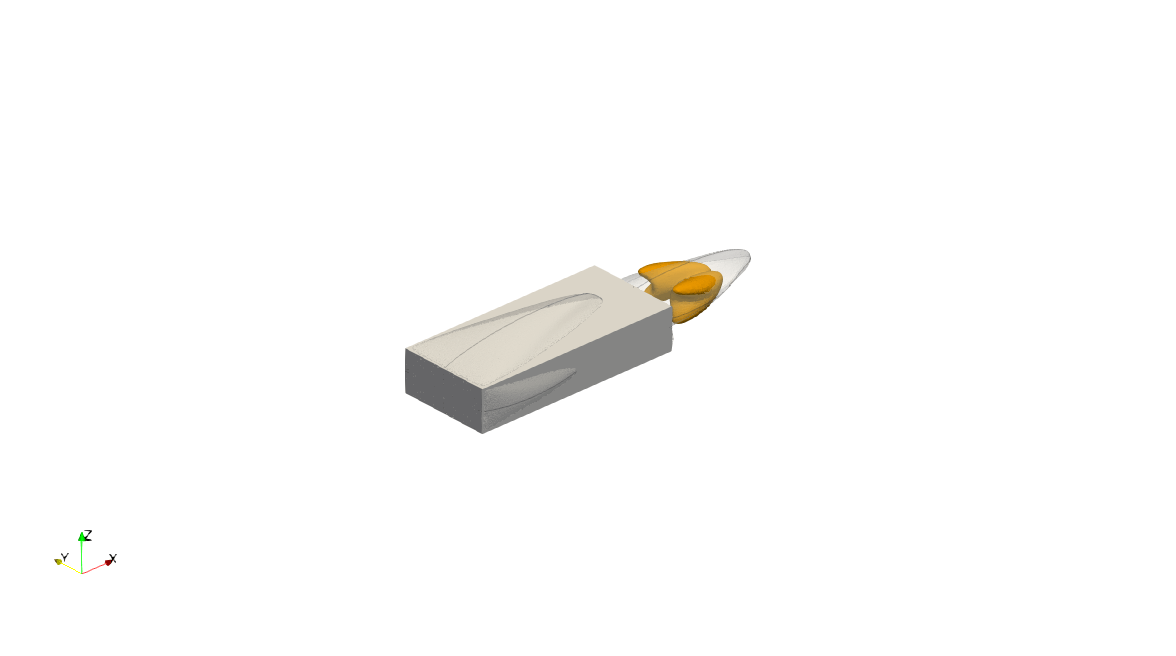}  
       \put(60,5){\textcolor{red}{$A_yS_z$}}  
 	\end{overpic} 
\\
\raisebox{2\normalbaselineskip}[0pt][0pt]{\rotatebox[origin=c]{90}{$W=1.2$}}
&
    \begin{overpic}[height=\hh cm, trim=100mm 40mm 130mm 40mm, clip=true]{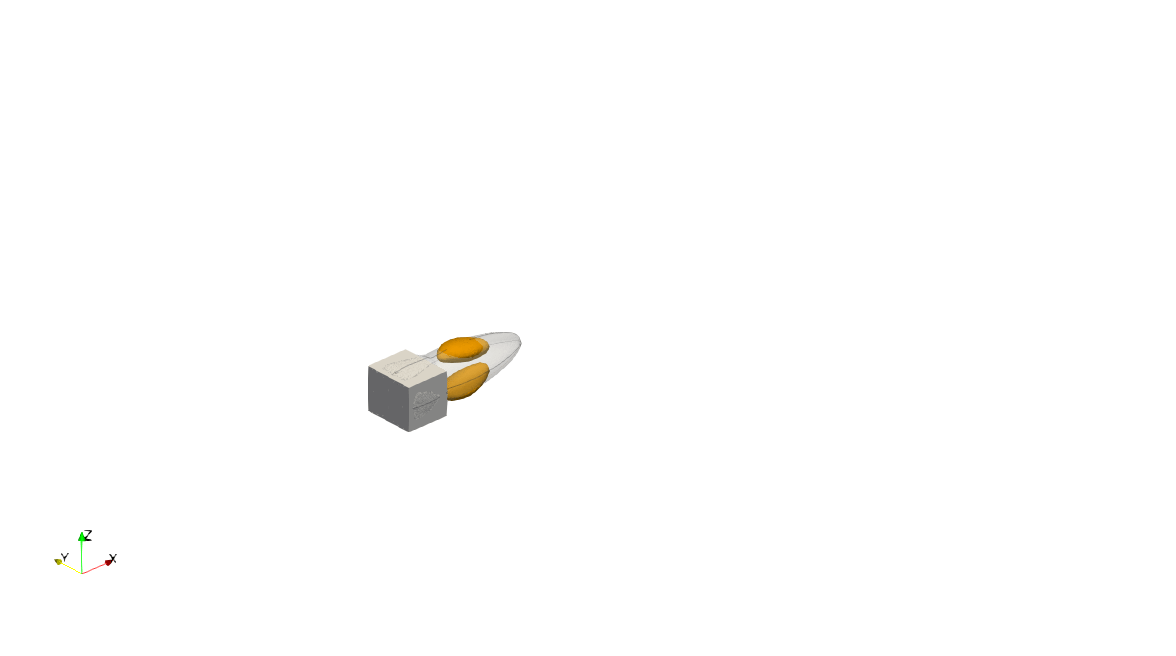}   
       \put(50,20){$S_yA_z$}
 	\end{overpic} 
 	\hspace{0.1cm}
    \begin{overpic}[height=\hh cm, trim=150mm 55mm 80mm 25mm, clip=true]{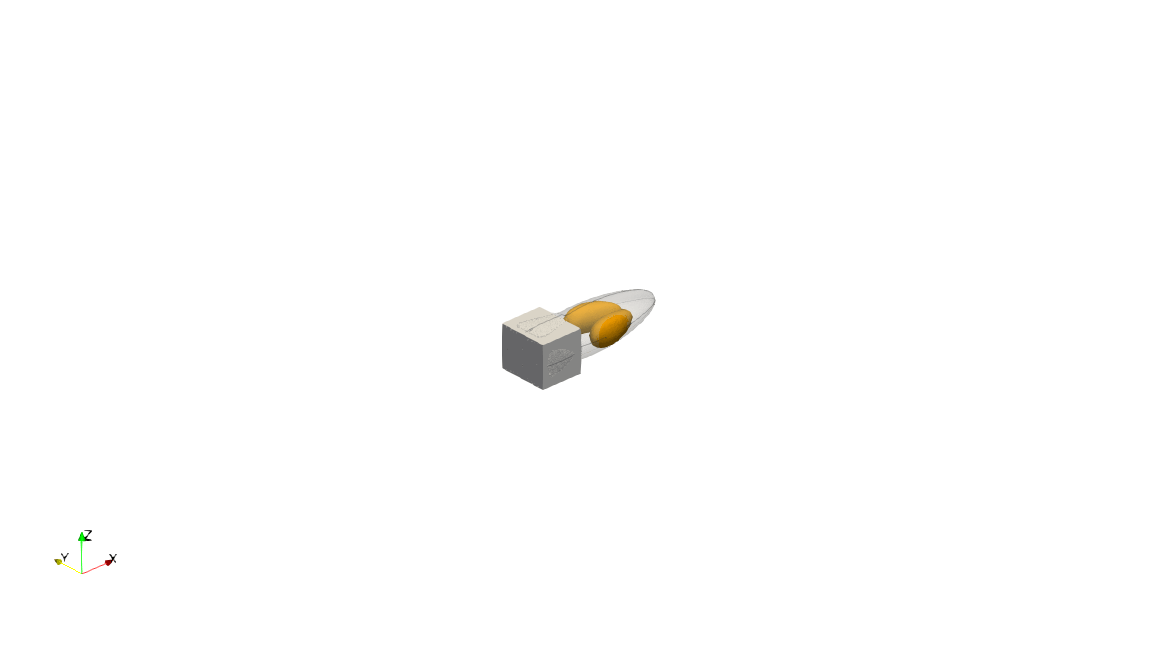}  
       \put(50,20){\textcolor{red}{$A_yS_z$}} 
 	\end{overpic} 
&
    \hspace{-0.3cm}
    \begin{overpic}[height=\hh cm, trim=75mm 33mm 155mm 47mm, clip=true]{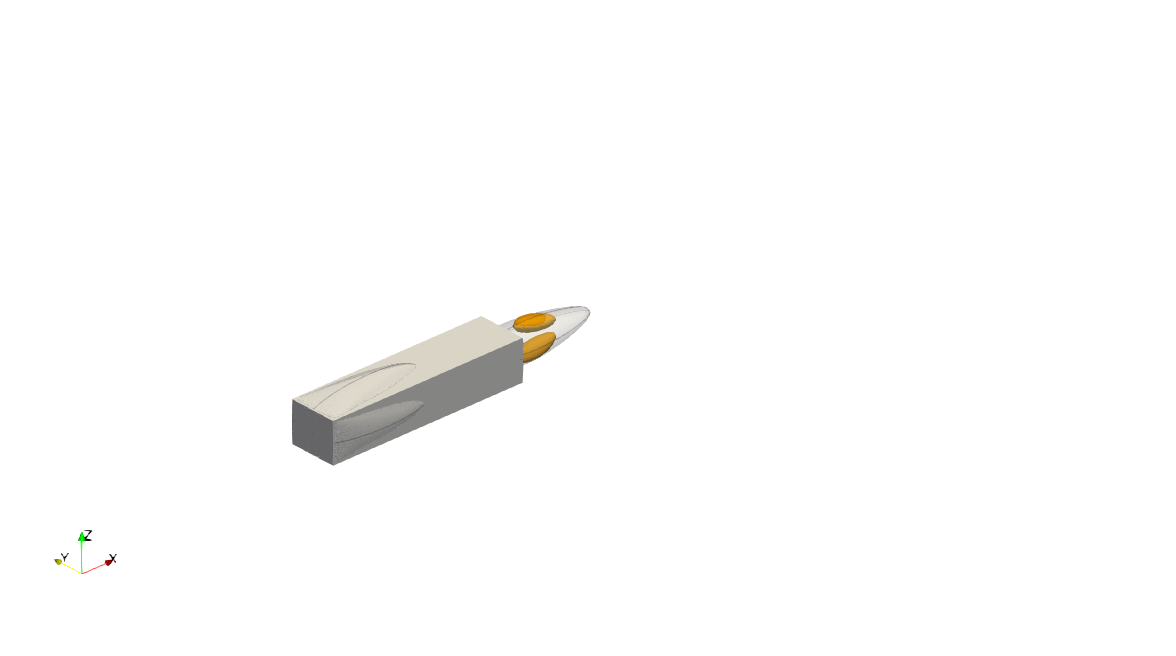}   
       \put(60,20){$S_yA_z$}
 	\end{overpic}
 	\hspace{0.1cm}
    \begin{overpic}[height=\hh cm, trim=120mm 49mm 110mm 31mm, clip=true]{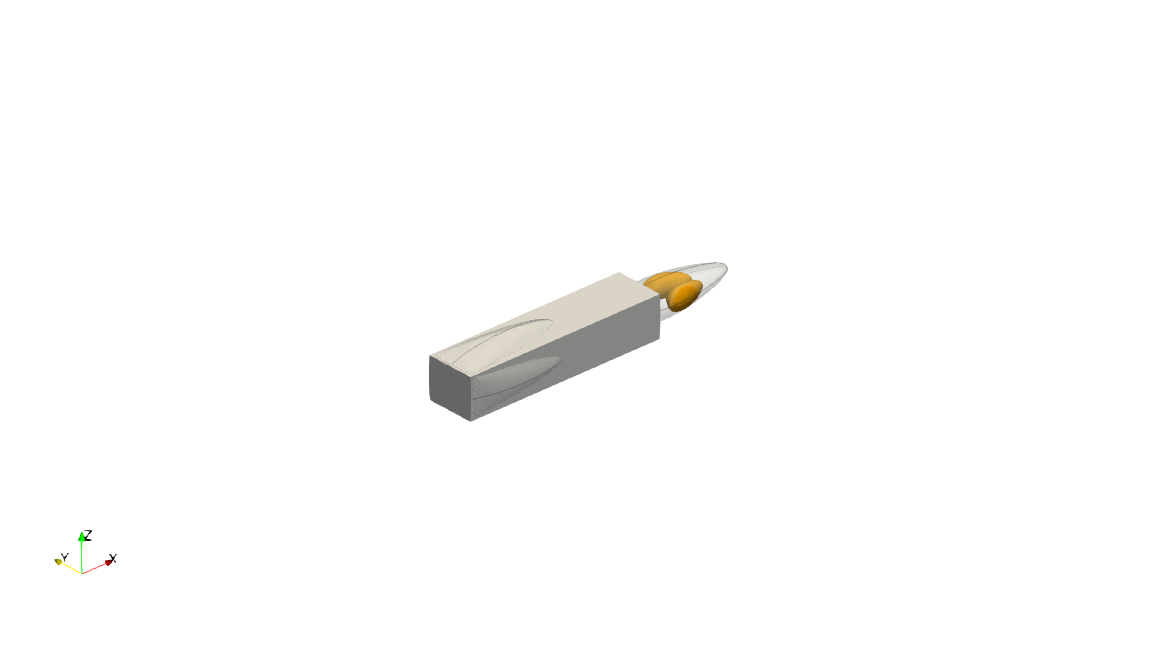}  
       \put(60,20){\textcolor{red}{$A_yS_z$}}  
 	\end{overpic}  
\\
\end{tabular} 
\vspace{-0.3cm}
\caption{
Structural sensitivity of the first (left) and second (right) bifurcating eigenmodes: representative isosurfaces of $||\hat{\boldsymbol{u}}_1||\,||\hat{\boldsymbol{u}}_1^{\dag}||$ (opaque orange).
The isosurface $u_0=0$ is reproduced from figure~\ref{tab:BF} (translucent grey).
}
\label{tab:first_two_modes_sensitivity}
\end{centering}
\end{figure}

\subsection{Weakly nonlinear analysis}
\label{sec:WNL}

In \S\ref{sec:stability} we have investigated the linear stability of the $S_yS_z$ steady base flow, characterising the different eigenmodes that become unstable.
It is worth emphasising that, when $Re$ increases past the first bifurcation, the linear stability of the base flow is not necessarily relevant. 
Rigorously speaking, to detect secondary bifurcations one should study the linear stability of the nonlinearly bifurcated state itself (e.g. static deflected wake or periodic vortex shedding) or use nonlinear simulations. 
For instance, the onset of the 3D instability in the 2D cylinder wake can be predicted using Floquet analysis of the 2D limit cycle for $Re>47$ \citep{barkley-henderson-1996}. 
In general, when the base flow is unstable to at least two eigenmodes, the nonlinear flow cannot be predicted  only by comparing the growth rates or the critical $Re$ of these modes. 
In some cases, weakly nonlinear (WNL) analysis is able to capture the dominant nonlinear interactions between competing modes, thus providing a very useful reduced-order model (a set of scalar amplitude equations) able to inform about the types of bifurcations (e.g. subcritical or supercritical), the different bifurcated flows and their linear stability, ranges of multi-stability and hysteresis, etc. 
For example, \cite{zampogna_boujo_2023} found that for the flow past an Ahmed body (3D rectangular prism with $(L,W)=(3,1.2)$ and rounded leading edges), where a steady $S_y A_z$ mode becomes linearly unstable before a steady $A_y S_z$ mode, the weakly nonlinear analysis correctly predicts that the vertically and horizontally deflected wakes exchange their stability as $Re$ increases.

In this section, we perform a WNL analysis to study the flow behaviour in the vicinity of codimension-two points, where two eigenmodes become unstable simultaneously. 
Anticipating on the DNS (\S\ref{sec:simulations}), we focus on $(L,W)=(5,1.2)$ and $(5,2.25)$. 
For each of these two geometries, the top/bottom ($S_y A_z$) and left/right ($A_y S_z$) symmetry-breaking eigenmodes undergo a stationary bifurcation at close $Re$ values.
The method is the same as in \cite{zampogna_boujo_2023} 
(see Appendix~\ref{sec:app-WNL} for details about the derivation).
In short, we use the technique of multiple scales. We introduce a small parameter $\epsilon^2$ quantifying the departure from criticality $Re_c^{-1}-Re^{-1}$, expand the flow field as a power series in $\epsilon$, introduce slow time scales $\epsilon^2 t$, $\epsilon^4 t \ldots$, and use a small-amplitude shift operator meant to bring the two bifurcations at the same critical Reynolds number $Re_c$. 
By injecting all this in the Navier-Stokes equations and collecting like-order terms in $\epsilon$, we obtain a series of linear problems to be solved successively. A set of two coupled equations for the slowly-varying amplitudes of the $S_yA_z$ and $A_yS_z$ modes, i.e. $A$ and $B$ respectively, is then obtained by imposing non-resonance conditions (Fredholm alternative).
Remarkably, although their coefficients are computed at $Re=Re_c$ only, these ordinary differential equations can be solved very easily and yield stable and unstable solutions $(A,B)$ as a function of $Re$ (assumed sufficient close to $Re_c$). 

\begin{figure}
\centerline{   
    \begin{overpic}[height=5.6cm, trim=35mm 90mm 40mm 95mm, clip=true]{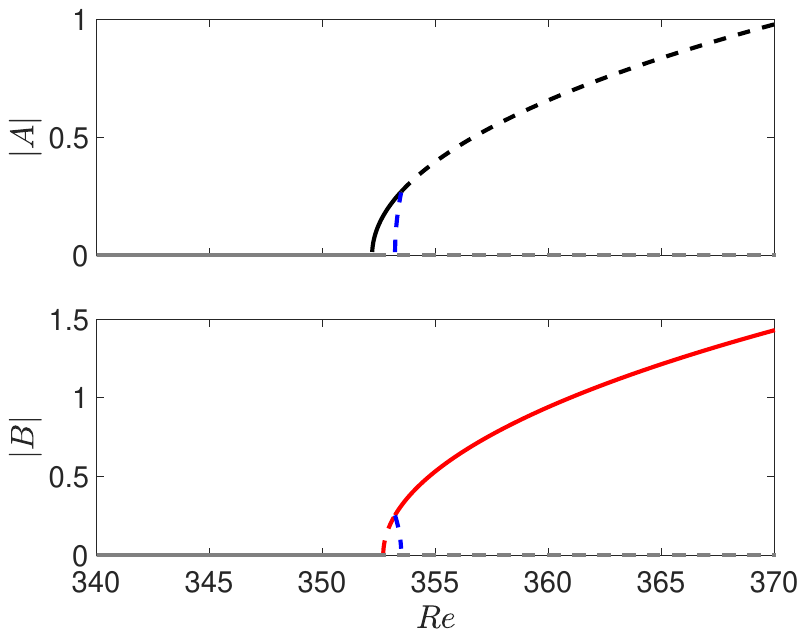}
       \put(-2,78){$(a)$}
       \put(41,62)                               {$(A,0)$ }  
       \put(50,52)  {\textcolor{blue}            {$(A,B)$}}  
       \put(20,51)  {\textcolor[rgb]{0.5,0.5,0.5}{$(0,0)$}}  
       \put(41,23)  {\textcolor{red}             {$(0,B)$}}  
       \put(50,14)  {\textcolor{blue}            {$(A,B)$}}  
       \put(20,13.5){\textcolor[rgb]{0.5,0.5,0.5}{$(0,0)$}}  
 	\end{overpic}
    \begin{overpic}[height=5.6cm, trim=35mm 90mm 40mm 95mm, clip=true]{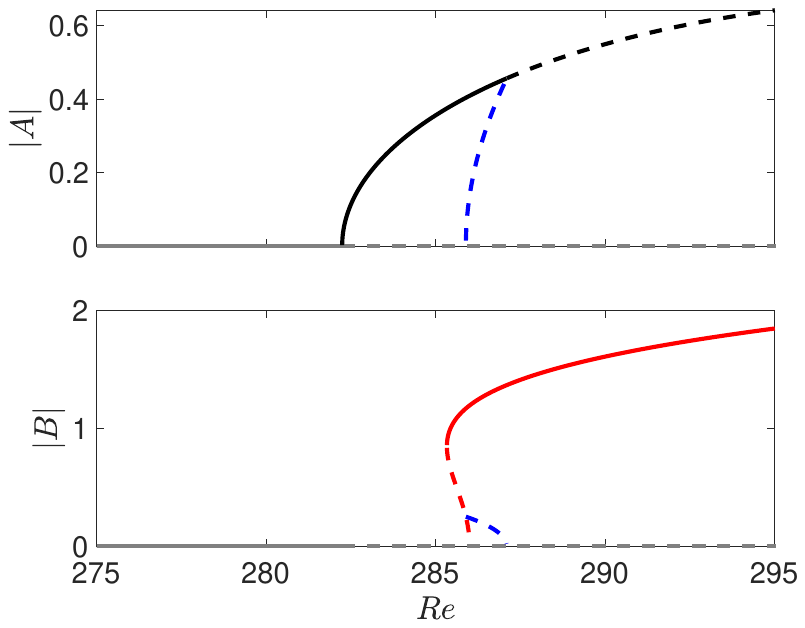}
       \put(-2,78){$(b)$}
       \put(43,68)                               {$(A,0)$ }  
       \put(59,52)  {\textcolor{blue}            {$(A,B)$}}  
       \put(20,51)  {\textcolor[rgb]{0.5,0.5,0.5}{$(0,0)$}}  
       \put(43,27)  {\textcolor{red}             {$(0,B)$}}  
       \put(60,16)  {\textcolor{blue}            {$(A,B)$}}  
       \put(20,13.5){\textcolor[rgb]{0.5,0.5,0.5}{$(0,0)$}}  
 	\end{overpic}
}
\caption{
Bifurcation diagrams obtained from weakly nonlinear analysis in the vicinity of codimension-two steady bifurcations.
$A$ and $B$ are the amplitudes of the top/bottom ($S_y A_z$) and left/right ($A_y S_z$) symmetry-breaking eigenmodes, respectively.
Solid and dashed lines denote stable and unstable branches, respectively.
$(a)$~$L=5, W=1.2$;
$(b)$~$L=5, W=2.25$.
}
\label{fig:WNL}
\end{figure}

For  $(L,W)=(5,1.2)$, we obtain the third-order amplitude equations
\begin{align}
\mathrm{d}_t A &=  \lambda_A A 
- \chi_A A^3
- \eta_A A B^2,
\label{eq:A}
\\
\mathrm{d}_t B &=  \lambda_B B 
- \chi_B B^3
- \eta_B A^2 B.
\label{eq:B}
\end{align}
See Appendix~\ref{sec:app-WNL} for the expressions of the $\lambda, \chi$ and $\eta$ coefficients.
The above system has four possible equilibrium solutions ($\mathrm{d}_t A=\mathrm{d}_t B=0$): (i)~symmetric base flow $(A,B)=(0,0)$ described in \S\ref{sec:baseflow}, (ii)~pure vertical deflection $(A,0)$, (iii)~pure horizontal deflection $(0,B)$, and (iv)~mixed state $(A,B)$. 
Note that (\ref{eq:A})-(\ref{eq:B}) is invariant under reflections $A\rightarrow-A$ and $B \rightarrow -B$, so we discuss only $A \geq 0$ and $B\geq 0$.
The linear stability of each state is determined by computing the eigenvalues of the $2\times 2$ Jacobian of (\ref{eq:A})-(\ref{eq:B}) linearised about the state of interest.
Here, we obtain  the bifurcation diagram of figure~\ref{fig:WNL}$(a)$.
The symmetric state $(A,B)=(0,0)$ is stable up to $Re \approx 352$, when the $S_yA_z$ eigenmode becomes unstable and the flow settles to a vertically deflected state $(A,0)$. 
Almost at the same $Re$, the $A_y S_z$ eigenmode becomes unstable, but the horizontally deflected state $(0,B)$ is initially unstable. 
Near $Re \approx 353$, there is an exchange of stability between states $(A,0)$ and $(0,B)$, i.e. the wake leaves the vertically deflected state and enters a horizontally deflected state. 
In the very narrow range of bistability between these two states, a mixed state $(A,B)$ exists, but is unstable. 
Qualitatively, the bifurcation scenario is similar to that found by \cite{zampogna_boujo_2023} for Ahmed bodies of dimensions $(L, W)=(3, 1.2)$ and a range of close-by geometries.

For $(L,W)=(5,2.25)$ we find a subcritical pitchfork bifurcation of the $A_y S_z$ mode. 
In this case, leading-order nonlinearities do not lead to saturation but instead make the system more unstable, so (\ref{eq:A})-(\ref{eq:B}) does not yield a stable solution.
This calls for the inclusion of higher-order nonlinearities to capture saturation, so we derive the fifth-order amplitude equations
\begin{align}
\mathrm{d}_t A
&= 
  \lambda_A A 
- \chi_A A^3 
- \eta_A A B^2
+ \kappa_A A B^4
+ \beta_A  A^3 B^2
+ \gamma_A A^5,
\label{AE_dAdt_2}
\\
\mathrm{d}_t B
&= 
 \lambda_B B 
-\chi_B B^3 
-\eta_B A^2 B
+  \kappa_B A^4 B
+  \beta_B  A^2 B^3
+  \gamma_B B^5.
\label{AE_dBdt_2}
\end{align}
The above system has the same type of equilibrium solutions as previously described. However, the pure deflected states may each have one or two different amplitudes $(A_1,0)$ and $(A_2,0)$, and $(0,B_1)$ and $(0,B_2)$;  similarly, up to four mixed states may exist, $(A_i,B_i)$, $i=1,2$.
Here, we obtain the bifurcation diagram of figure~\ref{fig:WNL}$(b)$.
The symmetric state $(A,B)=(0,0)$ is stable up to $Re \approx 282$, when the $S_yA_z$ eigenmode becomes unstable and the flow settles to a vertically deflected state $(A,0)$. This first bifurcation is supercritical, so there exists only one such state.
At $Re \approx 286$ the $A_yS_z$ mode becomes unstable, and this second bifurcation is subcritical: the branch $(0,B)$ is initially unstable and moves towards lower Reynolds numbers, before becoming stable again and folding back to larger Reynolds number. One mixed state $(A,B)$ exists but is unstable. 
The state $(A,0)$ becomes unstable at $Re \approx 287$, leaving $(0,B)$ as the only stable state.  Therefore, there is a narrow interval of bistability between the vertically and horizontally deflected states.

\section{Three-dimensional nonlinear simulations}
\label{sec:simulations}

In this section, we perform 3D, unsteady, fully nonlinear simulations to explore the successive bifurcations observed in the flow past elongated prisms with $L=5$ and $W=1.2$, $2.25$ and $5$, up to $Re=700$.
The specific length $L=5$ is chosen as it defines the international BARC benchmark \citep{bruno-salvetti-ricciardelli-2014}. 
To distinguish the different regimes, we use the notation $xY_yZ_z$:
the lower case $x$ determines whether the regime is steady ($s$), periodic ($p$) or aperiodic ($a$); the upper case $Y$ and $Z$ determine the planar symmetries of the flow, i.e. whether the flow is symmetric ($S$) or antisymmetric ($A$) with respect to the $y=0$ and $z=0$ planes. 
For unsteady flows, the symmetry refers to the time-averaged flow. 
For example, the $sS_yS_z$ regime refers to a steady regime that retains the left/right and top/bottom planar symmetries (like the symmetric base flow of \S\ref{sec:baseflow}). 
For $W=2.25$ and $5$, the periodic regimes $pS_yS_zl$ and $pS_yS_zt$ are characterised by vortex shedding  from the leading-edge ($l$) and trailing-edge ($t$) shear layers, respectively. 

We use the aerodynamic forces 
as driving quantities to detect the different flow regimes. Indeed, the frequency spectra and the instantaneous/mean/root-mean-square values of the aerodynamics forces provide immediate information about the spatial and temporal symmetries of the flow, i.e. whether the flow is steady/periodic/aperiodic and whether the instantaneous and mean fields exhibit top/bottom or right/left planar symmetries. However, we have verified that the same information regarding the temporal symmetries of the flow is obtained by inspecting the velocity field and probing time signals at different locations of the flow. To further characterise the flow regimes, we also report visualisations of mean and instantaneous fields, phase-space diagrams and POD modes.

\subsection{Narrow prism: $W=1.2$}
\begin{figure}
\centering
\begin{tikzpicture}

\definecolor{clr1}{RGB}{18 78 128}
\definecolor{clr2}{RGB}{89 165 216}
\definecolor{clr3}{RGB}{145 229 246}
\definecolor{clr4}{RGB}{255 143 163}
\definecolor{clr5}{RGB}{255 77 109}
\definecolor{clr6}{RGB}{201 24 74}
\definecolor{clr7}{RGB}{128 15 47}
\definecolor{clr8}{RGB}{174 32 18}
\definecolor{clr9}{RGB}{155 34 38}
\definecolor{clr10}{RGB}{46 139 87}

\begin{axis}[%
width=0.9\textwidth,
height=0.2\textwidth,
scale only axis,
xmin=250,
xmax=750,
ymin=0.75,
ymax=1.05,
xtick={300,400,500,600,700},
xticklabel=\empty,
ytick={0.75,0.8,0.85,0.9,0.95,1,1.05},
yticklabels={$ 0.75$,$0.80$,$ 0.85$,$0.90$,$0.95$,$1.00$,$1.05$},
y tick label style={/pgf/number format/zerofill},
ylabel={$F_x$},
ylabel style={at={(0.02,0.5)}},
axis background/.style={fill=white},
legend columns=3,transpose legend,
legend style={at={(0.99,0.82)}, anchor=east, legend cell align=left, align=left, fill=none, draw=none}
]

\addplot [color=black,solid,draw=none,mark=*,mark options={scale=1.4,black,fill=clr1}]
  table[row sep=crcr]{%
  300.0000    0.9778 \\
  350.0000    0.9136 \\
  352.0000    0.9115 \\
};
\addplot [color=black,solid,draw=none,mark=*,mark options={scale=1.4,black,fill=clr2}]
  table[row sep=crcr]{%
  355.0000    0.9085 \\
  357.0000    0.9066 \\
  360.0000    0.9037 \\
  365.0000    0.8990 \\
};
\addplot [color=black,solid,draw=none,mark=*,mark options={scale=1.4,black,fill=clr3}]
  table[row sep=crcr]{%
  370.0000    0.8948 \\
  390.0000    0.8780 \\
  450.0000    0.8341 \\
  500.0000    0.8010 \\
};

\addplot [color=black,solid,draw=none,mark=*,mark options={scale=1.4,black,fill=clr10}]
  table[row sep=crcr]{%
  510.0000    0.79500 \\
};

\addplot [color=black,solid,draw=none,mark=*,mark options={scale=1.4,black,fill=clr6}]
  table[row sep=crcr]{%
  515.0000    0.7931 \\
  520.0000    0.7907 \\
  535.0000    0.7831 \\
  550.0000    0.7759 \\
  575.0000    0.7703 \\
  590.0000    0.8253 \\
  600.0000    0.8319 \\
  625.0000    0.8187 \\ 
  650.0000    0.8229 \\
  700.0000    0.8185 \\
};

\addplot [color=black,solid,mark=none, line width=0.5, mark options={scale=1.4,black,fill=red!80!black}]
  table[row sep=crcr]{%
  510.0000    0.7949 \\
  510.0000    0.7952 \\
};

\addplot [color=black,solid,mark=none, line width=0.5, mark options={scale=1.4,black,fill=red!80!black}]
  table[row sep=crcr]{%
  515.0000    0.7926 \\
  515.0000    0.7936 \\
};

\addplot [color=black,solid,mark=none, line width=0.5, mark options={scale=1.4,black,fill=red!80!black}]
  table[row sep=crcr]{%
  520.0000    0.7900 \\
  520.0000    0.7914 \\
};

\addplot [color=black,solid,mark=none, line width=0.5, mark options={scale=1.4,black,fill=red!80!black}]
  table[row sep=crcr]{%
  535.0000    0.7822 \\
  535.0000    0.7840 \\
};

\addplot [color=black,solid,mark=none, line width=0.5, mark options={scale=1.4,black,fill=red!80!black}]
  table[row sep=crcr]{%
  550.0000    0.7740 \\
  550.0000    0.7778 \\
};

\addplot [color=black,solid,mark=none, line width=0.5, mark options={scale=1.4,black,fill=red!80!black}]
  table[row sep=crcr]{%
  575.0000    0.7656 \\
  575.0000    0.7759 \\
};

\addplot [color=black,solid,mark=none, line width=0.5, mark options={scale=1.4,black,fill=red!80!black}]
  table[row sep=crcr]{%
  590.0000    0.8135 \\
  590.0000    0.8329 \\
};

\addplot [color=black,solid,mark=none, line width=0.5, mark options={scale=1.4,black,fill=red!80!black}]
  table[row sep=crcr]{%
  600.0000    0.8210 \\
  600.0000    0.8436 \\
};

\addplot [color=black,solid,mark=none, line width=0.5,mark options={scale=1.4,black,fill=red!80!black}]
  table[row sep=crcr]{%
  625.0000    0.8042 \\
  625.0000    0.8332 \\
};

\addplot [color=black,solid,mark=none, line width=0.5,mark options={scale=1.4,black,fill=red!80!black}]
  table[row sep=crcr]{%
  650.0000    0.8059 \\
  650.0000    0.8400 \\
};

\addplot [color=black,solid,mark=none, line width=0.5,mark options={scale=1.4,black,fill=red!80!black}]
  table[row sep=crcr]{%
  700.0000    0.8033 \\
  700.0000    0.8337 \\
};

\addplot [color=black,dashed,mark=none, line width=0.5,mark options={scale=1.4,black,fill=red!80!black}]
  table[row sep=crcr]{%
  353    0.75 \\
  353    1.05 \\
};

\addplot [color=black,dashed,mark=none, line width=0.5,mark options={scale=1.4,black,fill=red!80!black}]
  table[row sep=crcr]{%
  367    0.75 \\
  367    1.05 \\
};

\addplot [color=black,dashed,mark=none, line width=0.5,mark options={scale=1.4,black,fill=red!80!black}]
  table[row sep=crcr]{%
  507    0.75 \\
  507    1.05 \\
};

\addplot [color=black,dashed,mark=none, line width=0.5,mark options={scale=1.4,black,fill=red!80!black}]
  table[row sep=crcr]{%
  512    0.75 \\
  512    1.05 \\
};
\addplot[red, solid, domain=250:575] {8.47*x^(-0.3799)};

\end{axis}

\node[] at (1.00,2.4) {$sS_yS_z$};
\draw[->] (3.05,2.4) -- (2.6,2.4);
\node[] at (3.5,2.4) {$sS_yA_z$};
\node[] at (5.1,2.4) {$sA_yS_z$};
\draw[->] (6.7,2.4) -- (6.3,2.4);
\node[] at (7.2,2.4) {$pA_yS_z$};
\node[] at (9.2,2.4) {$aA_yS_z$};

\end{tikzpicture}%
\vspace{-0.1cm}
\begin{tikzpicture}

\definecolor{clr1}{RGB}{18 78 128}
\definecolor{clr2}{RGB}{89 165 216}
\definecolor{clr3}{RGB}{145 229 246}
\definecolor{clr4}{RGB}{255 143 163}
\definecolor{clr5}{RGB}{255 77 109}
\definecolor{clr6}{RGB}{201 24 74}
\definecolor{clr7}{RGB}{128 15 47}
\definecolor{clr8}{RGB}{174 32 18}
\definecolor{clr9}{RGB}{155 34 38}
\definecolor{clr10}{RGB}{46 139 87}

\begin{axis}[%
scaled ticks=false,
width=0.9\textwidth,
height=0.2\textwidth,
scale only axis,
xmin=250,
xmax=750,
ymin=-0.04,
ymax=0.06,
xtick={300,400,500,600,700},
ytick={-0.04,-0.02,0,0.02,0.04,0.06},
yticklabels={-0.04,-0.02,0.00,0.02,0.04,0.06},
ylabel style={at={(0.02,0.5)}},
    y tick label style={
        /pgf/number format/.cd,
            fixed,
            fixed zerofill,
            precision=2,
        /tikz/.cd
    },
xlabel={$Re$},
ylabel={$F_y,F_z$},
axis background/.style={fill=white},
legend columns=3,transpose legend,
legend style={at={(0.13,0.15)}, anchor=east, legend cell align=left, align=left, fill=none, draw=none}
]

\addplot [color=black,solid,draw=none,mark=square,mark options={scale=1.3,black,fill=}]
  table[row sep=crcr]{%
  390.0000         0 \\
  450.0000         0 \\
  500.0000         0 \\
  550.0000         0 \\
  625.0000    0.0003 \\
  700.0000         0 \\
};
\addlegendentry{$F_z$};

\addplot [color=black,solid,draw=none,mark=diamond,mark options={scale=1.8,black,fill=red!80!black}]
  table[row sep=crcr]{%
  390.0000    0.0185 \\
  450.0000    0.0201 \\
  500.0000    0.0192 \\
  535.0000    0.0201 \\
  550.0000    0.0200 \\
  625.0000    0.0235 \\
  700.0000    0.0201 \\
};
\addlegendentry{$F_y$};

\addplot [color=black,solid,draw=none,mark=square*,mark options={scale=1.3,black,fill=clr1}]
  table[row sep=crcr]{%
  300.0000         0 \\
  350.0000         0 \\
  352.0000         0 \\
};

\addplot [color=black,solid,draw=none,mark=square*,mark options={scale=1.3,black,fill=clr2}]
  table[row sep=crcr]{
  355.0000    0.00691 \\
  357.0000    0.00848 \\
  360.0000    0.01023 \\
  365.0000    0.01232 \\
};

\addplot [color=black,solid,draw=none,mark=square*,mark options={scale=1.3,black,fill=clr3}]
  table[row sep=crcr]{%
  370.0000         0 \\
  390.0000         0 \\
  450.0000         0 \\
  500.0000         0 \\
};
\addplot [color=black,solid,draw=none,mark=square*,mark options={scale=1.3,black,fill=clr10}]
  table[row sep=crcr]{%
  510.0000         0 \\
};
\addplot [color=black,solid,draw=none,mark=square*,mark options={scale=1.3,black,fill=clr6}]
  table[row sep=crcr]{%
  515.0000         0 \\
  520.0000         0 \\
  535.0000         0 \\
  550.0000         0 \\
  575.0000         0 \\
  590.0000         0 \\
  600.0000         0 \\
  625.0000    0.0003 \\
  650.0000    0.0009 \\
  700.0000         0 \\
};
\addplot [color=black,solid,draw=none,mark=diamond*,mark options={scale=1.8,black,fill=clr1}]
  table[row sep=crcr]{%
  300.0000         0 \\
  350.0000         0 \\
};

\addplot [color=black,solid,draw=none,mark=diamond*,mark options={scale=1.8,black,fill=clr2}]
  table[row sep=crcr]{%
  352.0000         0 \\
  355.0000         0 \\
  357.0000         0 \\
  360.0000         0 \\
  365.0000         0 \\
};

\addplot [color=black,solid,draw=none,mark=diamond*,mark options={scale=1.8,black,fill=clr3}]
  table[row sep=crcr]{%
  370.0000    0.0149 \\
  390.0000    0.0185 \\
  450.0000    0.0201 \\
  500.0000    0.0192 \\
};
\addplot [color=black,solid,draw=none,mark=diamond*,mark options={scale=1.8,black,fill=clr10}]
  table[row sep=crcr]{
  510.0000    0.0191 \\
};
\addplot [color=black,solid,draw=none,mark=diamond*,mark options={scale=1.8,black,fill=clr6}]
  table[row sep=crcr]{
  515.0000    0.0195 \\
  520.0000    0.0198 \\
  535.0000    0.0201 \\  
  550.0000    0.0200 \\
  575.0000    0.0204 \\
  590.0000    0.0270 \\
  600.0000    0.0267 \\
  625.0000    0.0235 \\
  650.0000    0.0245 \\
  700.0000    0.0201 \\
};

\addplot [color=red,solid,mark=none, line width=0.5, mark options={scale=1.4,black,fill=red!80!black}]
  table[row sep=crcr]{%
  510.0000    0.0191 \\
  510.0000    0.0191 \\
};

\addplot [color=red,solid,mark=none, line width=0.5, mark options={scale=1.4,black,fill=red!80!black}]
  table[row sep=crcr]{%
  515.0000    0.0181 \\
  515.0000    0.0209 \\
};

\addplot [color=red,solid,mark=none, line width=0.5, mark options={scale=1.4,black,fill=red!80!black}]
  table[row sep=crcr]{%
  520.0000    0.0179 \\
  520.0000    0.0217 \\
};

\addplot [color=red,solid,mark=none, line width=0.5, mark options={scale=1.4,black,fill=red!80!black}]
  table[row sep=crcr]{%
  535.0000    0.0172 \\
  535.0000    0.0229 \\
};

\addplot [color=red,solid,mark=none, line width=0.5, mark options={scale=1.4,black,fill=red!80!black}]
  table[row sep=crcr]{%
  550.0000    0.0166 \\
  550.0000    0.0234 \\
};

\addplot [color=red,solid,mark=none, line width=0.5, mark options={scale=1.4,black,fill=red!80!black}]
  table[row sep=crcr]{%
  575.0000    0.0142 \\
  575.0000    0.0266 \\
};

\addplot [color=red,solid,mark=none, line width=0.5, mark options={scale=1.4,black,fill=red!80!black}]
  table[row sep=crcr]{%
  590.0000    0.0362 \\
  590.0000    0.0155 \\
};

\addplot [color=black,solid,mark=none, line width=0.5, mark options={scale=1.4,black,fill=red!80!black}]
  table[row sep=crcr]{%
  590.0000   -0.0163 \\
  590.0000    0.0163 \\
};

\addplot [color=red,solid,mark=none, line width=0.5, mark options={scale=1.4,black,fill=red!80!black}]
  table[row sep=crcr]{%
  600.0000    0.0139 \\
  600.0000    0.0395 \\
};

\addplot [color=red,solid,mark=none, line width=0.5,mark options={scale=1.4,black,fill=red!80!black}]
  table[row sep=crcr]{%
  625.0000    0.0014 \\
  625.0000    0.0408 \\
};

\addplot [color=red,solid,mark=none, line width=0.5,mark options={scale=1.4,black,fill=red!80!black}]
  table[row sep=crcr]{%
  650.0000    0.0063 \\
  650.0000    0.0428 \\
};

\addplot [color=black,solid,mark=none, line width=0.5,mark options={scale=1.4,black,fill=red!80!black}]
  table[row sep=crcr]{%
  650.0000    -0.0202 \\
  650.0000     0.0202 \\
};

\addplot [color=red,solid,mark=none, line width=0.5,mark options={scale=1.4,black,fill=red!80!black}]
  table[row sep=crcr]{%
  700.0000   -0.0103 \\
  700.0000    0.0505 \\
};

\addplot [color=black,dashed,mark=none, line width=0.5,mark options={scale=1.4,black,fill=red!80!black}]
  table[row sep=crcr]{%
  353    -0.04 \\
  353    0.06 \\
};

\addplot [color=black,dashed,mark=none, line width=0.5,mark options={scale=1.4,black,fill=red!80!black}]
  table[row sep=crcr]{%
  367    -0.04 \\
  367     0.06 \\
};

\addplot [color=black,dashed,mark=none, line width=0.5,mark options={scale=1.4,black,fill=red!80!black}]
  table[row sep=crcr]{%
  507    -0.04 \\
  507     0.06 \\
};

\addplot [color=black,dashed,mark=none, line width=0.5,mark options={scale=1.4,black,fill=red!80!black}]
  table[row sep=crcr]{%
  512    -0.04 \\
  512     0.06 \\
};

\addplot [color=black,solid,mark=none, line width=0.5,mark options={scale=1.4,black,fill=red!80!black}]
  table[row sep=crcr]{%
  535    -0.0143 \\
  535     0.0143 \\
};

\addplot [color=black,solid,mark=none, line width=0.5, mark options={scale=1.4,black,fill=red!80!black}]
  table[row sep=crcr]{%
  510.0000   -0.0056 \\
  510.0000    0.0056 \\
};

\addplot [color=black,solid,mark=none, line width=0.5, mark options={scale=1.4,black,fill=red!80!black}]
  table[row sep=crcr]{%
  515.0000   -0.0104 \\
  515.0000    0.0104 \\
};

\addplot [color=black,solid,mark=none, line width=0.5, mark options={scale=1.4,black,fill=red!80!black}]
  table[row sep=crcr]{%
  520.0000   -0.0123 \\
  520.0000    0.0123 \\
};

\addplot [color=black,solid,mark=none, line width=0.5, mark options={scale=1.4,black,fill=red!80!black}]
  table[row sep=crcr]{%
  550.0000   -0.0147 \\
  550.0000    0.0147 \\
};

\addplot [color=black,solid,mark=none, line width=0.5, mark options={scale=1.4,black,fill=red!80!black}]
  table[row sep=crcr]{%
  575.0000   -0.0116 \\
  575.0000    0.0116 \\
};

\addplot [color=black,solid,mark=none, line width=0.5, mark options={scale=1.4,black,fill=red!80!black}]
  table[row sep=crcr]{%
  600.0000   -0.0158 \\
  600.0000    0.0158 \\
};

\addplot [color=black,solid,mark=none, line width=0.5,mark options={scale=1.4,black,fill=red!80!black}]
  table[row sep=crcr]{%
  625.0000   -0.0137 \\
  625.0000    0.0225 \\
};

\addplot [color=black,solid,mark=none, line width=0.5,mark options={scale=1.4,black,fill=red!80!black}]
  table[row sep=crcr]{%
  700.0000   -0.0241 \\
  700.0000    0.0241 \\
};

\end{axis}

\node[] at (1.00,2.4) {$sS_yS_z$};
\draw[->] (3.05,2.4) -- (2.6,2.4);
\node[] at (3.5,2.4) {$sS_yA_z$};
\node[] at (5.1,2.4) {$sA_yS_z$};
\draw[->] (6.7,2.4) -- (6.3,2.4);
\node[] at (7.2,2.4) {$pA_yS_z$};
\node[] at (9.2,2.4) {$aA_yS_z$};

\end{tikzpicture}%
\includegraphics[width=0.325\textwidth]{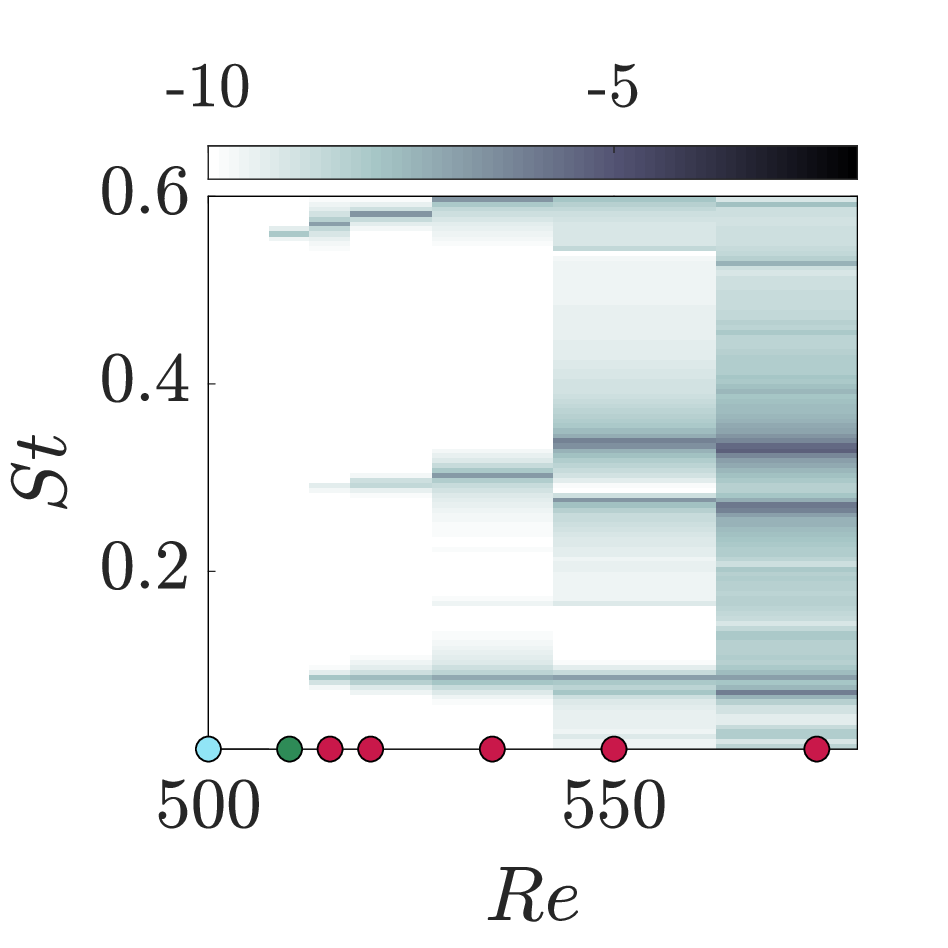}
\includegraphics[width=0.325\textwidth]{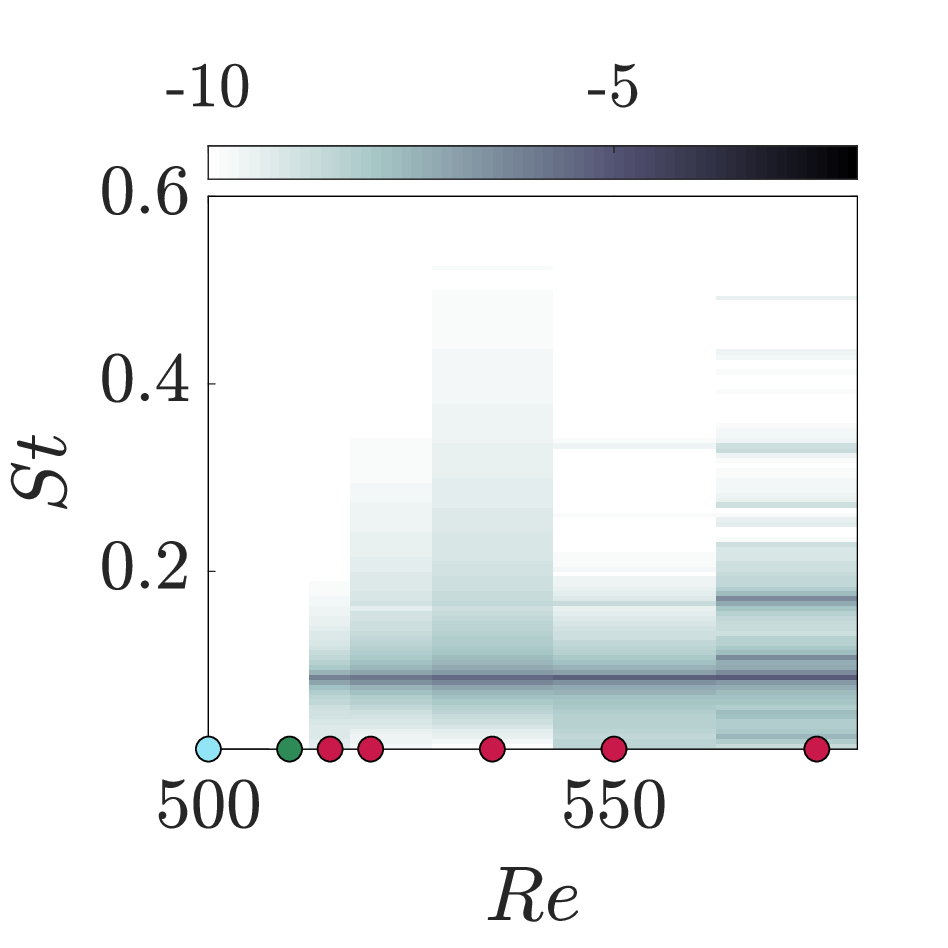}
\includegraphics[width=0.325\textwidth]{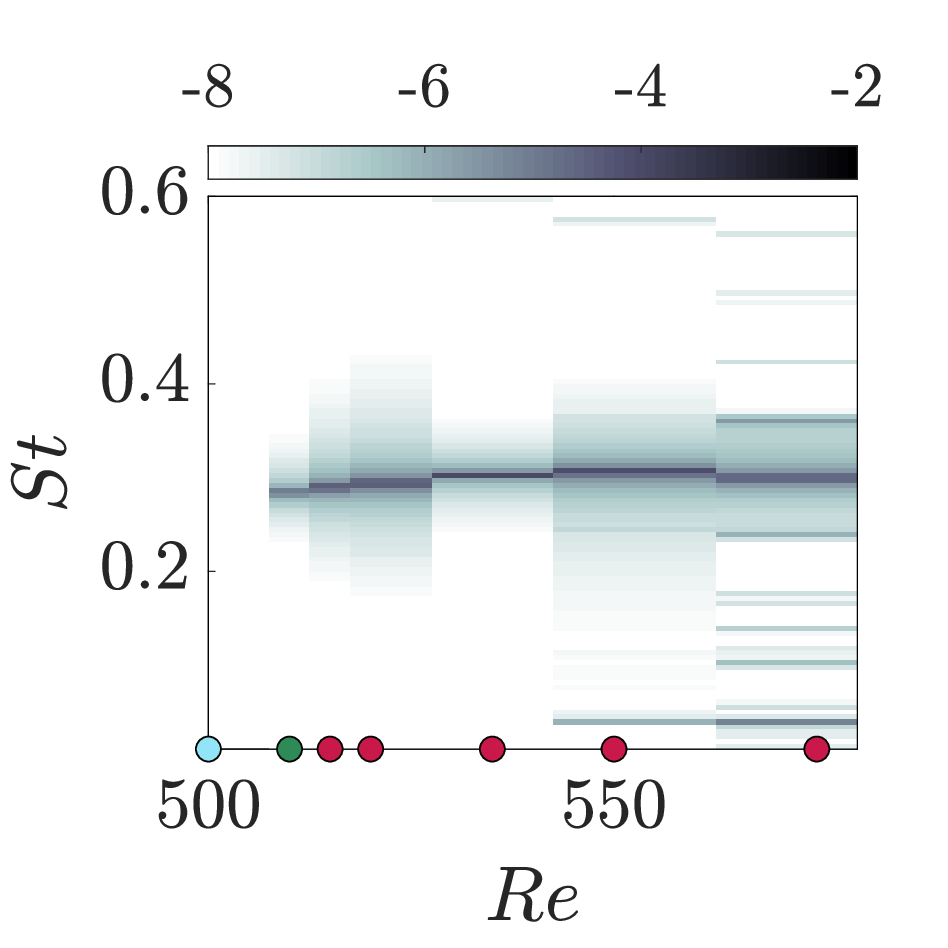}
\caption{Dependence of the aerodynamic forces on the Reynolds number for $L=5$ and $W=1.2$. 
In the top panels, circles refer to the average value, and bars to the root mean square  of the fluctuations. 
Top: streamwise force $F_x$. Centre: lateral forces $F_y$ and $F_z$. 
Red line in the top panel: $8.47 \times Re^{-0.3799}$.
Bottom:  frequency spectra of $F_x$ (left), $F_y$ (centre) and $F_z$ (right) in logarithmic scale.}
\label{fig:3D_Ary1_forces}
\end{figure}

For $L=5$ and $W=1.2$ five different regimes are identified when the Reynolds number is increased up to $Re=700$. This is conveniently visualised in figure \ref{fig:3D_Ary1_forces}, where the aerodynamic forces are shown as a function of $Re$. 
For $Re \lessapprox 352$ the flow is steady and retains the $S_yS_z$ symmetry, in agreement with the LSA (\S\ref{sec:stability}). 
At $Re \approx 355$ the $sS_yS_z$ regime is unstable, and the flow experiences a pitchfork bifurcation towards the $sS_yA_z$ regime. The flow loses the top/bottom planar symmetry, but retains the left/right one. For $Re \gtrapprox 370$ 
the flow enters the $sA_yS_z$ regime, i.e. recovers the top/bottom planar symmetry and loses the left/right one. 
This regime remains stable up to $Re \approx 500$, when the flow becomes unsteady. 
The flow first approaches the periodic $pA_yS_z$ regime, 
characterised by an alternating shedding of hairpin vortices from the top/bottom LE shear layers. 
For $Re \ge 515$ the wake  oscillates and the flow is in the aperiodic $aA_yS_z$ regime. 

\subsubsection{Low $Re$: Steady $sS_yA_z$ and $sA_yS_z$ regimes}

The nonlinear simulations show that the primary bifurcation breaks the top/bottom planar symmetry leading to the steady $sS_yA_z$ regime, similarly to  shorter bodies \citep{MarquetLarsson2015,zampogna_boujo_2023}. This bifurcation occurs at $Re \approx 355$, in agreement with the critical Reynolds number $Re_{c}=353$ found in \S\ref{sec:stability}. The simulations show that for $Re \gtrapprox 365$, the steady $A_yS_z$ mode dominates, and the flow recovers the top/bottom planar symmetry but loses the left/right one, as predicted by the WNL analysis. 
We observe a small discrepancy between the linear and WNL stability analyses ($Re_{c} \approx 355$) and the fully nonlinear simulations 
($Re_{c} \approx 365$).

\begin{figure}
\centering
\includegraphics[trim={0 180 0 180},clip,width=0.49\textwidth]{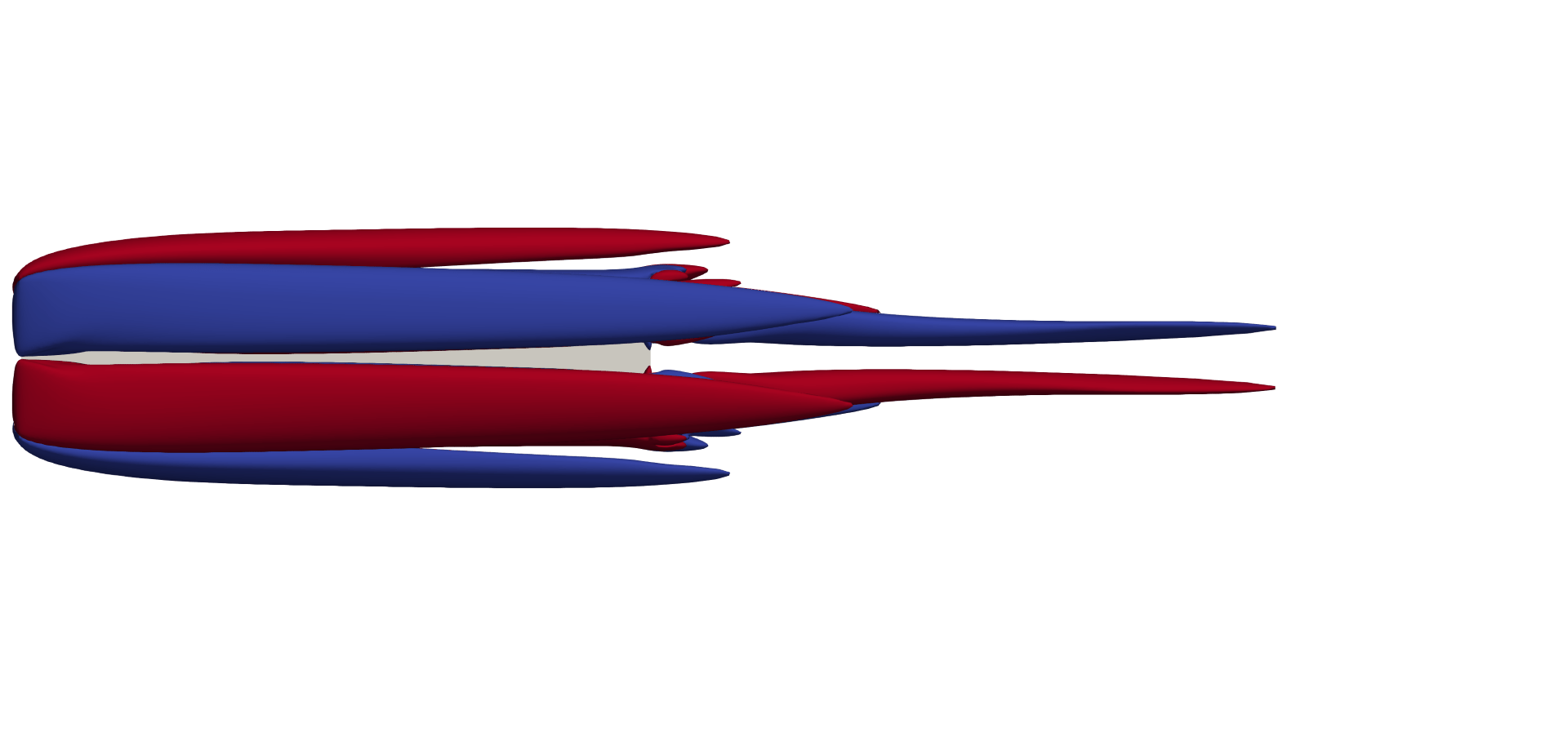}
\includegraphics[trim={0 180 0 180},clip,width=0.49\textwidth]{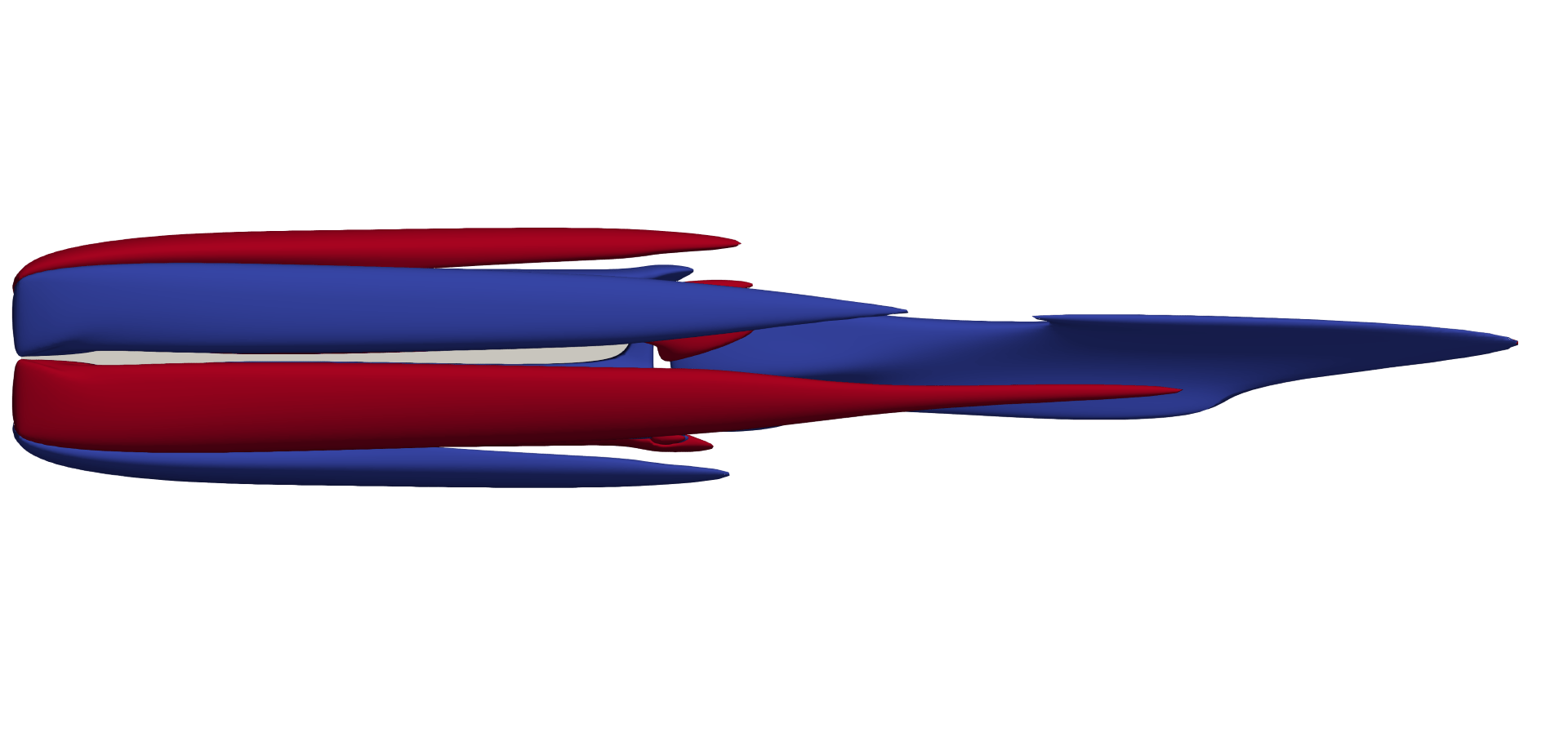}
\includegraphics[trim={0 180 0 180},clip,width=0.49\textwidth]{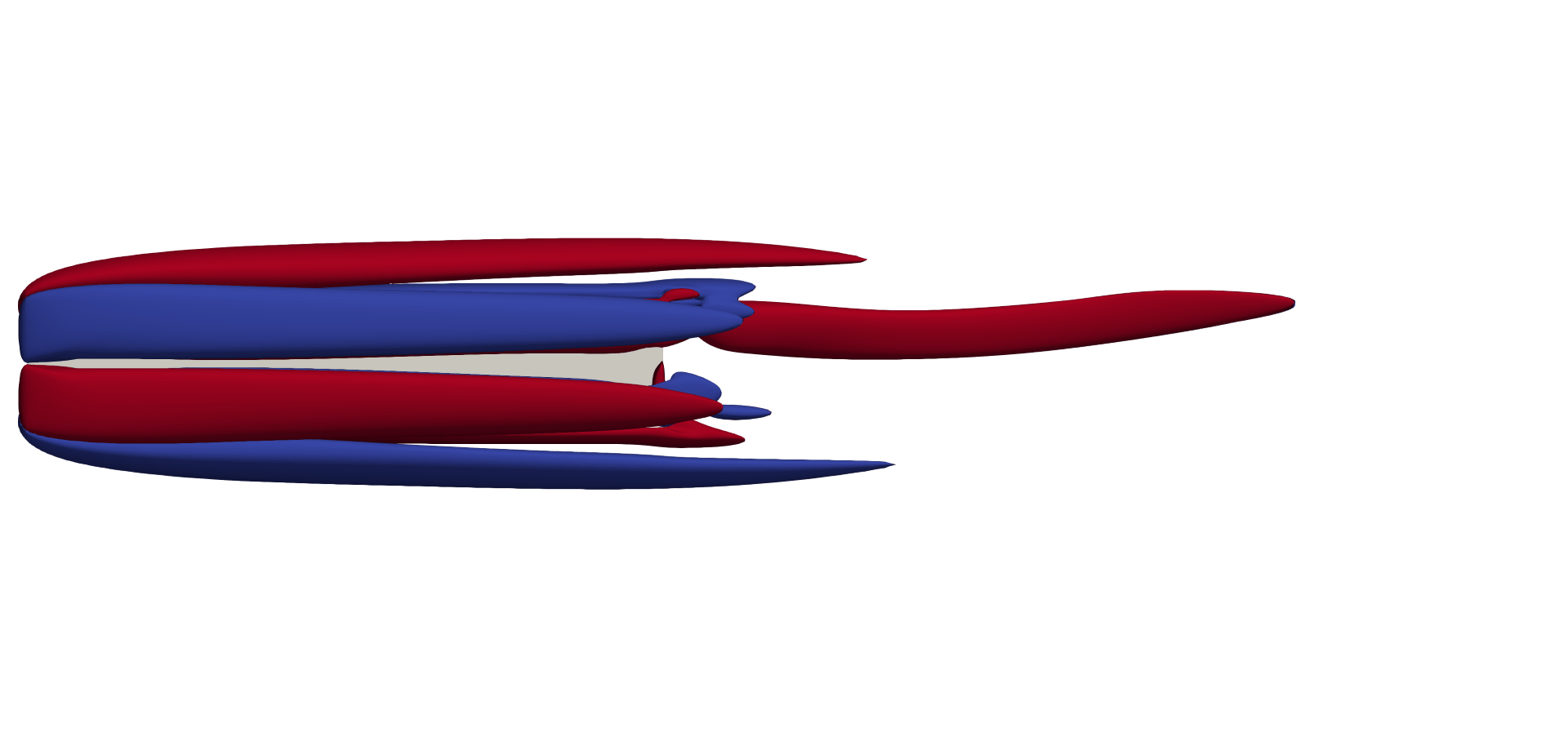}
\includegraphics[trim={0 180 0 180},clip,width=0.49\textwidth]{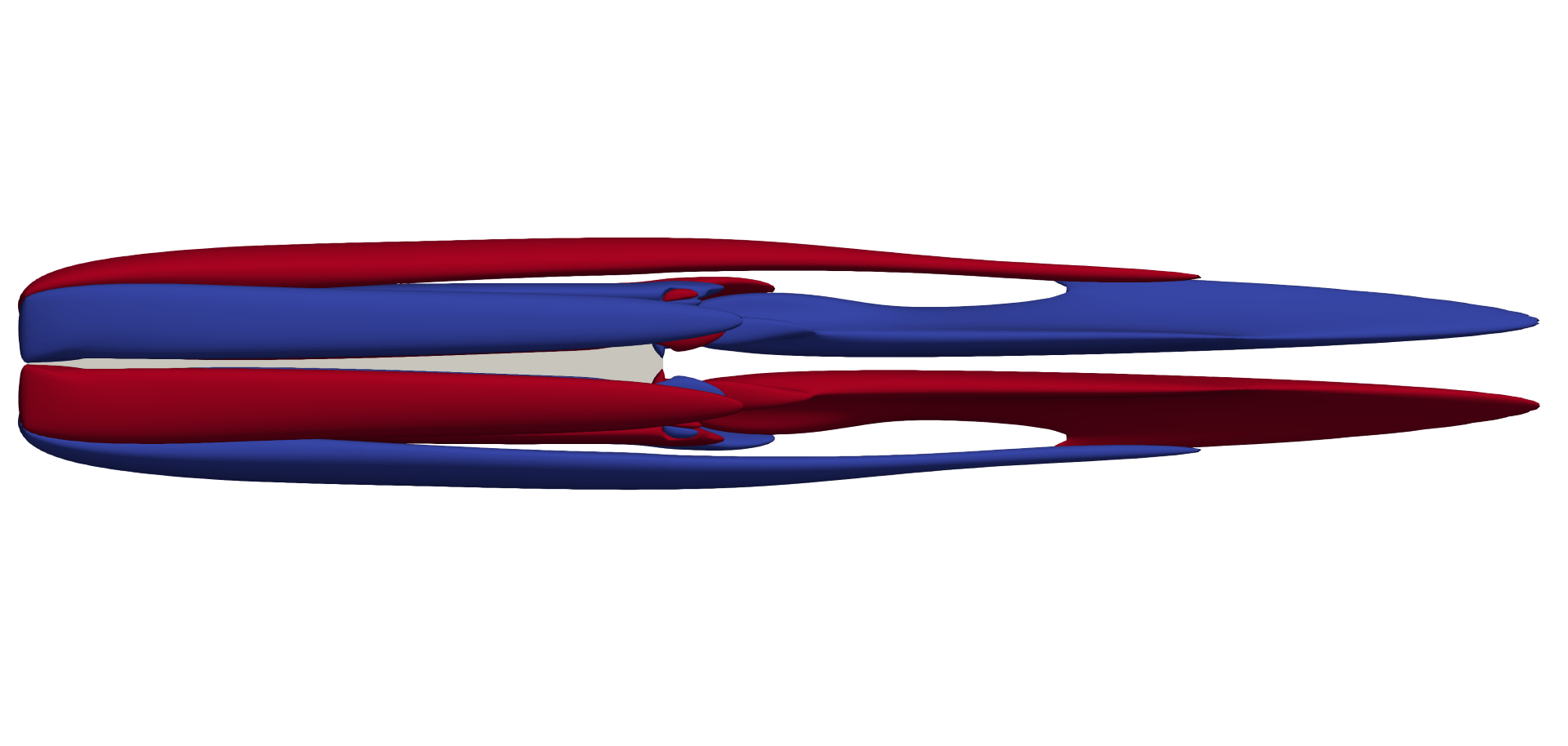}
\includegraphics[trim={350 0 350 0},clip,width=0.23\textwidth]{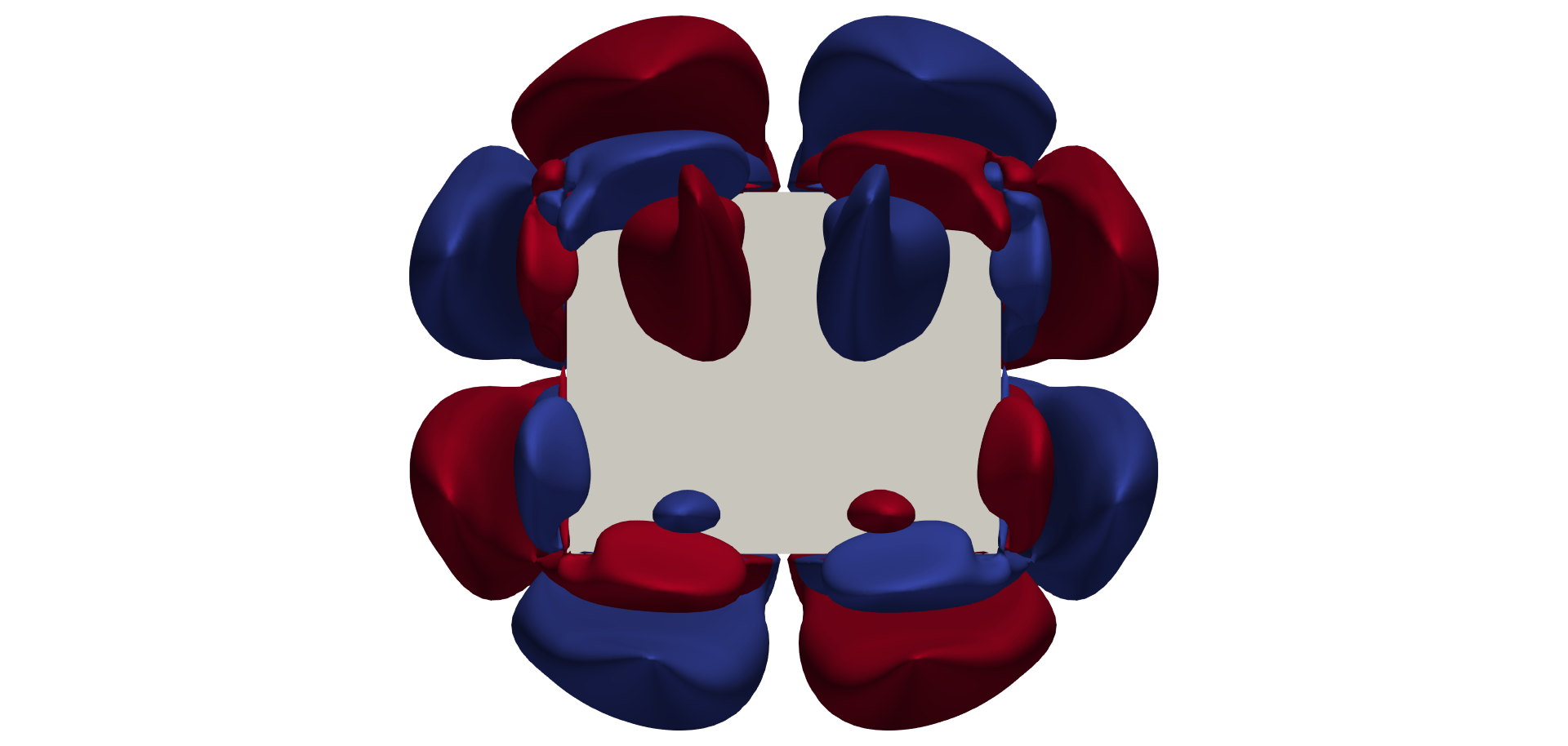} \hspace{0.2\textwidth}
\includegraphics[trim={350 0 350 0},clip,width=0.23\textwidth]{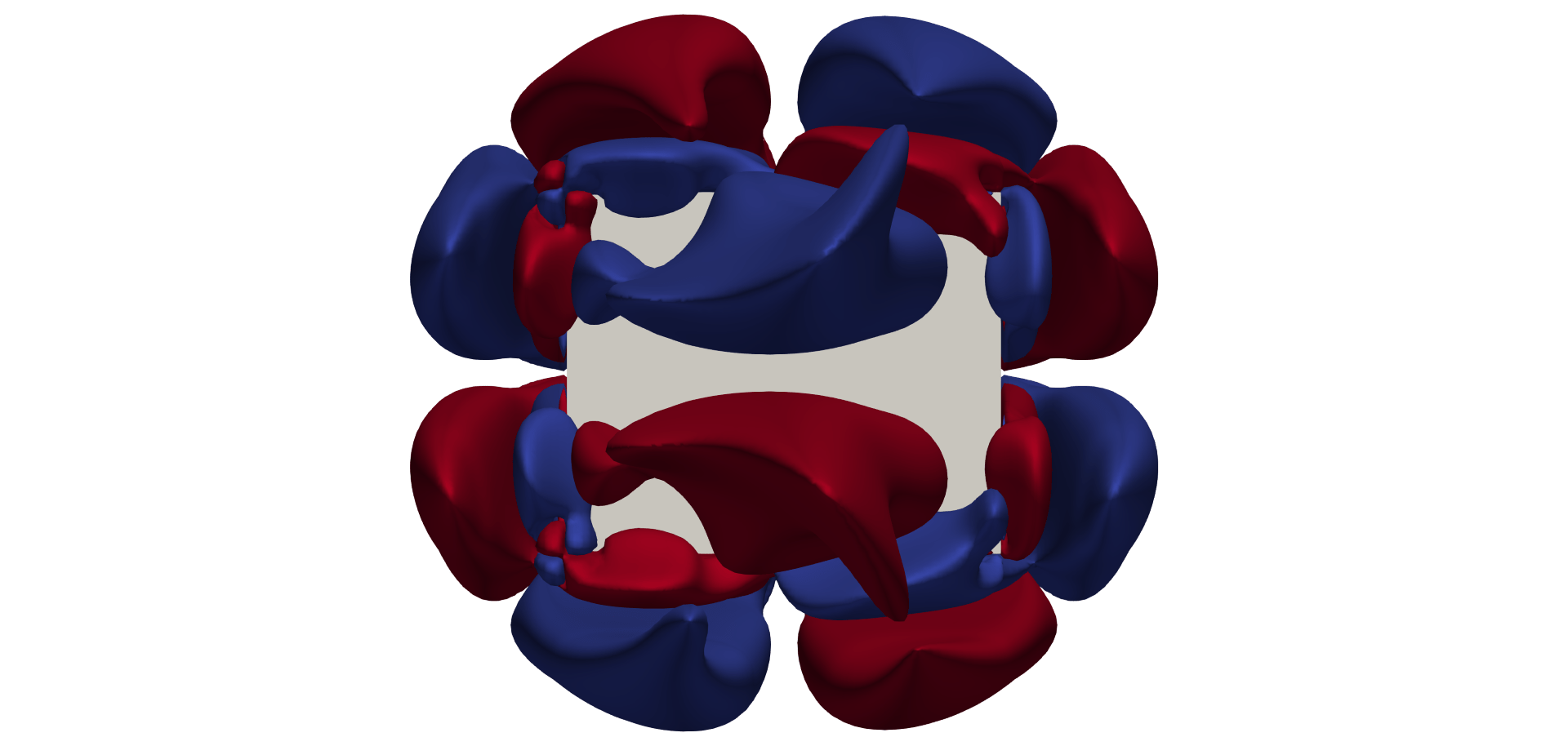}
\caption{
Steady regimes for $L=5$ and $W=1.2$.
Isosurfaces of streamwise vorticity, with red/blue  indicating $\omega_x=\pm 0.075$. 
Left: $sS_yA_z$ regime at $Re=355$. 
Right: $sA_yS_z$ regime at $Re=370$. 
Top: horizontal $x-y$ plane. Centre: vertical $x-z$ plane. Bottom $y-z$ plane.}
\label{fig:omegax_P1a_P1b}
\end{figure}

To characterise the $sA_yS_z$ and $sS_yA_z$ regimes, figure \ref{fig:omegax_P1a_P1b} shows  isosurfaces of positive (red) and negative (blue) values of the streamwise vorticity, and illustrates the loss of the planar symmetries. 
This is similar to the wakes past disks, spheres and bullet-shaped bodies at Reynolds numbers larger than $Re \approx 115, 210$ and $350$ \citep{johnson-patel-1999,tomboulides-orszag-2000,fabre-etal-2008,bohorquez-etal-2011}. 
In these steady regimes,  
the wake features a pair of counter-rotating  streamwise vortices that is absent at lower $Re$. 
These vortices exhibit an eccentricity that increases with $x$.
The eccentricity is in the $z$ direction in the $sS_yA_z$ regime, and in the $y$ direction in the $sA_yS_z$ regime. 

In both the $sS_yA_z$ and $sA_yS_z$ regimes we find that the flow asymmetry is confined in the wake, confirming that the $A_yS_z$ and $S_zA_y$ global modes are unstable modes of the wake. Indeed, the recirculating regions that arise along the sides of the prism retain the top/bottom and left/right 
planar symmetries in both regimes (not shown for brevity). These regions do not play a relevant role in the triggering mechanism of the primary instability, as confirmed by the results of the structural sensitivity (figure \ref{tab:first_two_modes_sensitivity}). 
A similar conclusion has been drawn for the primary, Hopf instability of the flow past elongated 2D rectangular cylinders \citep{chiarini-quadrio-auteri-2021a}.

\begin{figure}
\centering
\begin{tikzpicture}

\definecolor{clr1}{RGB}{18 78 128}
\definecolor{clr2}{RGB}{89 165 216}
\definecolor{clr3}{RGB}{145 229 246}
\definecolor{clr4}{RGB}{255 143 163}
\definecolor{clr5}{RGB}{255 77 109}
\definecolor{clr6}{RGB}{201 24 74}
\definecolor{clr7}{RGB}{128 15 47}
\definecolor{clr8}{RGB}{174 32 18}
\definecolor{clr9}{RGB}{155 34 38}
\definecolor{clr10}{RGB}{46 139 87}

\begin{axis}[%
width=0.36\textwidth,
height=0.15\textwidth,
scale only axis,
xmin=350,
xmax=500,
ymin=0.8,
ymax=0.92,
xtick={350,370,400,450,500,600,700},
ytick={0.75,0.8,0.85,0.9,0.95,1,1.05},
xlabel={$Re$},
ylabel={$F_x$},
ylabel style={at={(0.05,0.5)}},
axis background/.style={fill=white},
legend columns=3,transpose legend,
legend style={at={(0.99,0.82)}, anchor=east, legend cell align=left, align=left, fill=none, draw=none}
]

\addplot [color=black,solid,draw=none,mark=*,mark options={scale=1.4,black,fill=clr1}]
  table[row sep=crcr]{%
  300.0000    0.9778 \\
  350.0000    0.9136 \\
  352.0000    0.9115 \\
};
\addplot[red, solid, domain=350:500] {8.47*x^(-0.3799)};
\addplot [color=black,solid,draw=none,mark=*,mark options={scale=1.4,black,fill=clr2}]
  table[row sep=crcr]{%
  355.0000    0.9085 \\
  357.0000    0.9066 \\
  360.0000    0.9037 \\
  365.0000    0.8990 \\
};
\addplot [color=black,solid,draw=none,mark=*,mark options={scale=1.4,black,fill=clr3}]
  table[row sep=crcr]{%
  370.0000    0.8948 \\
  390.0000    0.8780 \\
  450.0000    0.8341 \\
  500.0000    0.8010 \\
};

\addplot [color=black,solid,draw=none,mark=*,mark options={scale=1.4,black,fill=clr10}]
  table[row sep=crcr]{%
  510.0000    0.79500 \\
};

\addplot [color=black,solid,draw=none,mark=*,mark options={scale=1.4,black,fill=clr6}]
  table[row sep=crcr]{%
  515.0000    0.7931 \\
  520.0000    0.7907 \\
  535.0000    0.7831 \\
  550.0000    0.7759 \\
  575.0000    0.7683 \\
  600.0000    0.8323 \\
  625.0000    0.8187 \\ 
  700.0000    0.8185 \\
};

\addplot [color=black,solid,mark=none, line width=0.5, mark options={scale=1.4,black,fill=red!80!black}]
  table[row sep=crcr]{%
  510.0000    0.7949 \\
  510.0000    0.7952 \\
};

\addplot [color=black,solid,mark=none, line width=0.5, mark options={scale=1.4,black,fill=red!80!black}]
  table[row sep=crcr]{%
  515.0000    0.7926 \\
  515.0000    0.7936 \\
};

\addplot [color=black,solid,mark=none, line width=0.5, mark options={scale=1.4,black,fill=red!80!black}]
  table[row sep=crcr]{%
  520.0000    0.7900 \\
  520.0000    0.7914 \\
};

\addplot [color=black,solid,mark=none, line width=0.5, mark options={scale=1.4,black,fill=red!80!black}]
  table[row sep=crcr]{%
  535.0000    0.7822 \\
  535.0000    0.7840 \\
};

\addplot [color=black,solid,mark=none, line width=0.5, mark options={scale=1.4,black,fill=red!80!black}]
  table[row sep=crcr]{%
  550.0000    0.7740 \\
  550.0000    0.7778 \\
};

\addplot [color=black,solid,mark=none, line width=0.5, mark options={scale=1.4,black,fill=red!80!black}]
  table[row sep=crcr]{%
  575.0000    0.7629 \\
  575.0000    0.7738 \\
};

\addplot [color=black,solid,mark=none, line width=0.5, mark options={scale=1.4,black,fill=red!80!black}]
  table[row sep=crcr]{%
  600.0000    0.8210 \\
  600.0000    0.8436 \\
};

\addplot [color=black,solid,mark=none, line width=0.5,mark options={scale=1.4,black,fill=red!80!black}]
  table[row sep=crcr]{%
  625.0000    0.8042 \\
  625.0000    0.8332 \\
};

\addplot [color=black,solid,mark=none, line width=0.5,mark options={scale=1.4,black,fill=red!80!black}]
  table[row sep=crcr]{%
  700.0000    0.8033 \\
  700.0000    0.8337 \\
};

\addplot [color=black,dashed,mark=none, line width=0.5,mark options={scale=1.4,black,fill=red!80!black}]
  table[row sep=crcr]{%
  353    0.75 \\
  353    1.05 \\
};

\addplot [color=black,dashed,mark=none, line width=0.5,mark options={scale=1.4,black,fill=red!80!black}]
  table[row sep=crcr]{%
  367    0.75 \\
  367    1.05 \\
};

\addplot [color=black,dashed,mark=none, line width=0.5,mark options={scale=1.4,black,fill=red!80!black}]
  table[row sep=crcr]{%
  507    0.75 \\
  507    1.05 \\
};

\addplot [color=black,dashed,mark=none, line width=0.5,mark options={scale=1.4,black,fill=red!80!black}]
  table[row sep=crcr]{%
  512    0.75 \\
  512    1.05 \\
};
\end{axis}
\end{tikzpicture}%
\begin{tikzpicture}

\definecolor{clr1}{RGB}{18 78 128}
\definecolor{clr2}{RGB}{89 165 216}
\definecolor{clr3}{RGB}{145 229 246}
\definecolor{clr4}{RGB}{255 143 163}
\definecolor{clr5}{RGB}{255 77 109}
\definecolor{clr6}{RGB}{201 24 74}
\definecolor{clr7}{RGB}{128 15 47}
\definecolor{clr8}{RGB}{174 32 18}
\definecolor{clr9}{RGB}{155 34 38}
\definecolor{clr10}{RGB}{46 139 87}

\begin{axis}[%
width=0.36\textwidth,
height=0.15\textwidth,
scale only axis,
xmin=350,
xmax=500,
ymin=-0.001,
ymax=0.021,
xtick={350,370,400,450,500,600,700},
ytick={0,0.01,0.02,0.04,0.06},
xlabel={$Re$},
ylabel={$F_y,F_z$},
ylabel style={at={(0.05,0.5)}},
axis background/.style={fill=white},
legend columns=3,transpose legend,
legend style={at={(0.99,0.5)}, anchor=east, legend cell align=left, align=left, fill=none, draw=none}
]

\addplot [color=black,solid,draw=none,mark=square,mark options={scale=1.3,black,fill=}]
  table[row sep=crcr]{%
  390.0000         0 \\
  450.0000         0 \\
  500.0000         0 \\
  550.0000         0 \\
  625.0000    0.0044 \\
  700.0000         0 \\
};
\addlegendentry{$F_z$};

\addplot [color=black,solid,draw=none,mark=diamond,mark options={scale=1.8,black,fill=red!80!black}]
  table[row sep=crcr]{%
  390.0000    0.0185 \\
  450.0000    0.0201 \\
  500.0000    0.0192 \\
  535.0000    0.0201 \\
  550.0000    0.0200 \\
  625.0000    0.0211 \\
  700.0000    0.0201 \\
};
\addlegendentry{$F_y$};

\addplot [color=black,solid,draw=none,mark=square*,mark options={scale=1.3,black,fill=clr1}]
  table[row sep=crcr]{%
  300.0000         0 \\
  350.0000         0 \\
  352.0000         0 \\
};

\addplot [color=black,solid,draw=none,mark=square*,mark options={scale=1.3,black,fill=clr2}]
  table[row sep=crcr]{
  355.0000    0.00691 \\
  357.0000    0.00848 \\
  360.0000    0.01023 \\
  365.0000    0.01232 \\
};

\addplot [color=black,solid,draw=none,mark=square*,mark options={scale=1.3,black,fill=clr3}]
  table[row sep=crcr]{%
  370.0000         0 \\
  390.0000         0 \\
  450.0000         0 \\
  500.0000         0 \\
};
\addplot [color=black,solid,draw=none,mark=square*,mark options={scale=1.3,black,fill=clr10}]
  table[row sep=crcr]{%
  510.0000         0 \\
};
\addplot [color=black,solid,draw=none,mark=square*,mark options={scale=1.3,black,fill=clr6}]
  table[row sep=crcr]{%
  515.0000         0 \\
  520.0000         0 \\
  535.0000         0 \\
  550.0000         0 \\
  575.0000         0 \\
  600.0000         0 \\
  625.0000    0.0044 \\
  700.0000         0 \\
};
\addplot [color=black,solid,draw=none,mark=diamond*,mark options={scale=1.8,black,fill=clr1}]
  table[row sep=crcr]{%
  300.0000         0 \\
  350.0000         0 \\
};

\addplot [color=black,solid,draw=none,mark=diamond*,mark options={scale=1.8,black,fill=clr2}]
  table[row sep=crcr]{%
  352.0000         0 \\
  355.0000         0 \\
  357.0000         0 \\
  360.0000         0 \\
  365.0000         0 \\
};

\addplot [color=black,solid,draw=none,mark=diamond*,mark options={scale=1.8,black,fill=clr3}]
  table[row sep=crcr]{%
  370.0000    0.0149 \\
  390.0000    0.0185 \\
  450.0000    0.0201 \\
  500.0000    0.0192 \\
};
\addplot [color=black,solid,draw=none,mark=diamond*,mark options={scale=1.8,black,fill=clr10}]
  table[row sep=crcr]{
  510.0000    0.0191 \\
};
\addplot [color=black,solid,draw=none,mark=diamond*,mark options={scale=1.8,black,fill=clr6}]
  table[row sep=crcr]{
  515.0000    0.0195 \\
  520.0000    0.0198 \\
  535.0000    0.0201 \\  
  550.0000    0.0200 \\
  575.0000    0.0200 \\
  600.0000    0.0267 \\
  625.0000    0.0211 \\
  700.0000    0.0201 \\
};

\addplot [color=red,solid,mark=none, line width=0.5, mark options={scale=1.4,black,fill=red!80!black}]
  table[row sep=crcr]{%
  510.0000    0.0191 \\
  510.0000    0.0191 \\
};

\addplot [color=red,solid,mark=none, line width=0.5, mark options={scale=1.4,black,fill=red!80!black}]
  table[row sep=crcr]{%
  515.0000    0.0181 \\
  515.0000    0.0209 \\
};

\addplot [color=red,solid,mark=none, line width=0.5, mark options={scale=1.4,black,fill=red!80!black}]
  table[row sep=crcr]{%
  520.0000    0.0179 \\
  520.0000    0.0217 \\
};

\addplot [color=red,solid,mark=none, line width=0.5, mark options={scale=1.4,black,fill=red!80!black}]
  table[row sep=crcr]{%
  535.0000    0.0172 \\
  535.0000    0.0229 \\
};

\addplot [color=red,solid,mark=none, line width=0.5, mark options={scale=1.4,black,fill=red!80!black}]
  table[row sep=crcr]{%
  550.0000    0.0166 \\
  550.0000    0.0234 \\
};

\addplot [color=red,solid,mark=none, line width=0.5, mark options={scale=1.4,black,fill=red!80!black}]
  table[row sep=crcr]{%
  575.0000    0.0148 \\
  575.0000    0.0251 \\
};

\addplot [color=red,solid,mark=none, line width=0.5, mark options={scale=1.4,black,fill=red!80!black}]
  table[row sep=crcr]{%
  600.0000    0.0139 \\
  600.0000    0.0395 \\
};

\addplot [color=red,solid,mark=none, line width=0.5,mark options={scale=1.4,black,fill=red!80!black}]
  table[row sep=crcr]{%
  625.0000    0.0014 \\
  625.0000    0.0408 \\
};

\addplot [color=red,solid,mark=none, line width=0.5,mark options={scale=1.4,black,fill=red!80!black}]
  table[row sep=crcr]{%
  700.0000   -0.0103 \\
  700.0000    0.0505 \\
};

\addplot [color=black,dashed,mark=none, line width=0.5,mark options={scale=1.4,black,fill=red!80!black}]
  table[row sep=crcr]{%
  353    -0.04 \\
  353    0.06 \\
};

\addplot [color=black,dashed,mark=none, line width=0.5,mark options={scale=1.4,black,fill=red!80!black}]
  table[row sep=crcr]{%
  367    -0.04 \\
  367     0.06 \\
};

\addplot [color=black,dashed,mark=none, line width=0.5,mark options={scale=1.4,black,fill=red!80!black}]
  table[row sep=crcr]{%
  507    -0.04 \\
  507     0.06 \\
};

\addplot [color=black,dashed,mark=none, line width=0.5,mark options={scale=1.4,black,fill=red!80!black}]
  table[row sep=crcr]{%
  512    -0.04 \\
  512     0.06 \\
};

\addplot [color=black,solid,mark=none, line width=0.5,mark options={scale=1.4,black,fill=red!80!black}]
  table[row sep=crcr]{%
  535    -0.0143 \\
  535     0.0143 \\
};

\addplot [color=black,solid,mark=none, line width=0.5, mark options={scale=1.4,black,fill=red!80!black}]
  table[row sep=crcr]{%
  510.0000   -0.0056 \\
  510.0000    0.0056 \\
};

\addplot [color=black,solid,mark=none, line width=0.5, mark options={scale=1.4,black,fill=red!80!black}]
  table[row sep=crcr]{%
  515.0000   -0.0104 \\
  515.0000    0.0104 \\
};

\addplot [color=black,solid,mark=none, line width=0.5, mark options={scale=1.4,black,fill=red!80!black}]
  table[row sep=crcr]{%
  520.0000   -0.0123 \\
  520.0000    0.0123 \\
};

\addplot [color=black,solid,mark=none, line width=0.5, mark options={scale=1.4,black,fill=red!80!black}]
  table[row sep=crcr]{%
  550.0000   -0.0147 \\
  550.0000    0.0147 \\
};

\addplot [color=black,solid,mark=none, line width=0.5, mark options={scale=1.4,black,fill=red!80!black}]
  table[row sep=crcr]{%
  575.0000   -0.0116 \\
  575.0000    0.0116 \\
};

\addplot [color=black,solid,mark=none, line width=0.5, mark options={scale=1.4,black,fill=red!80!black}]
  table[row sep=crcr]{%
  600.0000   -0.0158 \\
  600.0000    0.0158 \\
};

\addplot [color=black,solid,mark=none, line width=0.5,mark options={scale=1.4,black,fill=red!80!black}]
  table[row sep=crcr]{%
  625.0000   -0.0137 \\
  625.0000    0.0225 \\
};

\addplot [color=black,solid,mark=none, line width=0.5,mark options={scale=1.4,black,fill=red!80!black}]
  table[row sep=crcr]{%
  700.0000   -0.0241 \\
  700.0000    0.0241 \\
};
\end{axis}

\end{tikzpicture}%
\begin{tikzpicture}

\definecolor{clr1}{RGB}{18 78 128}
\definecolor{clr2}{RGB}{89 165 216}
\definecolor{clr3}{RGB}{145 229 246}
\definecolor{clr4}{RGB}{255 143 163}
\definecolor{clr5}{RGB}{255 77 109}
\definecolor{clr6}{RGB}{201 24 74}
\definecolor{clr7}{RGB}{128 15 47}
\definecolor{clr8}{RGB}{174 32 18}
\definecolor{clr9}{RGB}{155 34 38}

\begin{axis}[%
width=0.36\textwidth,
height=0.15\textwidth,
scale only axis,
xmin=350,
xmax=500,
ymin=0.05,
ymax=0.73,
xtick={300,350,400,450,500},
ytick={0.2, 0.4, 0.6},
xlabel={$Re$},
ylabel={$F_{x,f}, F_{x,p}$},
ylabel style={at={(0.05,0.5)}},
axis background/.style={fill=white},
legend columns=3,transpose legend,
legend style={at={(0.99,0.5)}, anchor=east, legend cell align=left, align=left, fill=none, draw=none}
]

\addplot [color=black,solid,draw=none,mark=diamond,mark options={scale=1.9,black,fill=}]
  table[row sep=crcr]{%
  300.0000    0.2571 \\
  350.0000    0.1985 \\
};
\addlegendentry{$F_{x,f}$};

\addplot [color=black,solid,draw=none,mark=square,mark options={scale=1.4,black,fill=}]
  table[row sep=crcr]{%
  300.0000    0.7236 \\
};
\addlegendentry{$F_{x,p}$};

\addplot [color=black,solid,draw=none,mark=diamond*,mark options={scale=1.9,black,fill=clr1}]
  table[row sep=crcr]{%
  300.0000    0.2571 \\
  350.0000    0.1985 \\
  352.0000    0.1965 \\
};

\addplot [color=black,solid,draw=none,mark=diamond*,mark options={scale=1.9,black,fill=clr2}]
  table[row sep=crcr]{%
  355.0000    0.1934 \\
  357.0000    0.1914 \\
  360.0000    0.1884 \\
  365.0000    0.1835 \\
};
\addplot [color=black,solid,draw=none,mark=diamond*,mark options={scale=1.9,black,fill=clr3}]
  table[row sep=crcr]{%
  370.0000    0.1786 \\
  390.0000    0.1605 \\
  450.0000    0.1123 \\
  500.0000    0.0779 \\
};

\addplot [color=black,solid,draw=none,mark=square*,mark options={scale=1.4,black,fill=clr1}]
  table[row sep=crcr]{%
  300.0000    0.7236 \\
  350.0000    0.7151 \\
  352.0000    0.7151 \\  
};
\addplot [color=black,solid,draw=none,mark=square*,mark options={scale=1.4,black,fill=clr2}]
  table[row sep=crcr]{%
  355.0000    0.7152 \\
  357.0000    0.7153 \\
  360.0000    0.7154 \\
  365.0000    0.7156 \\
};
\addplot [color=black,solid,draw=none,mark=square*,mark options={scale=1.4,black,fill=clr3}]
  table[row sep=crcr]{%
  370.0000    0.7162 \\
  390.0000    0.7175 \\
  450.0000    0.7158 \\
  500.0000    0.7207 \\
};

\addplot [color=black,dashed,mark=none, line width=0.5,mark options={scale=1.4,black,fill=red!80!black}]
  table[row sep=crcr]{%
  353    0 \\
  353    1 \\
};

\addplot [color=black,dashed,mark=none, line width=0.5,mark options={scale=1.4,black,fill=red!80!black}]
  table[row sep=crcr]{%
  367    0 \\
  367    1 \\
};

\end{axis}

\end{tikzpicture}%
\begin{tikzpicture}

\definecolor{clr1}{RGB}{18 78 128}
\definecolor{clr2}{RGB}{89 165 216}
\definecolor{clr3}{RGB}{145 229 246}
\definecolor{clr4}{RGB}{255 143 163}
\definecolor{clr5}{RGB}{255 77 109}
\definecolor{clr6}{RGB}{201 24 74}
\definecolor{clr7}{RGB}{128 15 47}
\definecolor{clr8}{RGB}{174 32 18}
\definecolor{clr9}{RGB}{155 34 38}

\begin{axis}[%
width=0.36\textwidth,
height=0.15\textwidth,
scale only axis,
xmin=350,
xmax=500,
ymin=-0.001,
ymax=0.0158,
xtick={300,350,370,400,450,500},
ytick={0,0.005,0.01,0.015},
xlabel={$Re$},
ylabel={$F_{\perp,f}, F_{\perp,p}$},
ylabel style={at={(0.05,0.5)}},
axis background/.style={fill=white},
legend columns=3,transpose legend,
legend style={at={(0.99,0.5)}, anchor=east, legend cell align=left, align=left, fill=none, draw=none}
]

\addplot [color=black,solid,draw=none,mark=diamond,mark options={scale=1.9,black,fill=}]
  table[row sep=crcr]{%
  300.0000    0 \\
  350.0000    0 \\
};
\addlegendentry{$F_{\perp,f}$};

\addplot [color=black,solid,draw=none,mark=square,mark options={scale=1.4,black,fill=}]
  table[row sep=crcr]{%
  300.0000    0 \\
};
\addlegendentry{$F_{\perp,p}$};

\addplot [color=black,solid,draw=none,mark=diamond*,mark options={scale=1.9,black,fill=clr1}]
  table[row sep=crcr]{%
  300.0000    0.0 \\
  350.0000    0.0 \\
  352.0000    0.0 \\
};

\addplot [color=black,solid,draw=none,mark=diamond*,mark options={scale=1.9,black,fill=clr2}]
  table[row sep=crcr]{%
  355.0000    0.0013 \\
  357.0000    0.0015 \\
  360.0000    0.0019 \\
  365.0000    0.0022 \\
};
\addplot [color=black,solid,draw=none,mark=diamond*,mark options={scale=1.9,black,fill=clr3}]
  table[row sep=crcr]{%
  370.0000    0.0032 \\
  390.0000    0.0040 \\
  450.0000    0.0043 \\
  500.0000    0.0040 \\
};

\addplot [color=black,solid,draw=none,mark=square*,mark options={scale=1.4,black,fill=clr1}]
  table[row sep=crcr]{%
  300.0000    0.0 \\
  350.0000    0.0 \\
  352.0000    0.0 \\  
};
\addplot [color=black,solid,draw=none,mark=square*,mark options={scale=1.4,black,fill=clr2}]
  table[row sep=crcr]{%
  355.0000    0.0057 \\
  357.0000    0.0069 \\
  360.0000    0.0084 \\
  365.0000    0.0101 \\
};
\addplot [color=black,solid,draw=none,mark=square*,mark options={scale=1.4,black,fill=clr3}]
  table[row sep=crcr]{%
  370.0000    0.0118 \\
  390.0000    0.0145 \\
  450.0000    0.0158 \\
  500.0000    0.0152 \\
};

\addplot [color=black,dashed,mark=none, line width=0.5,mark options={scale=1.4,black,fill=red!80!black}]
  table[row sep=crcr]{%
  353    -1 \\
  353    1 \\
};

\addplot [color=black,dashed,mark=none, line width=0.5,mark options={scale=1.4,black,fill=red!80!black}]
  table[row sep=crcr]{%
  367    -1 \\
  367    1 \\
};

\end{axis}

\end{tikzpicture}%
\caption{Dependence of the aerodynamic forces on $Re$ in the $sS_yS_z$, $sS_yA_z$ and $sA_yS_z$ regimes for $L=5$ and $W=1.2$. 
Top panels: total forces. 
Bottom left: viscous and  pressure components of the drag force. 
Bottom right: viscous and pressure components of $F_\perp$, the single non-zero force perpendicular to the incoming flow ($F_z$ in the $sS_yA_z$ regime and $F_y$ in the $sA_yS_z$ regime). 
Red line in the top left panel: $8.47 \times Re^{-0.3799}$.}
\label{fig:W12_forces}
\end{figure}
Figure \ref{fig:W12_forces} shows that when 
the flow bifurcates 
to the $sS_yA_z$ and $sA_yS_z$ regimes, the drag decreases according to a power law,  $F_x \sim Re^{\alpha}$ with $\alpha=-0.3799$. 
This is similar to  the flow past a sphere \citep{johnson-patel-1999}, a short circular cylinder \citep{yang-etal-2022} and a cube \citep{saha-2004,meng_an_cheng_kimiaei_2021}; for the latter \cite{saha-2004} report  $\alpha=-0.372$, and \cite{meng_an_cheng_kimiaei_2021} found a value between $-0.4$ and $-0.3$. 
The decrease of $F_x$ is entirely due to a decrease of the friction contribution $F_{x,f}$ (see the bottom left panel of figure \ref{fig:W12_forces}), as the pressure contribution $F_{x,p}$ does not change with $Re$. 
\begin{figure}
\centering
\begin{tikzpicture}
\definecolor{clr1}{RGB}{18 78 128}
\definecolor{clr2}{RGB}{89 165 216}
\definecolor{clr3}{RGB}{145 229 246}
\definecolor{clr4}{RGB}{255 143 163}
\definecolor{clr5}{RGB}{255 77 109}
\definecolor{clr6}{RGB}{201 24 74}
\definecolor{clr7}{RGB}{128 15 47}
\definecolor{clr8}{RGB}{174 32 18}
\definecolor{clr9}{RGB}{155 34 38}
\definecolor{clr10}{RGB}{46 139 87}

\begin{axis}[%
width=0.7\textwidth,
height=0.15\textwidth,
scale only axis,
xmin=350,
xmax=601,
ymin=2,
ymax=3.7,
xtick={300,350,400,450,500,550,600},
ytick={1.5,2,2.5,3,3.5},
xlabel={$Re$},
ylabel={$x_r$},
ylabel style={at={(0.05,0.5)}},
axis background/.style={fill=white},
legend columns=3,transpose legend,
legend style={at={(0.99,0.25)}, anchor=east, legend cell align=left, align=left, fill=none, draw=none}
]

\addplot [color=black,solid,draw=none,mark=diamond,mark options={scale=1.9,black,fill=}]
  table[row sep=crcr]{%
  300.0000    2.02073 \\
};
\addlegendentry{top/bottom};

\addplot [color=black,solid,draw=none,mark=square,mark options={scale=1.4,black,fill=}]
  table[row sep=crcr]{%
  300.0000    1.75284 \\
};
\addlegendentry{right/left};

\addplot [color=black,solid,draw=none,mark=diamond*,mark options={scale=1.9,black,fill=clr1}]
  table[row sep=crcr]{%
  300.0000    2.02073 \\
  350.0000    2.46680 \\
  352.0000    2.48360 \\  
};
\addplot[red, solid, domain=350:601] {0.0023*x^(1.1651)};
\addplot[red, dashed, domain=350:601] {0.0095*x^(0.9485)};
\addplot [color=black,solid,draw=none,mark=diamond*,mark options={scale=1.9,black,fill=clr2}]
  table[row sep=crcr]{%
  355.0000    2.51440 \\
  357.0000    2.52380 \\
  360.0000    2.54930 \\
  365.0000    2.58170 \\
};
\addplot [color=black,solid,draw=none,mark=diamond*,mark options={scale=1.9,black,fill=clr3}]
  table[row sep=crcr]{%
  370.0000    2.62220 \\
  390.0000    2.73795 \\
  450.0000    3.08612 \\
  500.0000    3.36465 \\
};  
\addplot [color=black,solid,draw=none,mark=diamond*,mark options={scale=1.9,black,fill=clr10}]
  table[row sep=crcr]{%
  510.0000    3.48000 \\
};

\addplot [color=black,solid,draw=none,mark=diamond*,mark options={scale=1.9,black,fill=clr6}]
  table[row sep=crcr]{%
  520.0000    3.49000 \\
  535.0000    3.49000 \\
  575.0000    3.52000 \\
  590.0000    3.30000 \\
  600.000     3.21000 \\
};

\addplot [color=black,solid,draw=none,mark=square*,mark options={scale=1.4,black,fill=clr1}]
  table[row sep=crcr]{%
  300.0000    1.75284 \\
  350.0000    2.18040 \\
  352.0000    2.19110 \\
};
\addplot [color=black,solid,draw=none,mark=square*,mark options={scale=1.4,black,fill=clr2}]
  table[row sep=crcr]{%
  355.0000    2.20790 \\
  357.0000    2.22310 \\
  360.0000    2.25310 \\
  365.0000    2.28340 \\
};
\addplot [color=black,solid,draw=none,mark=square*,mark options={scale=1.4,black,fill=clr3}]
  table[row sep=crcr]{%
  370.0000    2.31650 \\
  390.0000    2.47184\\
  450.0000    2.8895 \\
  500.0000    3.22507 \\
};
\addplot [color=black,solid,draw=none,mark=square*,mark options={scale=1.4,black,fill=clr10}]
  table[row sep=crcr]{%
  510.0000    3.29000 \\
};
\addplot [color=black,solid,draw=none,mark=square*,mark options={scale=1.4,black,fill=clr6}]
  table[row sep=crcr]{%
  520.0000    3.31000 \\
  535.0000    3.36000 \\
  575.0000    3.50000 \\
  590.0000    3.18000 \\
  600.0000    3.09000 \\
};

\end{axis}

\end{tikzpicture}%
\caption{Position of the reattachment point $x_r$ on the top/bottom and right/left sides (i.e. size of the recirculating regions on the prism walls) as a function of $Re$, for $L=5$ and $W=1.2$.
Red solid line: $0.0023 \times Re^{1.1651}$.
Red dashed line: $0.0095 \times Re^{0.9485}$. 
In the unsteady regimes ($Re>500$), $x_r$ is the reattachment point of the time averaged flow. 
}
\label{fig:xr_position} 
\end{figure}
An increase of $Re$, indeed, results into a downstream shift of the flow reattachment point $x_r$ over the sides of the prism, leading to a longer recirculating region and to a decrease of $F_{x,f}$ (figure \ref{fig:xr_position}). 
We note that a 
power function fits well the relation between $x_r$ and $Re$, being $x_r \sim Re^{1.1651}$ for the left and right sides and $x_r \sim Re^{0.9485}$ for the top and bottom sides. In contrast, the non-zero cross-stream component of the aerodynamic forces increases with $Re$.
Unlike for $F_x$, this is due to an increase of both the friction and pressure contributions (see the bottom right panel of figure \ref{fig:W12_forces}), and is associated with the increasing flow asymmetry. 

The regular bifurcation towards an asymmetric steady flow is found for prisms with small $W$ only, irrespective of $L$ (\S\ref{sec:stability}). 
This is the case of many other 3D bluff bodies: besides the largely studied axisymmetric sphere and disk, the existence of a regular bifurcation in non-axisymmetric flows has been reported by \cite{MarquetLarsson2015} for thin plates with $W/D<2$, by \cite{yang-etal-2022} for short circular cylinders with $W/D<1.75$, and by \cite{sheard-etal-2008} for short cylinders with hemispherical ends. 
The main difference is that the bifurcated wake retains a symmetry plane that is randomly oriented \citep[selected by 
perturbations;][]{tomboulides-orszag-2000} for axisymmetric bodies, but dictated by the planar symmetries of the geometry for non-axisymmetric bodies.

\subsubsection{Large $Re$: $pA_yS_z$ and $aA_yS_z$ regimes}
\label{sec:simW12-per-aper}

\begin{figure}
\centering
\includegraphics[width=0.49\textwidth]{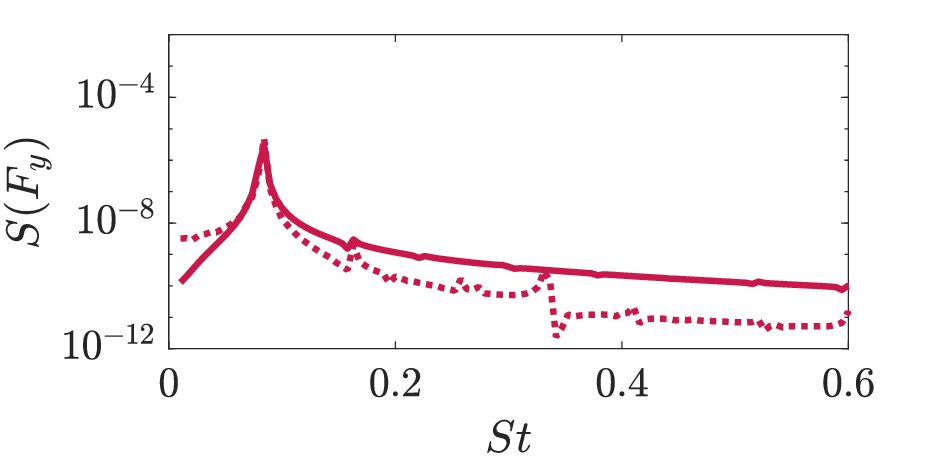}
\includegraphics[width=0.49\textwidth]{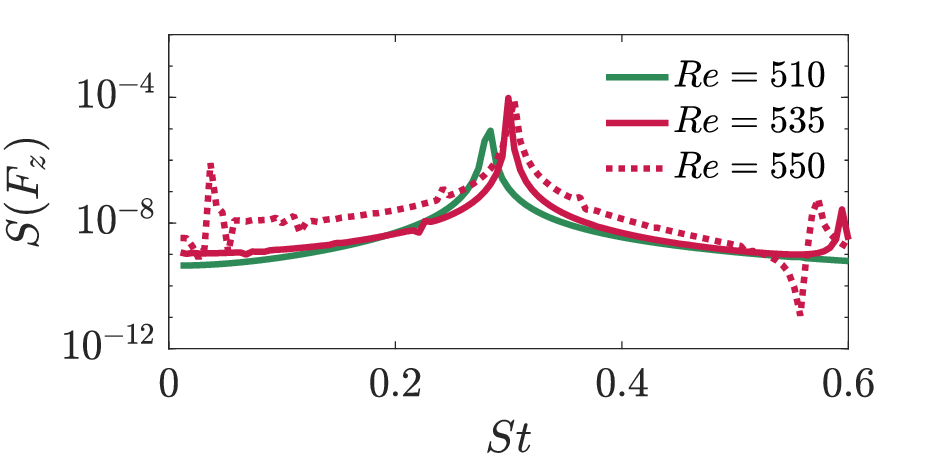}
%
\includegraphics[width=0.45\textwidth]{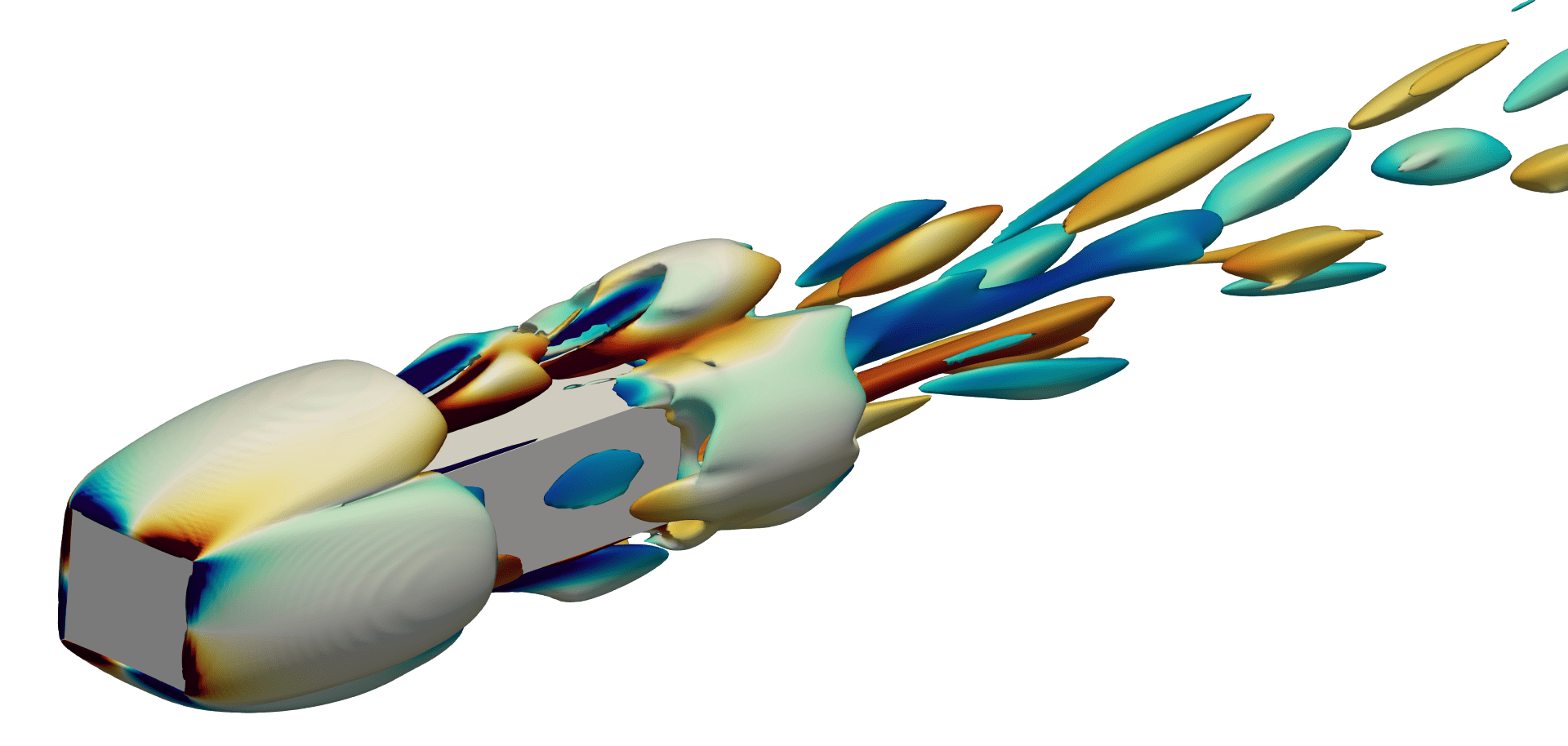}
\includegraphics[width=0.24\textwidth]{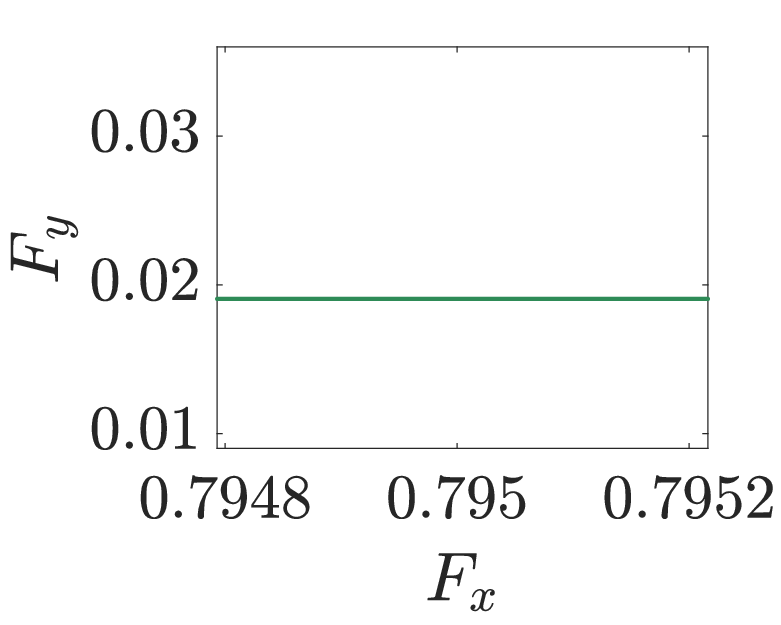}
\includegraphics[width=0.24\textwidth]{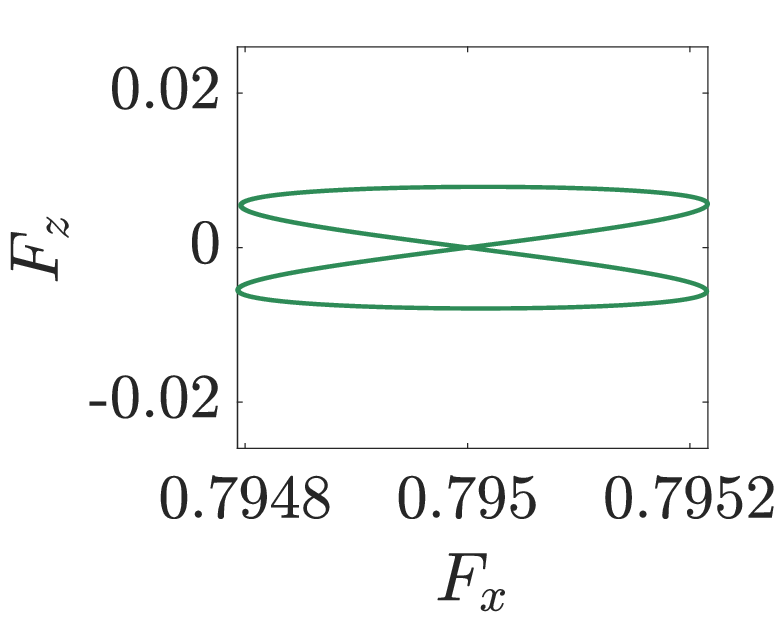}
\includegraphics[width=0.45\textwidth]{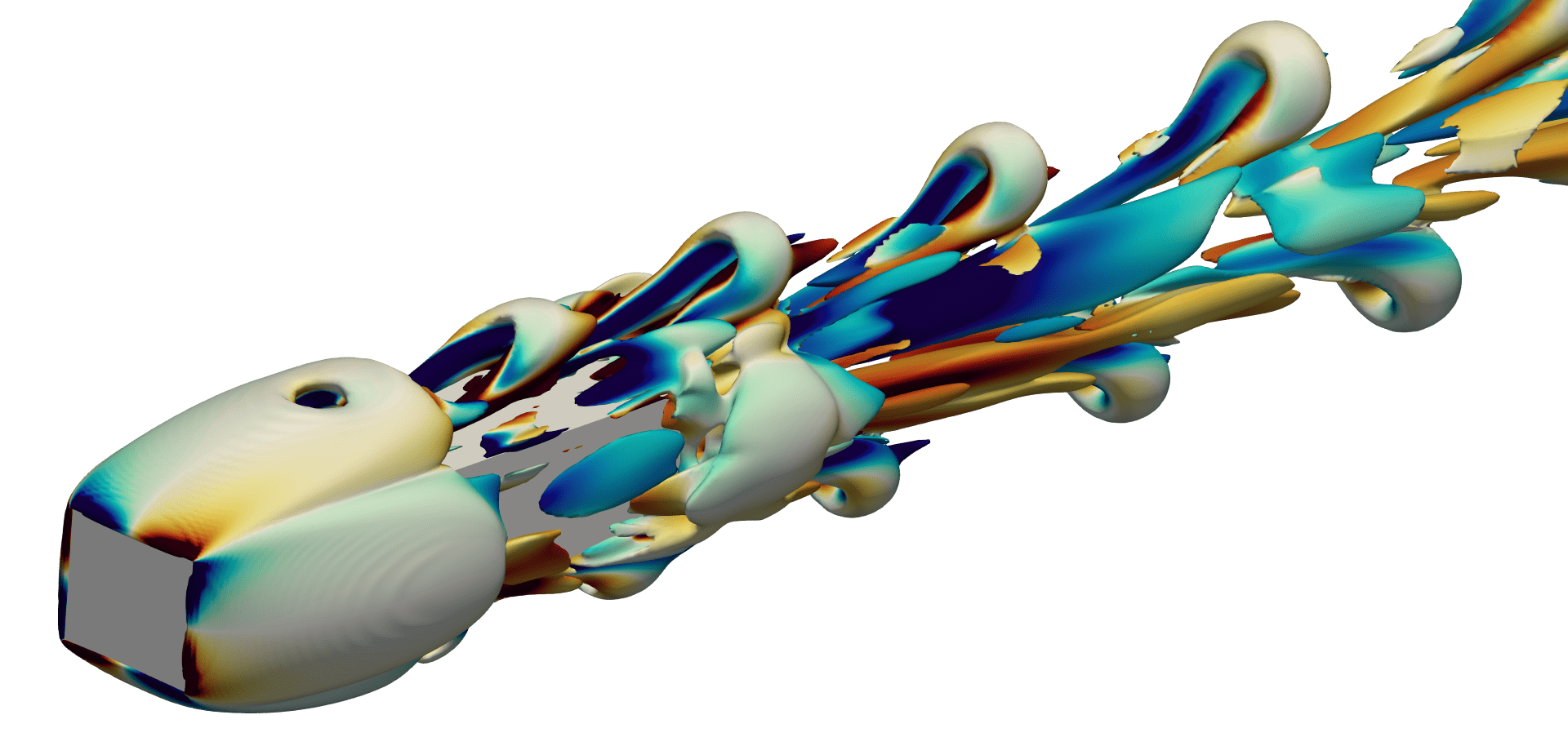}
\includegraphics[width=0.24\textwidth]{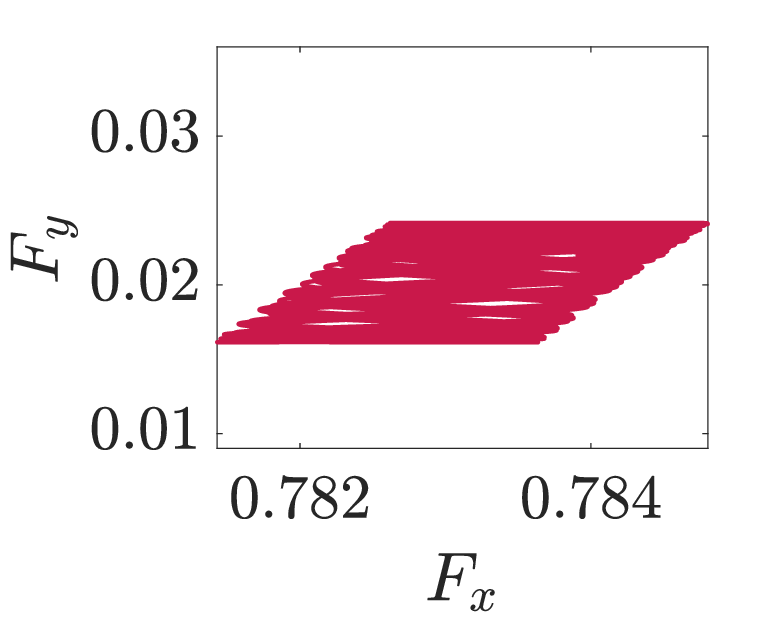}
\includegraphics[width=0.24\textwidth]{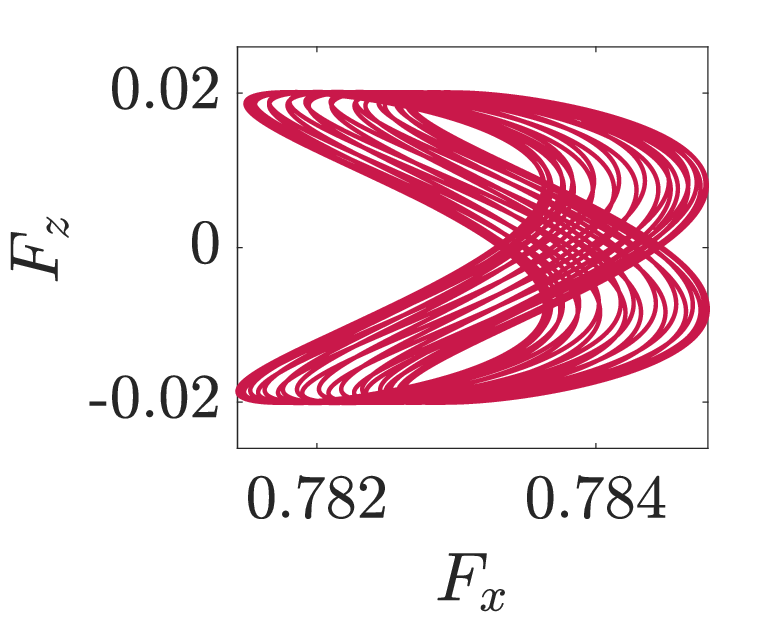}
\includegraphics[width=0.45\textwidth]{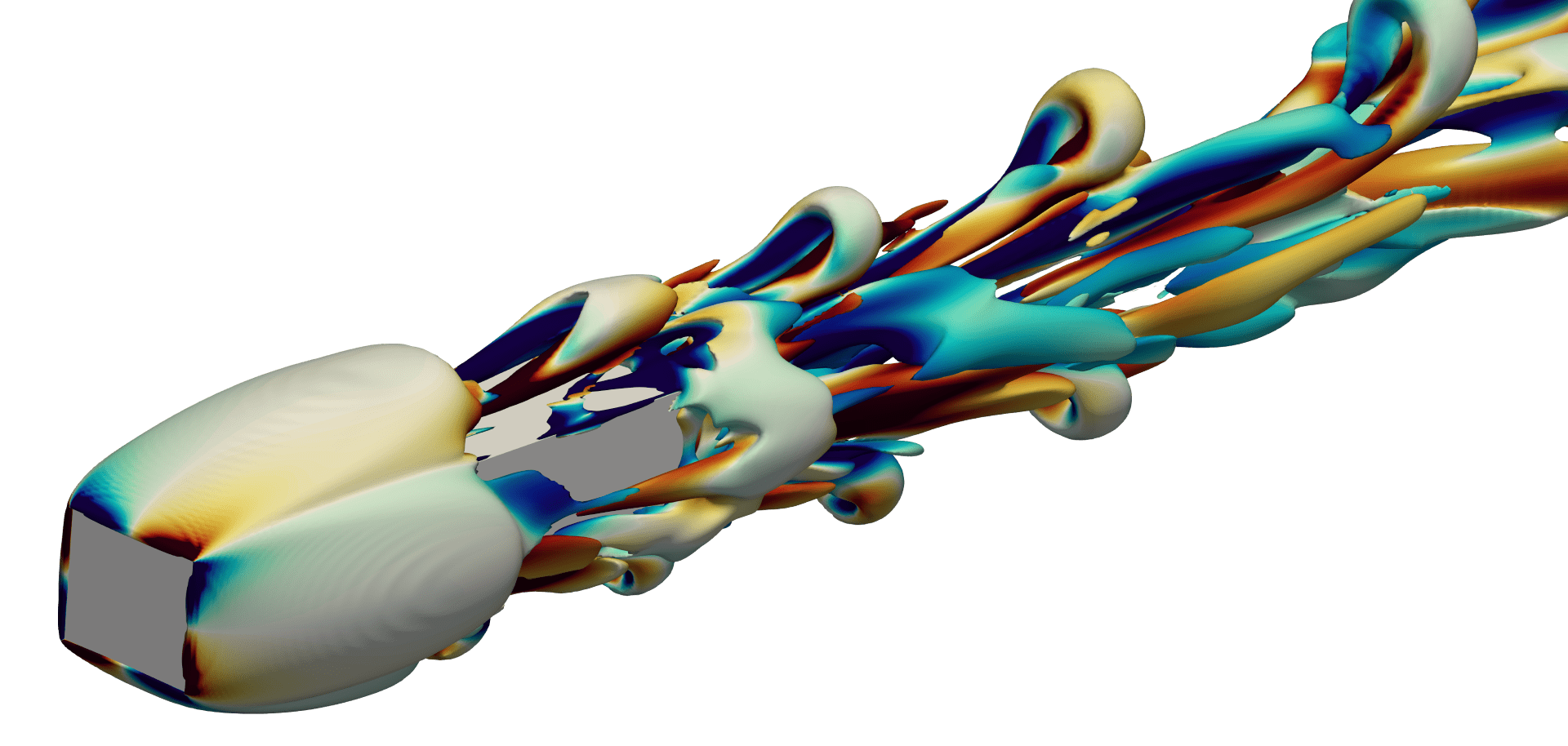}
\includegraphics[width=0.24\textwidth]{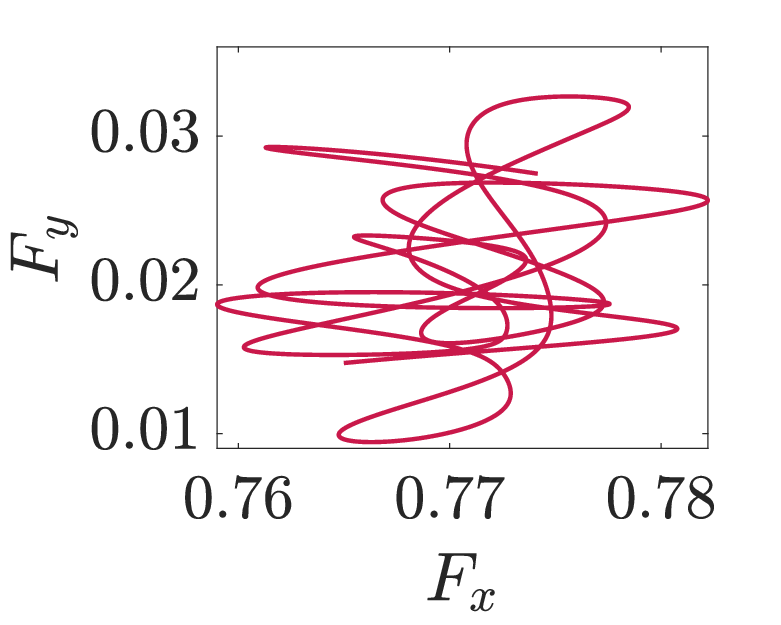}
\includegraphics[width=0.24\textwidth]{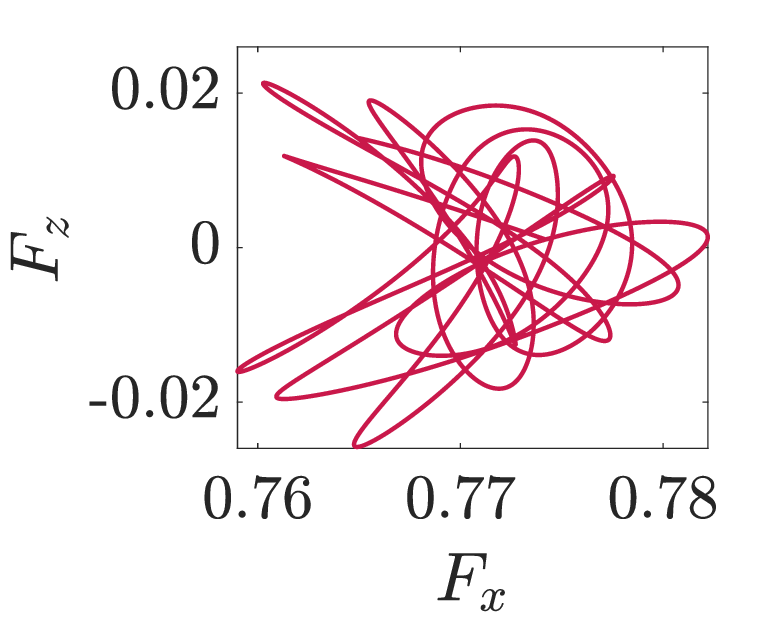}
\caption{
Unsteady regimes for $L=5$ and $W=1.2$.
Top panels: frequency spectra  of $F_y$ (left) and $F_z$ (right), $510 \le Re \le 550$. 
Bottom panels: structure of the flow for $Re=510$, $535$ and $575$ (top to bottom). 
Left: instantaneous isosurfaces $\lambda_2=-0.05$ coloured by streamwise vorticity in the range $-1 \le \omega_x \le 1$. 
Right: force diagrams.}
\label{fig:W12_lambda2_1}
\end{figure}
A further bifurcation occurs at $Re \approx 510$. 
It consists of a Hopf bifurcation of the $sA_yS_z$ deflected wake;  
the flow becomes unsteady and starts oscillating about the deflected $sA_yS_z$ regime. 
A similar regime, with the flow oscillating about an asymmetric steady state has been observed for the flow past a sphere \citep{ johnson-patel-1999}, a cube \citep{saha-2004}, a disk \citep{meliga-etal-2009} and a short cylinder \citep{pierson-etal-2019}. 
Interestingly, when considering the time-average forces, the decreasing power law that fits $F_x$ and $Re$ in the $sS_yS_z$, $sS_yA_z$ and $sA_yS_z$ regimes holds also in the unsteady regime for intermediate Reynolds numbers up to $Re=575$ (figure \ref{fig:3D_Ary1_forces}). 
For these $Re$, indeed, the time average flow field --- and its dependence on $Re$ --- resembles the steady $sA_yS_z$ regime. The average size of the recirculating regions over the lateral sides increases with the Reynolds number, as shown in figure \ref{fig:xr_position} by means of the reattaching point $x_r$. For larger $Re$ the structure of the time average flow changes, and $F_x$ increases. For $Re \ge 575$, indeed, the average $x_r(Re)$ deviates from the power law found at smaller $Re$: the lateral recirculating regions shrink as $Re$ increases (figure \ref{fig:xr_position}).

The dynamics of the flow progressively changes with $Re$. 
At $Re=510$ the flow enters the periodic $pA_yS_z$ regime, which is characterised by an alternating shedding of hairpin vortices (HVs) from the top and bottom LE shear layers (figure \ref{fig:W12_lambda2_1}). 
At this $Re$, a single peak $St \approx 0.277$ is found in the frequency spectrum $S(F_z)$, while $F_y$ remains practically constant in time. 
The attractor draws a limit cycle in the phase space. 

\begin{figure}
\centering
\begin{tikzpicture}

\definecolor{clr1}{RGB}{18 78 128}
\definecolor{clr2}{RGB}{89 165 216}
\definecolor{clr3}{RGB}{145 229 246}
\definecolor{clr4}{RGB}{255 143 163}
\definecolor{clr5}{RGB}{255 77 109}
\definecolor{clr6}{RGB}{201 24 74}
\definecolor{clr7}{RGB}{128 15 47}
\definecolor{clr8}{RGB}{174 32 18}
\definecolor{clr9}{RGB}{155 34 38}
\definecolor{clr10}{RGB}{46 139 87}

\begin{axis}[%
width=0.35\textwidth,
height=0.15\textwidth,
scale only axis,
ymode = log,
xmin=1,
xmax=8,
ymin=0.0,
ymax=0.3,
xtick={1,2,3,4,5,6,7,8},
xlabel={$mode$},
ylabel={$\lambda/\sum \lambda$},
ylabel style={at={(0.05,0.5)}},
axis background/.style={fill=white},
legend columns=3,transpose legend,
legend style={at={(0.99,0.25)}, anchor=east, legend cell align=left, align=left, fill=none, draw=none}
]

\addplot [color=black,solid,draw=none,mark=*,mark options={scale=1.4,black,fill=clr6}]
  table[row sep=crcr]{%
  1.0000 0.2999 \\
  2.0000 0.2967 \\
  3.0000 0.1909 \\
  4.0000 0.1744 \\
  5.0000 0.0091 \\
  6.0000 0.0091 \\
  7.0000 0.0044 \\
  8.0000 0.0043 \\
};

\end{axis}

\end{tikzpicture}%
\begin{tikzpicture}

\definecolor{clr1}{RGB}{18 78 128}
\definecolor{clr2}{RGB}{89 165 216}
\definecolor{clr3}{RGB}{145 229 246}
\definecolor{clr4}{RGB}{255 143 163}
\definecolor{clr5}{RGB}{255 77 109}
\definecolor{clr6}{RGB}{201 24 74}
\definecolor{clr7}{RGB}{128 15 47}
\definecolor{clr8}{RGB}{174 32 18}
\definecolor{clr9}{RGB}{155 34 38}
\definecolor{clr10}{RGB}{46 139 87}

\begin{axis}[%
width=0.35\textwidth,
height=0.15\textwidth,
scale only axis,
xmin=1,
xmax=8,
ymin=0.0,
ymax=0.6,
xtick={1,2,3,4,5,6,7,8},
xlabel={$mode$},
ylabel={$St$},
ylabel style={at={(0.05,0.5)}},
axis background/.style={fill=white},
legend columns=3,transpose legend,
legend style={at={(0.99,0.25)}, anchor=east, legend cell align=left, align=left, fill=none, draw=none}
]

\addplot[red, dashed, domain=1:8] {0.28155};
\addplot[red, dashed, domain=1:8] {0.07769};

\addplot [color=black,solid,draw=none,mark=*,mark options={scale=1.4,black,fill=clr6}]
  table[row sep=crcr]{%
  1.0000 0.28 \\
  2.0000 0.28 \\
  3.0000 0.078 \\
  4.0000 0.078 \\
  5.0000 0.56 \\
  6.0000 0.56 \\
  7.0000 0.15715 \\
  8.0000 0.15715 \\
};

\end{axis}

\end{tikzpicture}%
\includegraphics[trim={0 20 0 20},clip,width=0.49\textwidth]{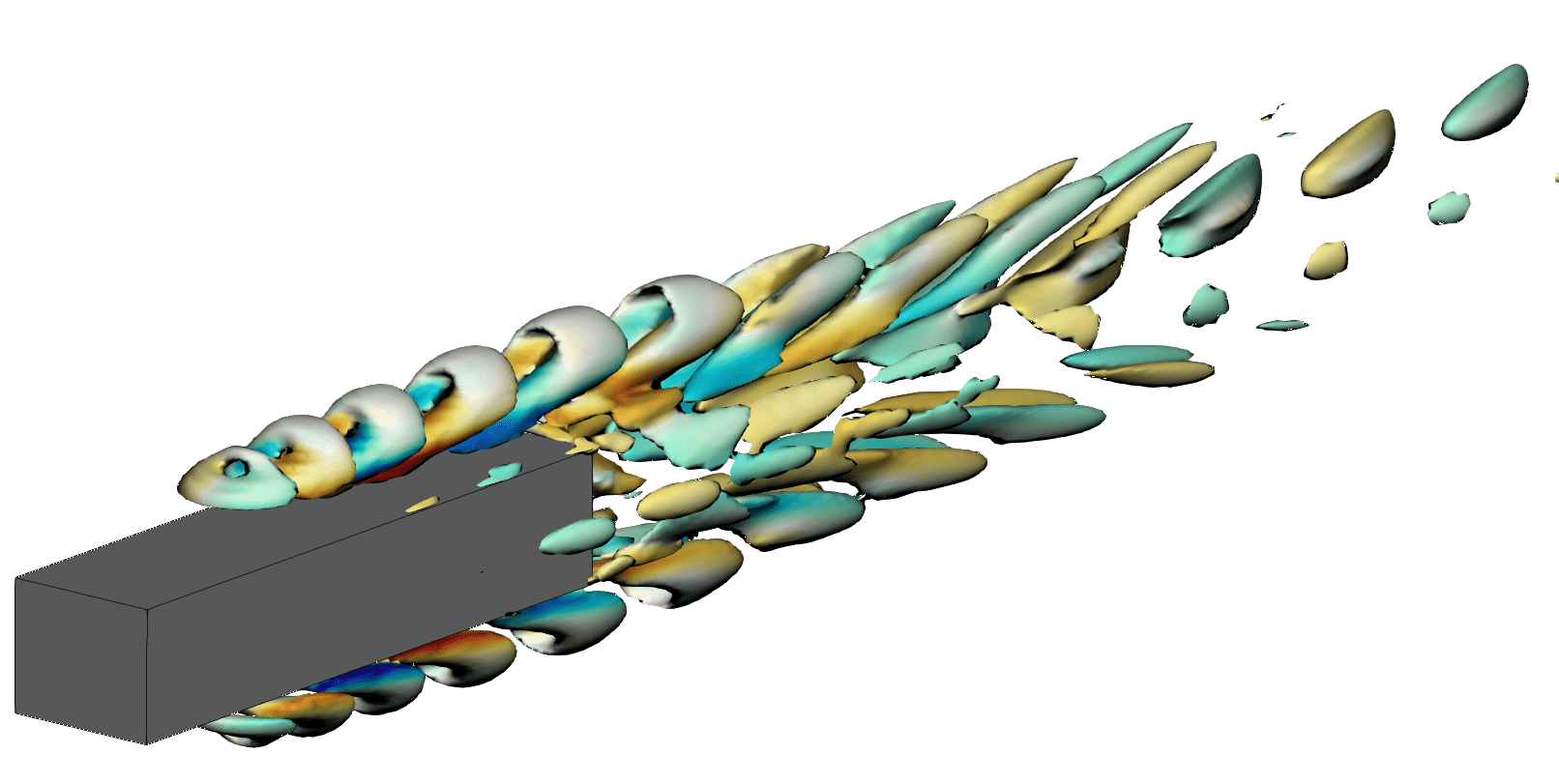}
\includegraphics[trim={0 20 0 20},clip,width=0.49\textwidth]{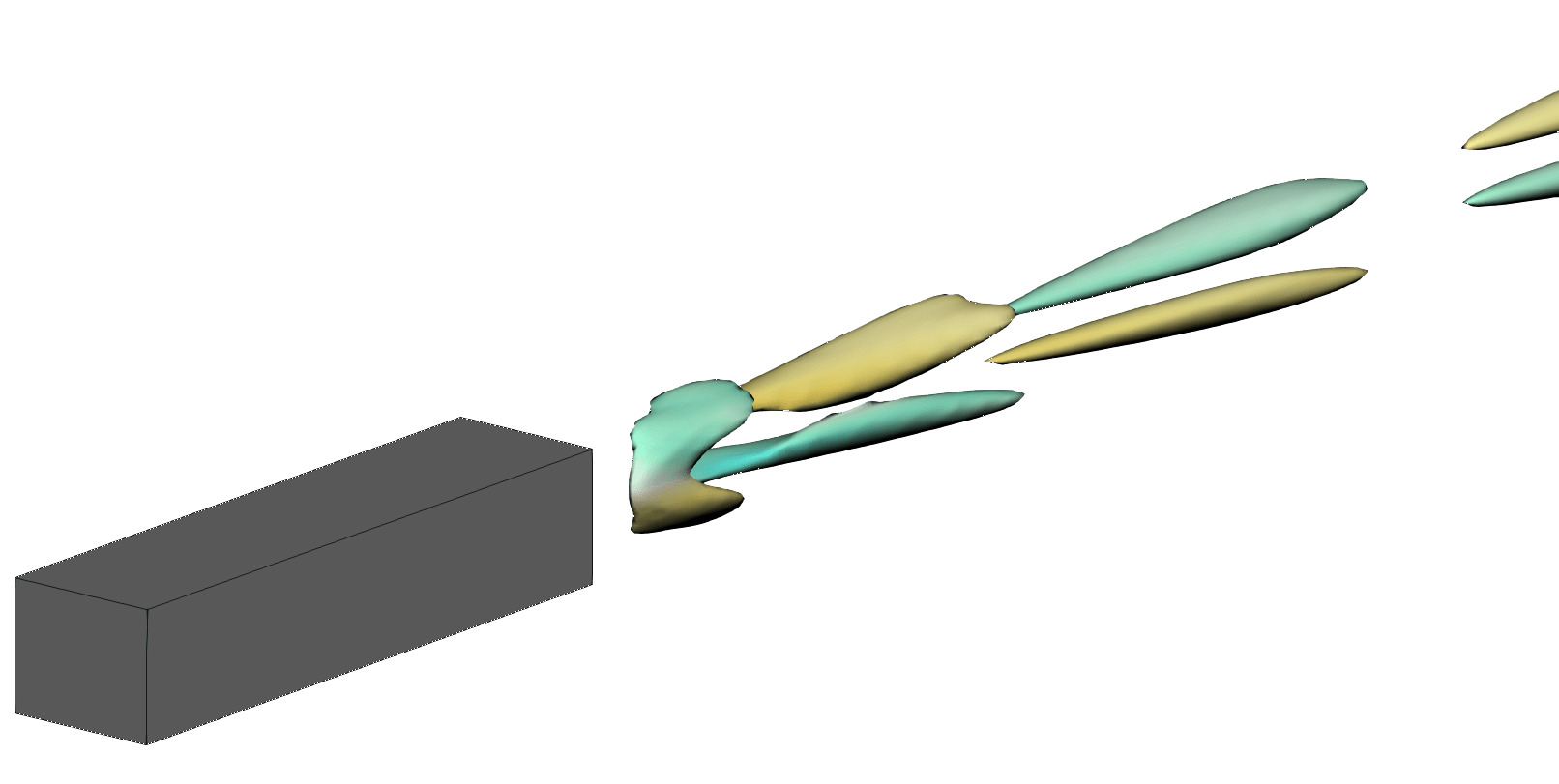}
\caption{POD analysis for $L=5$, $W=1.2$ and $Re=515$. 
Top left: energy fraction $E_j = \lambda_j/\sum\lambda_j$ of the first $8$ POD modes. 
$\lambda_j$ denotes the $j_{th}$ eigenvalue of the snapshot covariance matrix, and is related with the $j_{th}$ singular value by $\sigma_j = \lambda_j^2$ (see appendix \ref{sec:POD}). 
Top right: frequency associated with the first $8$ POD modes, ordered 
by decreasing $\sigma_j$ (i.e. decreasing $E_j$). 
Dashed lines: main DNS frequencies, also shown in figure \ref{fig:3D_Ary1_forces}. 
Bottom: visualisation of the POD modes 1 (left) and 3 (right). 
Isosurfaces of $\lambda_2$ (arbitrary value) 
coloured by streamwise vorticity (symmetric blue-to-red colourmap  from negative to positive values). }
\label{fig:W12_POD}
\end{figure}

For $Re\gtrapprox 515$ a frequency $St \approx 0.0776$ appears in  $S(F_y)$, and the flow becomes aperiodic. Here it is characterised by a superposition of two different modes, (i) the shedding of HVs from the top and bottom LE shear layers (with $St \approx 0.28$), and (ii) the oscillating motion of the wake in the $y$ direction (with $St \approx 0.08$); see figure \ref{fig:W12_POD} and the related discussion. 
The two frequencies are not commensurate and, as shown in the force diagrams of figure \ref{fig:W12_lambda2_1}, 
a torus replaces the limit cycle in the phase space. 
At this $Re$, the peak at $St \approx 0.28$  is much larger than the other one, indicating that the shedding of HVs dominates.

For $Re \ge 550$, the wake oscillates also in the $z$ direction, and a further frequency $St \approx 0.03$ appears in $S(F_z)$ (figure \ref{fig:W12_lambda2_1}). 
%
The different $Re$ at which the two oscillating modes arise agree with the results of the LSA, which reveals that for $L \le 2$ the unsteady $A_yS_z$ wake mode becomes unstable at smaller $Re$ compared to the unsteady $S_yA_z$ mode. 
As mentioned above, the presence of this additional mode leads to an increase of the average $F_x$ (figure \ref{fig:3D_Ary1_forces}). 
When $Re$ is further increased the different modes nonlinearly interact and the flow becomes progressively more chaotic (figure \ref{fig:W12_lambda2_1}).
The relative height of the peaks at $St \approx 0.28$ and $St \approx 0.03$ in $S(F_z)$ indicates that the flow dynamics is mainly driven by the LE vortex shedding at low $Re$, while the vertical wake oscillating motion takes over at large $Re$ (not shown).  

To examine the spatial structure of the different modes and relate them with the flow frequencies, we use proper orthogonal decomposition (POD); see appendix \ref{sec:POD} for details. 
Figure \ref{fig:W12_POD} considers $Re=515$. 
As shown in the top panels, the dominant POD modes are associated with the same frequencies found in the frequency spectra (figure \ref{fig:W12_lambda2_1}) and with their multiples. 
The POD modes come in pairs (i.e. with same single values and frequencies), as typical for oscillating structures. 
Mode 1 is associated with $St \approx 0.28$, and a shedding of vortices from the top and bottom sides is clearly visible in its spatial structure. Mode 3, instead, is associated with $St \approx 0.07$. Its spatial structure is confined in the wake and breaks the left/right symmetry consistently with an oscillating mode of the wake in the $y$ direction about a $A_yS_z$ state. 
At this $Re$ the energy fraction associated with the LE vortex shedding is approximately $1.5$ times larger than that associated with the lateral wake oscillation  (see  top left panel), consistently with the 
peaks heights in the frequency spectra.

For elongated prisms with $W=1.2$ and $L=5$, 
the characteristic frequency of the wake oscillation ($St \approx 0.07$) is smaller 
than for a sphere ($St \approx 0.129$ at $Re=270$; \cite{tomboulides-orszag-2000}), 
a cube ($St \approx 0.091-0.095$ at $Re= 270-300$; \cite{saha-2004}), 
and a short cylinder ($St \approx 0.125$ at $Re=283$; \cite{yang-etal-2022}). 
This is consistent with the LSA results (figure \ref{fig:neutral_curves_and_omc_vs_L}), that point out a decrease of $St$ with $L$. 
\subsection{Intermediate width: $W=2.25$}
\label{sec:sim_L5_W225}

\begin{figure}
\centering
\begin{tikzpicture}

\definecolor{clr1}{RGB}{18 78 128}
\definecolor{clr2}{RGB}{89 165 216}
\definecolor{clr3}{RGB}{145 229 246}
\definecolor{clr4}{RGB}{34 139 34}
\definecolor{clr5}{RGB}{255 77 109}
\definecolor{clr6}{RGB}{201 24 74}
\definecolor{clr7}{RGB}{128 15 47}
\definecolor{clr8}{RGB}{174 32 18}
\definecolor{clr9}{RGB}{155 34 38}
\definecolor{clr10}{RGB}{46 139 87}
\definecolor{clr11}{RGB}{240 128 128}
\definecolor{clr12}{RGB}{152 251 152}
\definecolor{clr13}{RGB}{143 188 143}
\definecolor{clr14}{RGB}{154 205 50}

\begin{axis}[%
width=0.9\textwidth,
height=0.2\textwidth,
scale only axis,
xmin=250,
xmax=750,
ymin=0.74,
ymax=0.95,
xtick={300,400,500,600,700},
xticklabel=\empty,
ytick={0.75,0.8,0.85,0.9,0.95,1},
yticklabels={$ 0.75$,$ 0.80$,$ 0.85$,$0.90$,$0.95$,$1.00$},
ylabel={$F_x$},
ylabel style={at={(0.02,0.5)}},
axis background/.style={fill=white},
legend columns=3,transpose legend,
legend style={at={(0.99,0.82)}, anchor=east, legend cell align=left, align=left, fill=none, draw=none}
]

\addplot[red, solid, domain=250:450] {4.3313*x^(-0.2905)};

\addplot [color=black,solid,draw=none,mark=*,mark options={scale=1.4,black,fill=clr1}]
  table[row sep=crcr]{%
  250.0000    0.88665 \\
  270.0000    0.85853 \\
  360.0000    0.78739 \\
  370.0000    0.78297 \\
};
\addplot [color=black,solid,draw=none,mark=*,mark options={scale=1.4,black,fill=clr2}]
  table[row sep=crcr]{
  290.0000    0.83512 \\
  300.0000    0.82605 \\
};
\addplot [color=black,solid,draw=none,mark=*,mark options={scale=1.4,black,fill=clr3}]
  table[row sep=crcr]{%
  300.0000    0.82596 \\
  310.0000    0.81706 \\
  320.0000    0.81042 \\
  330.0000    0.80345 \\
  340.0000    0.79682 \\
  345.0000    0.79362 \\
  350.0000    0.79053 \\
};

\addplot [color=black,solid,draw=none,mark=*,mark options={scale=1.4,black,fill=clr11}]
  table[row sep=crcr]{%
  375.0000    0.7814 \\
  380.0000    0.7791 \\
};

\addplot [color=black,solid,draw=none,mark=*,mark options={scale=1.4,black,fill=clr12}]
  table[row sep=crcr]{%
  385.0000    0.7766 \\
  390.0000    0.7748 \\
};

\addplot [color=black,solid,draw=none,mark=*,mark options={scale=1.4,black,fill=clr4}]
  table[row sep=crcr]{%
  395.0000    0.7734 \\
  400.0000    0.7725 \\
  420.0000    0.7682 \\
};

\addplot [color=black,solid,draw=none,mark=*,mark options={scale=1.4,black,fill=clr13}]  
  table[row sep=crcr]{%
  440.0000    0.7670 \\
  450.0000    0.7658 \\
  460.0000    0.7645 \\
};

\addplot [color=black,solid,draw=none,mark=*,mark options={scale=1.4,black,fill=clr14}]
  table[row sep=crcr]{
  465.000     0.7652 \\
  470.000     0.7659 \\
};

\addplot [color=black,solid,draw=none,mark=*,mark options={scale=1.4,black,fill=clr7}]
  table[row sep=crcr]{%
  485.000     0.7670 \\  
  500.0000    0.7660 \\
  550.0000    0.7634 \\
  625.0000    0.7722 \\
  700.0000    0.7862 \\
};

\addplot [color=black,solid,mark=none, line width=0.5, mark options={scale=1.4,black,fill=red!80!black}]
  table[row sep=crcr]{%
  400.0000    0.7554 \\
  400.0000    0.7896 \\
};

\addplot [color=black,solid,mark=none, line width=0.5, mark options={scale=1.4,black,fill=red!80!black}]
  table[row sep=crcr]{%
  460.0000    0.7614 \\
  460.0000    0.7656 \\
};

\addplot [color=black,solid,mark=none, line width=0.5,mark options={scale=1.4,black,fill=red!80!black}]
  table[row sep=crcr]{%
  500.0000    0.7562 \\
  500.0000    0.7758 \\
};

\addplot [color=black,solid,mark=none, line width=0.5,mark options={scale=1.4,black,fill=red!80!black}]
  table[row sep=crcr]{%
  550.0000    0.7473 \\
  550.0000    0.7787 \\
};

\addplot [color=black,dashed,mark=none, line width=0.5,mark options={scale=1.4,black,fill=red!80!black}]
  table[row sep=crcr]{%
  280    0.91 \\
  280    0.78 \\
};

\addplot [color=black,dashed,mark=none, line width=0.5,mark options={scale=1.4,black,fill=red!80!black}]
  table[row sep=crcr]{%
  295    0.74 \\
  295    1 \\
};

\addplot [color=black,dashed,mark=none, line width=0.5,mark options={scale=1.4,black,fill=red!80!black}]
  table[row sep=crcr]{%
  305    0.74 \\
  305    1 \\
};

\addplot [color=black,dashed,mark=none, line width=0.5,mark options={scale=1.4,black,fill=red!80!black}]
  table[row sep=crcr]{%
  355    0.74 \\
  355    1 \\
};

\addplot [color=black,dashed,mark=none, line width=0.5,mark options={scale=1.4,black,fill=red!80!black}]
  table[row sep=crcr]{%
  372    0.74 \\
  372    1 \\
};

\addplot [color=black,dashed,mark=none, line width=0.5,mark options={scale=1.4,black,fill=red!80!black}]
  table[row sep=crcr]{%
  382    0.74 \\
  382    1 \\
};

\addplot [color=black,dashed,mark=none, line width=0.5,mark options={scale=1.4,black,fill=red!80!black}]
  table[row sep=crcr]{%
  392    0.74 \\
  392    0.92 \\
};

\addplot [color=black,dashed,mark=none, line width=0.5,mark options={scale=1.4,black,fill=red!80!black}]
  table[row sep=crcr]{%
  477    0.74 \\
  477    1 \\
};

\addplot [color=black,dashed,mark=none, line width=0.5,mark options={scale=1.4,black,fill=red!80!black}]
  table[row sep=crcr]{%
  462.5    0.74 \\
  462.5    1 \\
};

\addplot [color=black,dashed,mark=none, line width=0.5,mark options={scale=1.4,black,fill=red!80!black}]
  table[row sep=crcr]{%
  435    0.74 \\
  435    1 \\
};

\addplot [color=black,solid,mark=none, line width=0.5,mark options={scale=1.4,black,fill=red!80!black}]
  table[row sep=crcr]{%
  700.0000    0.7678 \\
  700.0000    0.8046 \\
};

\addplot [color=black,solid,mark=none, line width=0.5,mark options={scale=1.4,black,fill=red!80!black}]
  table[row sep=crcr]{%
  375.0000    0.7710 \\
  375.0000    0.7918 \\
};

\addplot [color=black,solid,mark=none, line width=0.5,mark options={scale=1.4,black,fill=red!80!black}]
  table[row sep=crcr]{%
  380.0000    0.7652 \\
  380.0000    0.7929 \\
};

\addplot [color=black,solid,mark=none, line width=0.5,mark options={scale=1.4,black,fill=red!80!black}]
  table[row sep=crcr]{%
  385.0000    0.7613 \\
  385.0000    0.7920 \\
};

\addplot [color=black,solid,mark=none, line width=0.5,mark options={scale=1.4,black,fill=red!80!black}]
  table[row sep=crcr]{%
  390.0000    0.7586 \\
  390.0000    0.7910 \\
};

\addplot [color=black,solid,mark=none, line width=0.5,mark options={scale=1.4,black,fill=red!80!black}]
  table[row sep=crcr]{%
  395.0000    0.7566 \\
  395.0000    0.7901 \\
};

\addplot [color=black,solid,mark=none, line width=0.5,mark options={scale=1.4,black,fill=red!80!black}]
  table[row sep=crcr]{%
  420.0000    0.7507 \\
  420.0000    0.7857 \\
};

\addplot [color=black,solid,mark=none, line width=0.5,mark options={scale=1.4,black,fill=red!80!black}]
  table[row sep=crcr]{%
  440.0000    0.7661 \\
  440.0000    0.7698 \\
};

\addplot [color=black,solid,mark=none, line width=0.5,mark options={scale=1.4,black,fill=red!80!black}]
  table[row sep=crcr]{%
  450.0000    0.7638 \\
  450.0000    0.7678 \\
};

\addplot [color=black,solid,mark=none, line width=0.5,mark options={scale=1.4,black,fill=red!80!black}]
  table[row sep=crcr]{%
  465.0000    0.7630 \\
  465.0000    0.7673 \\
};

\addplot [color=black,solid,mark=none, line width=0.5,mark options={scale=1.4,black,fill=red!80!black}]
  table[row sep=crcr]{%
  470.0000    0.7638 \\
  470.0000    0.7678 \\
};

\addplot [color=black,solid,mark=none, line width=0.5,mark options={scale=1.4,black,fill=red!80!black}]
  table[row sep=crcr]{%
  485.0000    0.7644 \\
  485.0000    0.7696 \\
};

\addplot [color=black,solid,mark=none, line width=0.5,mark options={scale=1.4,black,fill=red!80!black}]
  table[row sep=crcr]{%
  625.0000    0.7564 \\
  625.0000    0.7881 \\
};

\end{axis}

\node[] at (0.45,2.4) {$sS_yS_z$};
\draw[->] (1.6,1.5) -- (1.2,1.5);
\node[] at (1.9,1.8) {$sS_yA_z$};
\node[] at (1.9,1.5) {$\&$};
\node[] at (1.9,1.2) {$sA_yS_z$};
\node[] at (0.6,0.3) {$sS_yA_z$};
\node[] at (1.9,2.4) {$sA_yS_z$};
\node[] at (3.9,2.4) {$pS_yS_zla$};
\draw[->] (5.6,1.7) -- (5,1.7);
\node[] at (6.2,1.7) {$pS_yS_zlb$};
\draw[->] (3.6,1.7) -- (2.8,1.7);
\node[] at (4.1,1.7) {$sS_yS_z$};
\draw[->] (3.6,1.4) -- (3.1,1.4);
\node[] at (4.1,1.4) {$aS_yA_z$};
\draw[->] (3.6,1.1) -- (3.3,1.1);
\node[] at (4.1,1.1) {$pS_yA_z$};
\node[] at (8.1,2.4) {$aS_yS_z$};
\draw[->] (5.6,1.1) -- (5.3,1.1);
\node[] at (6.1,1.1) {$pS_yS_z$};

\end{tikzpicture}%
\input{./figure/Ary225/FyFz-Re-2.tex}
\includegraphics[width=0.325\textwidth]{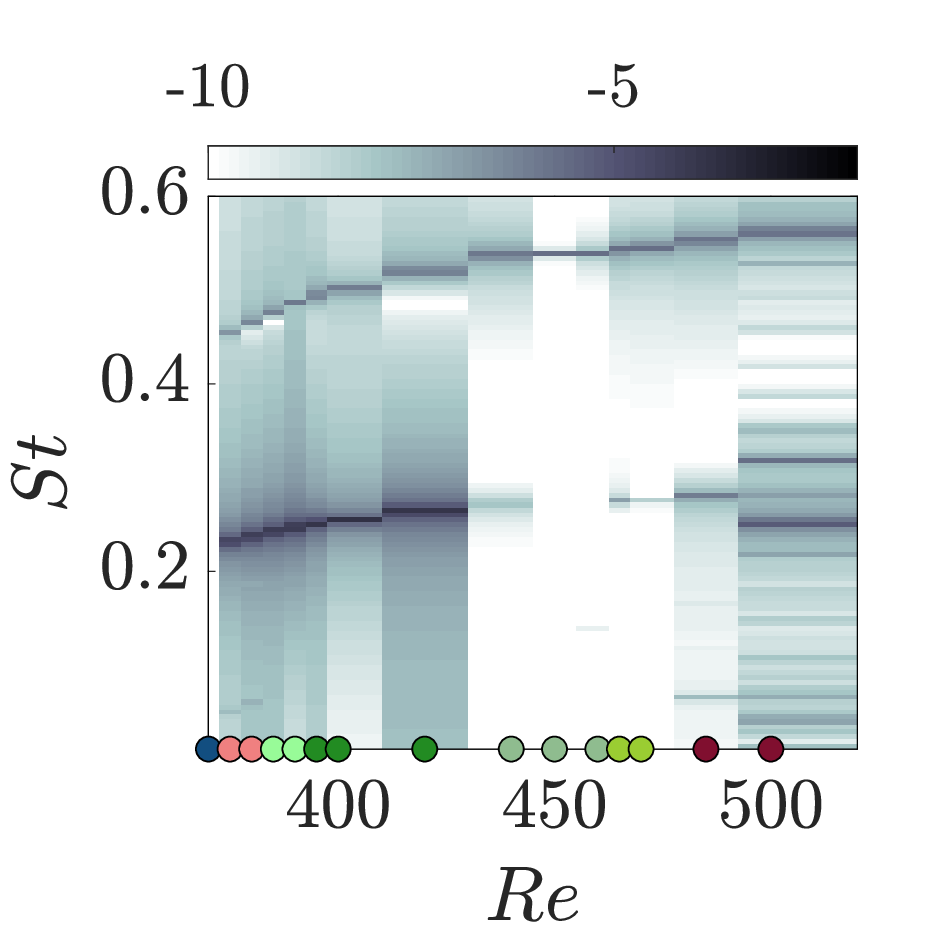}
\includegraphics[width=0.325\textwidth]{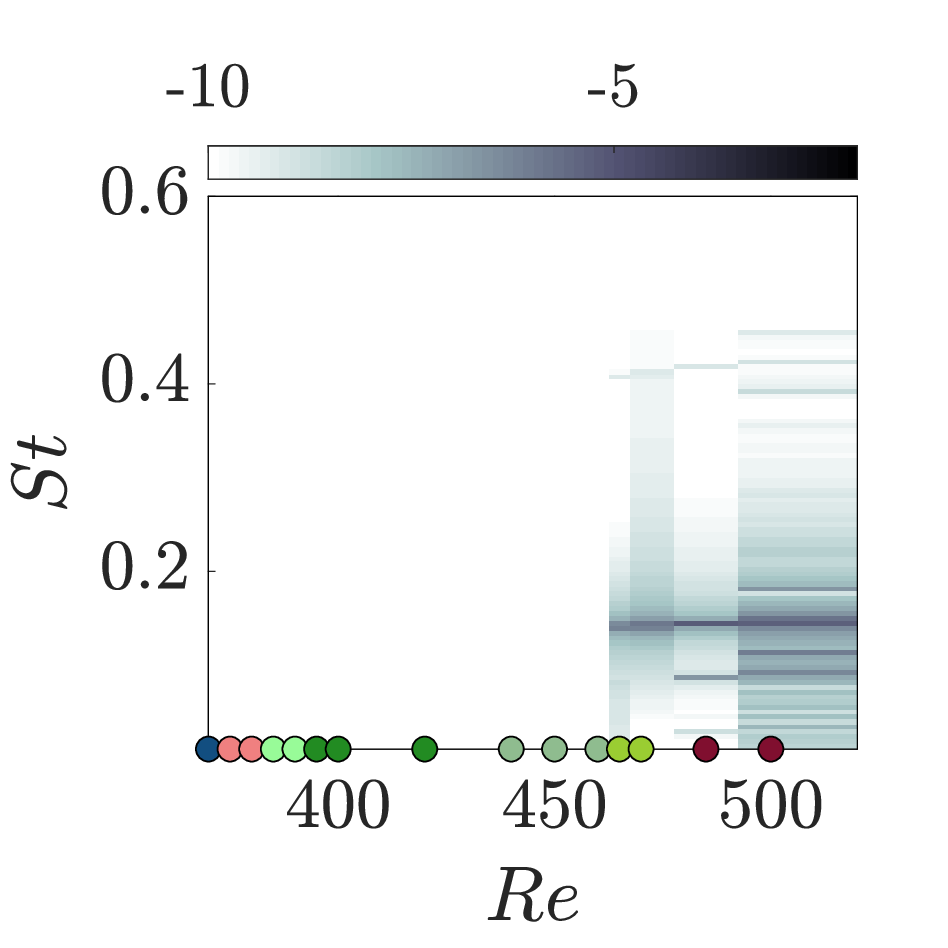}
\includegraphics[width=0.325\textwidth]{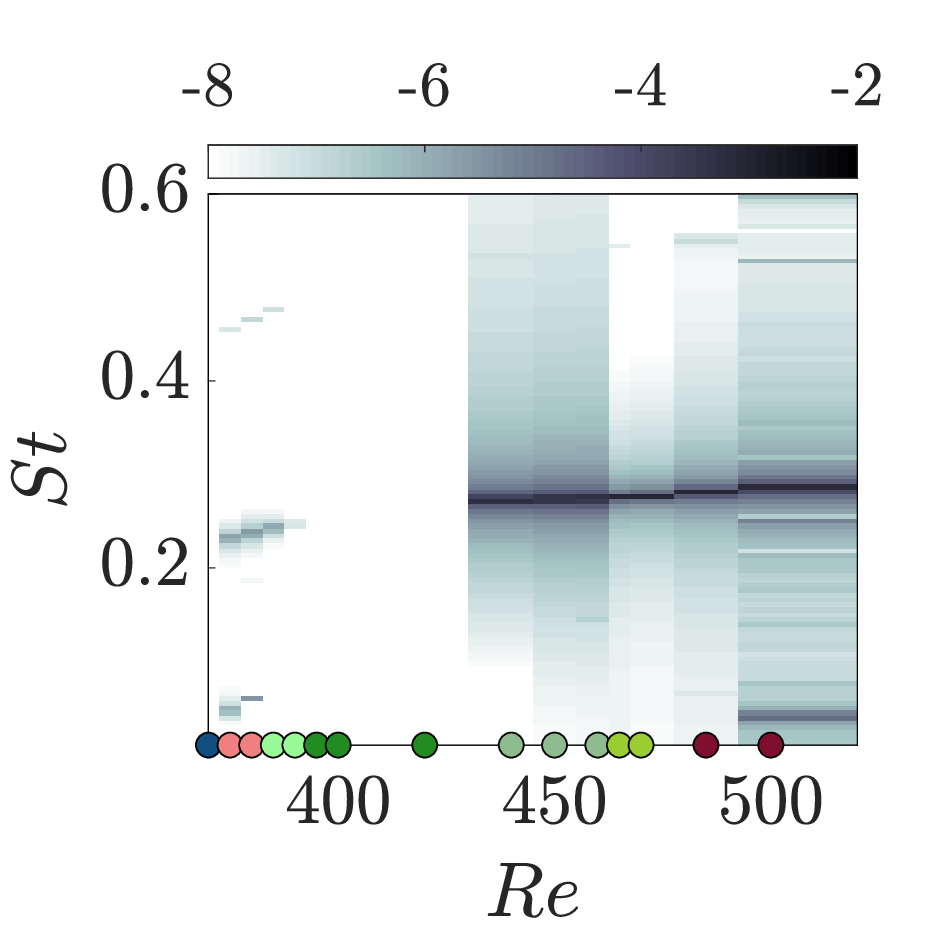}
\caption{As figure \ref{fig:3D_Ary1_forces},  for $L=5$ and $W=2.25$. 
Red line in the top panel: $4.33 \times Re^{-0.2905}$.}
\label{fig:AR225_forces}
\end{figure}
For $L=5$ and $W=2.25$ the scenario becomes more complicated, with many regimes identified up to $Re = 700$ (figure \ref{fig:AR225_forces}). 
In agreement with the LSA (\S\ref{sec:stability}), the critical $Re$ corresponding to the onset of the first 
bifurcation decreases with  the width of the prism. 
For $W=2.25$ the primary instability consists of a pitchfork bifurcation towards the steady $sS_yA_z$ regime,  for $Re$ between $270$ and $290$.
At $Re \approx 300$, another bifurcation restores the top/bottom planar symmetry and breaks the left/right one, thus leading to the $sA_yS_z$ regime, similar to smaller $W$. 
The simulations show an hysteresis in the transition from $sS_yA_z$ to $sA_yS_z$. 
Bistability is detected for some values of $Re$, in agreement with the subcritical nature of the bifurcation (see the WNL analysis in \S\ref{sec:WNL}). 
Interestingly, at larger $Re$ the flow recovers both planar symmetries and settles again in a $sS_yS_z$ regime,
before becoming unsteady at $Re \approx 375$ with a combination of two different modes: (i) oscillation of the wake, and (ii) symmetric shedding of HVs from the top and bottom LE shear layers. 
As $Re$ further increases, the wake oscillation mode stabilises, leading to the periodic $pS_yS_zla$ and $pS_yS_zlb$ regimes, and then destabilises again at larger $Re$, leading eventually to the aperiodic $aS_yS_z$ regime.

\subsubsection{Low $Re$: Steady regimes $sS_yA_z$ and $sA_yS_z$}

\begin{figure}
\centering
  \begin{tikzpicture}
\definecolor{clr1}{RGB}{18 78 128}  
\definecolor{clr2}{RGB}{89 165 216}
\definecolor{clr3}{RGB}{145 229 246}
\node at (-3.0,4.5)   {\includegraphics[width=0.24\textwidth]{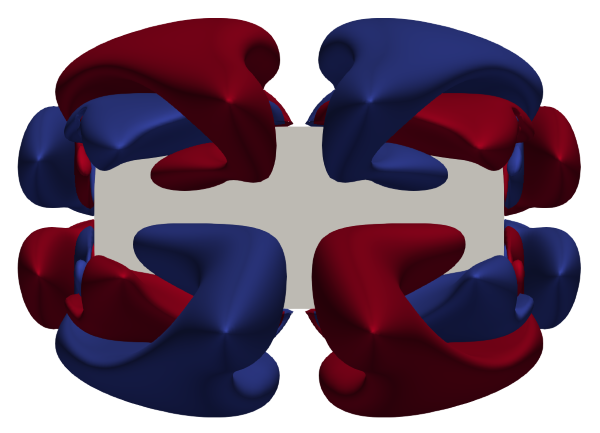}};
\node at (0.4,4.5) {\includegraphics[width=0.24\textwidth]{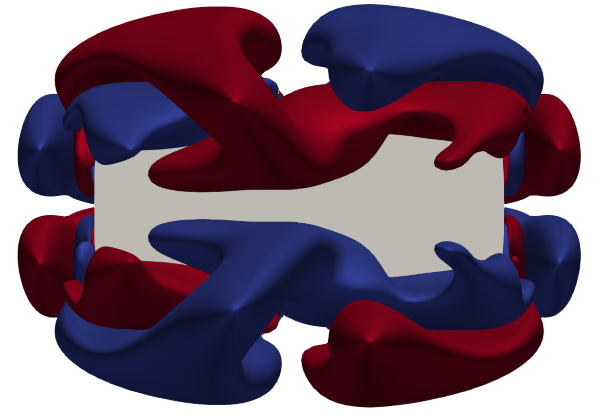}};
\node at (3.8,4.5)   {\includegraphics[width=0.24\textwidth]{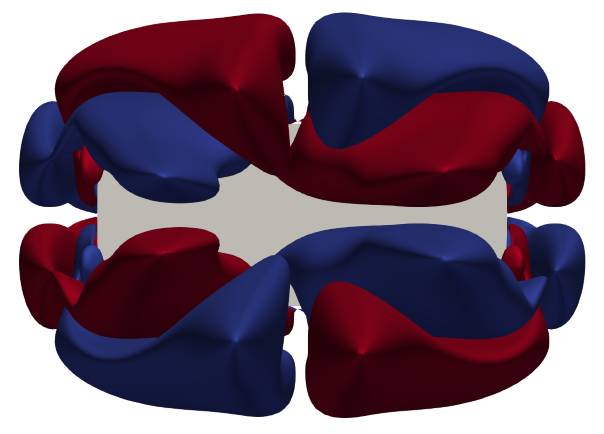}};
\node at (7.2,4.5) {\includegraphics[width=0.24\textwidth]{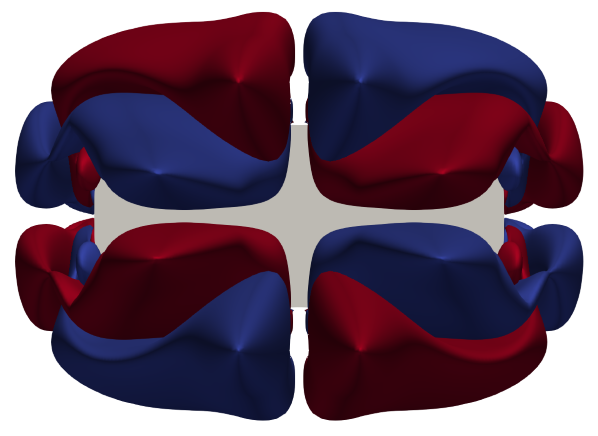}};   
\node at (-4.6,5.3){({\color{clr2}$a$})};
\node at (-1.2,5.3){({\color{clr3}$b$})};
\node at ( 2.2,5.3){({\color{clr3}$c$})};
\node at ( 5.6,5.3){({\color{clr1}$d$})};    
  \end{tikzpicture}
\begin{tikzpicture}
\definecolor{clr1}{RGB}{18 78 128}
\definecolor{clr2}{RGB}{89 165 216}
\definecolor{clr3}{RGB}{145 229 246}
\definecolor{clr4}{RGB}{34 139 34}
\definecolor{clr5}{RGB}{255 77 109}
\definecolor{clr6}{RGB}{201 24 74}
\definecolor{clr7}{RGB}{128 15 47}
\definecolor{clr8}{RGB}{174 32 18}
\definecolor{clr9}{RGB}{155 34 38}
\definecolor{clr10}{RGB}{46 139 87}
\definecolor{clr11}{RGB}{240 128 128}
\definecolor{clr12}{RGB}{152 251 152}

\begin{axis}[%
scaled ticks=false,
width=0.7\textwidth,
height=0.15\textwidth,
scale only axis,
xmin=280,
xmax=370,
ymin=-0.001,
ymax=0.02,
xtick={280,300,320,340,360},
ytick={0,0.01,0.02},
yticklabels={0.00,0.01,0.02},
xlabel={$Re$},
ylabel style={at={(0.1,0.5)}},
axis background/.style={fill=white},
legend columns=3,transpose legend,
legend style={at={(0.99,0.7)}, anchor=east, legend cell align=left, align=left, fill=none, draw=none}
]

\addplot [color=black,solid,draw=none,mark=square,mark options={scale=1.3,black,fill=}]
  table[row sep=crcr]{%
  250.0000         0 \\
  300.0000    0.0133 \\
  350.0000         0 \\
  400.0000         0 \\
  500.0000    0.0023 \\
  550.0000         0 \\
  700.0000    0.0022 \\
};
\addlegendentry{$F_z$};

\addplot [color=black,solid,draw=none,mark=diamond,mark options={scale=1.8,black,fill=red!80!black}]
  table[row sep=crcr]{%
   250.0000000000000   0.0000004680000 \\
   300.0000000000000   0.0000010000000 \\
   400.0000000000000   0.0000019397000 \\
   500.0000000000000                 0 \\
   550.0000000000000   0.0016000000000 \\
   700.0000000000000                 0 \\
};
\addlegendentry{$F_y$};

\addplot [color=black,solid,draw=none,mark=square*,mark options={scale=1.3,black,fill=clr1}]
  table[row sep=crcr]{%
  250.0000         0 \\
  360.0000         0 \\
  370.0000         0 \\
};

\addplot [color=black,solid,draw=none,mark=square*,mark options={scale=1.3,black,fill=clr2}]
  table[row sep=crcr]{
  290.000     0.0106 \\
  300.0000    0.0133 \\
};

\addplot [color=black,solid,draw=none,mark=square*,mark options={scale=1.3,black,fill=clr3}]
  table[row sep=crcr]{
  300.0000         0 \\
  310.0000         0 \\
  320.0000         0 \\
  330.0000         0 \\
  340.0000         0 \\
  345.0000         0 \\
  350.0000         0 \\
};

\addplot [color=black,solid,draw=none,mark=square*,mark options={scale=1.3,black,fill=clr11}]
  table[row sep=crcr]{
  375.0000         0.0148 \\
  380.0000         0.0152 \\
};

\addplot [color=black,solid,draw=none,mark=diamond*,mark options={scale=1.8,black,fill=clr11}]
  table[row sep=crcr]{
  375.0000         0.0000 \\
  380.0000         0.0000 \\
};

\addplot [color=black,solid,draw=none,mark=square*,mark options={scale=1.3,black,fill=clr12}]
  table[row sep=crcr]{
  385.0000         0.0130 \\
  390.0000         0.0084 \\
};

\addplot [color=black,solid,draw=none,mark=diamond*,mark options={scale=1.8,black,fill=clr12}]
  table[row sep=crcr]{
  385.0000         0.0000 \\
  390.0000         0.0000 \\
};

\addplot [color=black,solid,draw=none,mark=square*,mark options={scale=1.3,black,fill=clr4}]
  table[row sep=crcr]{
  395.000          0 \\
  400.0000         0 \\
  420.0000         0 \\
  450.0000         0 \\
  470.0000         0.0017 \\
  485.0000         0.0033 \\
};

\addplot [color=black,solid,draw=none,mark=square*,mark options={scale=1.3,black,fill=clr7}]
  table[row sep=crcr]{%
  500.0000    0.0023 \\
  550.0000         0 \\
  625.0000         0 \\
  700.0000    0.0022 \\
};

\addplot [color=black,solid,draw=none,mark=diamond*,mark options={scale=1.8,black,fill=clr1}]
  table[row sep=crcr]{%
   250.0000000000000   0.0000004680000 \\
   360.0000000000000   0 \\
   370.0000000000000   0 \\   
};

\addplot [color=black,solid,draw=none,mark=diamond*,mark options={scale=1.8,black,fill=clr2}]
  table[row sep=crcr]{%
   290.0000000000000   0.0000000000000 \\
   300.0000000000000   0.0000010000000 \\
};

\addplot [color=black,solid,draw=none,mark=diamond*,mark options={scale=1.8,black,fill=clr3}]
  table[row sep=crcr]{
   300.0000000000000   0.0127 \\
   310.0000000000000   0.013773 \\
   320.0000000000000   0.013432 \\
   330.0000000000000   0.012071 \\
   340.0000000000000   0.0087 \\
   345.0000000000000   0.0067 \\
   350.0000000000000   0.0007658 \\
};

\addplot [color=black,solid,draw=none,mark=diamond*,mark options={scale=1.8,black,fill=clr4}]
  table[row sep=crcr]{%
   395.0000000000000   0.0000000000000 \\
   400.0000000000000   0.0000019397000 \\
   420.0000000000000   0.0000000000000 \\
   450.0000000000000   0.0000000000000 \\
   470.0000000000000   0.0000000000000 \\
   485.0000000000000   0.0000000000000 \\
};

\addplot [color=black,solid,draw=none,mark=diamond*,mark options={scale=1.8,black,fill=clr7}]
  table[row sep=crcr]{%
   500.0000000000000                 0 \\
   550.0000000000000   0.0016000000000 \\
   625.0000000000000   0.0014000000000 \\
   700.0000000000000                 0 \\
};

\addplot [color=black,solid,mark=none, line width=0.5,mark options={scale=1.4,black,fill=red!80!black}]
  table[row sep=crcr]{%
  500.0000    0.0436 \\
  500.0000   -0.0390 \\
};

\addplot [color=red,solid,mark=none, line width=0.5,mark options={scale=1.4,black,fill=red!80!black}]
  table[row sep=crcr]{%
  500.0000    0.0045 \\
  500.0000   -0.0045 \\
};

\addplot [color=black,solid,mark=none, line width=0.5,mark options={scale=1.4,black,fill=red!80!black}]
  table[row sep=crcr]{%
  550.0000    0.0432 \\
  550.0000   -0.0432 \\
};

\addplot [color=red,solid,mark=none, line width=0.5,mark options={scale=1.4,black,fill=red!80!black}]
  table[row sep=crcr]{%
  550.0000    0.0159 \\
  550.0000   -0.0135 \\
};

\addplot [color=black,solid,mark=none, line width=0.5,mark options={scale=1.4,black,fill=red!80!black}]
  table[row sep=crcr]{%
  700.0000    0.0524 \\
  700.0000   -0.048 \\
};

\addplot [color=red,solid,mark=none, line width=0.5,mark options={scale=1.4,black,fill=red!80!black}]
  table[row sep=crcr]{%
  700.0000    0.0196 \\
  700.0000   -0.0106 \\
};

\addplot [color=black,solid,mark=none, line width=0.5,mark options={scale=1.4,black,fill=red!80!black}]
  table[row sep=crcr]{%
  375    0.0117 \\
  375    0.0178 \\
};

\addplot [color=black,solid,mark=none, line width=0.5,mark options={scale=1.4,black,fill=red!80!black}]
  table[row sep=crcr]{%
  380    0.0114 \\
  380    0.0190 \\
};

\addplot [color=black,solid,mark=none, line width=0.5,mark options={scale=1.4,black,fill=red!80!black}]
  table[row sep=crcr]{%
  385    0.0109 \\
  385    0.0151 \\
};

\addplot [color=black,solid,mark=none, line width=0.5,mark options={scale=1.4,black,fill=red!80!black}]
  table[row sep=crcr]{%
  390    0.0079 \\
  390    0.0089 \\
};

\addplot [color=black,solid,mark=none, line width=0.5,mark options={scale=1.4,black,fill=red!80!black}]
  table[row sep=crcr]{%
  450    -0.0410 \\
  450     0.0410\\
};

\addplot [color=black,solid,mark=none, line width=0.5,mark options={scale=1.4,black,fill=red!80!black}]
  table[row sep=crcr]{%
  470    -0.0416 \\
  470     0.0416\\
};

\addplot [color=red,solid,mark=none, line width=0.5,mark options={scale=1.4,black,fill=red!80!black}]
  table[row sep=crcr]{%
  470    -0.0021 \\
  470     0.0021\\
};

\addplot [color=black,solid,mark=none, line width=0.5,mark options={scale=1.4,black,fill=red!80!black}]
  table[row sep=crcr]{%
  485    -0.0414 \\
  485     0.0414\\
};

\addplot [color=red,solid,mark=none, line width=0.5,mark options={scale=1.4,black,fill=red!80!black}]
  table[row sep=crcr]{%
  485    -0.0036 \\
  485     0.0036\\
};

\addplot [color=black,solid,mark=none, line width=0.5,mark options={scale=1.4,black,fill=red!80!black}]
  table[row sep=crcr]{%
  625    -0.0443 \\
  625     0.0443\\
};

\addplot [color=red,solid,mark=none, line width=0.5,mark options={scale=1.4,black,fill=red!80!black}]
  table[row sep=crcr]{%
  625    -0.0158 \\
  625     0.0158\\
};

\addplot [color=black,dashed,mark=none, line width=0.5,mark options={scale=1.4,black,fill=red!80!black}]
  table[row sep=crcr]{%
  280   -0.03 \\
  280    0.04 \\
};

\addplot [color=black,dashed,mark=none, line width=0.5,mark options={scale=1.4,black,fill=red!80!black}]
  table[row sep=crcr]{%
  295    -0.045 \\
  295     0.060 \\
};

\addplot [color=black,dashed,mark=none, line width=0.5,mark options={scale=1.4,black,fill=red!80!black}]
  table[row sep=crcr]{%
  305    -0.045 \\
  305     0.060 \\
};

\addplot [color=black,dashed,mark=none, line width=0.5,mark options={scale=1.4,black,fill=red!80!black}]
  table[row sep=crcr]{%
  355    -0.045 \\
  355     0.060 \\
};

\addplot [color=black,dashed,mark=none, line width=0.5,mark options={scale=1.4,black,fill=red!80!black}]
  table[row sep=crcr]{%
  372    -0.045 \\
  372    0.06 \\
};

\addplot [color=black,dashed,mark=none, line width=0.5,mark options={scale=1.4,black,fill=red!80!black}]
  table[row sep=crcr]{%
  382    -0.045 \\
  382     0.06 \\
};

\addplot [color=black,dashed,mark=none, line width=0.5,mark options={scale=1.4,black,fill=red!80!black}]
  table[row sep=crcr]{%
  392   -0.045 \\
  392    0.06 \\
};

\addplot [color=black,dashed,mark=none, line width=0.5,mark options={scale=1.4,black,fill=red!80!black}]
  table[row sep=crcr]{%
  492    -0.045 \\
  492     0.06 \\
};


\end{axis}

\end{tikzpicture}%
\caption{
Steady regimes for $L=5$ and $W=2.25$.
Back view of the $y-z$ plane. The red and blue isosurfaces are for $\omega_x= \pm 0.1$. 
Top row: 
$(a)$ $Re=300$, regime $sS_yA_z$; 
$(b)$ $Re=300$, regime $sA_yS_z$; 
$(c)$ $Re=330$, regime $sA_yS_z$;
$(d)$ $Re=360$, regime $sS_yS_z$.
Bottom panel: zoom of the central panel of figure \ref{fig:AR225_forces} in the range $280 \le Re \le 370$.}
\label{fig:W225_omegax_Re300_Re360}
\end{figure}

The nonlinear simulations show that the flow is steady and retains the $S_yS_z$ symmetry for $Re \lessapprox 290$, in agreement with the results of the LSA. 
At $Re \approx 290$ the flow undergoes a regular bifurcation towards the steady $sS_yA_z$ regime, qualitatively similar to $W=1.2$. 
Further increasing $Re$, the flow remains steady but recovers the top/bottom symmetry and loses the left/right one, i.e. enters the $sA_yS_z$ regime. Unlike for smaller $W$, this bifurcation is hysteretic, and both the $sS_yA_Z$ and $sA_yS_z$ regimes are simultaneously stable for a small range of $Re \approx 300$; see the top panels of figure \ref{fig:W225_omegax_Re300_Re360} and the WNL analysis in \S\ref{sec:WNL}. 

By increasing the Reynolds number in the $320 \le Re \le 350$ range, the flow asymmetry in the $y$ direction decreases, and at $Re=350$ the flow recovers the left/right planar symmetry; for $Re \approx 360$ the flow is steady and $F_y \approx F_z \approx 0$. 
This symmetrisation is shown in figure \ref{fig:W225_omegax_Re300_Re360}.
Like for $W=1.2$ (figure \ref{fig:omegax_P1a_P1b}), at $Re=300$  a pair of counter-rotating streamwise vortices is found in the wake, with eccentricity in the $y$ direction 
(panel $b$). 
As $Re$ increases, a new pair of counter rotating vortices arises, 
becomes stronger and progressively pushes the other pair away (panel $c$). 
Eventually, at $Re \approx 360$ the two pairs of vortices are placed symmetrically with respect to the $z$ axis, and  have the same intensity (panel $d$). 
At this $Re$ the flow exhibits again both top/bottom and left/right planar symmetries ($sS_yS_z$ regime). 
This does not happen for $W=1.2$, as two pairs of streamwise vortices of characteristic size $H=1$ cannot be simultaneously accommodated at the TE when $W<2$.

In passing, we note that in the steady $sS_yA_z$, $sA_yS_z$ and $sS_yS_z$ regimes a decreasing power law fits 
well the relation between $F_x$ and $Re$, being $F_x \sim Re^{\alpha}$ with $\alpha=-0.2905$. However, the agreement does not extend to the unsteady regime, unlike $W=1.2$. 
Also, like for smaller $W$, the decrease of $F_x$ with $Re$ is mainly due to the enlargement of the side recirculating regions (not shown).

\subsubsection{Intermediate $Re$: aperiodic $aS_yA_z$ and periodic $pS_yA_z$ regimes} 
\label{sec:ARy225P2a}
\begin{figure}
\centering
\includegraphics[width=0.49\textwidth]{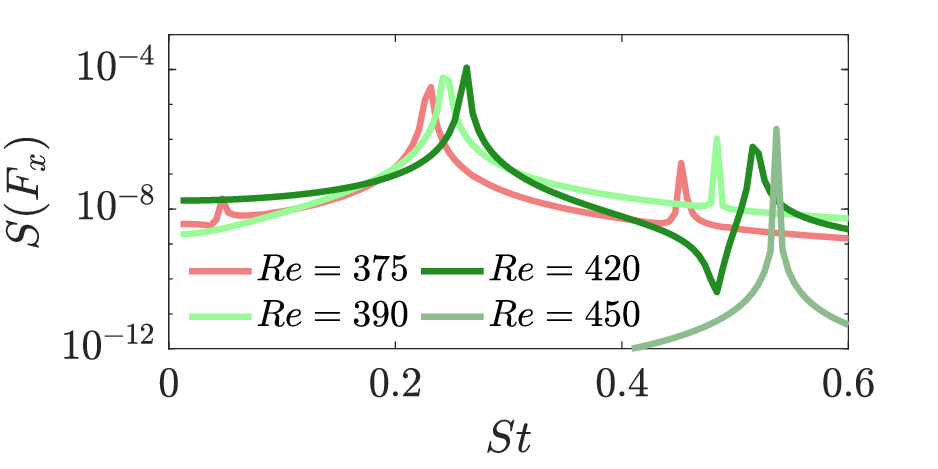}
\includegraphics[width=0.49\textwidth]{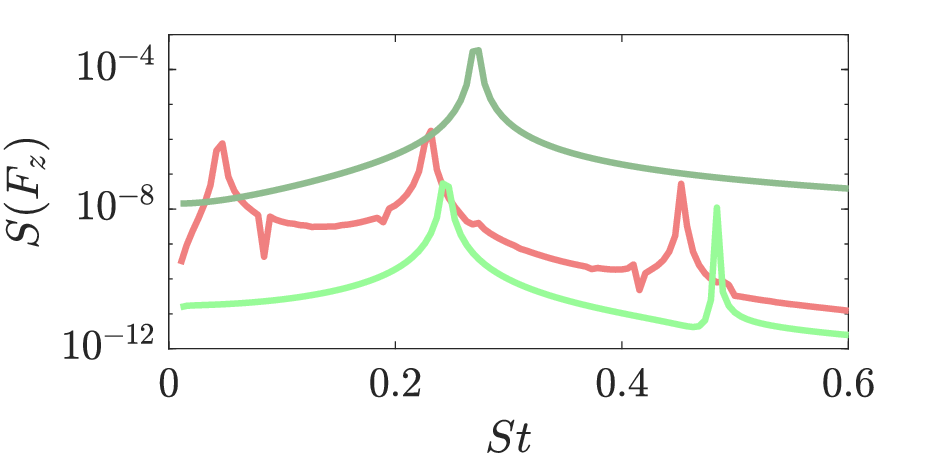}\\
\hspace{-0.75cm}
\begin{tikzpicture}

\definecolor{clr1}{RGB}{18 78 128}
\definecolor{clr2}{RGB}{89 165 216}
\definecolor{clr3}{RGB}{145 229 246}
\definecolor{clr4}{RGB}{34 139 34}
\definecolor{clr5}{RGB}{255 77 109}
\definecolor{clr6}{RGB}{201 24 74}
\definecolor{clr7}{RGB}{128 15 47}
\definecolor{clr8}{RGB}{174 32 18}
\definecolor{clr9}{RGB}{155 34 38}
\definecolor{clr10}{RGB}{46 139 87}
\definecolor{clr11}{RGB}{240 128 128}
\definecolor{clr12}{RGB}{152 251 152}
\definecolor{clr13}{RGB}{143 188 143}
\definecolor{clr14}{RGB}{154 205 50}

\begin{axis}[%
width=0.355\textwidth,
height=0.15\textwidth,
scale only axis,
xmin=374,
xmax=461,
ymin=0.0,
ymax=0.54,
xtick={380,400,420,440,460},
xticklabels=\empty,
ytick={0.00,0.20,0.40},
yticklabels={0.00,0.20,0.40},
ylabel={$St_{F_x}$},
ylabel style={at={(0.05,0.5)}},
axis background/.style={fill=white},
legend columns=3,transpose legend,
legend style={at={(0.99,0.25)}, anchor=east, legend cell align=left, align=left, fill=none, draw=none}
]

\addplot [color=black,solid,draw=none,mark=*,mark options={scale=1.4,black,fill=clr11}]
  table[row sep=crcr]{%
  375.0000    0.22391\\
  380.0000    0.23080\\
  375.0000    0.03924\\
  380.0000    0.05077 \\
  375.0000    0.45233\\
  380.0000    0.4616 \\    
};

\addplot [color=black,solid,draw=none,mark=*,mark options={scale=1.4,black,fill=clr12}]
  table[row sep=crcr]{%
  385.0000    0.2346 \\
  390.0000    0.2395 \\
  385.0000    0.4692 \\
  390.0000    0.4790 \\  
};

\addplot [color=black,solid,draw=none,mark=*,mark options={scale=1.4,black,fill=clr4}]
  table[row sep=crcr]{%
  395.0000    0.24327 \\
  400.0000    0.24668 \\
  420.0000    0.25571 \\
  395.0000    0.48650 \\
  400.0000    0.49330 \\
  420.0000    0.51147 \\  
};

\addplot [color=black,solid,draw=none,mark=*,mark options={scale=1.4,black,fill=clr13}]
  table[row sep=crcr]{%
  440.0000    0.53085 \\
  450.0000    0.53107 \\
  460.0000    0.53123 \\
};

\end{axis}

\end{tikzpicture}%
\hspace{0.45cm}
\begin{tikzpicture}

\definecolor{clr1}{RGB}{18 78 128}
\definecolor{clr2}{RGB}{89 165 216}
\definecolor{clr3}{RGB}{145 229 246}
\definecolor{clr4}{RGB}{34 139 34}
\definecolor{clr5}{RGB}{255 77 109}
\definecolor{clr6}{RGB}{201 24 74}
\definecolor{clr7}{RGB}{128 15 47}
\definecolor{clr8}{RGB}{174 32 18}
\definecolor{clr9}{RGB}{155 34 38}
\definecolor{clr10}{RGB}{46 139 87}
\definecolor{clr11}{RGB}{240 128 128}
\definecolor{clr12}{RGB}{152 251 152}
\definecolor{clr13}{RGB}{143 188 143}
\definecolor{clr14}{RGB}{154 205 50}

\begin{axis}[%
width=0.355\textwidth,
height=0.15\textwidth,
scale only axis,
xmin=374,
xmax=461,
ymin=0.00,
ymax=0.54,
xtick={380,400,420,440,460},
xticklabels=\empty,
ytick={0.00,0.20,0.40},
yticklabels={0.00,0.20,0.40},
ylabel={$St_{F_z}$},
ylabel style={at={(0.05,0.5)}},
axis background/.style={fill=white},
legend columns=3,transpose legend,
legend style={at={(0.99,0.25)}, anchor=east, legend cell align=left, align=left, fill=none, draw=none}
]

\addplot [color=black,solid,draw=none,mark=*,mark options={scale=1.4,black,fill=clr11}]
  table[row sep=crcr]{
  375.0000    0.03924\\
  380.0000    0.05077 \\
  375.0000    0.22391\\
  380.0000    0.23080\\
  375.0000    0.45233\\
  380.0000    0.4616 \\   
};

\addplot [color=black,solid,draw=none,mark=*,mark options={scale=1.4,black,fill=clr12}]
  table[row sep=crcr]{%
  385.0000    0.2346 \\
  390.0000    0.2395 \\
  385.0000    0.4692 \\
  390.0000    0.4790 \\  
};

\addplot [color=black,solid,draw=none,mark=*,mark options={scale=1.4,black,fill=clr13}]
  table[row sep=crcr]{%
  440.0000    0.26543 \\
  450.0000    0.26554 \\
  460.0000    0.26562 \\
};

\end{axis}

\end{tikzpicture}%
\\
\hspace{-0.5cm}
\begin{tikzpicture}

\definecolor{clr1}{RGB}{18 78 128}
\definecolor{clr2}{RGB}{89 165 216}
\definecolor{clr3}{RGB}{145 229 246}
\definecolor{clr4}{RGB}{34 139 34}
\definecolor{clr5}{RGB}{255 77 109}
\definecolor{clr6}{RGB}{201 24 74}
\definecolor{clr7}{RGB}{128 15 47}
\definecolor{clr8}{RGB}{174 32 18}
\definecolor{clr9}{RGB}{155 34 38}
\definecolor{clr10}{RGB}{46 139 87}
\definecolor{clr11}{RGB}{240 128 128}
\definecolor{clr12}{RGB}{152 251 152}
\definecolor{clr13}{RGB}{143 188 143}
\definecolor{clr14}{RGB}{154 205 50}

\begin{axis}[%
scaled ticks=false,
width=0.355\textwidth,
height=0.15\textwidth,
scale only axis,
xmin=374,
xmax=461,
ymin=0.75,
ymax=0.8,
xtick={380,400,420,440,460},
ytick={0.74,0.76,0.78,0.8},
yticklabels={0.74,0.76,0.78,0.80},
xlabel={$Re$},
ylabel={$F_x$},
ylabel style={at={(0.05,0.5)}},
axis background/.style={fill=white},
legend columns=3,transpose legend,
legend style={at={(0.99,0.82)}, anchor=east, legend cell align=left, align=left, fill=none, draw=none}
]

\addplot [color=black,solid,draw=none,mark=*,mark options={scale=1.4,black,fill=clr1}]
  table[row sep=crcr]{%
  250.0000    0.88665 \\
  360.0000    0.78739 \\
  370.0000    0.78297 \\
};
\addplot [color=black,solid,draw=none,mark=*,mark options={scale=1.4,black,fill=clr2}]
  table[row sep=crcr]{
  290.0000    0.83512 \\
  300.0000    0.82605 \\
};
\addplot [color=black,solid,draw=none,mark=*,mark options={scale=1.4,black,fill=clr3}]
  table[row sep=crcr]{%
  300.0000    0.82596 \\
  310.0000    0.81706 \\
  320.0000    0.81042 \\
  330.0000    0.80345 \\
  340.0000    0.79682 \\
  345.0000    0.79362 \\
  350.0000    0.79053 \\
};

\addplot [color=black,solid,draw=none,mark=*,mark options={scale=1.4,black,fill=clr11}]
  table[row sep=crcr]{%
  375.0000    0.7814 \\
  380.0000    0.7791 \\
};

\addplot [color=black,solid,draw=none,mark=*,mark options={scale=1.4,black,fill=clr12}]
  table[row sep=crcr]{%
  385.0000    0.7766 \\
  390.0000    0.7748 \\
};

\addplot [color=black,solid,draw=none,mark=*,mark options={scale=1.4,black,fill=clr4}]
  table[row sep=crcr]{%
  395.0000    0.7734 \\
  400.0000    0.7725 \\
  420.0000    0.7682 \\
};

\addplot [color=black,solid,draw=none,mark=*,mark options={scale=1.4,black,fill=clr13}]  
  table[row sep=crcr]{%
  440.0000    0.7670 \\
  450.0000    0.7658 \\
  460.0000    0.7645 \\
};

\addplot [color=black,solid,draw=none,mark=*,mark options={scale=1.4,black,fill=clr14}]
  table[row sep=crcr]{
  465.000     0.7652 \\
  470.000     0.7659 \\
};

\addplot [color=black,solid,draw=none,mark=*,mark options={scale=1.4,black,fill=clr7}]
  table[row sep=crcr]{%
  485.000     0.7670 \\  
  500.0000    0.7660 \\
  550.0000    0.7634 \\
  625.0000    0.7722 \\
  700.0000    0.7862 \\
};

\addplot [color=black,solid,mark=none, line width=0.5, mark options={scale=1.4,black,fill=red!80!black}]
  table[row sep=crcr]{%
  400.0000    0.7554 \\
  400.0000    0.7896 \\
};

\addplot [color=black,solid,mark=none, line width=0.5, mark options={scale=1.4,black,fill=red!80!black}]
  table[row sep=crcr]{%
  460.0000    0.7614 \\
  460.0000    0.7656 \\
};

\addplot [color=black,solid,mark=none, line width=0.5,mark options={scale=1.4,black,fill=red!80!black}]
  table[row sep=crcr]{%
  500.0000    0.7562 \\
  500.0000    0.7758 \\
};

\addplot [color=black,solid,mark=none, line width=0.5,mark options={scale=1.4,black,fill=red!80!black}]
  table[row sep=crcr]{%
  550.0000    0.7473 \\
  550.0000    0.7787 \\
};

\addplot [color=black,dashed,mark=none, line width=0.5,mark options={scale=1.4,black,fill=red!80!black}]
  table[row sep=crcr]{%
  280    0.91 \\
  280    0.78 \\
};

\addplot [color=black,dashed,mark=none, line width=0.5,mark options={scale=1.4,black,fill=red!80!black}]
  table[row sep=crcr]{%
  295    0.74 \\
  295    1 \\
};

\addplot [color=black,dashed,mark=none, line width=0.5,mark options={scale=1.4,black,fill=red!80!black}]
  table[row sep=crcr]{%
  305    0.74 \\
  305    1 \\
};

\addplot [color=black,dashed,mark=none, line width=0.5,mark options={scale=1.4,black,fill=red!80!black}]
  table[row sep=crcr]{%
  355    0.74 \\
  355    1 \\
};

\addplot [color=black,dashed,mark=none, line width=0.5,mark options={scale=1.4,black,fill=red!80!black}]
  table[row sep=crcr]{%
  372    0.74 \\
  372    1 \\
};

\addplot [color=black,dashed,mark=none, line width=0.5,mark options={scale=1.4,black,fill=red!80!black}]
  table[row sep=crcr]{%
  382    0.74 \\
  382    1 \\
};

\addplot [color=black,dashed,mark=none, line width=0.5,mark options={scale=1.4,black,fill=red!80!black}]
  table[row sep=crcr]{%
  392    0.74 \\
  392    0.92 \\
};

\addplot [color=black,dashed,mark=none, line width=0.5,mark options={scale=1.4,black,fill=red!80!black}]
  table[row sep=crcr]{%
  477    0.74 \\
  477    1 \\
};

\addplot [color=black,dashed,mark=none, line width=0.5,mark options={scale=1.4,black,fill=red!80!black}]
  table[row sep=crcr]{%
  462.5    0.74 \\
  462.5    1 \\
};

\addplot [color=black,dashed,mark=none, line width=0.5,mark options={scale=1.4,black,fill=red!80!black}]
  table[row sep=crcr]{%
  435    0.74 \\
  435    1 \\
};

\addplot [color=black,solid,mark=none, line width=0.5,mark options={scale=1.4,black,fill=red!80!black}]
  table[row sep=crcr]{%
  700.0000    0.7678 \\
  700.0000    0.8046 \\
};

\addplot [color=black,solid,mark=none, line width=0.5,mark options={scale=1.4,black,fill=red!80!black}]
  table[row sep=crcr]{%
  375.0000    0.7710 \\
  375.0000    0.7918 \\
};

\addplot [color=black,solid,mark=none, line width=0.5,mark options={scale=1.4,black,fill=red!80!black}]
  table[row sep=crcr]{%
  380.0000    0.7652 \\
  380.0000    0.7929 \\
};

\addplot [color=black,solid,mark=none, line width=0.5,mark options={scale=1.4,black,fill=red!80!black}]
  table[row sep=crcr]{%
  385.0000    0.7613 \\
  385.0000    0.7920 \\
};

\addplot [color=black,solid,mark=none, line width=0.5,mark options={scale=1.4,black,fill=red!80!black}]
  table[row sep=crcr]{%
  390.0000    0.7586 \\
  390.0000    0.7910 \\
};

\addplot [color=black,solid,mark=none, line width=0.5,mark options={scale=1.4,black,fill=red!80!black}]
  table[row sep=crcr]{%
  395.0000    0.7566 \\
  395.0000    0.7901 \\
};

\addplot [color=black,solid,mark=none, line width=0.5,mark options={scale=1.4,black,fill=red!80!black}]
  table[row sep=crcr]{%
  420.0000    0.7507 \\
  420.0000    0.7857 \\
};

\addplot [color=black,solid,mark=none, line width=0.5,mark options={scale=1.4,black,fill=red!80!black}]
  table[row sep=crcr]{%
  440.0000    0.7661 \\
  440.0000    0.7698 \\
};

\addplot [color=black,solid,mark=none, line width=0.5,mark options={scale=1.4,black,fill=red!80!black}]
  table[row sep=crcr]{%
  450.0000    0.7638 \\
  450.0000    0.7678 \\
};

\addplot [color=black,solid,mark=none, line width=0.5,mark options={scale=1.4,black,fill=red!80!black}]
  table[row sep=crcr]{%
  465.0000    0.7630 \\
  465.0000    0.7673 \\
};

\addplot [color=black,solid,mark=none, line width=0.5,mark options={scale=1.4,black,fill=red!80!black}]
  table[row sep=crcr]{%
  470.0000    0.7638 \\
  470.0000    0.7678 \\
};

\addplot [color=black,solid,mark=none, line width=0.5,mark options={scale=1.4,black,fill=red!80!black}]
  table[row sep=crcr]{%
  485.0000    0.7644 \\
  485.0000    0.7696 \\
};

\addplot [color=black,solid,mark=none, line width=0.5,mark options={scale=1.4,black,fill=red!80!black}]
  table[row sep=crcr]{%
  625.0000    0.7564 \\
  625.0000    0.7881 \\
};

\end{axis}

\end{tikzpicture}%
\hspace{0.2cm}
\input{./figure/Ary225/FyFz-Re-p-2.tex}
\caption{Forces and frequency content for $L=5$, $W=2.25$ and $375 \le Re \le 450$. 
Top panels: frequency spectra for $F_x$ (left) and $F_z$ (right) for $Re=375, 390, 420$ and $450$. 
For $Re=420$, $S(F_z)$ is not visible as fluctuations in $F_z$ are almost null.
Central panels: main frequencies  for $F_x$ (left) and $F_z$ (right).
Bottom panels: zoom of  figure \ref{fig:AR225_forces} in the range $370 \le Re \le 460$. 
For $440 \le Re \le 460$, the bars are not visible as the oscillations of $F_x$ are small.}
\label{fig:W225_spec_1}
\end{figure}

\begin{figure}
\centering
\includegraphics[trim={0 60 0 60},clip,width=0.49\textwidth]{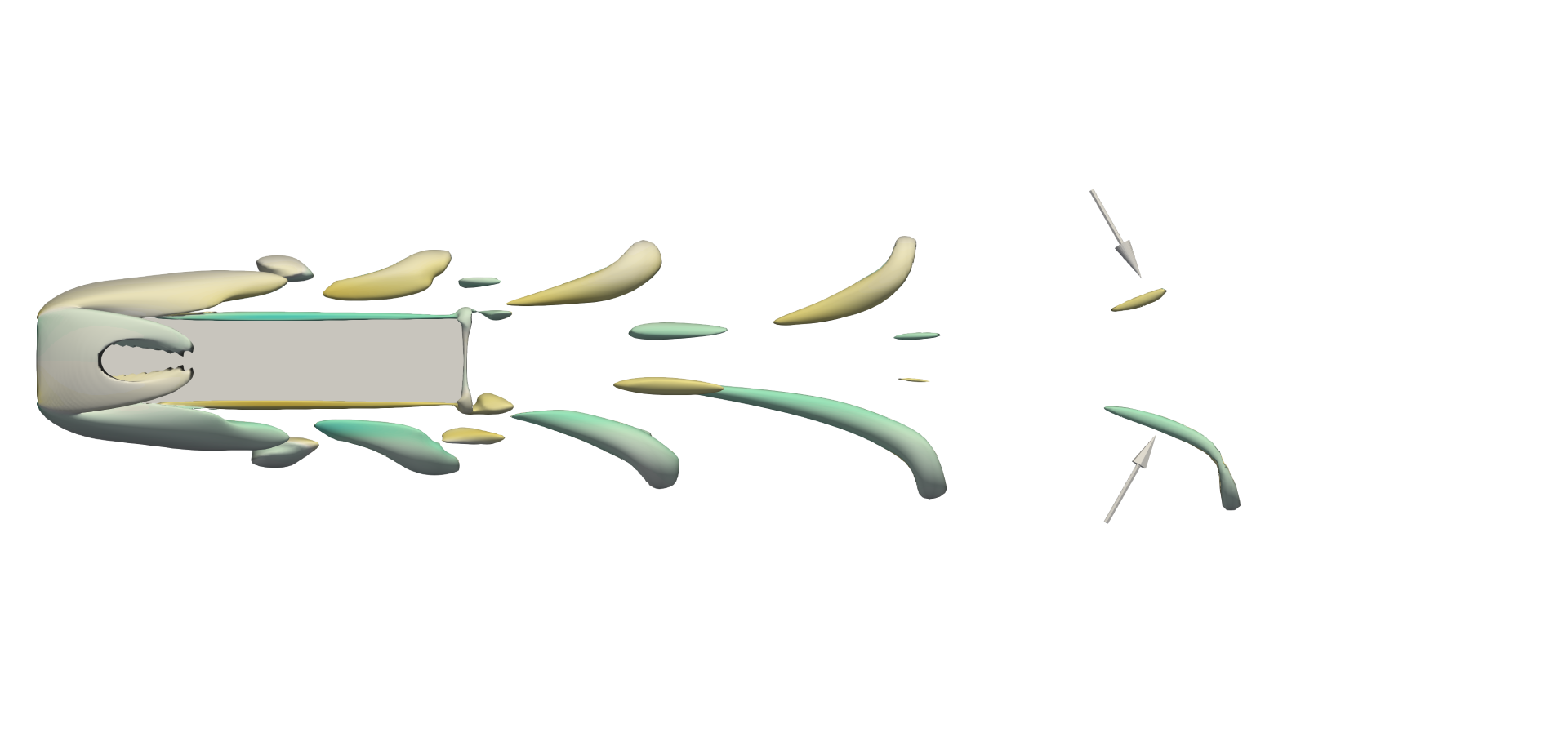}
\includegraphics[width=0.24\textwidth]{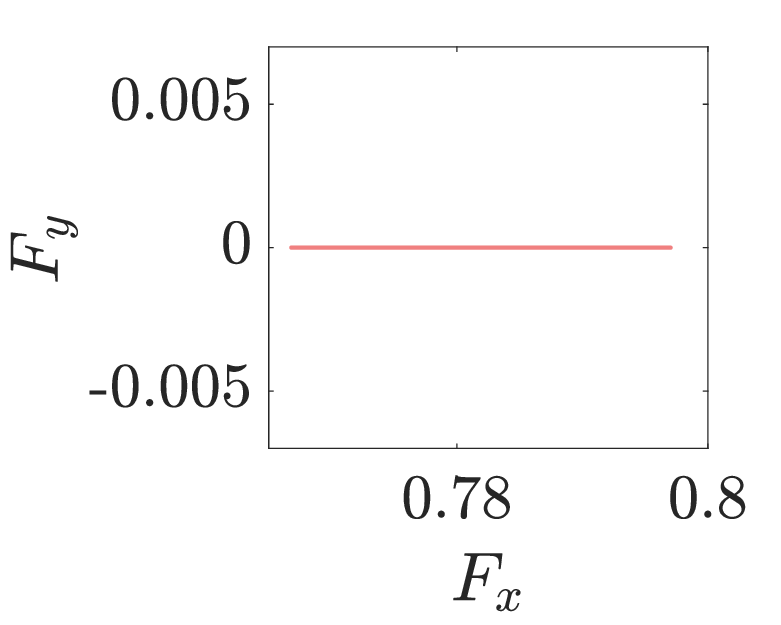}
\includegraphics[width=0.24\textwidth]{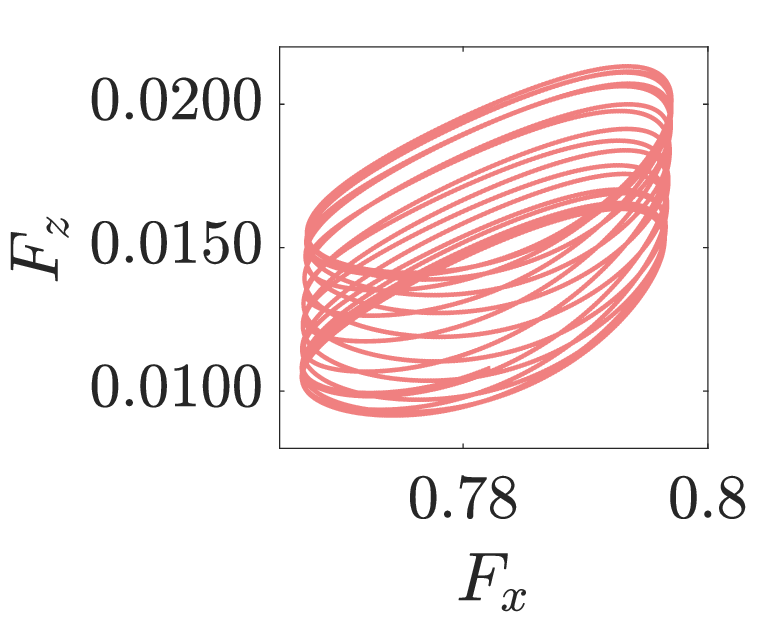}
\includegraphics[trim={0 60 0 60},clip,width=0.49\textwidth]{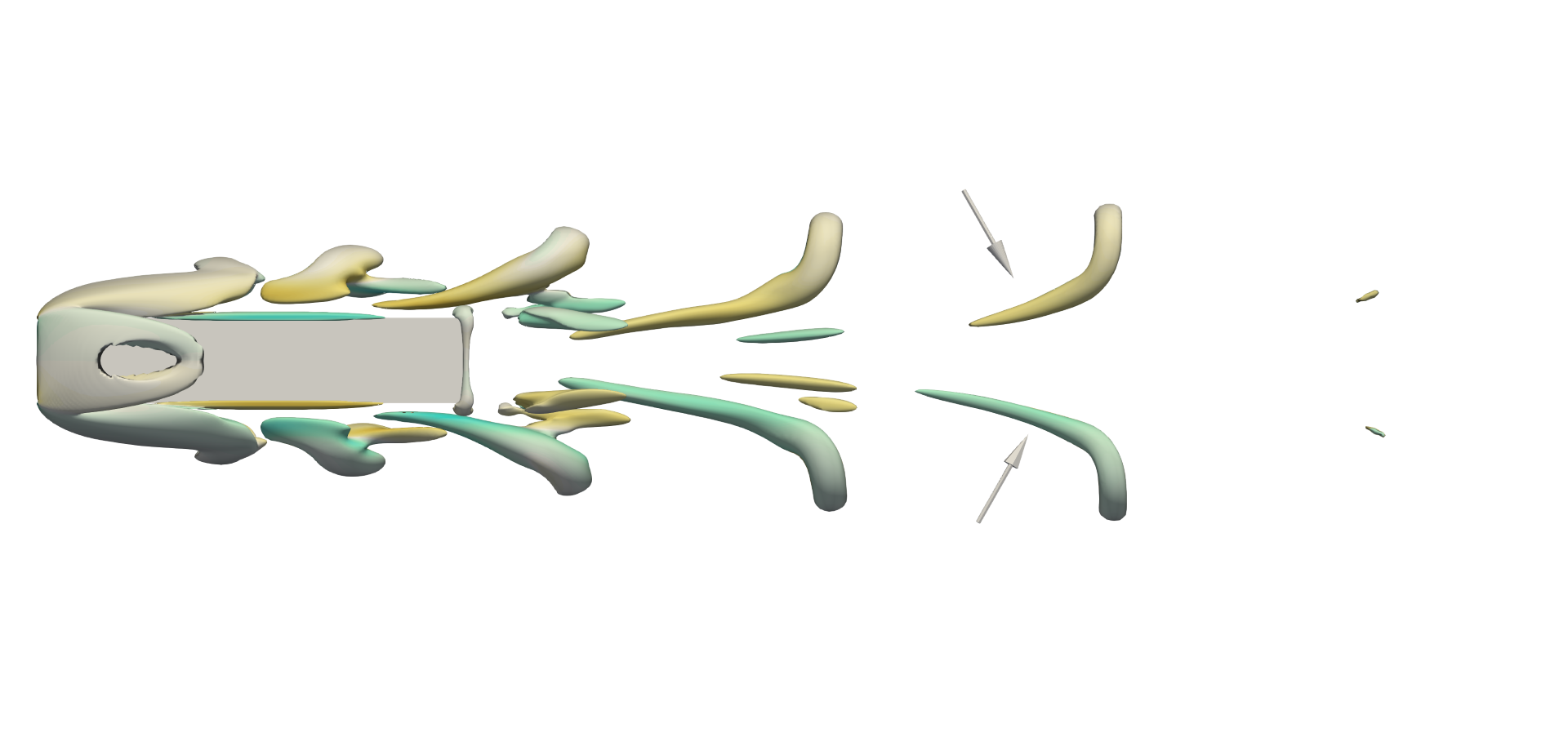}
\includegraphics[width=0.24\textwidth]{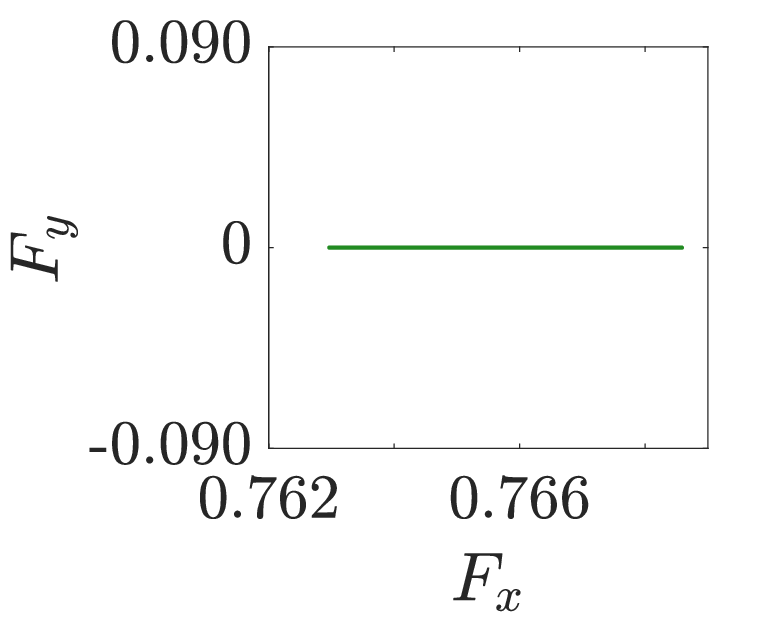}
\includegraphics[width=0.24\textwidth]{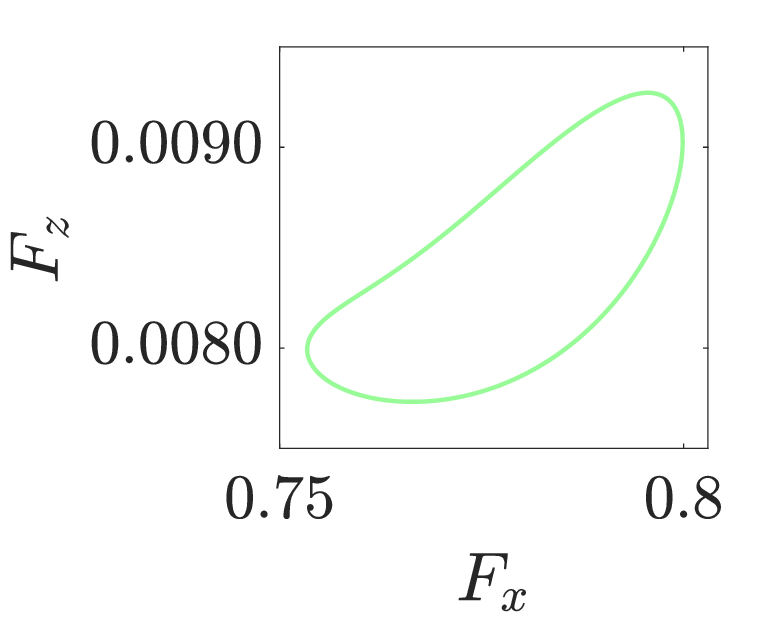}
\includegraphics[trim={0 60 0 60},clip,width=0.49\textwidth]{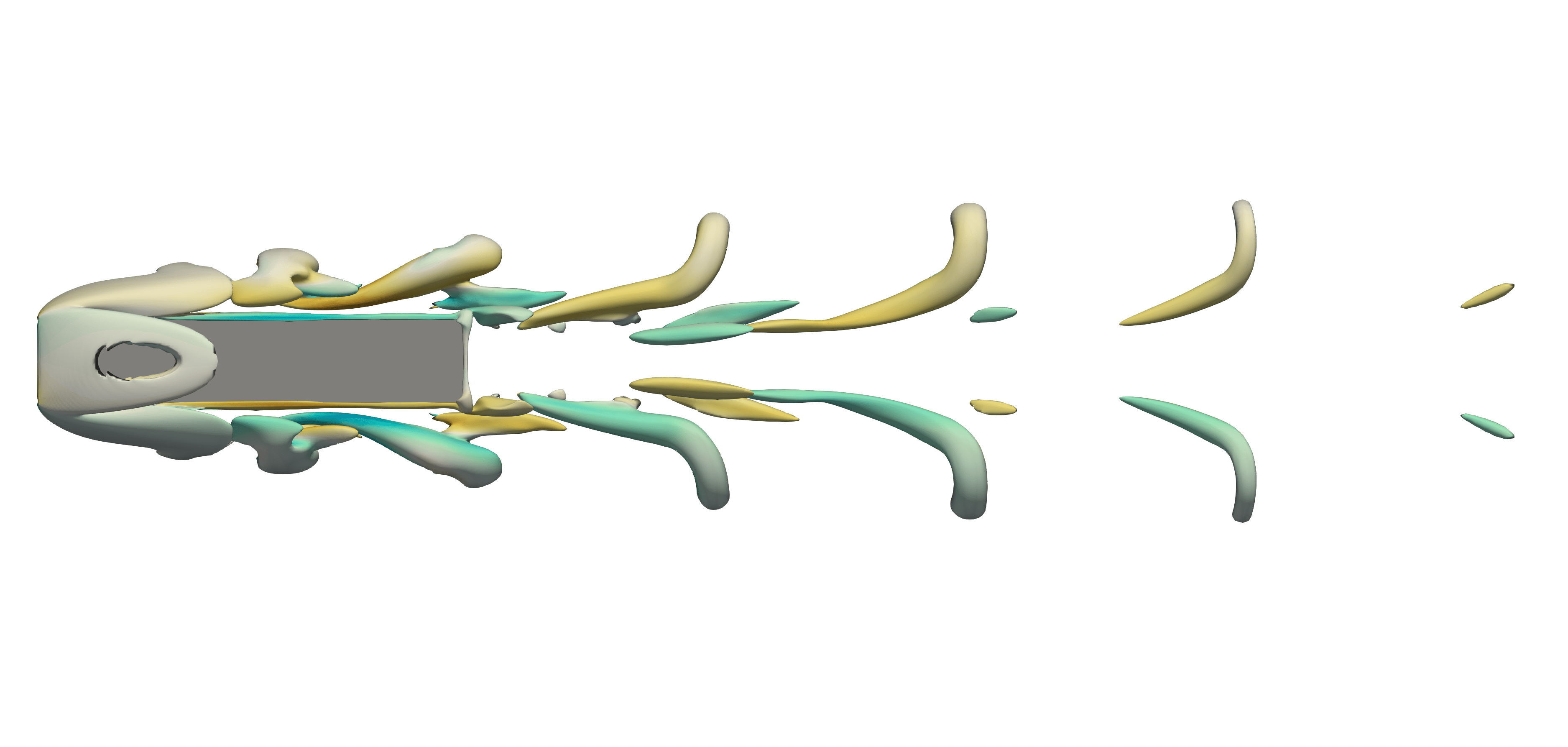}
\includegraphics[width=0.24\textwidth]{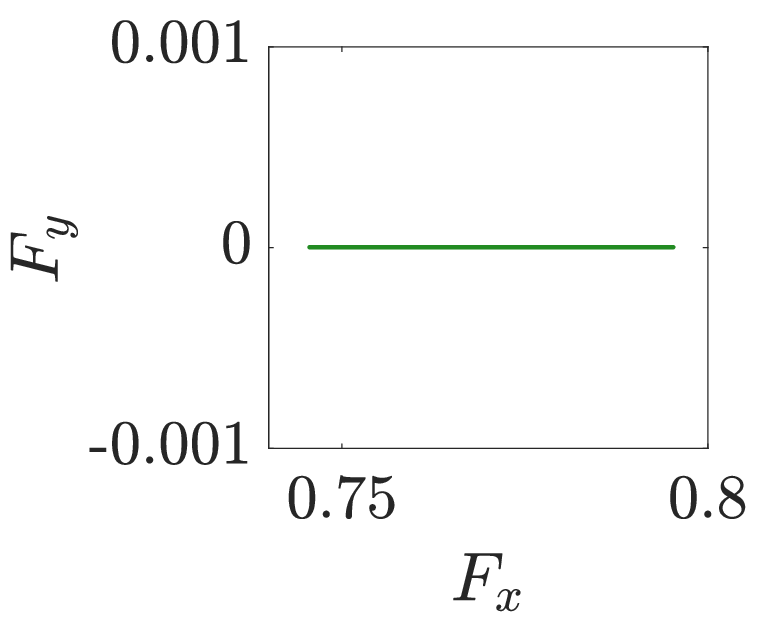}
\includegraphics[width=0.24\textwidth]{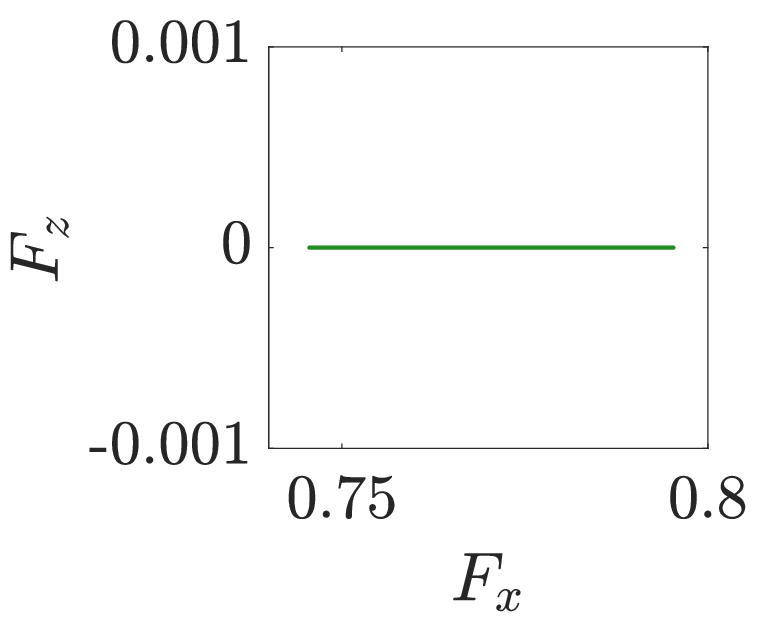}
\includegraphics[trim={0 60 0 60},clip,width=0.49\textwidth]{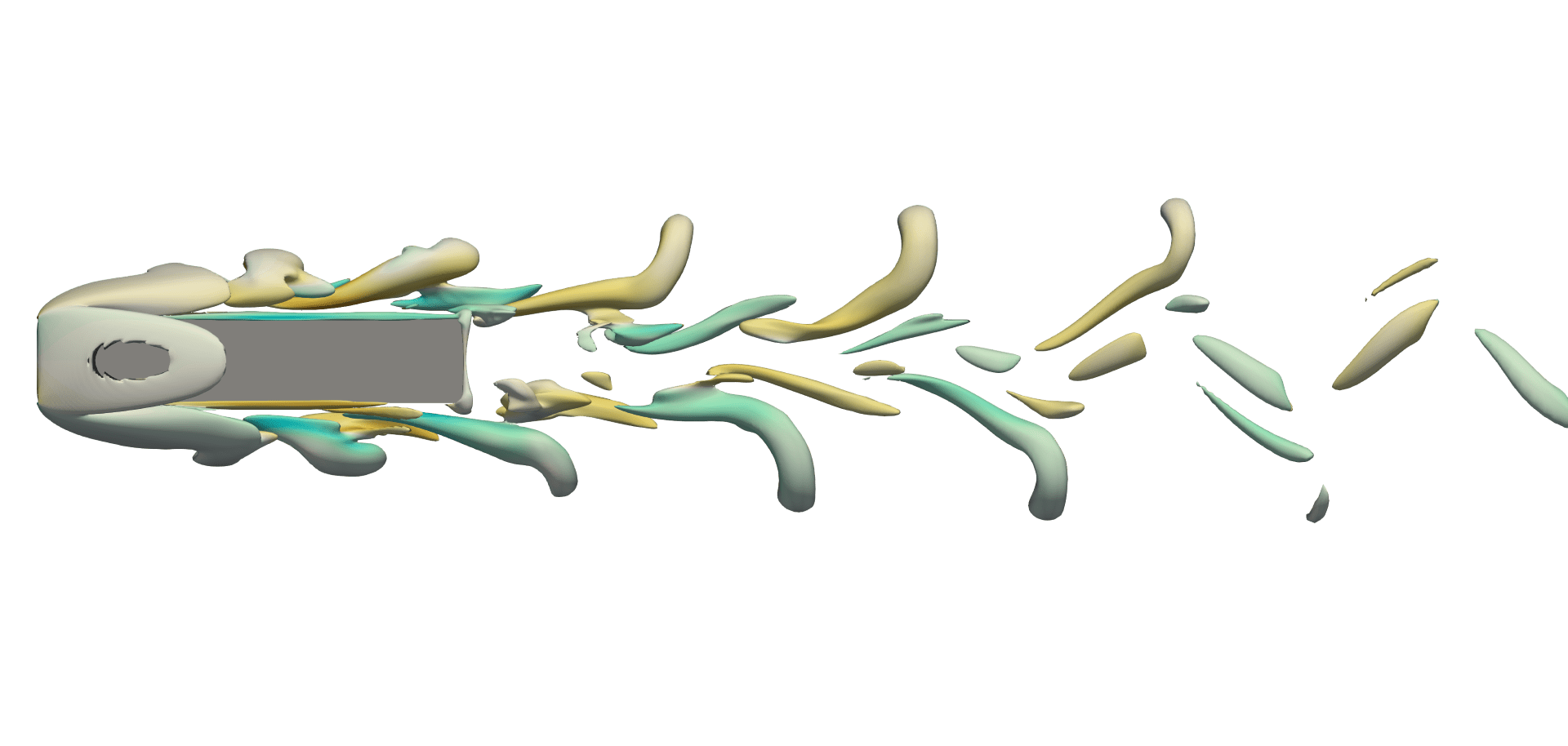}
\includegraphics[width=0.24\textwidth]{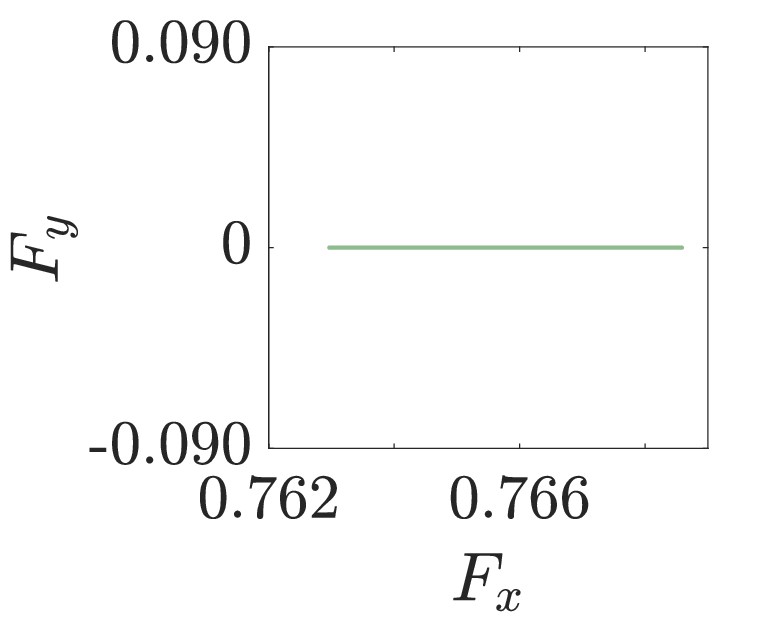}
\includegraphics[width=0.24\textwidth]{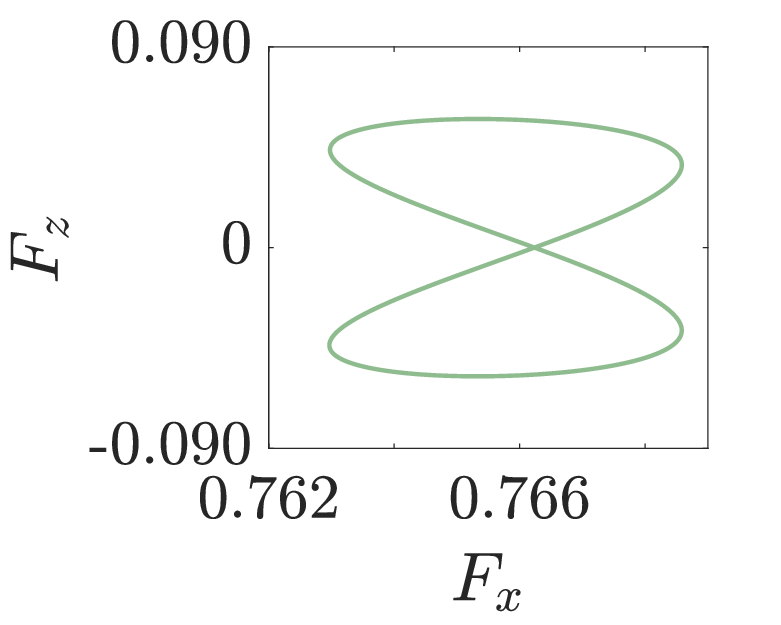}
\caption{Structure of the flow for $L=5$, $W=2.25$ and $Re=375$, $390$, $420$ and $450$ (top to bottom). 
Left: side view of instantaneous isosurfaces of $\lambda_2=-0.25$ coloured with $-1 \le \omega_x \le 1$. Right: force diagrams $F_y-F_x$ and $F_z-F_x$. 
The lack of top/bottom symmetry in the two top left panels is highlighted with arrows.}
\label{fig:W225_FlowStr_375_470}
\end{figure}

At $Re \approx 375$ the flow becomes unsteady and enters different regimes in short succession. 
This is conveniently visualised in figures \ref{fig:W225_spec_1} and \ref{fig:W225_FlowStr_375_470}, where the frequency content and the structure of the flow are shown for $375 \le Re \le 470$. 

For $375 \le Re \le 380$, the flow approaches the aperiodic $aS_yA_z$ regime: unlike for smaller $W$, the flow oscillates about a $S_yA_z$ state that retains the left/right symmetry and breaks the top/bottom one. 
With two non commensurate frequencies, $St \approx 0.05$ and $0.23$, the attractor draws a torus in the phase space. 
As shown in the top left panel of figure \ref{fig:W225_FlowStr_375_470}, at this $Re$ the flow behaviour is the result of the superposition of two different modes: 
(i) asymmetric oscillating mode of the wake in the $z$ direction ($St \approx 0.05$) and 
(ii) symmetric in-phase shedding of HVs from the top and bottom LE shear layers ($St \approx 0.23$). 
For $Re \approx 385$, instead, the wake oscillation mode stabilises and the flow approaches the periodic $pS_yA_z$ regime. The wake remains deflected in the $z$ direction and the unsteadiness is only due to the in-phase shedding of HVs, with a frequency $St \approx 0.24$ that slightly increases with $Re$ (see bottom panel of figure \ref{fig:W225_spec_1}). A limit cycle replaces the torus in the phase space (figure \ref{fig:W225_FlowStr_375_470}). 

\begin{figure}
\centering
\begin{tikzpicture}

\definecolor{clr1}{RGB}{18 78 128}
\definecolor{clr2}{RGB}{89 165 216}
\definecolor{clr3}{RGB}{145 229 246}
\definecolor{clr4}{RGB}{34 139 34}
\definecolor{clr5}{RGB}{255 77 109}
\definecolor{clr6}{RGB}{201 24 74}
\definecolor{clr7}{RGB}{128 15 47}
\definecolor{clr8}{RGB}{174 32 18}
\definecolor{clr9}{RGB}{155 34 38}
\definecolor{clr10}{RGB}{46 139 87}
\definecolor{clr11}{RGB}{240 128 128}
\definecolor{clr12}{RGB}{152 251 152}

\begin{axis}[%
width=0.35\textwidth,
height=0.15\textwidth,
scale only axis,
ymode = log,
xmin=1,
xmax=8,
ymin=0.0,
ymax=0.7,
xtick={1,2,3,4,5,6,7,8},
xlabel={$mode$},
ylabel={$\lambda/\sum \lambda$},
ylabel style={at={(0.05,0.5)}},
axis background/.style={fill=white},
legend columns=3,transpose legend,
legend style={at={(0.99,0.25)}, anchor=east, legend cell align=left, align=left, fill=none, draw=none}
]

\addplot [color=black,solid,draw=none,mark=*,mark options={scale=1.4,black,fill=clr11}]
  table[row sep=crcr]{%
  1.0000 0.4298 \\
  2.0000 0.4298 \\
  3.0000 0.0448 \\
  4.0000 0.0446 \\
  5.0000 0.0160 \\
  6.0000 0.0148 \\
  7.0000 0.0069 \\
  8.0000 0.0069 \\
};

\end{axis}

\end{tikzpicture}%
\begin{tikzpicture}

\definecolor{clr1}{RGB}{18 78 128}
\definecolor{clr2}{RGB}{89 165 216}
\definecolor{clr3}{RGB}{145 229 246}
\definecolor{clr4}{RGB}{34 139 34}
\definecolor{clr5}{RGB}{255 77 109}
\definecolor{clr6}{RGB}{201 24 74}
\definecolor{clr7}{RGB}{128 15 47}
\definecolor{clr8}{RGB}{174 32 18}
\definecolor{clr9}{RGB}{155 34 38}
\definecolor{clr10}{RGB}{46 139 87}
\definecolor{clr11}{RGB}{240 128 128}
\definecolor{clr12}{RGB}{152 251 152}

\begin{axis}[%
width=0.35\textwidth,
height=0.15\textwidth,
scale only axis,
xmin=1,
xmax=8,
ymin=0.0,
ymax=0.7,
xtick={1,2,3,4,5,6,7,8},
xlabel={$mode$},
ylabel={$St$},
ylabel style={at={(0.05,0.5)}},
axis background/.style={fill=white},
legend columns=3,transpose legend,
legend style={at={(0.99,0.25)}, anchor=east, legend cell align=left, align=left, fill=none, draw=none}
]

\addplot[red, dashed, domain=1:8] {0.23};
\addplot[red, dashed, domain=1:8] {0.055};

\addplot [color=black,solid,draw=none,mark=*,mark options={scale=1.4,black,fill=clr11}]
  table[row sep=crcr]{%
  1.0000 0.23 \\
  2.0000 0.23 \\
  3.0000 0.46 \\
  4.0000 0.46 \\
  5.0000 0.0508 \\
  6.0000 0.0508 \\
  7.0000 0.6861 \\
  8.0000 0.6861 \\
};

\end{axis}

\end{tikzpicture}%
\includegraphics[trim={0 20 0 20},clip,width=0.49\textwidth]{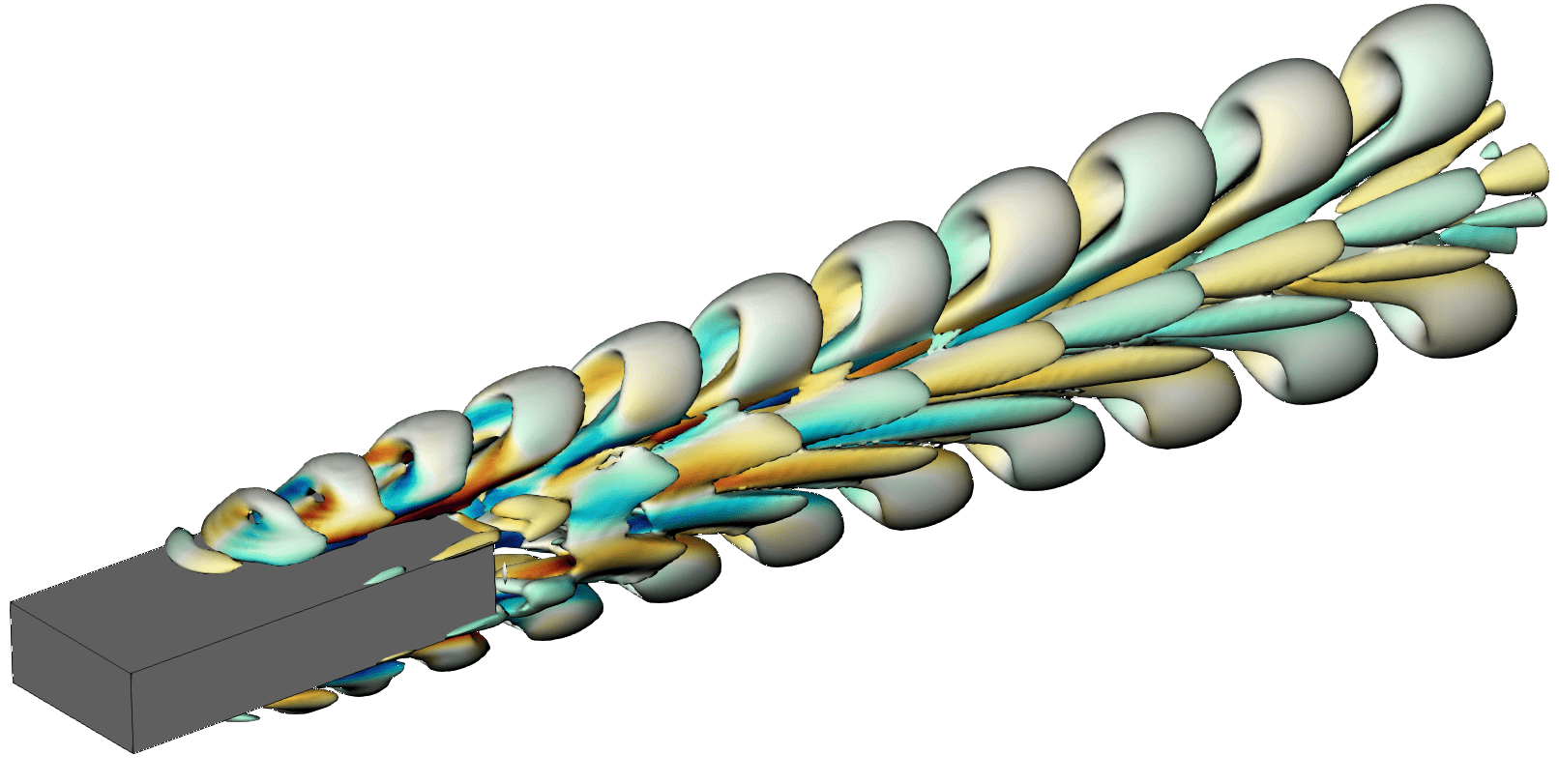}
\includegraphics[trim={0 20 0 20},clip,width=0.49\textwidth]{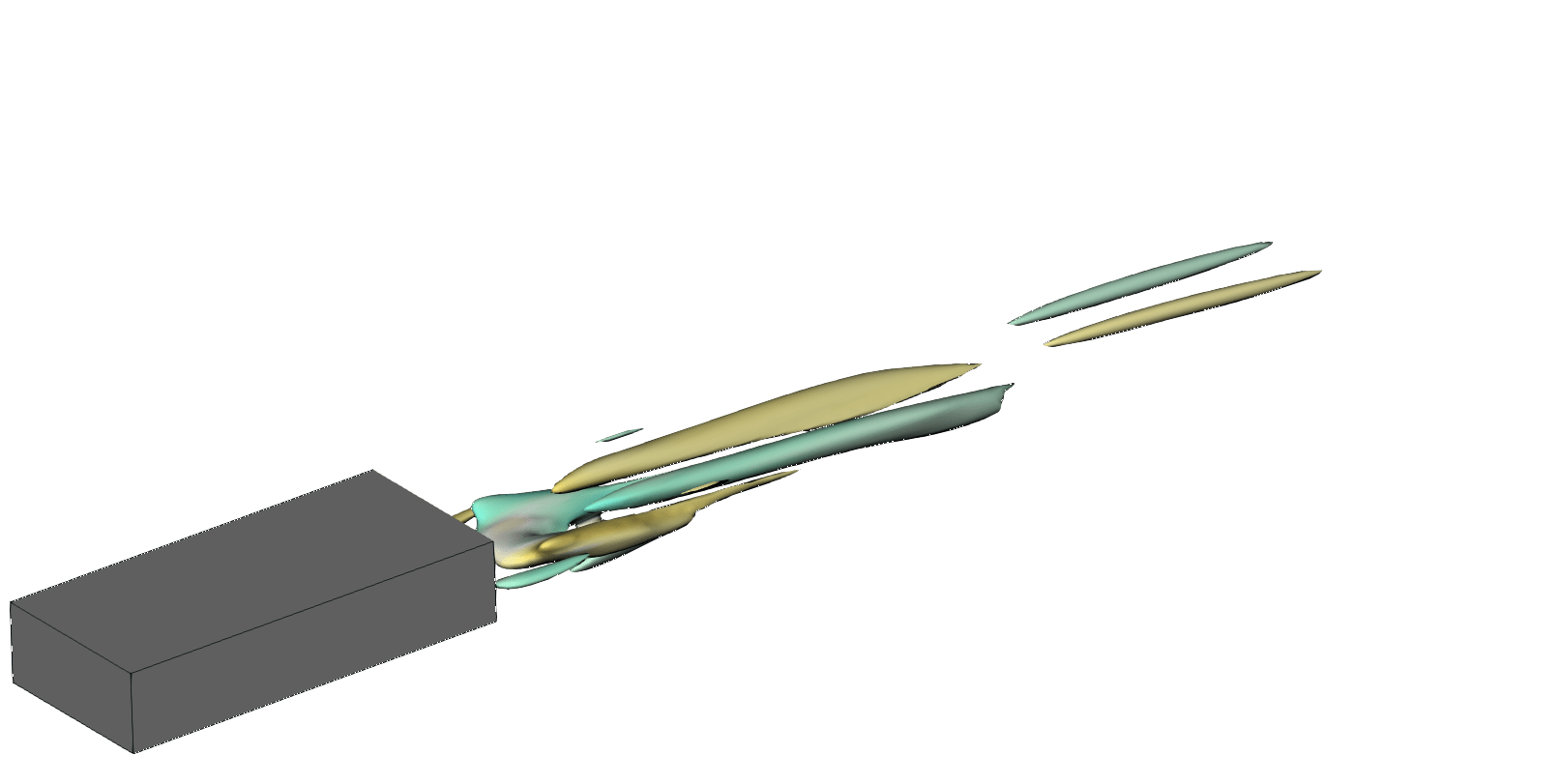}
\caption{POD analysis for $L=5$, $W=2.25$ and $Re=380$. 
Top left: energy fraction $E_j = \lambda_j/\sum\lambda_j$ of the first $8$ POD modes.
Top right: frequency associated with these modes.
Dashed lines: main DNS frequencies as in figure \ref{fig:W225_spec_1}.
Bottom: POD modes 1 (left) and 5 (right). Isosurfaces of $\lambda_2$ (arbitrary value) coloured by streamwise vorticity.
}
\label{fig:W225_POD}
\end{figure}
%
We use  POD to examine the structure of the flow at $Re=380$. 
In figure \ref{fig:W225_POD}, we separate the modes responsible for the two non commensurate frequencies 
detected in the frequency spectra. 
POD mode $1$ is associated with $St \approx 0.23$, and its spatial structure shows a $S_yS_z$ shedding of HVs from the top and bottom sides of the prism. 
Mode $5$, instead, matches the $St \approx 0.05$ frequency, with a $S_yA_z$ spatial structure that agrees with an asymmetric oscillating mode of the wake in the $z$ direction. 
This suggests that the aperiodic $aS_yA_z$ regime found at $Re \approx 380$ is the result of a superposition of an oscillatory $S_yA_z$ mode of the wake, 
and an unsteady $S_yS_z$ mode of the LE shear layers. Interestingly, this interpretation is supported by the LSA, which indeed predicts an unsteady $S_yS_z$ mode to become unstable for $Re \gtrapprox 350$ for this geometry, with a frequency that is compatible with $St \approx 0.2$ (figure \ref{fig:neutral_curves_and_omc_vs_L}). 
Also, the neutral curves (figure \ref{fig:neutral_curves_and_omc_vs_L}) show that for $L\le 4$ an unsteady $S_yA_z$ mode of the wake with $St \approx 0.06-0.07$ becomes unstable, but for a small range of $Re$ only. For $L \approx 4$, for example, the unstable range is between $330 \lessapprox Re \lessapprox 350$. Recalling that these neutral curves refer to the low-$Re$ $sS_yS_z$ steady base flow, it is possible that due to nonlinear effects, an increase of $Re$ extends this curve to $L=5$, explaining this sudden destabilisation/stabilisation of the wake  mode. 

Figure \ref{fig:W225_POD} shows that at $Re=380$ the energy fraction associated with LE vortex shedding is approximately $25$ times larger than that associated with the asymmetric oscillating mode of the wake; at this $Re$ the flow dynamics is mainly driven by the HVs shed by the LE shear layers. The effect of the Reynolds number on the spatial structure of the POD mode associated with the shedding of HVs from the LE shear layers is further discussed in appendix \S\ref{app:POD_LE_VS} (figure \ref{fig:W225_POD_Re420_Re450}).

\subsubsection{Intermediate $Re$: periodic $pS_yS_zl$ regime}
\label{sec:pSySzl}

As $Re$ is further increased, the flow 
 recovers once again the top/bottom planar symmetry, and for $390 < Re \le 470$ it is in the periodic and symmetric $pS_yS_zl$ regime (figure \ref{fig:W225_FlowStr_375_470}). In this regime, two distinct behaviours can be further distinguished, based on the fluctuations of $F_z$, characterised by a synchronisation/anti-synchronisation of the vortex shedding from the top and bottom LE shear layers.

For $390 < Re \le 420$ the flow shows an in-phase shedding of HVs from the top and bottom LE shear layers, and the flow \textit{instantaneously} retains the top/bottom and left/right planar symmetries (left panels of figure \ref{fig:W225_FlowStr_375_470}); this is regime $pS_yS_zla$. 
Accordingly, the unbalance of the viscous and pressure forces between the top/bottom and left/right sides is null at all times, leading to  $F_z \approx F_y \approx 0 $. For this regime, we detect the leading flow frequency by inspecting the $S(F_x)$ spectrum. 
To the best of our knowledge, such a synchronous vortex shedding that preserves all spatial symmetries at all times and generates neither lift force nor side force has not been reported.

For $420 < Re \le 470$, instead, the flow remains periodic with  $St \approx 0.24$, but the shedding of vortices from the top and bottom LE shear layers 
is in phase opposition (bottom panels in figure \ref{fig:W225_FlowStr_375_470}); this is regime $pS_yS_zlb$. In this regime the flow retains the top/bottom planar symmetry in an average sense only, and the oscillations of $F_z$ are not null. Accordingly, unlike for $380 \le Re \le 420$, at these $Re$ the POD mode associated with the LE vortex shedding is top/down antisymmetric (see appendix \S\ref{app:POD_LE_VS})

We now consider regime $pS_yS_zla$ and detail the dynamics of the HVs shed from the top and bottom LE shear layers at $Re \approx 400$.
\begin{figure}
\centering
\includegraphics[width=0.49\textwidth]{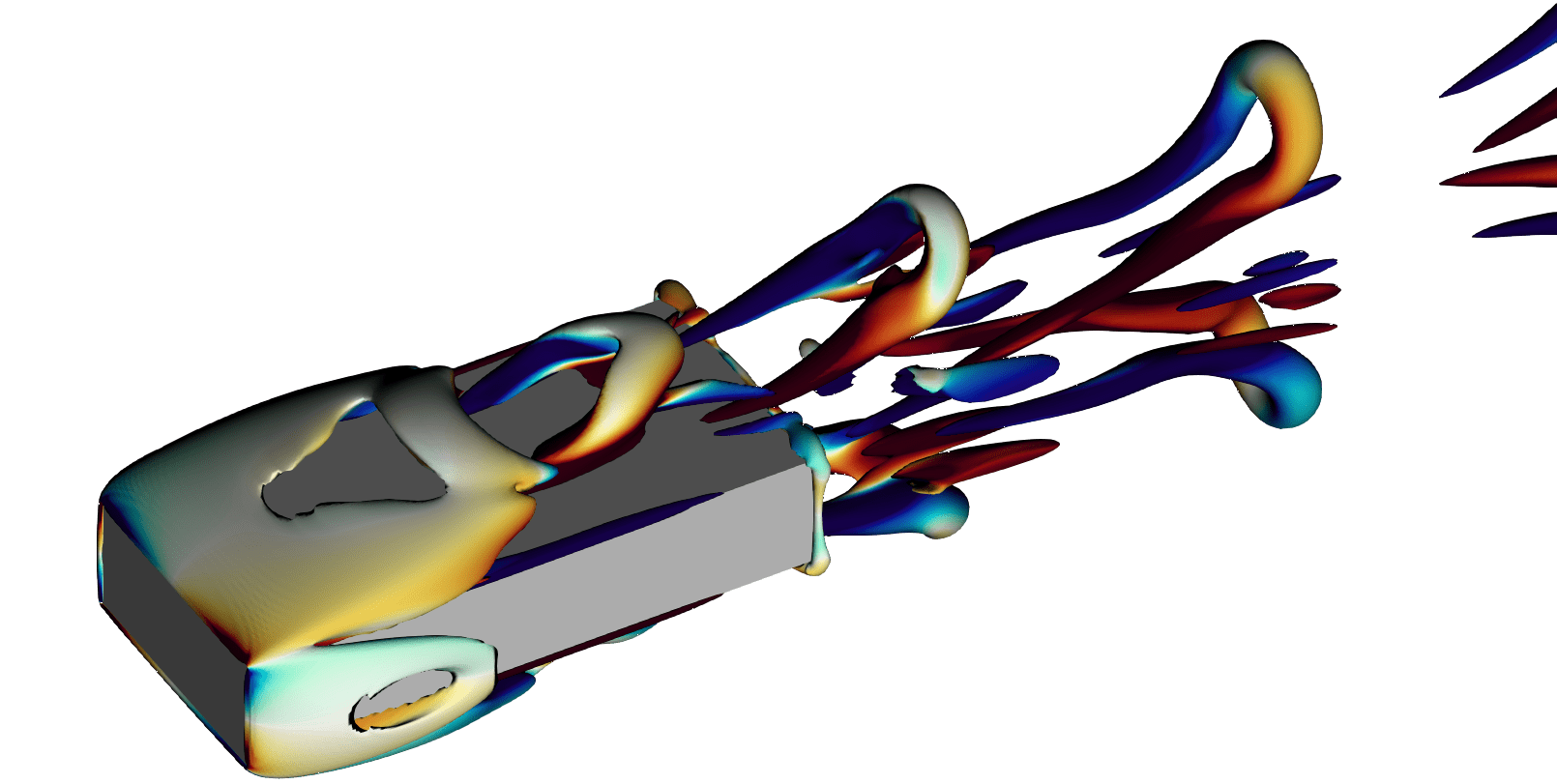}
\includegraphics[width=0.49\textwidth]{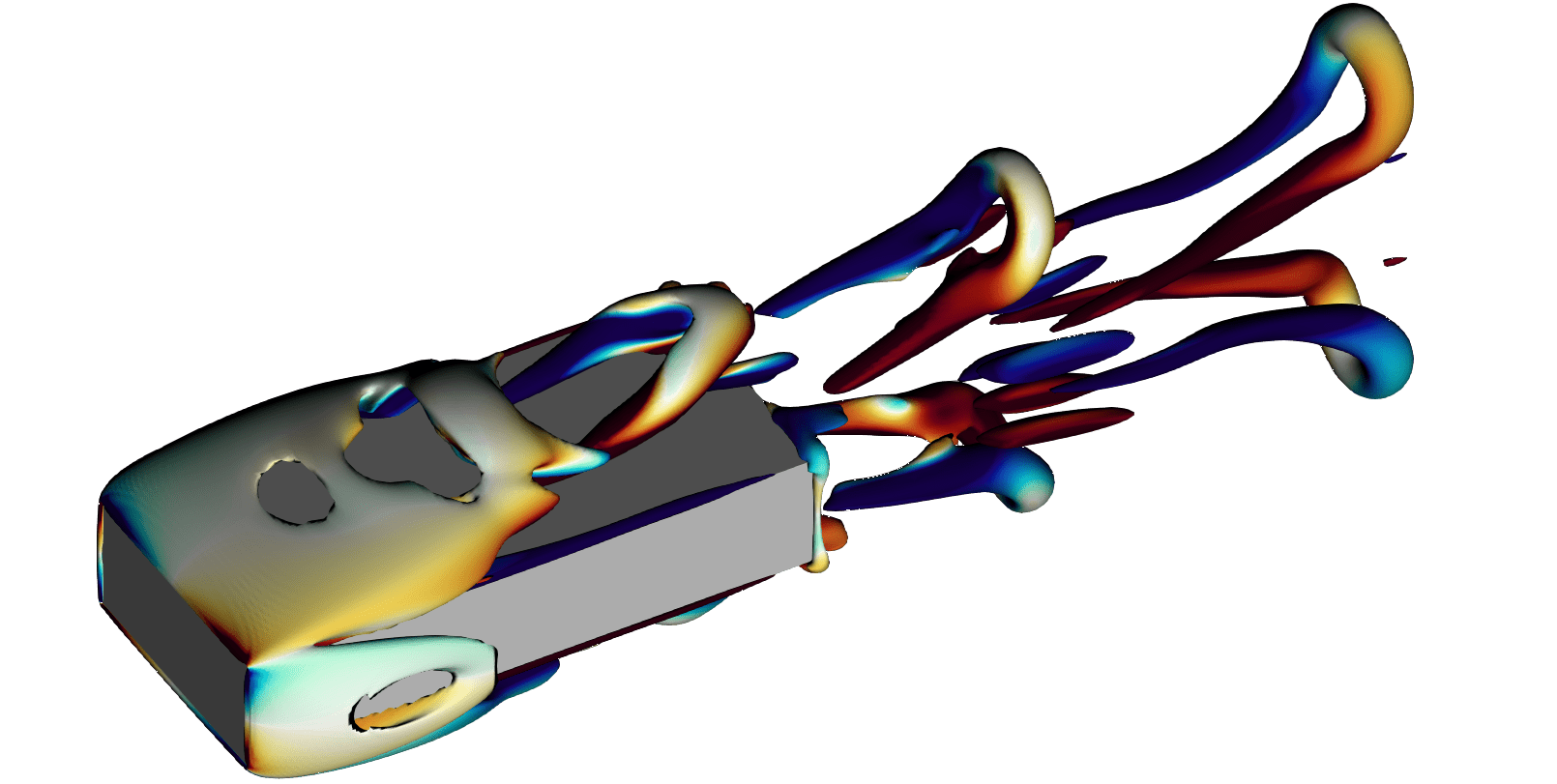}
\includegraphics[width=0.49\textwidth]{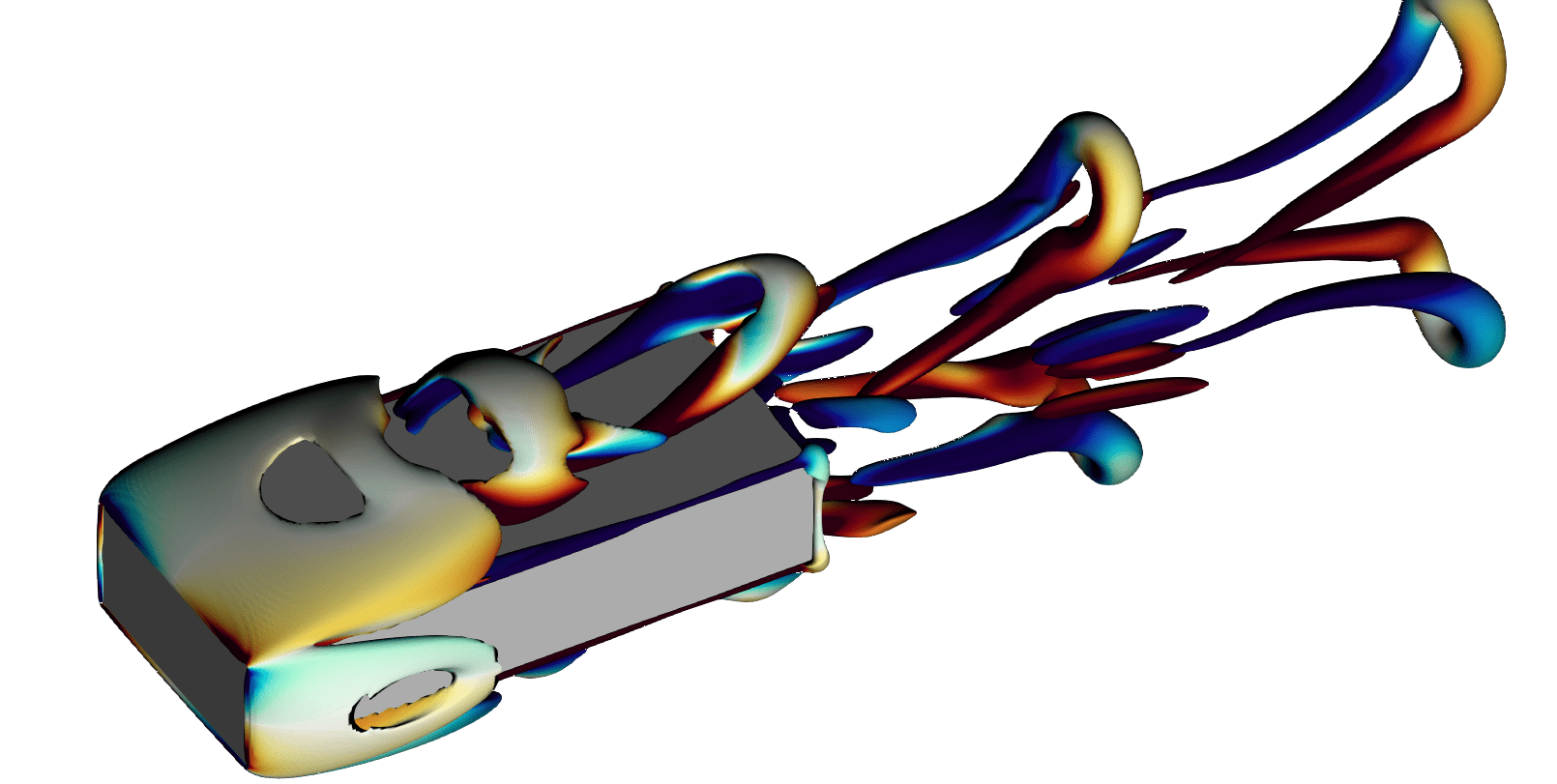}
\includegraphics[width=0.49\textwidth]{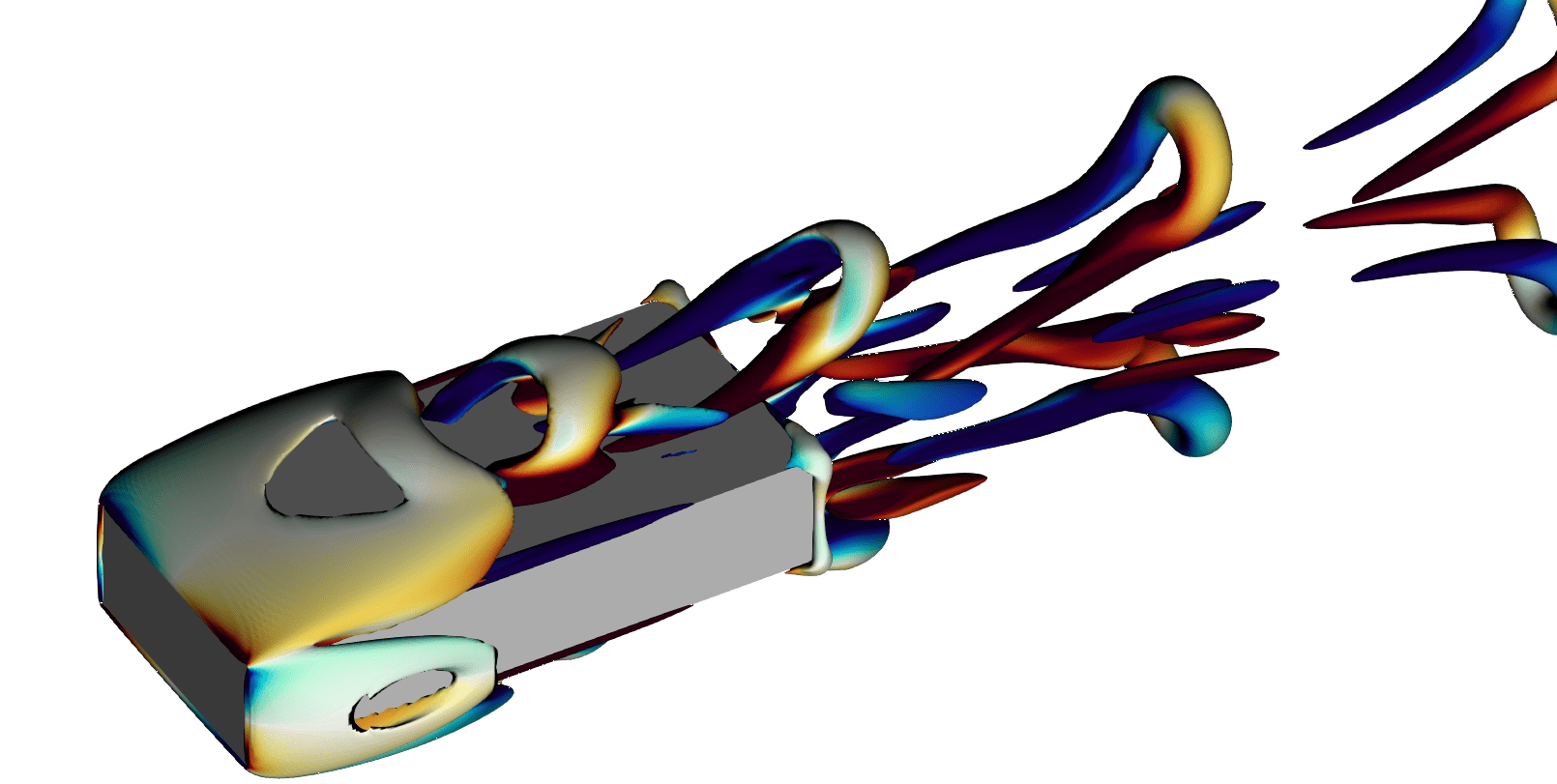}\\
\caption{Flow structures of the doubly symmetric $pS_yS_zla$  regime for $L=5$, $W=2.25$ and $Re=400$:
isosurfaces of $\lambda_2 = -0.05$ coloured by $-1 \le \omega_x \le 1$.
The four snapshots are separated in time by $T/4$, where $T$ is the period of the shedding of hairpin vortices. 
}
\label{fig:AR225_Re400_lambda2}
\end{figure}
Figure \ref{fig:AR225_Re400_lambda2} shows the flow structures at four different instants equispaced in one shedding period $T=1/St$. 
A pair of HVs of opposite vorticity  is shed in phase from the top and bottom LE shear layers, with a spanwise wavelength $\lambda_z$ dictated by the width of the prism $\lambda_z \approx W$. 
The top/bottom LE shear layers, indeed, periodically release tubes of negative/positive spanwise vorticity which, due to the finite width of the prism, are not exactly aligned with the spanwise direction. As such, they are modulated by the vertical and spanwise velocity gradients and form hairpin vortices: 
the central part of the tubes, farther from the wall, is convected downstream faster and forms the HV heads;
the lateral parts, closer to the wall, are convected more slowly and form the HV legs.
Accordingly, the HVs induce low-speed velocity (negative velocity fluctuations) between their legs, and high-speed velocity (positive velocity fluctuations) in the outer region. 
Once generated, the two HVs are shed downstream and continue moving as a pair in the wake, where they progressively move apart in the $z$ direction due to their mutual induction, until they are dissipated by viscosity. 
The shedding of HVs is accompanied by a periodic enlargement and shrinking of the top/bottom  recirculation regions (not shown). 
The recirculating regions enlarge while the leading-edge shear layers accumulate vorticity, 
and suddenly shrink when the HVs are shed.
%
%

The shedding of HVs from the LE shear layers is reminiscent of the flow past 2D rectangular cylinders:  spanwise vortices are periodically shed from the LE shear layers due to the so-called impinging-LE-vortex (ILEV) instability \citep{naudascher-rockwell-1994}. The ILEV instability is a resonant oscillation of the fluid.  
Vortices are shed periodically from the LE shear layer and, when a LE vortex passes over the TE it 
triggers the shedding of a new LE vortex \citep{chiarini-quadrio-auteri-2022}. 
The flow frequency thus changes with the length of the body, and the dynamics of the LE and TE vortex shedding are synchronised. In the two-dimensional case, therefore, the shedding of vortices from the LE is not the result of an absolute instability of the LE shear layer, but  is interconnected with the dynamics of the TE vortex shedding \citep{hourigan-thompson-tan-2001,chiarini-quadrio-auteri-2022}. 
%
%
\begin{figure}
\centering
\includegraphics[trim={30 240 80 170},clip,width=0.49\textwidth]{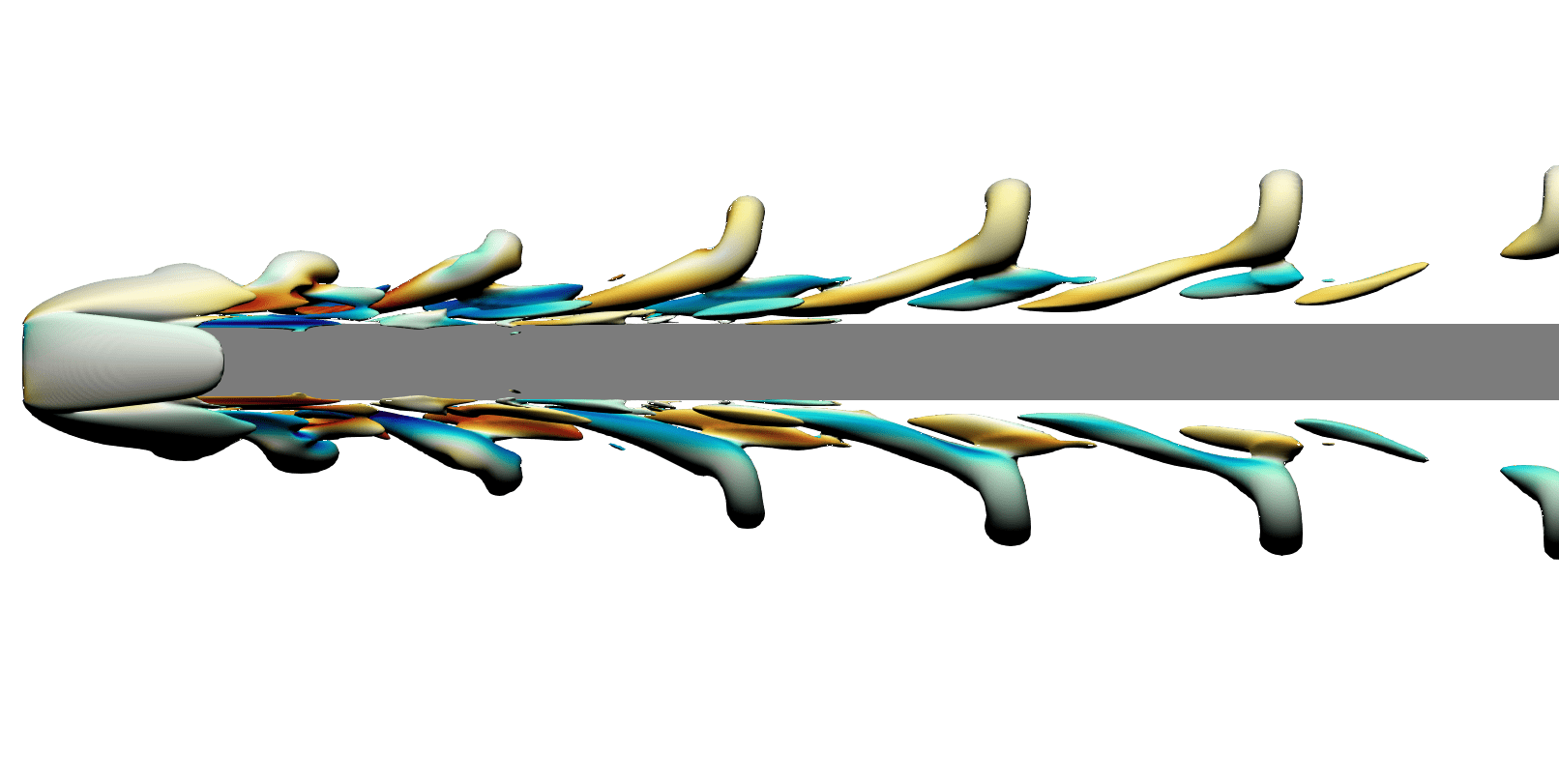}
\includegraphics[trim={30 240 80 170},clip,width=0.49\textwidth]{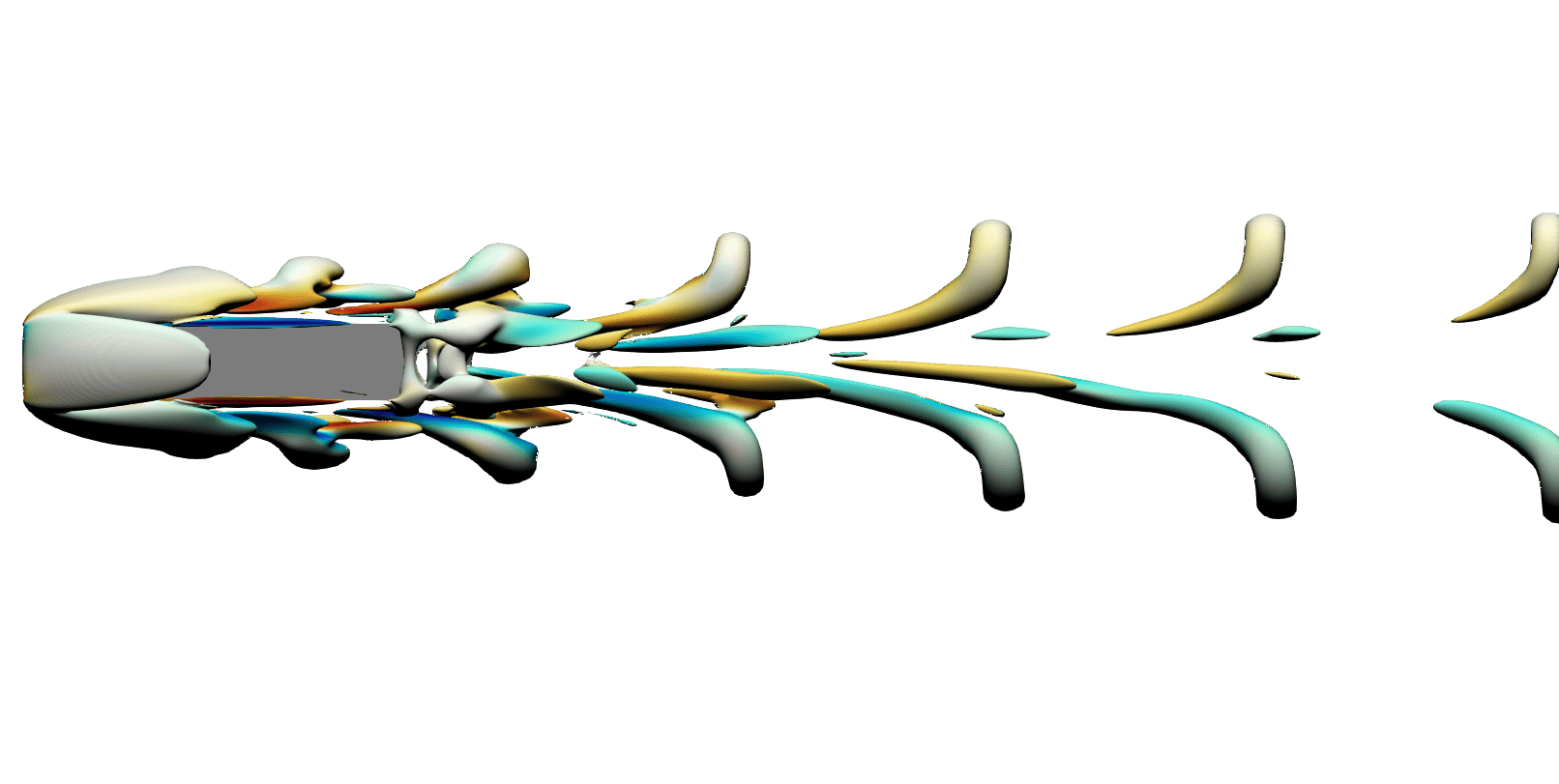}
\caption{
Lateral view of the flow structures for rectangular prisms with $(L,W)=(\infty,2.25)$ and sharp LE (left), and $(L,W)=(5,2.25)$ and rounded LE with $R=1/64$ (right),  at $Re=400$. Isosurfaces of $\lambda_2 = -0.05$ coloured by $-1 \le \omega_x \le 1$.
The  flow approaches the regimes $pS_yS_zla$  and $pS_yA_z$, respectively.}
\label{fig:plate225}
\end{figure}

To investigate the role of the LE and TE shear layers in the mechanism that sustains the $pS_yS_zla$ regime, two types of additional simulations have been carried out at $Re=400$ for modified prism geometries.
First, to isolate the LE shear layer from the interaction with the TE, we have considered a rectangular prism of infinite length and $W=2.25$.
As shown in the left panel of figure \ref{fig:plate225}, 
the flow enters the $pS_yS_zla$ regime in this case too, with pairs of HVs shed in phase from the top and bottom LE shear layers. 
Interestingly, the shedding frequency closely matches that found for $L=5$, $St \approx 0.24$. 
This indicates that for 3D prisms the triggering mechanism of the LE vortex shedding does not require the presence of a TE, as instead observed for 2D blunt bodies \citep{chaurasia-thompson-2011,thompson-2012}. 
A possible feedback mechanism that triggers the formation of the HVs may be therefore embedded within the recirculating regions that arise over the top and bottom sides of the prism. 

Second, to support this hypothesis we have considered rectangular prisms with $L=5$ and $W=2.25$, but with a rounded LE. Five curvature radii $R$ have been considered, ranging from $R=1/2$ (semicircular LE) to $R=1/64$. 
For $R=1/2$ the flow does not separate from the LE, and recirculating regions do not form over the top and bottom sides of the prism. 
In this case the flow approaches a steady and asymmetric $sS_yA_z$ regime and shedding of HVs from the LE is not detected. 
When a smaller $R$ is considered, the flow separates from the LE, and progressively larger recirculating regions 
form over the top and bottom sides of the prism. However, the shedding of HVs is 
only detected for the smallest radius $R=1/64$.
In this case the flow enters a periodic regime around the asymmetric $S_yA_z$ state, characterised by the in-phase shedding of HVs from the top and bottom LE shear layers (right panel of figure \ref{fig:plate225}). 
Again, the shedding frequency matches 
well that found for the sharp configuration,  $St \approx 0.238$. 

In summary, these additional simulations support the hypothesis that in the $pS_yS_zl$ regime the shedding of HVs from the LE shear layers is the result of a feedback mechanism that takes place in the recirculating regions over the top/bottom sides of the prism and does not require the presence of a TE. 

\subsubsection{Large $Re$: Aperiodic regime $aS_yS_z$}


\begin{figure}
\centering
\includegraphics[width=0.49\textwidth]{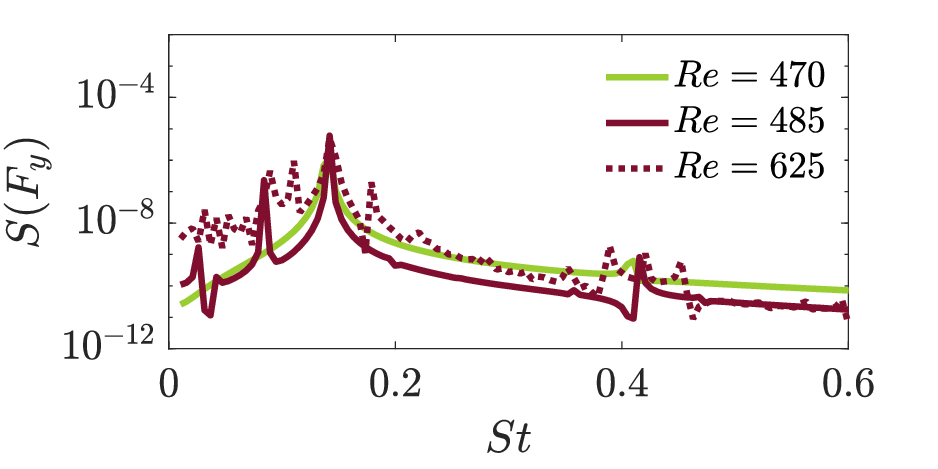}
\includegraphics[width=0.49\textwidth]{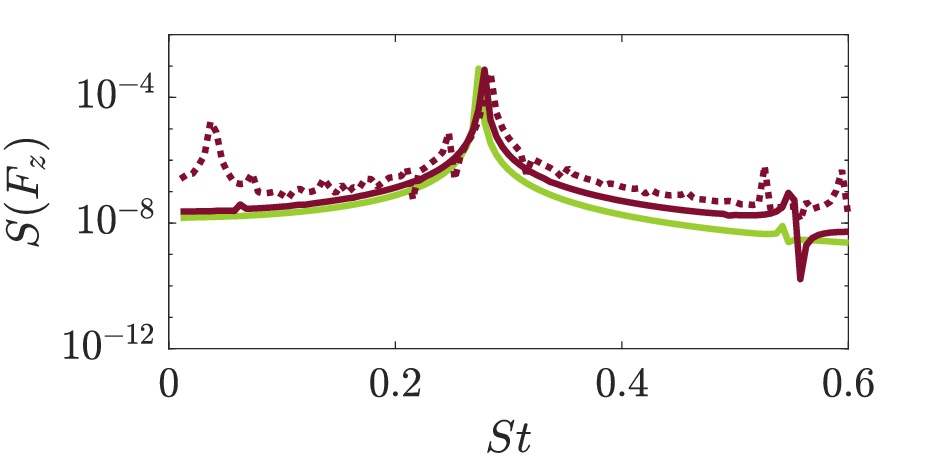}
%
\includegraphics[trim={0 0 0 0},clip,width=0.49\textwidth]{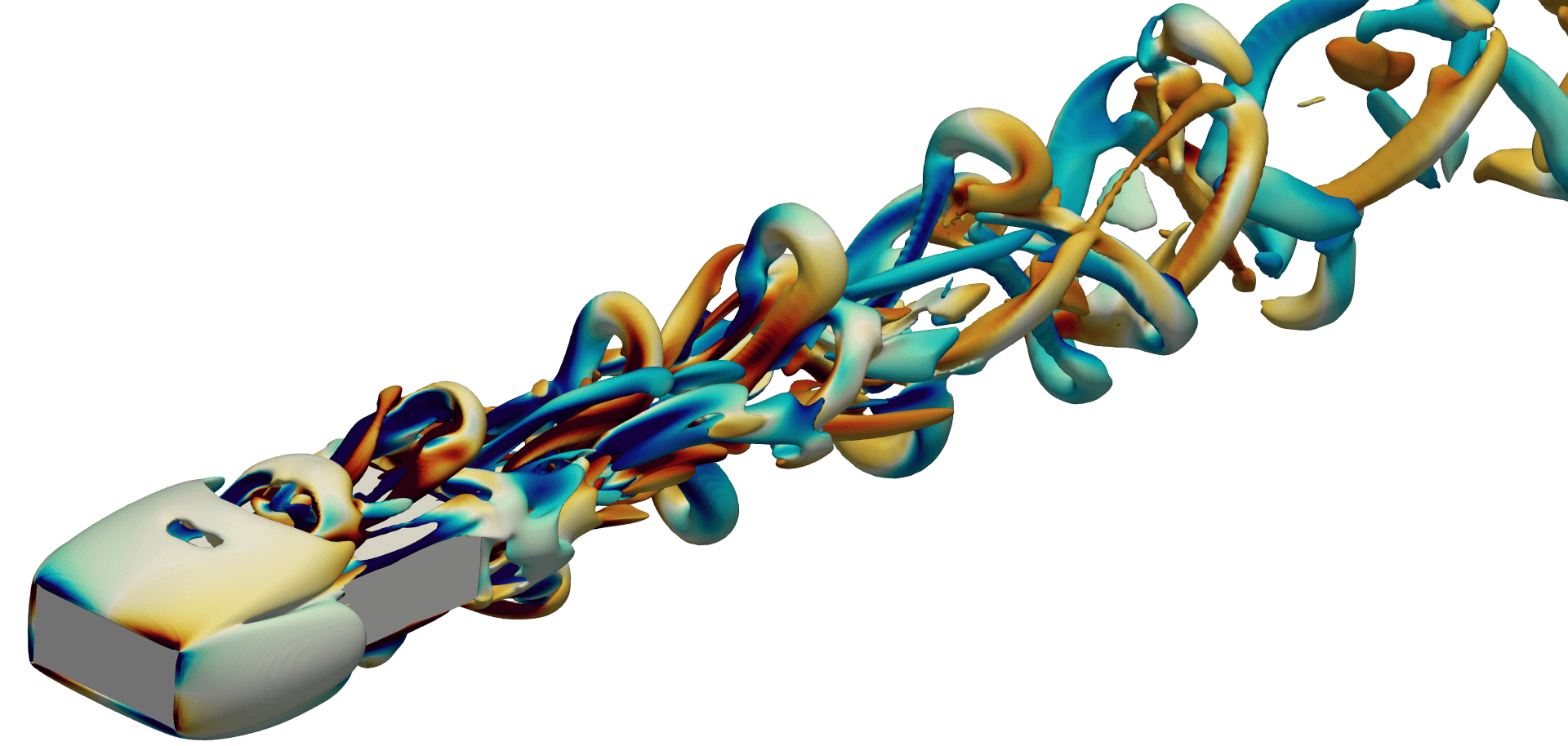}
\includegraphics[width=0.24\textwidth]{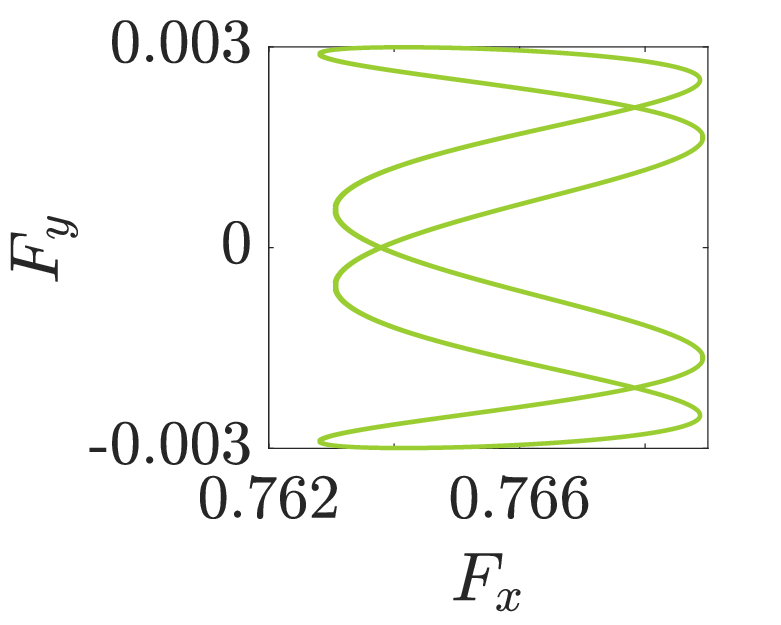}
\includegraphics[width=0.24\textwidth]{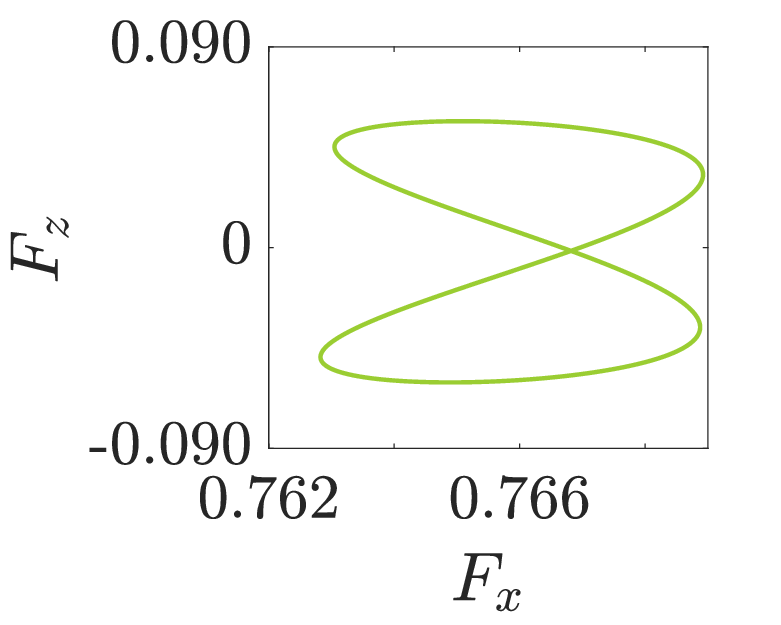}
\includegraphics[trim={0 0 0 0},clip,width=0.49\textwidth]{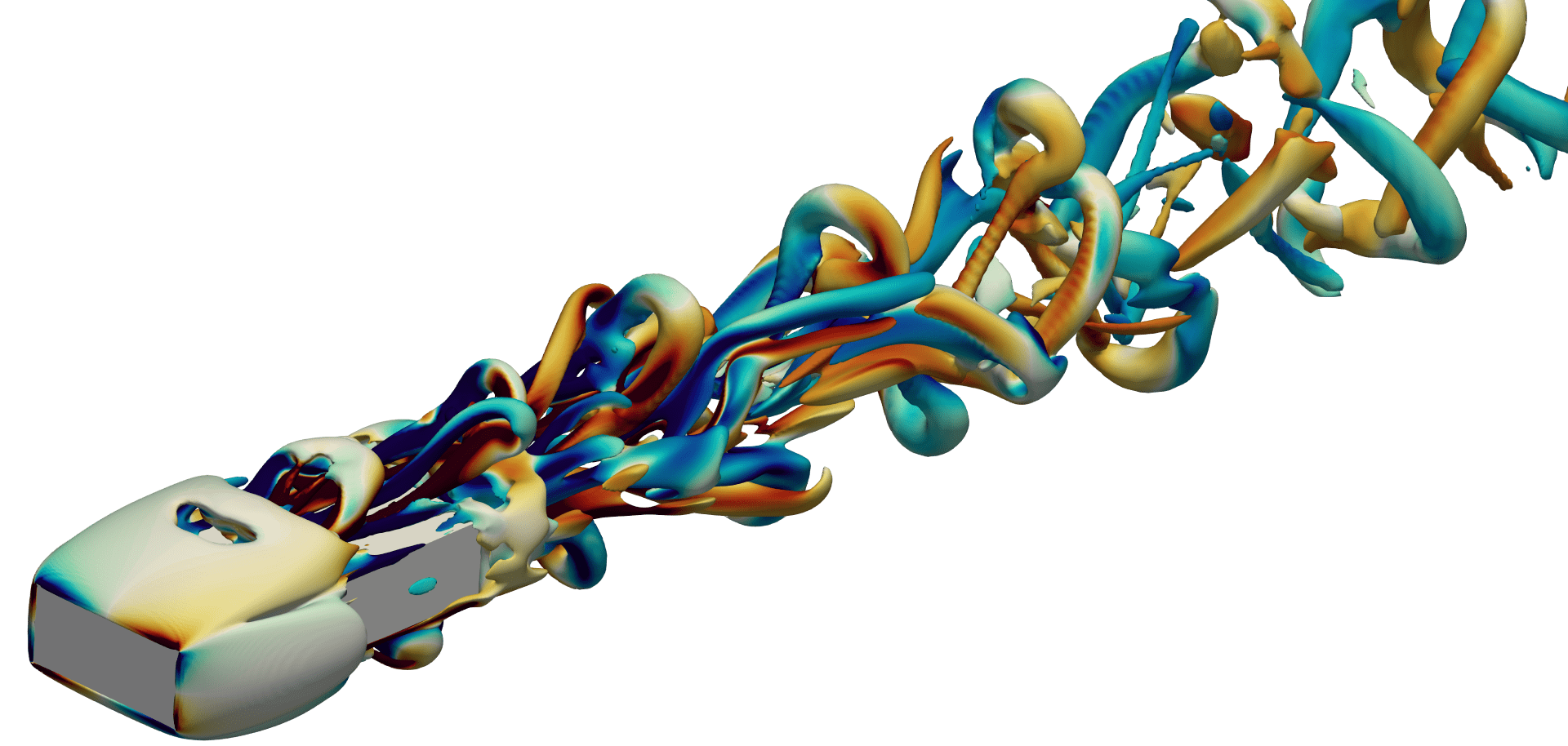}
\includegraphics[width=0.24\textwidth]{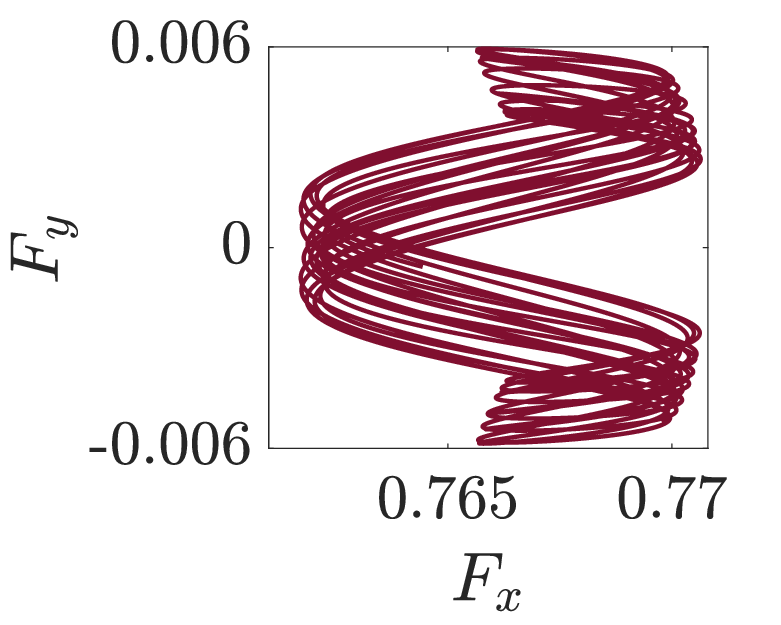}
\includegraphics[width=0.24\textwidth]{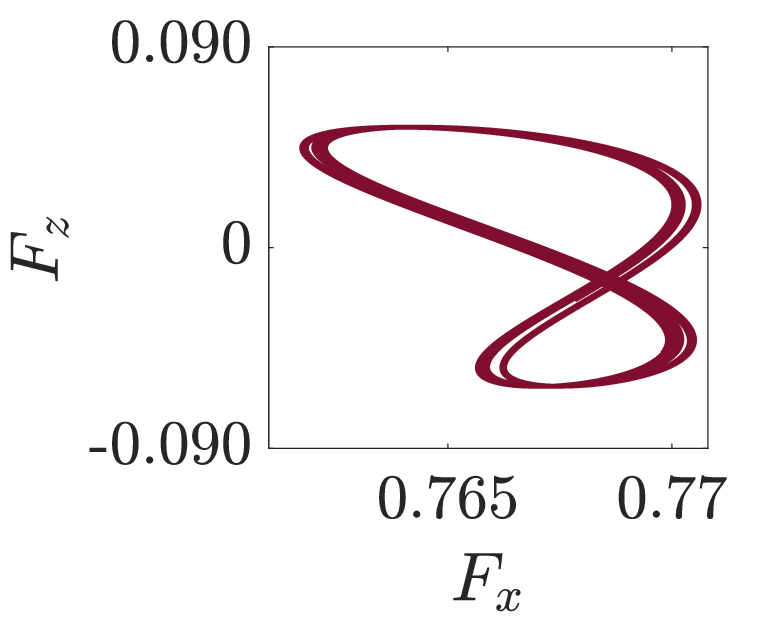}
\includegraphics[trim={0 0 0 0},clip,width=0.49\textwidth]{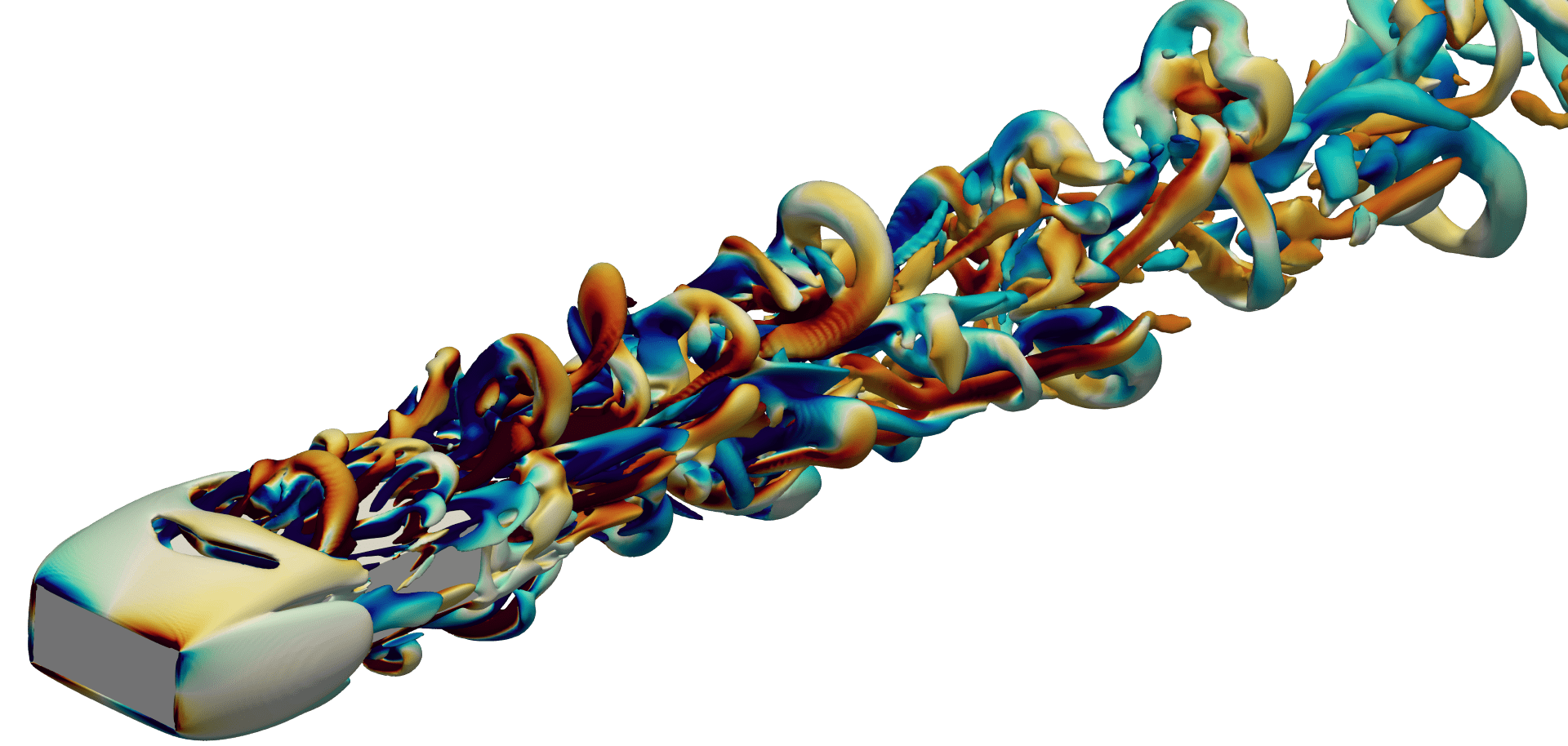}
\includegraphics[width=0.24\textwidth]{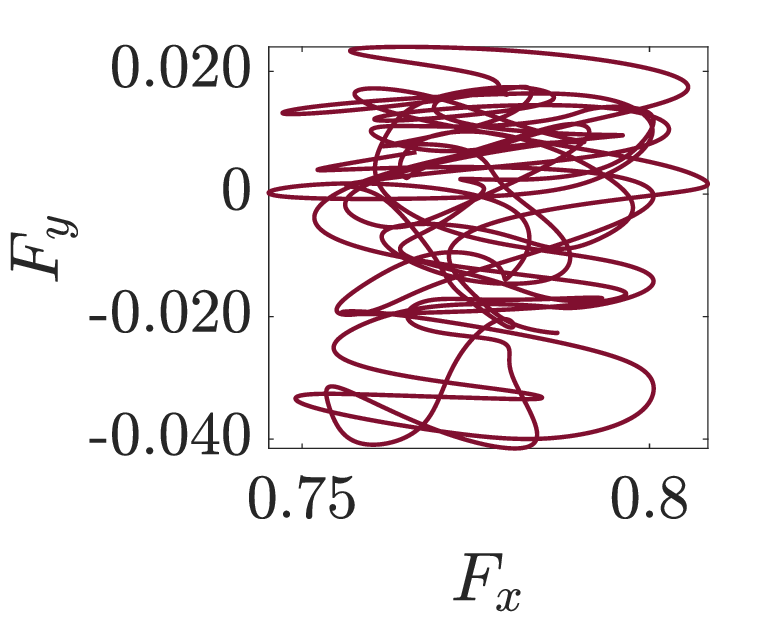}
\includegraphics[width=0.24\textwidth]{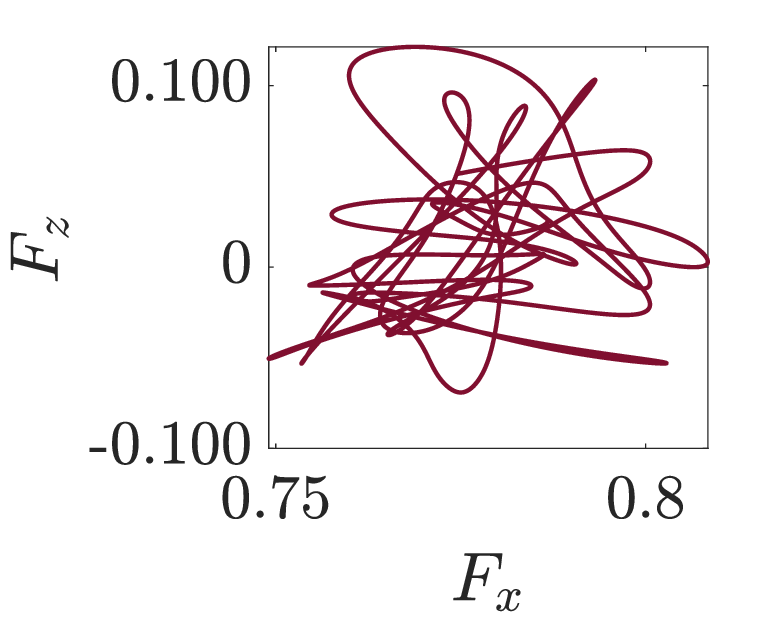}
\caption{Unsteady regimes at larger $Re$ for $L=5$ and $W=2.25$.
Top panels: frequency spectra of $F_y$ (left) and $F_z$ (right) for $470 \le Re \le 625$. 
Bottom panels: structure of the flow for $Re=470$, $485$ and $625$ (top to bottom). 
Left: instantaneous isosurfaces of $\lambda_2=-0.05$ coloured by $-1 \le \omega_x \le 1$. 
Right: force diagrams $F_y-F_x$ and $F_z-F_x$. }
\label{fig:W225_FlowStr_470_700}
\end{figure}

For larger Reynolds numbers $Re \ge 460$, the lateral and vertical oscillating modes of the wake become unstable, and the flow progressively enters a chaotic regime (figure \ref{fig:W225_FlowStr_470_700}). 
Unlike for smaller $W$, for $500 \le Re \le 700$ the flow oscillates around a $S_yS_z$ state, and the time average value of the cross-stream forces is $F_y=F_z=0$. This recalls the results of \cite{zdravkovic-etal-1989} and \cite{zdravkovich-etal-1998}, who investigated the flow past a finite circular cylinder at $Re \approx 10^5$, varying the width-to-height ratio between $0.25 \le W \le 10$. In agreement with our results, they found an asymmetric flow pattern for $1 \le W \le 2$ only.

According to our simulations, the flow loses the instantaneous left/right symmetry at $Re \approx 465$. 
For $465 \le Re \le 470$, $S(F_y)$ (figure \ref{fig:W225_FlowStr_470_700}) shows a single peak at  $St \approx 0.134$, which is half the frequency of 
the LE vortex shedding (see $S(F_z)$ in the same figure). The flow enters a $pS_yS_z$ regime.
In this range of $Re$ the flow remains periodic and the attractor is a limit cycle that draws a closed line in the phase space (see the force diagrams in figure \ref{fig:W225_FlowStr_470_700}). 
As $Re$ increases, the bifurcated limit cycle becomes unstable. 
At $Re=485$, a new incommensurate frequency $St \approx 0.079$ arises in $S(Fy)$;
the flow is therefore aperiodic 
and the attractor draws a torus in the phase space.

For $Re \ge 500$, additional frequencies appear and the flow becomes chaotic.
In the $aS_yS_z$ regime the flow dynamics is dominated by the nonlinear interaction of three different modes: 
(i)~shedding of HVs from top and bottom LE shear layers, and  
(ii)-(iii)~oscillating modes of the wake in the $z$ and $y$ directions. 
The frequency of the first two modes are detected with the peaks in $S(F_z)$, i.e. $St \approx 0.25-0.30$ and $St \approx 0.02-0.03$. The frequency $St \approx 0.04-0.05$ of the wake oscillating mode in the $y$ direction, instead, is visible in  $S(F_y)$. As $Re$ increases, the frequency of the LE vortex shedding increases and the corresponding peak in the frequency spectrum becomes less evident and more broad-banded (not shown).
Like for smaller $W$, indeed, at larger $Re$ the LE vortex shedding weakens, and the flow unsteadiness is mainly driven by the wake oscillating modes. Note that $St \approx 0.03$ is close to what was found for $W=1.2$, indicating that the frequency of the wake oscillating mode in the $z$ direction is only marginally influenced by the width of the prism for $1.2 \le W \le 2.25$. 
In contrast, the frequency of the wake oscillation mode in the $y$ direction clearly decreases with $W$.
Figure \ref{fig:W225_FlowStr_470_700} shows that for $Re > 500$ the shedding of HVs arises also over the lateral sides of the prism.

\subsection{Wide prism: $W=5$}
\label{subsec:w5}

\begin{figure}
\centering
\begin{tikzpicture}

\definecolor{clr1}{RGB}{18 78 128}
\definecolor{clr2}{RGB}{89 165 216}
\definecolor{clr3}{RGB}{145 229 246}
\definecolor{clr4}{RGB}{255 143 163}
\definecolor{clr5}{RGB}{50 205 50}
\definecolor{clr6}{RGB}{201 24 74}
\definecolor{clr7}{RGB}{128 15 47}
\definecolor{clr8}{RGB}{174 32 18}
\definecolor{clr9}{RGB}{155 34 38}
\definecolor{clr14}{RGB}{154 205 50}

\begin{axis}[%
width=0.9\textwidth,
height=0.2\textwidth,
scale only axis,
xmin=200,
xmax=725,
ymin=0.77,
ymax=0.89,
xtick={200,300,400,500,600,700},
xticklabels=\empty,
ytick={0.78,0.82,0.86},
yticklabels={$ 0.78$,$ 0.82$,$ 0.86$},
ylabel={$F_x$},
ylabel style={at={(0.02,0.5)}},
axis background/.style={fill=white},
legend columns=3,transpose legend,
legend style={at={(0.99,0.82)}, anchor=east, legend cell align=left, align=left, fill=none, draw=none}
]


\addplot [color=black,solid,draw=none,mark=*,mark options={scale=1.4,black,fill=clr1}]
  table[row sep=crcr]{%
  225.0000    0.828 \\
  240.0000    0.81299 \\
};
\addplot [color=black,solid,draw=none,mark=*,mark options={scale=1.4,black,fill=clr5}]
  table[row sep=crcr]{%
  250.0000    0.81084 \\
  255.0000    0.81130 \\
  265.0000    0.81670 \\
  275.0000    0.82460 \\
  280.0000    0.83250 \\
  300.0000    0.85310 \\
};

\addplot [color=black,solid,draw=none,mark=*,mark options={scale=1.3,black,fill=clr14}]
  table[row sep=crcr]{%
  353.0000    0.83660 \\
  355.0000    0.83420 \\
  357.0000    0.83250 \\
  373.0000    0.83420 \\
  374.0000    0.83240 \\
  375.0000    0.83010 \\  
};
\addplot [color=black,solid,draw=none,mark=*,mark options={scale=1.4,black,fill=clr7}]
  table[row sep=crcr]{%
  305.0000    0.84910 \\
  315.0000    0.85830 \\
  325.0000    0.86650 \\
  335.0000    0.87230 \\
  345.0000    0.85820 \\
  350.0000    0.85106 \\
  360.0000    0.82440 \\
  365.0000    0.82070 \\
  370.0000    0.81780 \\
  377.0000    0.81450 \\
  380.0000    0.81430 \\
  385.0000    0.81310 \\
  400.0000    0.80720 \\
  450.0000    0.80150 \\
  500.0000    0.78960 \\
  550.0000    0.79620 \\
  600.0000    0.79950 \\
  650.0000    0.80820 \\
  700.0000    0.81190 \\
};

\addplot [color=black,solid,mark=none, line width=0.5, mark options={scale=1.4,black,fill=red!80!black}]
  table[row sep=crcr]{%
  325.0000    0.8642 \\
  325.0000    0.8689 \\
};

\addplot [color=black,solid,mark=none, line width=0.5, mark options={scale=1.4,black,fill=red!80!black}]
  table[row sep=crcr]{%
  335.0000    0.8663 \\
  335.0000    0.8782 \\
};

\addplot [color=black,solid,mark=none, line width=0.5, mark options={scale=1.4,black,fill=red!80!black}]
  table[row sep=crcr]{%
  345.0000    0.8451 \\
  345.0000    0.8712 \\
};

\addplot [color=black,solid,mark=none, line width=0.5, mark options={scale=1.4,black,fill=red!80!black}]
  table[row sep=crcr]{%
  353.0000    0.8295 \\
  353.0000    0.8437 \\
};

\addplot [color=black,solid,mark=none, line width=0.5, mark options={scale=1.4,black,fill=red!80!black}]
  table[row sep=crcr]{%
  355.0000    0.8276 \\
  355.0000    0.8409 \\
};

\addplot [color=black,solid,mark=none, line width=0.5, mark options={scale=1.4,black,fill=red!80!black}]
  table[row sep=crcr]{%
  357.0000    0.8265 \\
  357.0000    0.8386 \\
};

\addplot [color=black,solid,mark=none, line width=0.5, mark options={scale=1.4,black,fill=red!80!black}]
  table[row sep=crcr]{%
  360.0000    0.8135 \\
  360.0000    0.8353 \\
};

\addplot [color=black,solid,mark=none, line width=0.5, mark options={scale=1.4,black,fill=red!80!black}]
  table[row sep=crcr]{%
  365.0000    0.8103 \\
  365.0000    0.8357 \\
};

\addplot [color=black,solid,mark=none, line width=0.5, mark options={scale=1.4,black,fill=red!80!black}]
  table[row sep=crcr]{%
  370.0000    0.8077 \\
  370.0000    0.8330 \\
};

\addplot [color=black,solid,mark=none, line width=0.5, mark options={scale=1.4,black,fill=red!80!black}]
  table[row sep=crcr]{%
  377.0000    0.8033 \\
  377.0000    0.8252 \\
};

\addplot [color=black,solid,mark=none, line width=0.5, mark options={scale=1.4,black,fill=red!80!black}]
  table[row sep=crcr]{%
  380.0000    0.8035 \\
  380.0000    0.8250 \\
};

\addplot [color=black,solid,mark=none, line width=0.5, mark options={scale=1.4,black,fill=red!80!black}]
  table[row sep=crcr]{%
  373.0000    0.8171 \\
  373.0000    0.8513 \\
};

\addplot [color=black,solid,mark=none, line width=0.5, mark options={scale=1.4,black,fill=red!80!black}]
  table[row sep=crcr]{%
  374.0000    0.8148 \\
  374.0000    0.8499 \\
};

\addplot [color=black,solid,mark=none, line width=0.5, mark options={scale=1.4,black,fill=red!80!black}]
  table[row sep=crcr]{%
  375.0000    0.8120 \\
  375.0000    0.8483 \\
};

\addplot [color=black,solid,mark=none, line width=0.5, mark options={scale=1.4,black,fill=red!80!black}]
  table[row sep=crcr]{%
  385.0000    0.8015 \\
  385.0000    0.8246 \\
};

\addplot [color=black,dashed,mark=none, line width=0.5, mark options={scale=1.4,black,fill=red!80!black}]
  table[row sep=crcr]{%
  245.0000    0.77 \\
  245.0000    0.89 \\
};

\addplot [color=black,dashed,mark=none, line width=0.5, mark options={scale=1.4,black,fill=red!80!black}]
  table[row sep=crcr]{%
  302.5000    0.77 \\
  302.5000    0.89 \\
};

\addplot [color=black,dotted,mark=none, line width=0.5, mark options={scale=1.4,black,fill=red!80!black}]
  table[row sep=crcr]{%
  352.0000    0.77 \\
  352.0000    0.89 \\
};

\addplot [color=black,dotted,mark=none, line width=0.5, mark options={scale=1.4,black,fill=red!80!black}]
  table[row sep=crcr]{%
  359.0000    0.77 \\
  359.0000    0.89 \\
};

\addplot [color=black,dotted,mark=none, line width=0.5, mark options={scale=1.4,black,fill=red!80!black}]
  table[row sep=crcr]{%
  372.0000    0.77 \\
  372.0000    0.89 \\
};

\addplot [color=black,dotted,mark=none, line width=0.5, mark options={scale=1.4,black,fill=red!80!black}]
  table[row sep=crcr]{%
  376.0000    0.77 \\
  376.0000    0.89 \\
};

\addplot [color=black,solid,mark=none, line width=0.5,mark options={scale=1.4,black,fill=red!80!black}]
  table[row sep=crcr]{%
  350.0000    0.8643 \\
  350.0000    0.8378 \\
};

\addplot [color=black,solid,mark=none, line width=0.5,mark options={scale=1.4,black,fill=red!80!black}]
  table[row sep=crcr]{%
  400.0000    0.7868 \\
  400.0000    0.8276 \\
};

\addplot [color=black,solid,mark=none, line width=0.5,mark options={scale=1.4,black,fill=red!80!black}]
  table[row sep=crcr]{%
  450.0000    0.7850 \\
  450.0000    0.8181 \\
};

\addplot [color=black,solid,mark=none, line width=0.5,mark options={scale=1.4,black,fill=red!80!black}]
  table[row sep=crcr]{%
  500.0000    0.7778 \\
  500.0000    0.8014 \\
};

\addplot [color=black,solid,mark=none, line width=0.5,mark options={scale=1.4,black,fill=red!80!black}]
  table[row sep=crcr]{%
  550.0000    0.7841 \\
  550.0000    0.8084 \\
};

\addplot [color=black,solid,mark=none, line width=0.5,mark options={scale=1.4,black,fill=red!80!black}]
  table[row sep=crcr]{%
  600.0000    0.7836 \\
  600.0000    0.8154 \\
};

\addplot [color=black,solid,mark=none, line width=0.5,mark options={scale=1.4,black,fill=red!80!black}]
  table[row sep=crcr]{%
  650.0000    0.7920 \\
  650.0000    0.8243 \\
};

\addplot [color=black,solid,mark=none, line width=0.5,mark options={scale=1.4,black,fill=red!80!black}]
  table[row sep=crcr]{%
  700.0000    0.7951 \\
  700.0000    0.8287 \\
};

\end{axis}

\node[] at (0.5,2.4) {$sS_yS_z$};
\node[] at (1.7,2.4) {$pS_yS_zt$};
\node[] at (6.4,2.4) {$aS_yS_z$};

\end{tikzpicture}%
\input{./figure/Ary5/FyFz-Re-2.tex}
\begin{tikzpicture}

\definecolor{clr1}{RGB}{18 78 128}
\definecolor{clr2}{RGB}{89 165 216}
\definecolor{clr3}{RGB}{145 229 246}
\definecolor{clr4}{RGB}{255 143 163}
\definecolor{clr5}{RGB}{50 205 50}
\definecolor{clr6}{RGB}{201 24 74}
\definecolor{clr7}{RGB}{128 15 47}
\definecolor{clr8}{RGB}{174 32 18}
\definecolor{clr9}{RGB}{155 34 38}
\definecolor{clr14}{RGB}{154 205 50}

\begin{axis}[%
width=0.38\textwidth,
height=0.12\textwidth,
scale only axis,
xmin=333,
xmax=383,
ymin=0.77,
ymax=0.89,
xlabel={$Re$},
ylabel={$F_x$},
ylabel style={at={(0.0,0.5)}},
axis background/.style={fill=white},
legend columns=3,transpose legend,
legend style={at={(0.99,0.82)}, anchor=east, legend cell align=left, align=left, fill=none, draw=none}
]


\addplot [color=black,solid,draw=none,mark=*,mark options={scale=1.4,black,fill=clr1}]
  table[row sep=crcr]{%
  225.0000    0.828 \\
  240.0000    0.81299 \\
};
\addplot [color=black,solid,draw=none,mark=*,mark options={scale=1.4,black,fill=clr5}]
  table[row sep=crcr]{%
  250.0000    0.81084 \\
  275.0000    0.82460 \\
  280.0000    0.83250 \\
  300.0000    0.85310 \\
};

\addplot [color=black,solid,draw=none,mark=*,mark options={scale=1.3,black,fill=clr14}]
  table[row sep=crcr]{%
  353.0000    0.83660 \\
  355.0000    0.83420 \\
  357.0000    0.83250 \\
  373.0000    0.83420 \\
  374.0000    0.83240 \\
  375.0000    0.83010 \\  
};
\addplot [color=black,solid,draw=none,mark=*,mark options={scale=1.4,black,fill=clr7}]
  table[row sep=crcr]{%
  305.0000    0.84910 \\
  315.0000    0.85830 \\
  325.0000    0.86650 \\
  335.0000    0.87230 \\
  345.0000    0.85820 \\
  350.0000    0.85106 \\
  360.0000    0.82440 \\
  365.0000    0.82070 \\
  370.0000    0.81780 \\
  377.0000    0.81450 \\
  380.0000    0.81430 \\
  385.0000    0.81310 \\
  400.0000    0.80720 \\
  450.0000    0.80150 \\
  500.0000    0.78960 \\
  550.0000    0.79620 \\
  600.0000    0.79950 \\
  650.0000    0.80820 \\
  700.0000    0.81190 \\
};

\addplot [color=black,solid,mark=none, line width=0.5, mark options={scale=1.4,black,fill=red!80!black}]
  table[row sep=crcr]{%
  325.0000    0.8642 \\
  325.0000    0.8689 \\
};

\addplot [color=black,solid,mark=none, line width=0.5, mark options={scale=1.4,black,fill=red!80!black}]
  table[row sep=crcr]{%
  335.0000    0.8663 \\
  335.0000    0.8782 \\
};

\addplot [color=black,solid,mark=none, line width=0.5, mark options={scale=1.4,black,fill=red!80!black}]
  table[row sep=crcr]{%
  345.0000    0.8451 \\
  345.0000    0.8712 \\
};

\addplot [color=black,solid,mark=none, line width=0.5, mark options={scale=1.4,black,fill=red!80!black}]
  table[row sep=crcr]{%
  353.0000    0.8295 \\
  353.0000    0.8437 \\
};

\addplot [color=black,solid,mark=none, line width=0.5, mark options={scale=1.4,black,fill=red!80!black}]
  table[row sep=crcr]{%
  355.0000    0.8276 \\
  355.0000    0.8409 \\
};

\addplot [color=black,solid,mark=none, line width=0.5, mark options={scale=1.4,black,fill=red!80!black}]
  table[row sep=crcr]{%
  357.0000    0.8265 \\
  357.0000    0.8386 \\
};

\addplot [color=black,solid,mark=none, line width=0.5, mark options={scale=1.4,black,fill=red!80!black}]
  table[row sep=crcr]{%
  360.0000    0.8135 \\
  360.0000    0.8353 \\
};

\addplot [color=black,solid,mark=none, line width=0.5, mark options={scale=1.4,black,fill=red!80!black}]
  table[row sep=crcr]{%
  365.0000    0.8103 \\
  365.0000    0.8357 \\
};

\addplot [color=black,solid,mark=none, line width=0.5, mark options={scale=1.4,black,fill=red!80!black}]
  table[row sep=crcr]{%
  370.0000    0.8077 \\
  370.0000    0.8330 \\
};

\addplot [color=black,solid,mark=none, line width=0.5, mark options={scale=1.4,black,fill=red!80!black}]
  table[row sep=crcr]{%
  377.0000    0.8033 \\
  377.0000    0.8252 \\
};

\addplot [color=black,solid,mark=none, line width=0.5, mark options={scale=1.4,black,fill=red!80!black}]
  table[row sep=crcr]{%
  380.0000    0.8035 \\
  380.0000    0.8250 \\
};

\addplot [color=black,solid,mark=none, line width=0.5, mark options={scale=1.4,black,fill=red!80!black}]
  table[row sep=crcr]{%
  373.0000    0.8171 \\
  373.0000    0.8513 \\
};

\addplot [color=black,solid,mark=none, line width=0.5, mark options={scale=1.4,black,fill=red!80!black}]
  table[row sep=crcr]{%
  374.0000    0.8148 \\
  374.0000    0.8499 \\
};

\addplot [color=black,solid,mark=none, line width=0.5, mark options={scale=1.4,black,fill=red!80!black}]
  table[row sep=crcr]{%
  375.0000    0.8120 \\
  375.0000    0.8483 \\
};

\addplot [color=black,solid,mark=none, line width=0.5, mark options={scale=1.4,black,fill=red!80!black}]
  table[row sep=crcr]{%
  385.0000    0.8015 \\
  385.0000    0.8246 \\
};

\addplot [color=black,dashed,mark=none, line width=0.5, mark options={scale=1.4,black,fill=red!80!black}]
  table[row sep=crcr]{%
  245.0000    0.77 \\
  245.0000    0.89 \\
};

\addplot [color=black,dashed,mark=none, line width=0.5, mark options={scale=1.4,black,fill=red!80!black}]
  table[row sep=crcr]{%
  302.5000    0.77 \\
  302.5000    0.89 \\
};

\addplot [color=black,dotted,mark=none, line width=0.5, mark options={scale=1.4,black,fill=red!80!black}]
  table[row sep=crcr]{%
  352.0000    0.77 \\
  352.0000    0.89 \\
};

\addplot [color=black,dotted,mark=none, line width=0.5, mark options={scale=1.4,black,fill=red!80!black}]
  table[row sep=crcr]{%
  359.0000    0.77 \\
  359.0000    0.89 \\
};

\addplot [color=black,dotted,mark=none, line width=0.5, mark options={scale=1.4,black,fill=red!80!black}]
  table[row sep=crcr]{%
  372.0000    0.77 \\
  372.0000    0.89 \\
};

\addplot [color=black,dotted,mark=none, line width=0.5, mark options={scale=1.4,black,fill=red!80!black}]
  table[row sep=crcr]{%
  376.0000    0.77 \\
  376.0000    0.89 \\
};

\addplot [color=black,solid,mark=none, line width=0.5,mark options={scale=1.4,black,fill=red!80!black}]
  table[row sep=crcr]{%
  350.0000    0.8643 \\
  350.0000    0.8378 \\
};

\addplot [color=black,solid,mark=none, line width=0.5,mark options={scale=1.4,black,fill=red!80!black}]
  table[row sep=crcr]{%
  400.0000    0.7868 \\
  400.0000    0.8276 \\
};

\addplot [color=black,solid,mark=none, line width=0.5,mark options={scale=1.4,black,fill=red!80!black}]
  table[row sep=crcr]{%
  450.0000    0.7850 \\
  450.0000    0.8181 \\
};

\addplot [color=black,solid,mark=none, line width=0.5,mark options={scale=1.4,black,fill=red!80!black}]
  table[row sep=crcr]{%
  500.0000    0.7778 \\
  500.0000    0.8014 \\
};

\addplot [color=black,solid,mark=none, line width=0.5,mark options={scale=1.4,black,fill=red!80!black}]
  table[row sep=crcr]{%
  550.0000    0.7841 \\
  550.0000    0.8084 \\
};

\addplot [color=black,solid,mark=none, line width=0.5,mark options={scale=1.4,black,fill=red!80!black}]
  table[row sep=crcr]{%
  600.0000    0.7836 \\
  600.0000    0.8154 \\
};

\addplot [color=black,solid,mark=none, line width=0.5,mark options={scale=1.4,black,fill=red!80!black}]
  table[row sep=crcr]{%
  650.0000    0.7920 \\
  650.0000    0.8243 \\
};

\addplot [color=black,solid,mark=none, line width=0.5,mark options={scale=1.4,black,fill=red!80!black}]
  table[row sep=crcr]{%
  700.0000    0.7951 \\
  700.0000    0.8287 \\
};

\end{axis}

\end{tikzpicture}%
\input{./figure/Ary5/FyFz-Re-zoom.tex}
\includegraphics[width=0.325\textwidth]{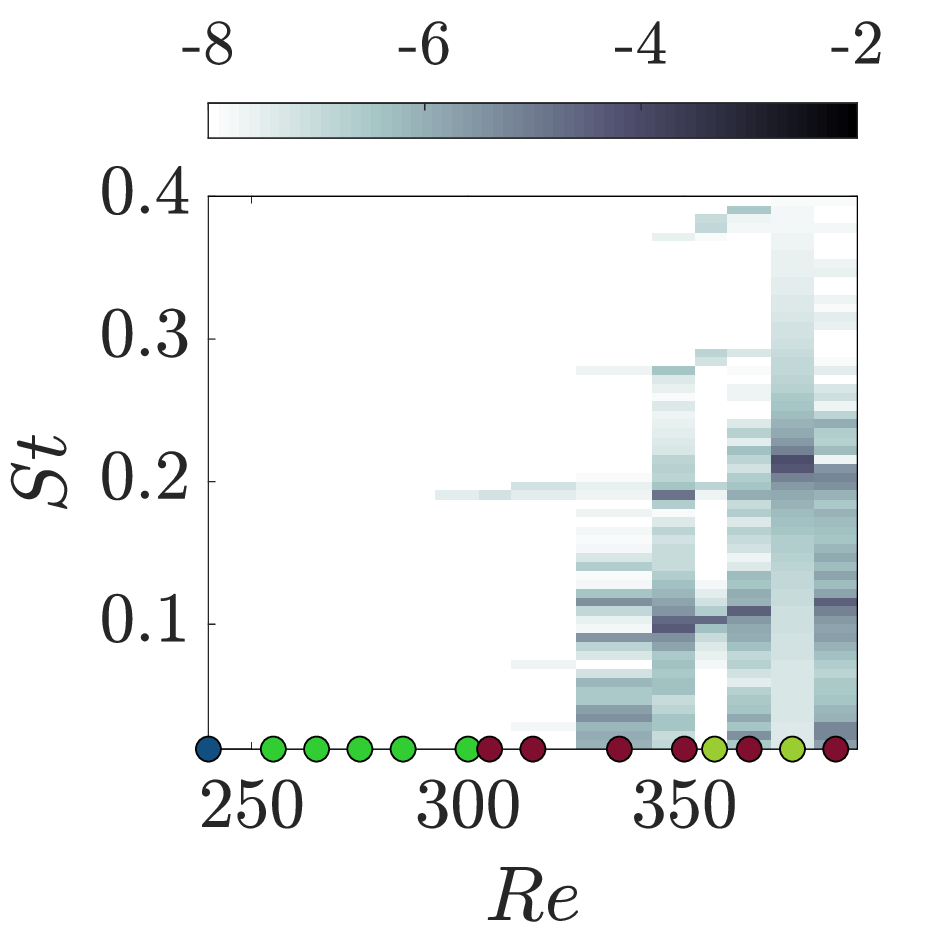}
\includegraphics[width=0.325\textwidth]{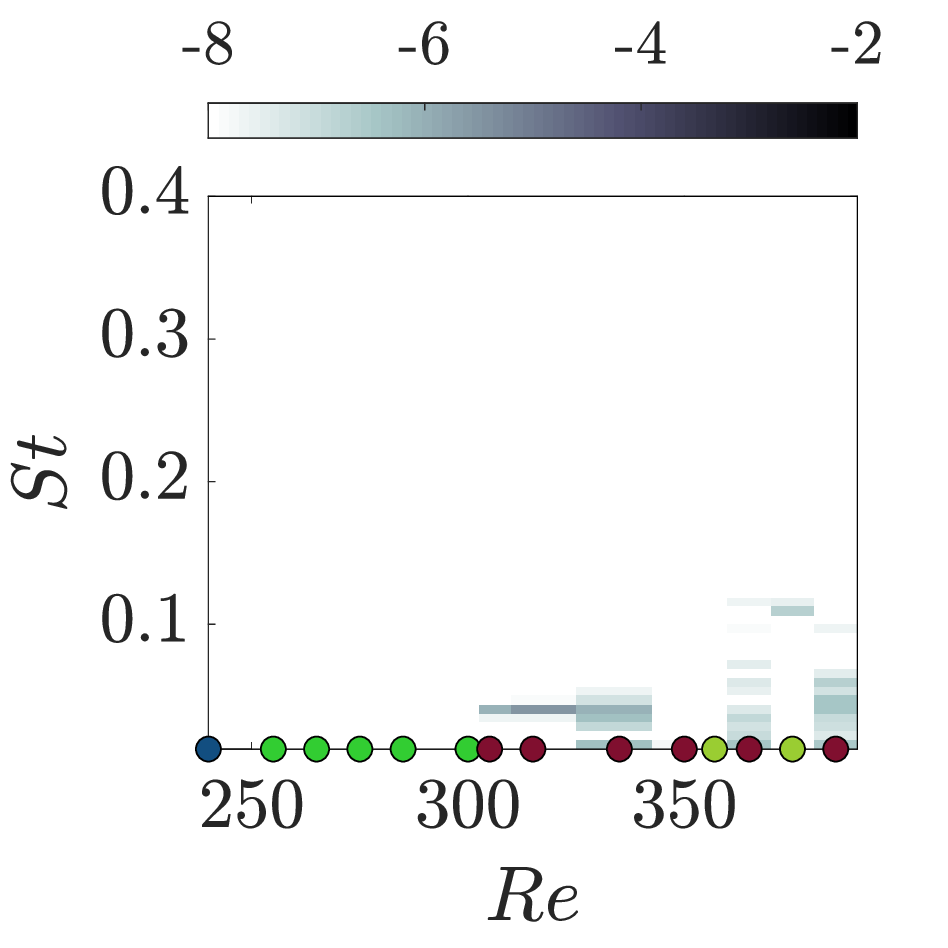}
\includegraphics[width=0.325\textwidth]{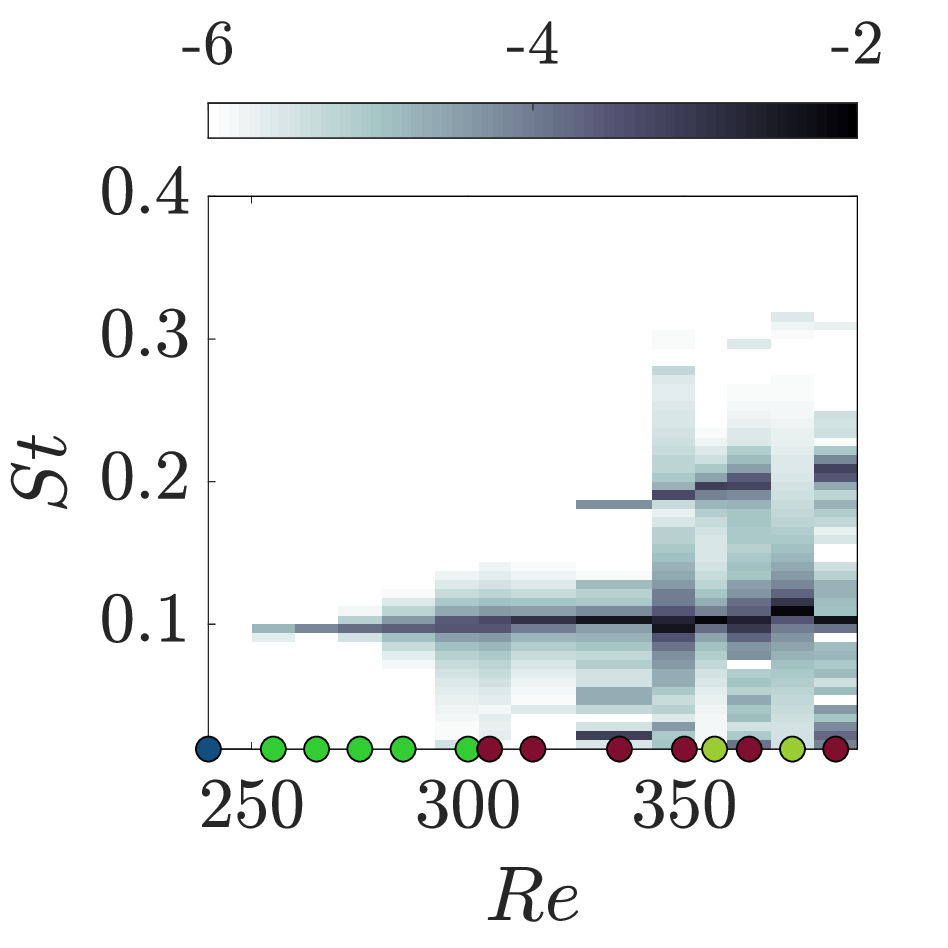}
\caption{As figure \ref{fig:3D_Ary1_forces},  for $L=5$ and $W=5$. The additional panels are a zoom in the range $330 \le Re \le 380$, where synchronisation occurs (see \S\ref{sec:lock-in_W5}).}
\label{fig:AR5_forces}
\end{figure}

For $L=5$ and $W=5$ three main  regimes have been identified for $Re < 700$ (figure \ref{fig:AR5_forces}). 
In agreement with the LSA, at $Re \approx 250$ the  steady $S_yS_z$  flow experiences a Hopf bifurcation towards the periodic $pS_yS_zt$ regime, which is characterised by the unsteady $S_yA_z$ mode of the wake. For $Re \gtrapprox 305$ the flow approaches the aperiodic $aS_yS_z$ regime. Here the wake oscillates in both the $y$ and $z$ directions, and vortices are shed from the top/bottom LE shear layers. 
The interaction between these modes largely changes with the Reynolds number.
For some $Re$,  wake oscillations and vortex shedding synchronise and the flow recovers periodicity.

\subsubsection{Low $Re$: Periodic regime $pS_yS_zt$}

\begin{figure}
\centering
\includegraphics[width=0.65\textwidth]{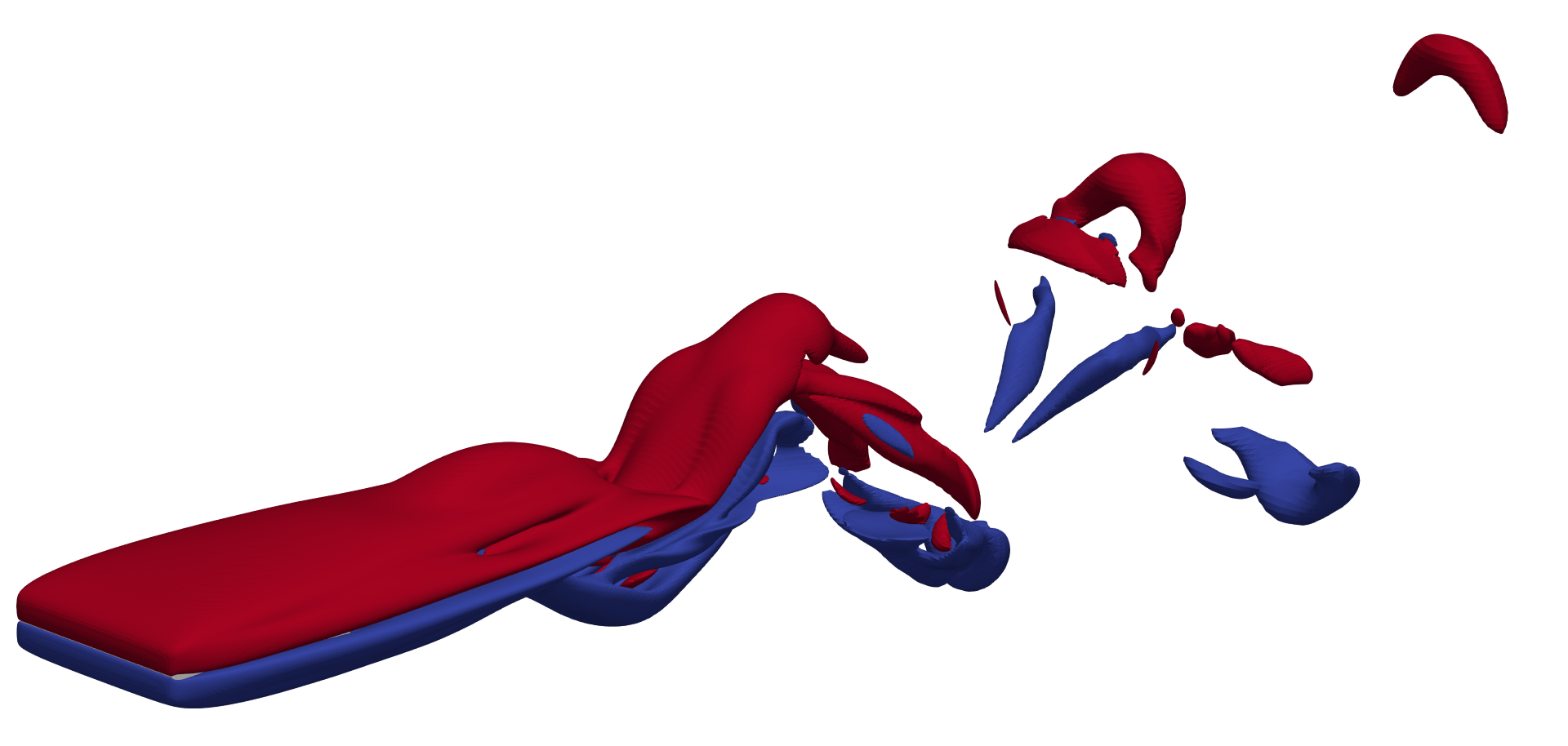} \\
\begin{tikzpicture}

\definecolor{clr1}{RGB}{18 78 128}
\definecolor{clr2}{RGB}{89 165 216}
\definecolor{clr3}{RGB}{145 229 246}
\definecolor{clr4}{RGB}{255 143 163}
\definecolor{clr5}{RGB}{50 205 50}
\definecolor{clr6}{RGB}{201 24 74}
\definecolor{clr7}{RGB}{128 15 47}
\definecolor{clr8}{RGB}{174 32 18}
\definecolor{clr9}{RGB}{155 34 38}
\definecolor{clr14}{RGB}{154 205 50}

\begin{axis}[%
width=0.2\textwidth,
height=0.15\textwidth,
scale only axis,
xmin=224,
xmax=301,
ymin=-0.015,
ymax= 0.045,
xtick={225,250,275,300},
xlabel={$Re$},
ylabel={$F_{x,f}$},
ylabel style={at={(0.05,0.5)}},
axis background/.style={fill=white},
legend columns=3,transpose legend,
legend style={at={(0.99,0.5)}, anchor=east, legend cell align=left, align=left, fill=none, draw=none}
]

\addplot [color=black,solid,draw=none,mark=diamond*,mark options={scale=1.9,black,fill=clr1}]
  table[row sep=crcr]{%
  225.0000   0.0411 \\
  240.0000   0.0217 \\
};

\addplot [color=black,solid,draw=none,mark=diamond*,mark options={scale=1.9,black,fill=clr5}]
  table[row sep=crcr]{%
  250.0000   0.012726 \\
  255.0000   0.008734 \\
  265.0000   0.001885 \\
  275.0000  -0.0045 \\
  285.000   -0.0101 \\
  300.000   -0.0092 \\
};

\end{axis}

\end{tikzpicture}%
\begin{tikzpicture}

\definecolor{clr1}{RGB}{18 78 128}
\definecolor{clr2}{RGB}{89 165 216}
\definecolor{clr3}{RGB}{145 229 246}
\definecolor{clr4}{RGB}{255 143 163}
\definecolor{clr5}{RGB}{50 205 50}
\definecolor{clr6}{RGB}{201 24 74}
\definecolor{clr7}{RGB}{128 15 47}
\definecolor{clr8}{RGB}{174 32 18}
\definecolor{clr9}{RGB}{155 34 38}

\begin{axis}[%
width=0.2\textwidth,
height=0.15\textwidth,
scale only axis,
xmin=224,
xmax=301,
ymin=0.785,
ymax=0.865,
xtick={225,250,275,300},
xlabel={$Re$},
ylabel={$F_{x,p}$},
ylabel style={at={(0.05,0.5)}},
axis background/.style={fill=white},
legend columns=3,transpose legend,
legend style={at={(0.99,0.5)}, anchor=east, legend cell align=left, align=left, fill=none, draw=none}
]

\addplot [color=black,solid,draw=none,mark=square*,mark options={scale=1.4,black,fill=clr1}]
  table[row sep=crcr]{%
  225.0000   0.7892 \\
  240.0000   0.7913 \\
};

\addplot [color=black,solid,draw=none,mark=square*,mark options={scale=1.4,black,fill=clr5}]
  table[row sep=crcr]{%
  250.0000   0.7981 \\
  255.0000   0.8026 \\
  265.0000   0.8148 \\
  275.0000   0.8291 \\
  285.0000   0.8426 \\
  300.0000   0.8623 \\
};

\end{axis}

\end{tikzpicture}%
\begin{tikzpicture}

\definecolor{clr1}{RGB}{18 78 128}
\definecolor{clr2}{RGB}{89 165 216}
\definecolor{clr3}{RGB}{145 229 246}
\definecolor{clr4}{RGB}{255 143 163}
\definecolor{clr5}{RGB}{50 205 50}
\definecolor{clr6}{RGB}{201 24 74}
\definecolor{clr7}{RGB}{128 15 47}
\definecolor{clr8}{RGB}{174 32 18}
\definecolor{clr9}{RGB}{155 34 38}

\begin{axis}[%
width=0.2\textwidth,
height=0.15\textwidth,
scale only axis,
xmin=224,
xmax=301,
ymin=0.0859,
ymax=0.0894,
xtick={225,250,275,300},
xlabel={$Re$},
ylabel={$St(F_z)$},
ylabel style={at={(0.05,0.5)}},
axis background/.style={fill=white},
legend columns=3,transpose legend,
legend style={at={(0.99,0.25)}, anchor=east, legend cell align=left, align=left, fill=none, draw=none}
]

\addplot [color=black,solid,draw=none,mark=diamond*,mark options={scale=1.9,black,fill=clr5}]
  table[row sep=crcr]{%
  250.000  0.08591 \\
  255.000  0.08601 \\
  265.000  0.087295 \\
  275.000  0.089157 \\
  285.000  0.089195 \\
  300.000  0.089318 \\ 
};

\end{axis}

\end{tikzpicture}%
\caption{Characterisation of the $pS_yS_zt$ regime for $L=5$ and $W=5$. 
Top: isosurfaces of spanwise vorticity (red and blue for $\omega_y= \pm 0.25$) at $Re=275$. 
Bottom panels: friction drag (left) and pressure drag (middle) for $225 \le Re \le 300$; flow frequency (right) for $250 \le Re \le 300$.}
\label{fig:W5_p_fre_f}
\end{figure}

The nonlinear simulations show that the unsteady $S_yA_z$ mode of the wake becomes unstable at $Re \approx 250$, and that for $250 \lessapprox Re \lessapprox 300$ the flow experiences a periodic oscillation of the wake in the $z$ direction around the steady $S_yS_z$ state (top panel in figure \ref{fig:W5_p_fre_f}). 
Unlike  for smaller $W$, here the flow unsteadiness is driven by the wake oscillation and HVs are not shed by the LE shear layers. 
The hairpin-like structures observed in the wake are thus the result of an interaction between the top and bottom TE shear layers;
this resembles the so-called hairpin shedding regime found for shorter bluff bodies of various shapes \citep{tomboulides-orszag-2000,saha-2004,yang-etal-2022}. 
The flow is periodic 
with a single frequency close to that found with LSA ($St=0.09$; \S\ref{sec:stability}).
As expected, the intensity of the wake oscillation increases with $Re$, as conveniently visualised in figure \ref{fig:AR5_forces} by means of the root mean square of $F'_z = F_z - \aver{F_z}_t$ ($\aver{\cdot}_t$ indicates average in time): 
$F'_{z,rms} \approx 0.004$ at $Re=250$ and $F'_{z,rms} \approx 0.036$ 
at $Re=300$. 
In this regime, an increase of $Re$ is accompanied by an increase of $F_x$, which is entirely ascribed to a monotonic increase of the pressure contribution $F_{x,p}$. 
In fact, similarly to $W=1.2$ (figure \ref{fig:xr_position}), an increase of $Re$ leads to larger side (time-average) recirculating regions and, thus, to a decrease of the friction contribution to $F_x$ (bottom left panel in figure \ref{fig:W5_p_fre_f}). 

\subsubsection{Intermediate $Re$: aperiodic and frequency-locking regimes}
\label{sec:lock-in_W5}

\begin{figure}
\centering
\includegraphics[width=0.49\textwidth]{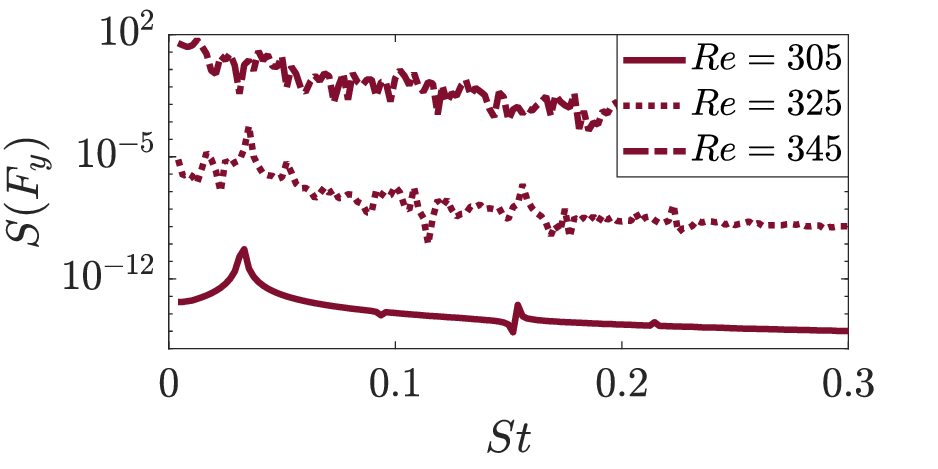}
\includegraphics[width=0.49\textwidth]{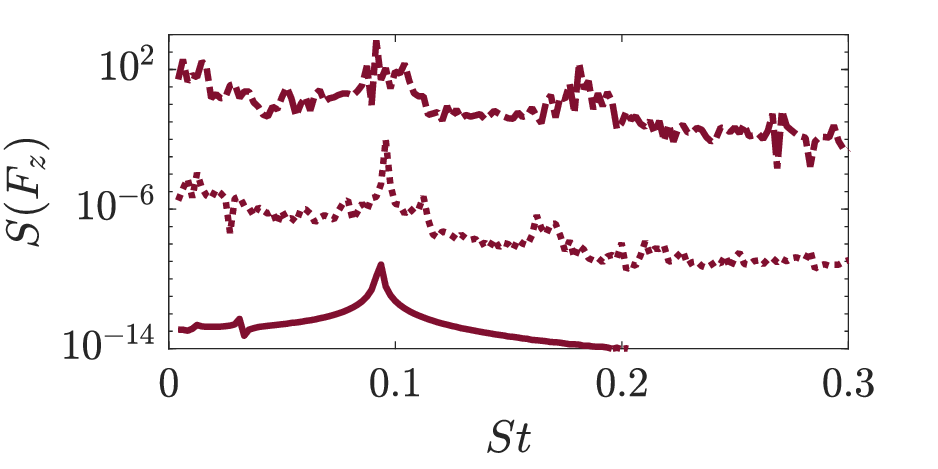}
\includegraphics[trim={0 0 0 0},clip,width=0.49\textwidth]{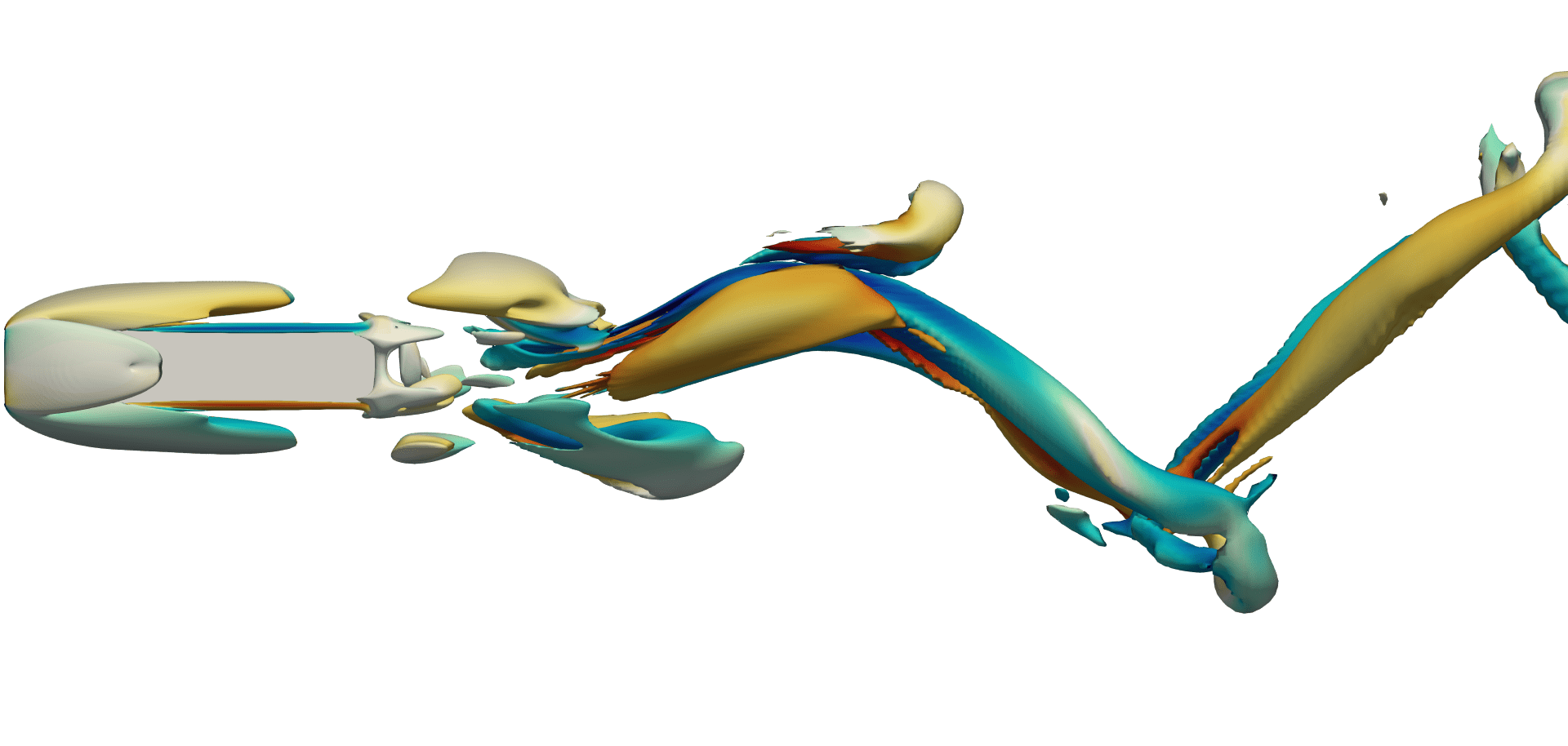}
\includegraphics[width=0.24\textwidth]{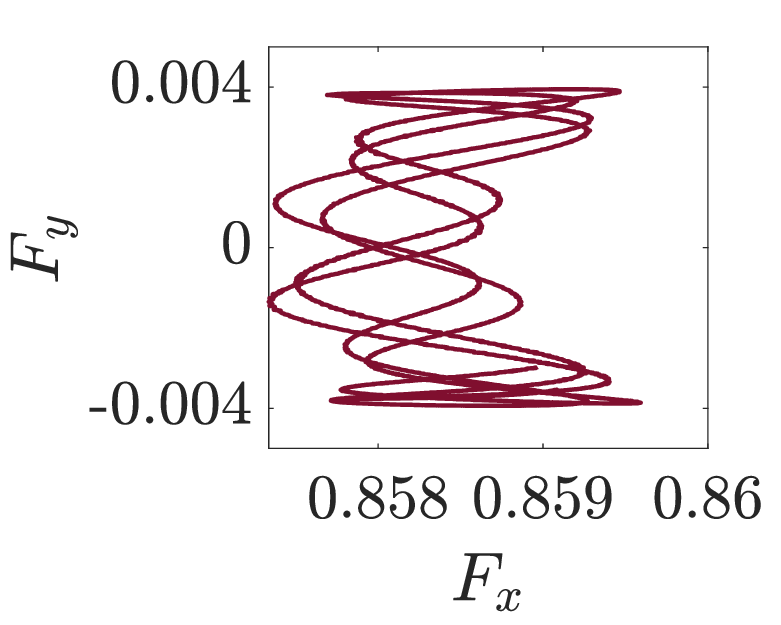}
\includegraphics[width=0.24\textwidth]{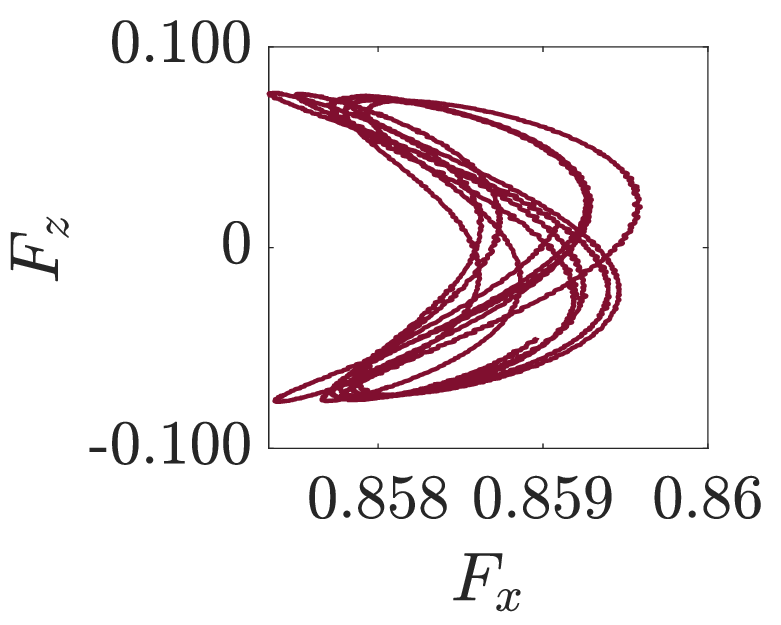}
\includegraphics[trim={0 0 0 0},clip,width=0.49\textwidth]{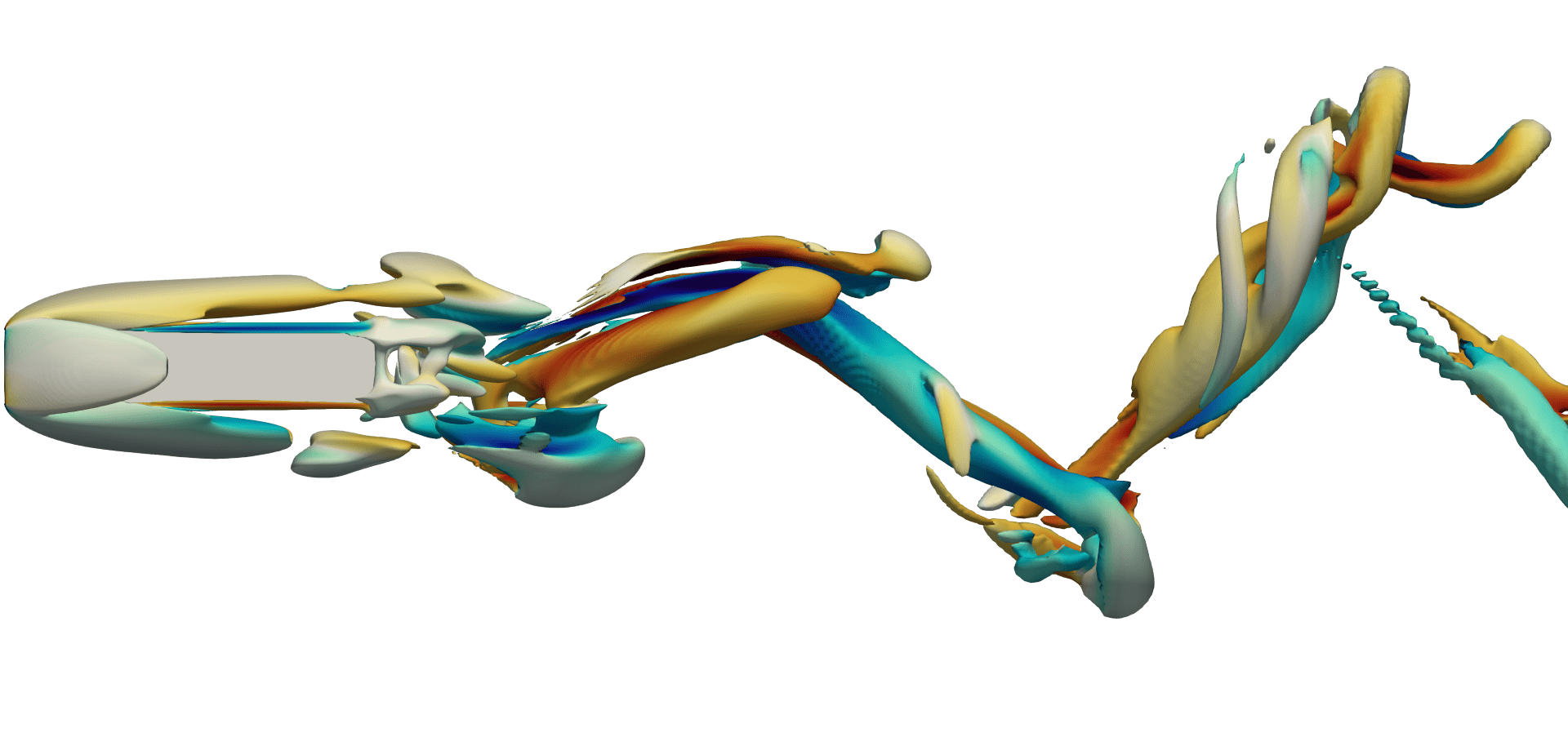}
\includegraphics[width=0.24\textwidth]{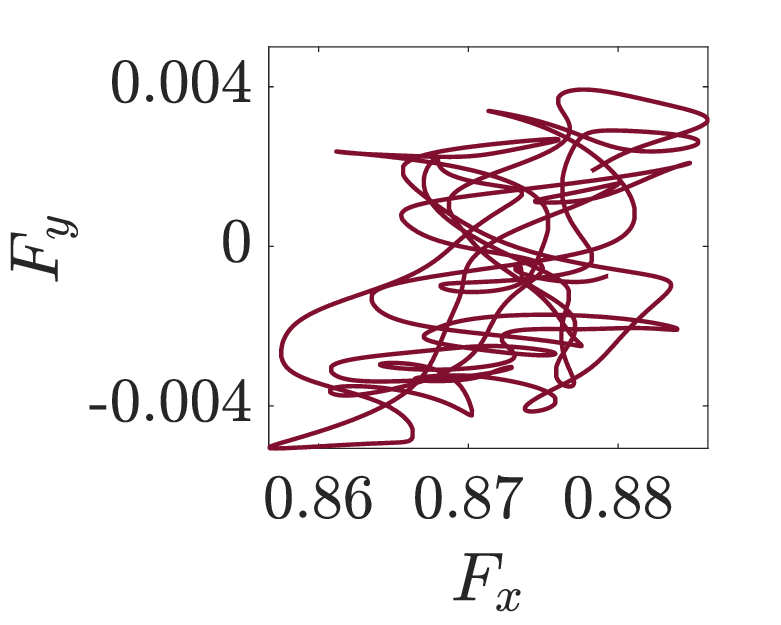}
\includegraphics[width=0.24\textwidth]{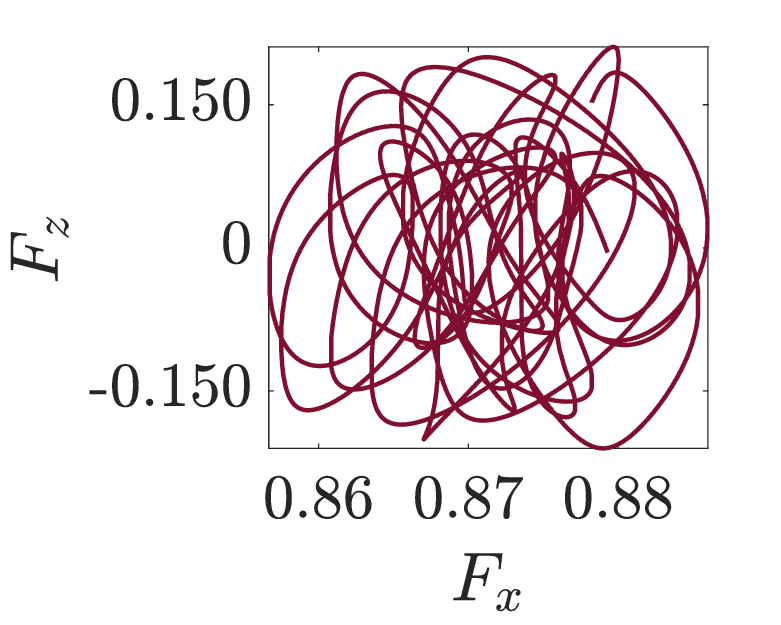}
\caption{First aperiodic regimes for $L=5$ and $W=5$. 
Top panels: frequency spectra of  $F_y$ (left) and $F_z$ (right) for $305 \le Re \le 345$. 
Bottom panels: structure of the flow for  $Re=315$ (top) and $Re=335$ (bottom).
Left: instantaneous isosurfaces $\lambda_2=-0.25$ coloured by $-1 \le \omega_x \le 1$.
Right: force diagrams $F_y-F_x$ and $F_z-F_x$.}
\label{fig:FlowStrucW5305335}
\end{figure}

As the Reynolds number increases, the limit cycle loses its stability via a Neimark-Sacker bifurcation, and new frequencies appear in the flow. 
At $Re \approx 305$ a torus replaces the limit cycle. The unsteady $A_yS_z$ wake mode predicted by the LSA
of the steady $S_yS_z$ base flow (\S\ref{sec:stability}) becomes unstable, and the flow experiences an oscillating motion in both $y$ and $z$ directions. 
The new frequency is $St \approx 0.03$, in agreement with the value predicted by the LSA.
As $Re$ increases, the two modes interact nonlinearly  and different frequencies arise in the spectra (figure \ref{fig:FlowStrucW5305335}), resulting into a progressive loss of coherency. 
However, 
the frequency spectra (and the POD, not shown) indicate that the horizontal oscillations at $St=0.03$ are less intense than the vertical oscillations at $St=0.09$.
As a result, up to $Re \approx 325$ the  dynamics is mainly driven by the vertical oscillation of the wake and the resulting hairpin-like structures.

For $Re \gtrapprox 325$ a new mode arises with a frequency in the range $0.16 \le St \le 0.2$  
associated with the shedding  of spanwise vorticity tubes from the top/bottom LE shear layers, that roll up and generate HVs in the wake, similarly to what was found for smaller $W$. 
The interaction of these vortices with the wake unsteadiness is analysed later in this subsection.
The LE vortex shedding frequency 
for $W=5$ is smaller than that for $W=1.2$ and $2.25$.
%
\begin{figure}
\centering
\includegraphics[width=0.49\textwidth]{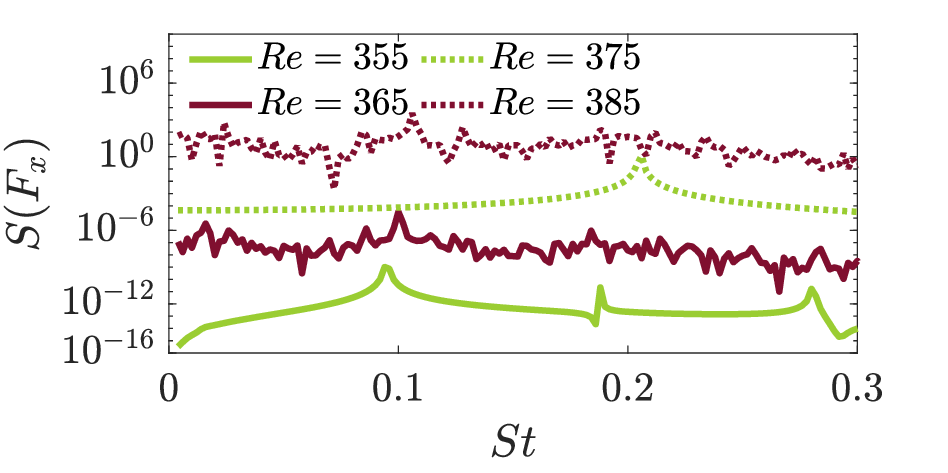}
\includegraphics[width=0.49\textwidth]{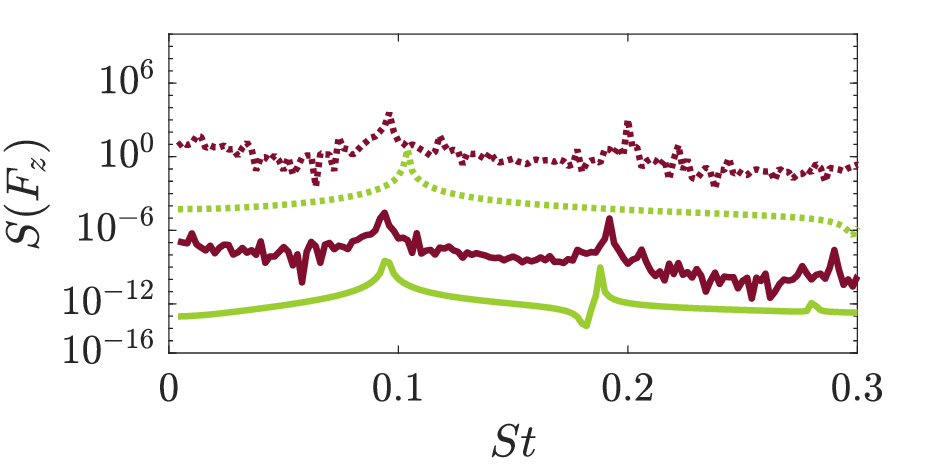}
\vspace{0.5cm}\\
\begin{tikzpicture}

\definecolor{clr1}{RGB}{18 78 128}
\definecolor{clr2}{RGB}{89 165 216}
\definecolor{clr3}{RGB}{145 229 246}
\definecolor{clr4}{RGB}{255 143 163}
\definecolor{clr5}{RGB}{50 205 50}
\definecolor{clr6}{RGB}{201 24 74}
\definecolor{clr7}{RGB}{128 15 47}
\definecolor{clr8}{RGB}{174 32 18}
\definecolor{clr9}{RGB}{155 34 38}
\definecolor{clr14}{RGB}{154 205 50}

\begin{axis}[%
width=0.85\textwidth,
height=0.15\textwidth,
scale only axis,
xmin=344,
xmax=386,
ymin=0.089,
ymax=0.21,
xlabel={$Re$},
ylabel={$St$},
ylabel style={at={(0.02,0.5)}},
axis background/.style={fill=white},
legend columns=3,transpose legend,
legend style={at={(0.99,0.25)}, anchor=east, legend cell align=left, align=left, fill=none, draw=none}
]

\addplot [color=black,solid,draw=none,mark=*,mark options={scale=1.4,black,fill=clr7}]
  table[row sep=crcr]{%
  345  0.17888 \\
  350  0.18035 \\
  360  0.18762 \\
  365  0.19003 \\
  380  0.19582 \\
  385  0.19851 \\
};

\addplot [color=black,solid,draw=none,mark=square*,mark options={scale=1.4,black,fill=clr7}]
  table[row sep=crcr]{%
  345  0.089442 \\
  350  0.090787 \\
  360  0.090534 \\
  365  0.090883 \\
  380  0.092634 \\
  385  0.093348 \\
};

\addplot [color=black,solid,draw=none,mark=*,mark options={scale=1.4,black,fill=clr14}]
  table[row sep=crcr]{%
  353  0.18521 \\
  355  0.18546 \\
  357  0.18960 \\
  373  0.20279 \\
  374  0.2028  \\
  375  0.20343 \\
};

\addplot [color=black,solid,draw=none,mark=square*,mark options={scale=1.4,black,fill=clr14}]
  table[row sep=crcr]{%
  353 0.092606 \\
  355 0.092729 \\
  357 0.092849 \\
  373 0.10139 \\
  374 0.10140 \\
  375 0.10171 \\
};

\addplot [color=black,dashed]
  table[row sep=crcr]{%
    351.5 0.089 \\
    351.5 0.21 \\
};

\addplot [color=black,dashed]
  table[row sep=crcr]{%
    358.5 0.089 \\
    358.5 0.21 \\
};

\addplot [color=black,dashed]
  table[row sep=crcr]{%
    370.5 0.089 \\
    370.5 0.21 \\
};

\addplot [color=black,dashed]
  table[row sep=crcr]{%
    377.5 0.089 \\
    377.5 0.21 \\
};

\node at (110,50) {{\color{clr14}lock-in I}};
\node at (300,50) {{\color{clr14}lock-in II}};

\end{axis}

\end{tikzpicture}%
\includegraphics[trim={0 0 0 0},clip,width=0.49\textwidth]{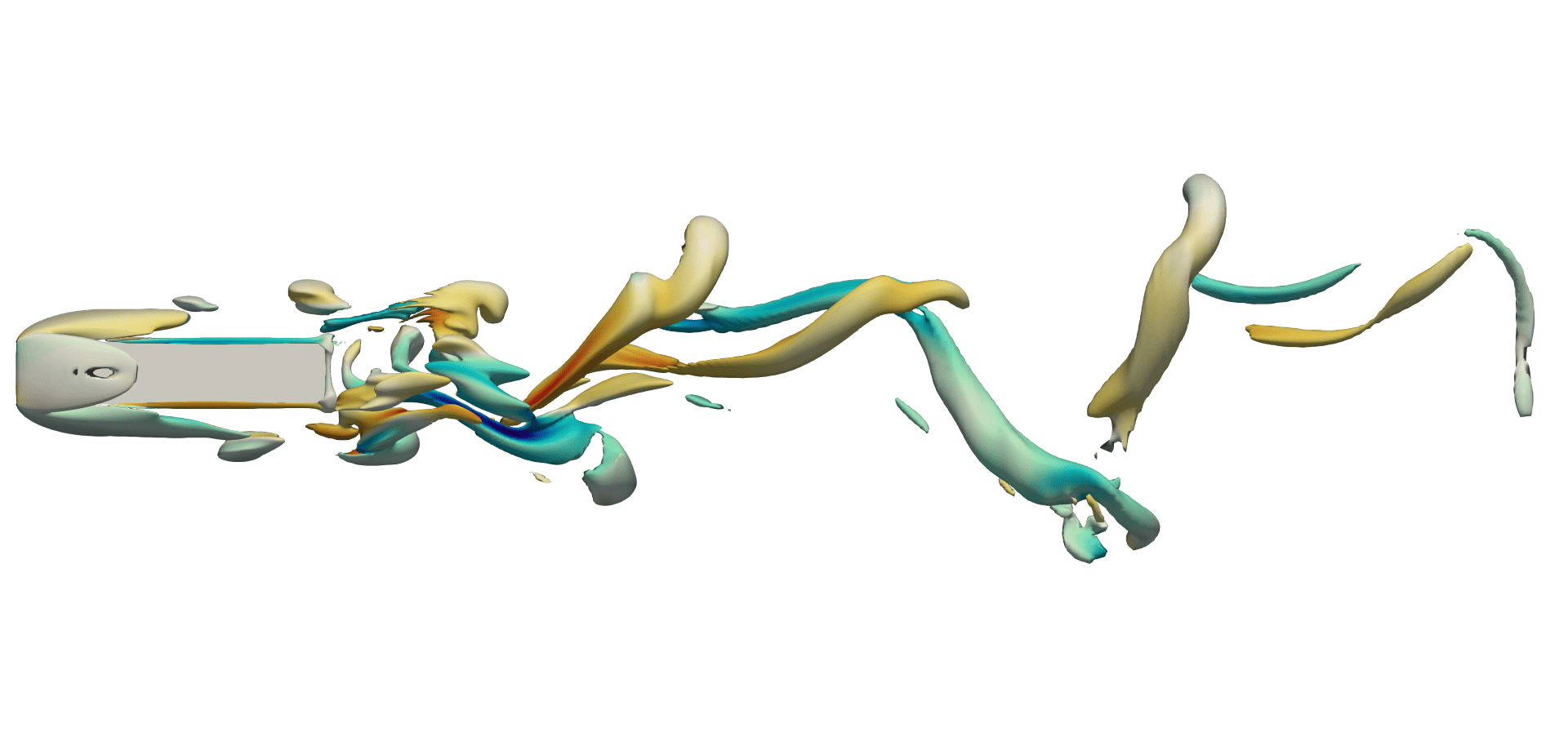}
\includegraphics[width=0.24\textwidth]{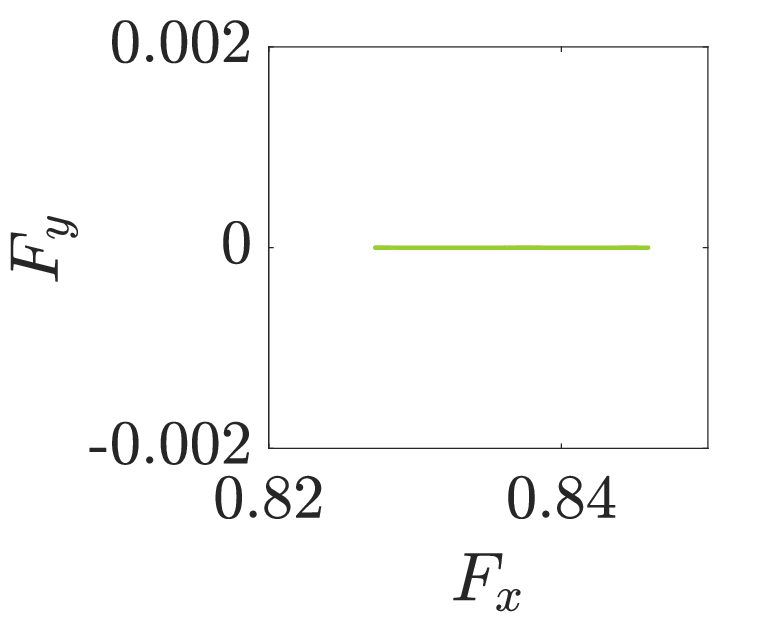}
\includegraphics[width=0.24\textwidth]{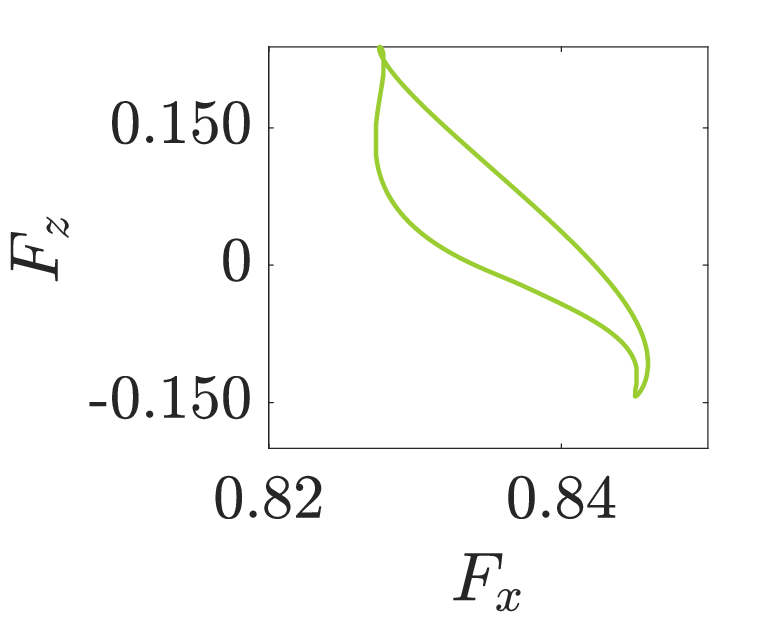}
\includegraphics[trim={0 0 0 0},clip,width=0.49\textwidth]{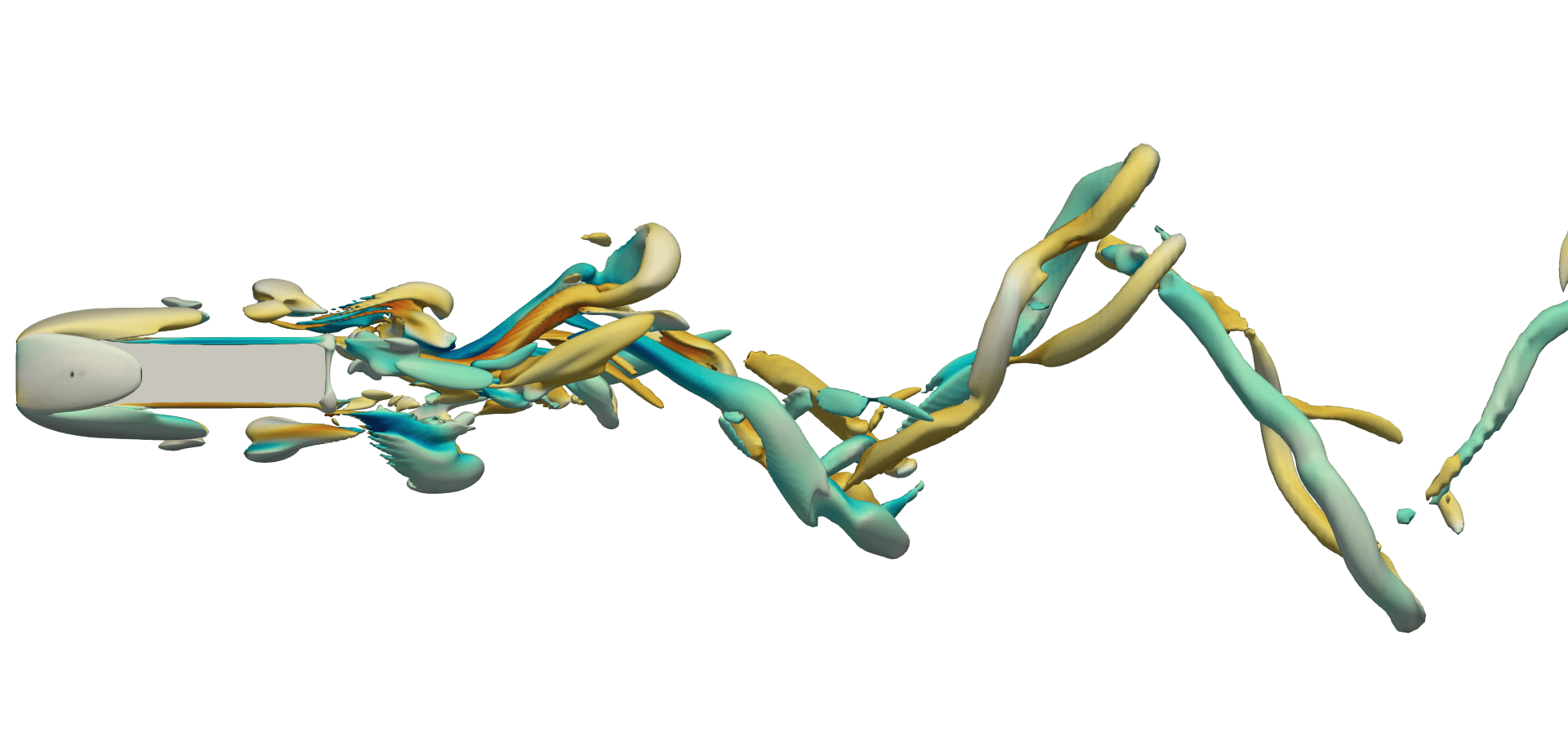}
\includegraphics[width=0.24\textwidth]{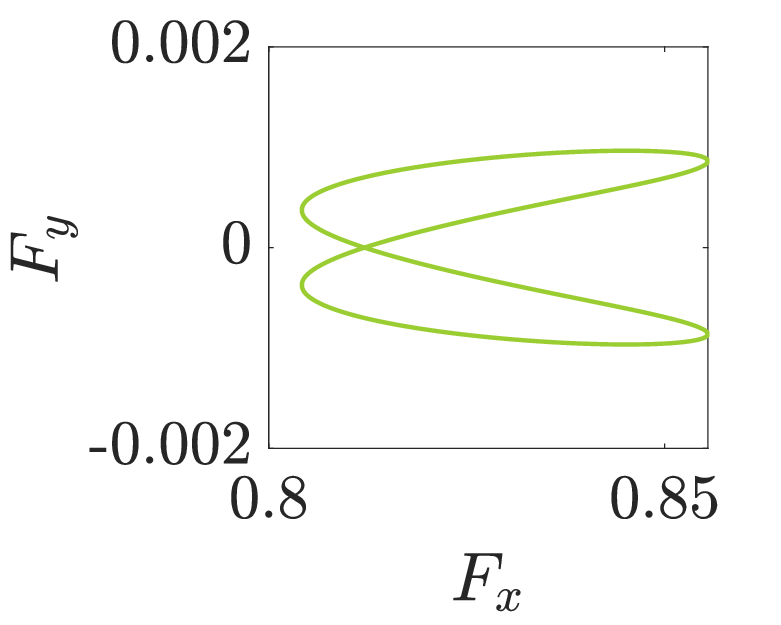}
\includegraphics[width=0.24\textwidth]{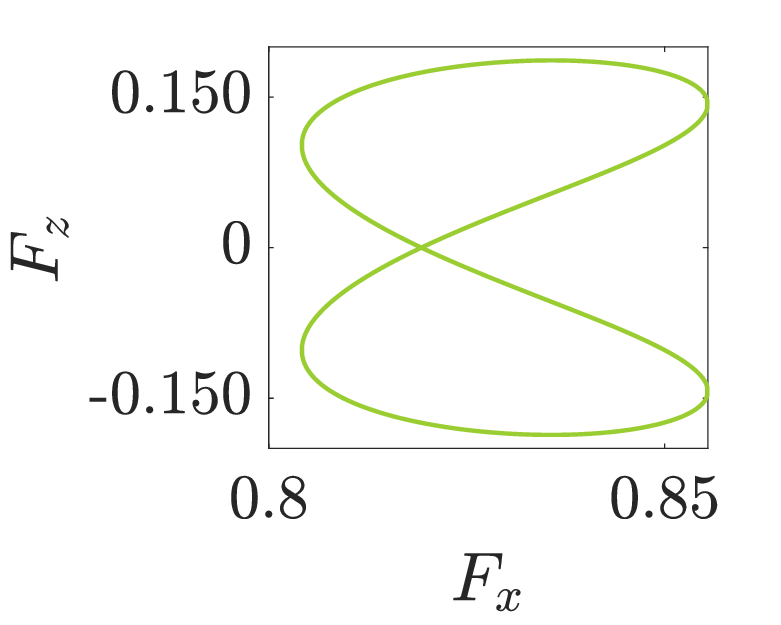}
\caption{
Aperiodic and frequency-locking regimes for $L=5$ and $W=5$ at larger Reynolds number  $345 \le Re \le 385$.
Same conventions as figure \ref{fig:FlowStrucW5305335}. 
Central panel: main flow frequencies. 
Bottom panels: $Re=353$ (top) and $Re=375$ (bottom). 
Note that vortices are shed from the LE shear layer in phase for $Re=353$ (lock-in region~I) and in phase opposition for $Re=375$ (lock-in region~II).
}
\label{fig:W5_freq_335_380}
\end{figure}
%
This LE vortex shedding progressively breaks the coherency of the flow (see the force diagrams and the emergence of several peaks in the frequency spectra in figure \ref{fig:FlowStrucW5305335}). 
As $Re$ increases, the importance of the LE vortex shedding on the flow dynamics progressively increases 
(see the peaks at $St \approx 0.09$ and $0.19$ in $S(F_z)$ and the following POD analysis)
and for $Re \gtrapprox 330$ its contribution 
is comparable to that of the wake oscillation.

For two narrow ranges of Reynolds numbers the flow oscillations influence each other in a way that produces synchronisation 
into a periodic regime. 
This phenomenon, known as frequency locking \citep{ioos-1990,kuznetsov-2004}, has already been observed by \cite{chiarini-quadrio-auteri-2022} for LE and TE vortex shedding in the flow past 2D rectangular cylinders.
The two sheddings, of periods $T_{LE}$ and $T_{TE}$, synchronise, and different stable cycles (periodic orbits) arise in the torus, with a long-time period $T_{lp}=p T_{TE}=q T_{LE}$, where $p,q \in \mathbb{N}$.
In the present case, the frequency locking is visualised in figure \ref{fig:W5_freq_335_380}.
The characteristic frequencies  of the vertical wake oscillation ($St \approx 0.09-0.10$; squares) and LE vortex shedding ($St \approx 0.18-0.20$; circles) increase weakly with $Re$.
The  frequency of the LE vortex shedding is approximately twice that of the wake oscillation, meaning that two LE vortices are shed from the top/bottom LE shear layers during  one wake oscillation period. 
The flow is periodic for $353 \le Re \le 357$ (lock-in region~I) and $ 373 \le Re \le 375$ (lock-in region~II): the LE vortex shedding synchronises with the vertical wake oscillation, and a  limit cycle arises with $T_{lp}=T_{TE} = 2 T_{LE}$ (i.e. $p=1$ and $q=2$).
%
Notably, the relative phase between the top and bottom vortex shedding is different in the two synchronised regimes. The LE vortices are indeed shed from the top and bottom LE shear layers in phase opposition 
in region I and in phase 
in region II
(see figure \ref{fig:W5_freq_335_380} and the following POD analysis).
This explains why the $St = 1/T_{LE}$ frequency is not visible in  $S(F_z)$ 
for the lock-in region II:
since the top and bottom LE vortices are shed in phase, they do not create any instantaneous unbalance of  vertical force.
By contrast, 
in region I
the closed trajectory in the $F_z-F_x$ force diagram is not symmetric with respect to $F_z=0$, meaning that the time-averaged $F_z$ is non-zero.
More generally, the time-averaged top/bottom 
symmetry is lost for $345 \le Re \le 360$ (figure \ref{fig:AR5_forces}).   
Regarding the lateral aerodynamic force, the different phase locking in the two synchronised regimes results in a different behaviour: 
$F_y \approx 0$ at all times 
in region I, while  $F_y$ oscillates with $St = 1/T_{TE} = 1/(2 T_{LE})$ 
in region II.


\begin{figure}
\centering
\includegraphics[trim={0 30 0 30},clip,width=0.49\textwidth]{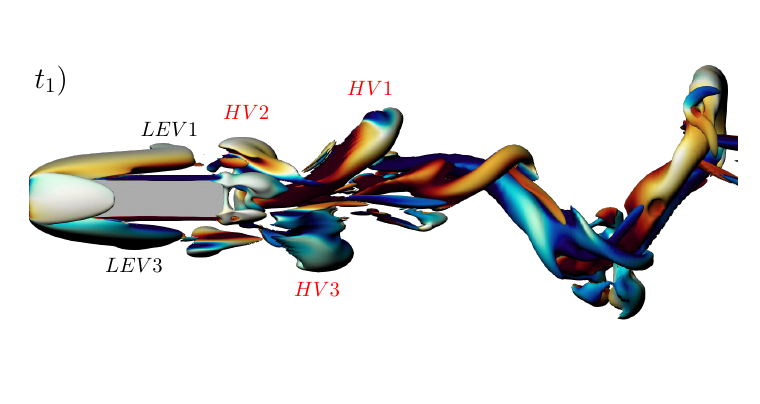}
\includegraphics[trim={0 30 0 30},clip,width=0.49\textwidth]{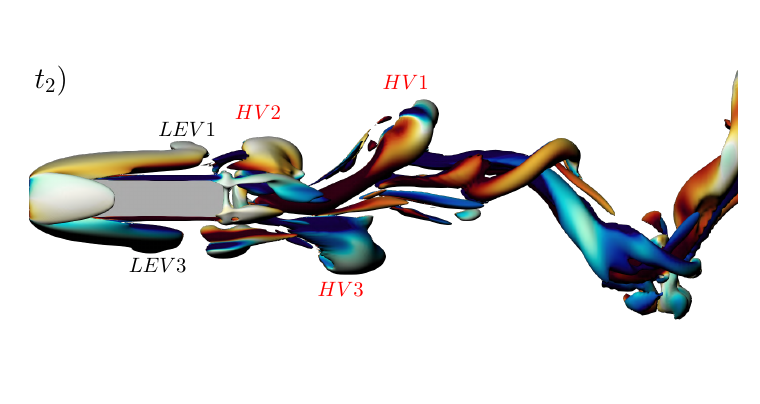}
\includegraphics[trim={0 30 0 30},clip,width=0.49\textwidth]{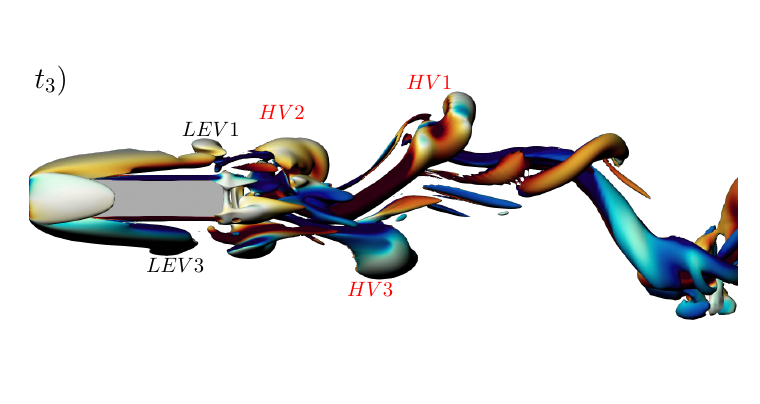}
\includegraphics[trim={0 30 0 30},clip,width=0.49\textwidth]{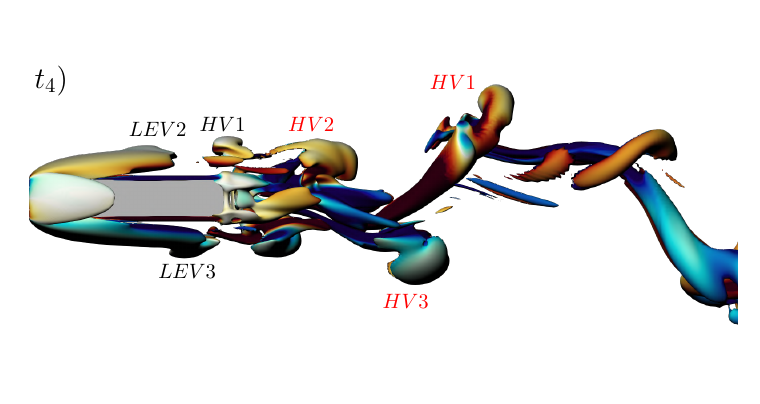}
\includegraphics[trim={0 30 0 30},clip,width=0.49\textwidth]{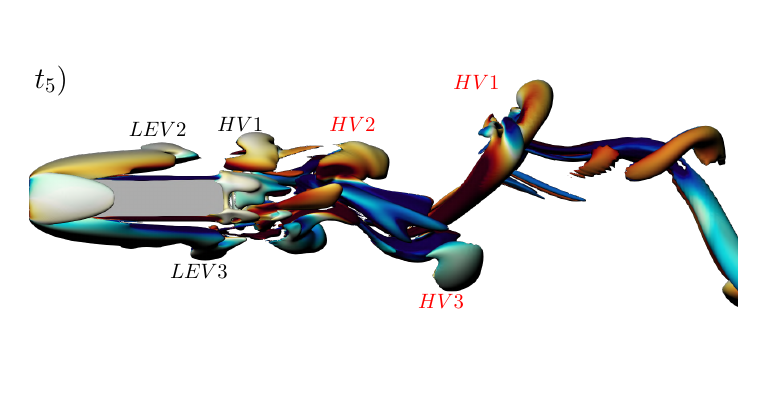}
\includegraphics[trim={0 30 0 30},clip,width=0.49\textwidth]{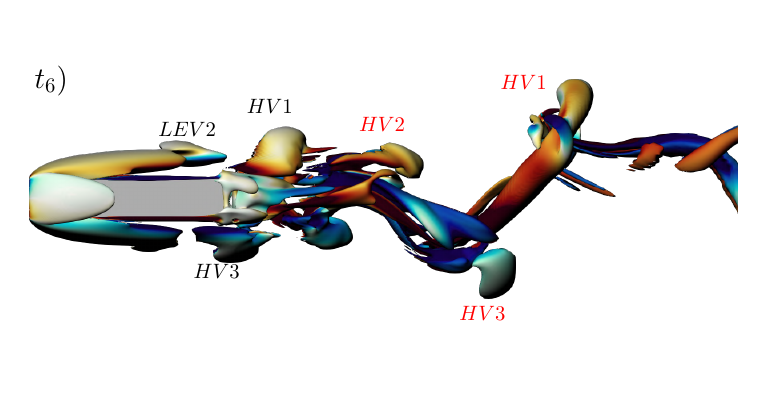}
\includegraphics[trim={0 30 0 30},clip,width=0.49\textwidth]{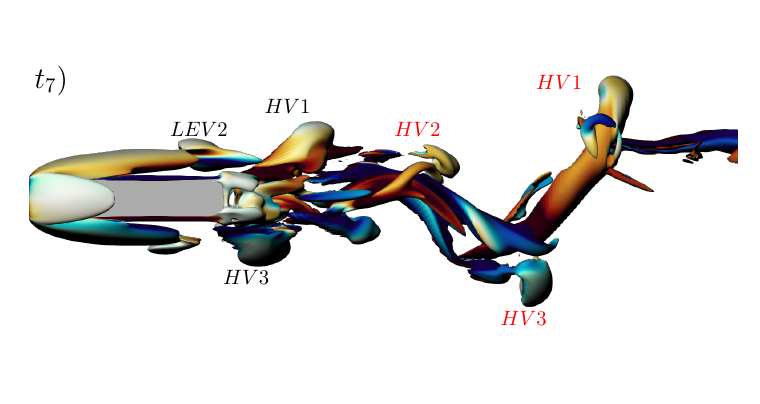}
\includegraphics[trim={0 30 0 30},clip,width=0.49\textwidth]{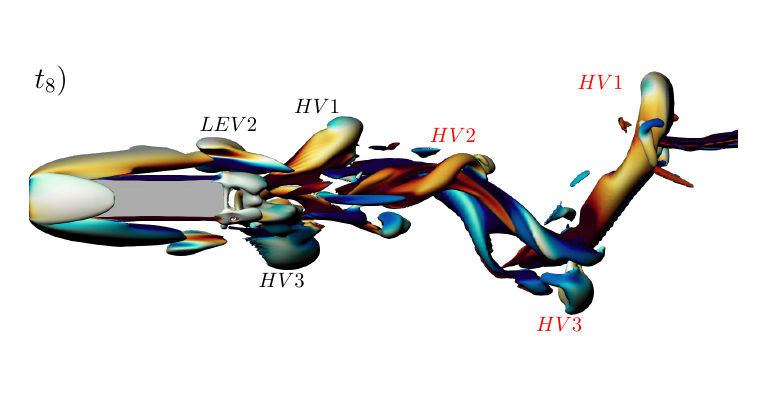}
\caption{
Frequency-locking regime for $L=5$, $W=5$ and $Re=353$ (lock-in region~I).
Lateral view of isosurfaces  $\lambda_2=-0.05$ coloured by  streamwise vorticity (blue-to-red colourmap  for $-1 \le \omega_x \le 1$). 
The snapshots are separated in time by $T/8$, where $T$ is the period of the wake oscillation. 
Black/red labels refer to vortices shed in the considered/previous period. 
LEV1-LEV2 and LEV3 refer to LE vortices shed from the top and bottom LE shear layers, respectively. 
HV1, HV2 and HV3 indicate  hairpin vortices that arise in the wake once LEV1, LEV2 and LEV3 cross the trailing edge.}
\label{fig:Ary5-lambda2-Re353-2D}
\end{figure}
We now illustrate the interaction between the LE and TE vortex shedding in 
region I.
Figure \ref{fig:Ary5-lambda2-Re353-2D} shows eight snapshots of the flow at $Re=353$, equispaced in one period $T_{TE}$. 
Black and red labels refer to vortices  shed from the prism in the current and previous periods, respectively.
Vortices shed from the top and bottom LE shear layers are in phase opposition. 
At $t = t_1$ a spanwise-aligned vortex, LEV1, is generated from the top LE shear layer, and  convected downstream (see $t=t_2$). 
At $t=t_3$, LEV1 crosses the TE corner in phase with the upwards motion of the wake, and rolls up ($t_4$ to $t_6$), generating a large head-up hairpin vortex (HV1) that is then shed downstream ($t_7$ and $t_8$). 
The same dynamics is observed over the bottom side. A spanwise-aligned vortex LEV3 
is shed from the LE shear layer at $t=t_3$. It crosses the bottom TE at $t=t_5$, and rolls up generating a head-down hairpin vortex HV3 at $t=t_6$. 
Moving again to the top side, at $t=t_5$ the second spanwise-aligned vortex LEV2 is shed from the top LE shear layer. 
It crosses the TE corner at $t=t_5$ in phase with the downwards motion of the wake, and generates a hairpin vortex HV2 that is then connected with the HV3 hairpin vortex.

In region II, LE vortices interact with the vertical wake oscillation in a similar way to generate the large HVs that characterise the wake dynamics. 
However, the top/bottom LE vortices are shed in phase, and thus cross the top/bottom TE in phase (bottom left panel in figure \ref{fig:W5_freq_335_380}). 
When a top/bottom LE vortex crosses the top/bottom TE in phase with an upward/downward motion of the wake, a head-up/head-down hairpin-like vortex arises in the wake.

In the aperiodic regimes
($Re \in  [345,350]$, $[360,365]$ and $[380,385]$),
a similar interaction between the LE vortex shedding and the wake unsteadiness as in the lock-in region I is observed, although not fully periodic.

\begin{figure}
  \centering
  \begin{tikzpicture}

\definecolor{clr1}{RGB}{18 78 128}
\definecolor{clr2}{RGB}{89 165 216}
\definecolor{clr3}{RGB}{145 229 246}
\definecolor{clr4}{RGB}{255 143 163}
\definecolor{clr5}{RGB}{50 205 50}
\definecolor{clr6}{RGB}{201 24 74}
\definecolor{clr7}{RGB}{128 15 47}
\definecolor{clr8}{RGB}{174 32 18}
\definecolor{clr9}{RGB}{155 34 38}
\definecolor{clr14}{RGB}{154 205 50}

\begin{axis}[%
width=0.35\textwidth,
height=0.15\textwidth,
scale only axis,
xmin=300,
xmax=390,
ymin=0.009,
ymax=0.235,
xlabel={$Re$},
ylabel={$St$},
ylabel style={at={(0.05,0.5)}},
axis background/.style={fill=white},
legend columns=3,transpose legend,
legend style={at={(0.99,0.25)}, anchor=east, legend cell align=left, align=left, fill=none, draw=none}
]

\addplot[red, dashed, domain=1:8] {0.1};
\addplot[red, dashed, domain=1:8] {0.2};

\addplot [color=black,solid,draw=none,mark=*,mark options={scale=1.6,black,fill=clr14}]
  table[row sep=crcr]{%
  357 0.096281 \\
  375 0.11655 \\
};

\addplot [color=black,solid,draw=none,mark=*,mark options={scale=1.6,black,fill=clr7}]
  table[row sep=crcr]{%
  305 0.0912 \\
  315 0.0912 \\
  325 0.0912 \\
  335 0.0912 \\
  345 0.0912 \\
  365 0.0912 \\
  385 0.0912 \\
};

\addplot [color=black,solid,draw=none,mark=square*,mark options={scale=1.4,black,fill=clr14}]
  table[row sep=crcr]{%
  357 0.18516 \\
  375 0.23315 \\
};

\addplot [color=black,solid,draw=none,mark=square*,mark options={scale=1.4,black,fill=clr7}]
  table[row sep=crcr]{%
  325 0.18867 \\
  335 0.18867 \\
  345 0.18169 \\
  365 0.193 \\
  385 0.2016 \\
};

\end{axis}

\end{tikzpicture}%
  \begin{tikzpicture}

\definecolor{clr1}{RGB}{18 78 128}
\definecolor{clr2}{RGB}{89 165 216}
\definecolor{clr3}{RGB}{145 229 246}
\definecolor{clr4}{RGB}{255 143 163}
\definecolor{clr5}{RGB}{50 205 50}
\definecolor{clr6}{RGB}{201 24 74}
\definecolor{clr7}{RGB}{128 15 47}
\definecolor{clr8}{RGB}{174 32 18}
\definecolor{clr9}{RGB}{155 34 38}
\definecolor{clr14}{RGB}{154 205 50}

\begin{axis}[%
width=0.35\textwidth,
height=0.15\textwidth,
scale only axis,
xmin=300,
xmax=390,
xlabel={$Re$},
ylabel={$\lambda/\sum \lambda$},
ylabel style={at={(0.05,0.5)}},
axis background/.style={fill=white},
legend columns=3,transpose legend,
legend style={at={(0.99,0.25)}, anchor=east, legend cell align=left, align=left, fill=none, draw=none}
]

\addplot[red, dashed, domain=1:8] {0.1};
\addplot[red, dashed, domain=1:8] {0.2};

\addplot [color=black,solid,draw=none,mark=*,mark options={scale=1.6,black,fill=clr14}]
  table[row sep=crcr]{%
  357 0.2768 \\
  375 0.3252 \\
};

\addplot [color=black,solid,draw=none,mark=*,mark options={scale=1.6,black,fill=clr7}]
  table[row sep=crcr]{%
  305 0.3649 \\
  315 0.3434 \\
  325 0.3062 \\
  335 0.2716 \\
  345 0.2133 \\
  365 0.1870 \\
  385 0.1670 \\
};

\addplot [color=black,solid,draw=none,mark=square*,mark options={scale=1.4,black,fill=clr14}]
  table[row sep=crcr]{%
  357 0.1402 \\
  375 0.1015 \\
};

\addplot [color=black,solid,draw=none,mark=square*,mark options={scale=1.4,black,fill=clr7}]
  table[row sep=crcr]{%
  325 0.0485 \\
  335 0.0511 \\
  345 0.0872 \\
  365 0.1058 \\
  385 0.1323 \\
};

\end{axis}

\end{tikzpicture}%
  \includegraphics[width=0.32\textwidth]{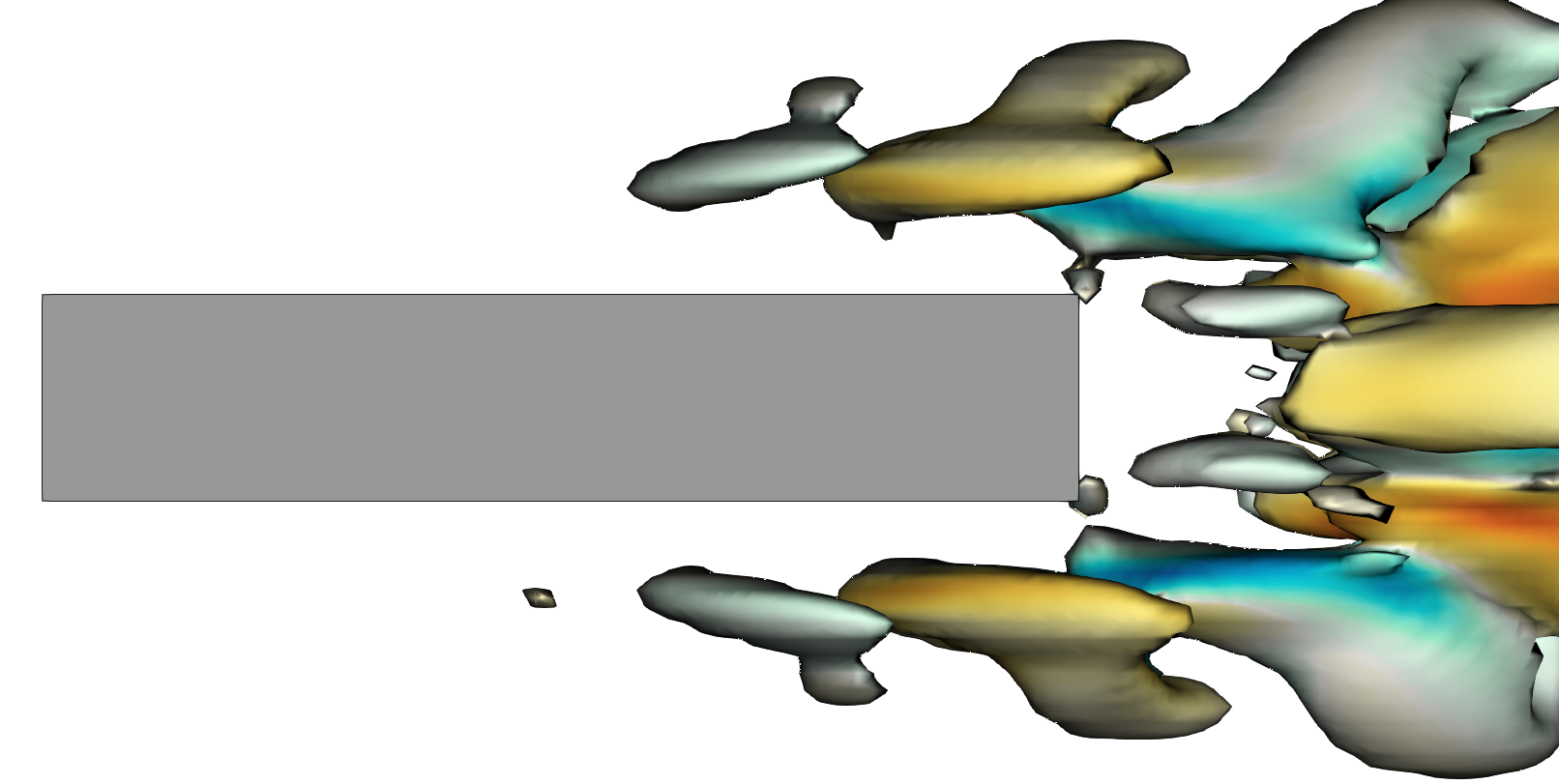}
  \includegraphics[width=0.32\textwidth]{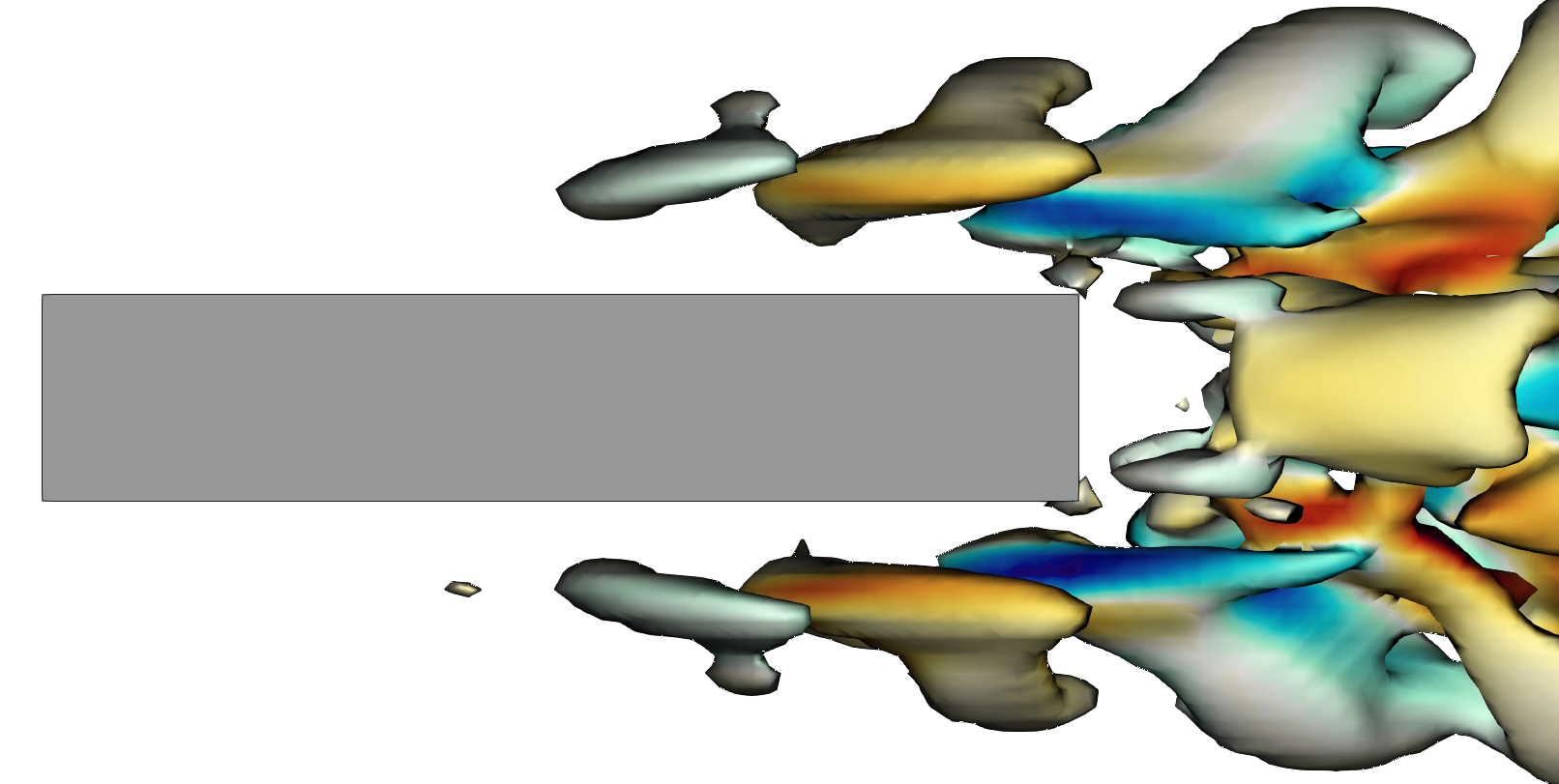}
  \includegraphics[width=0.32\textwidth]{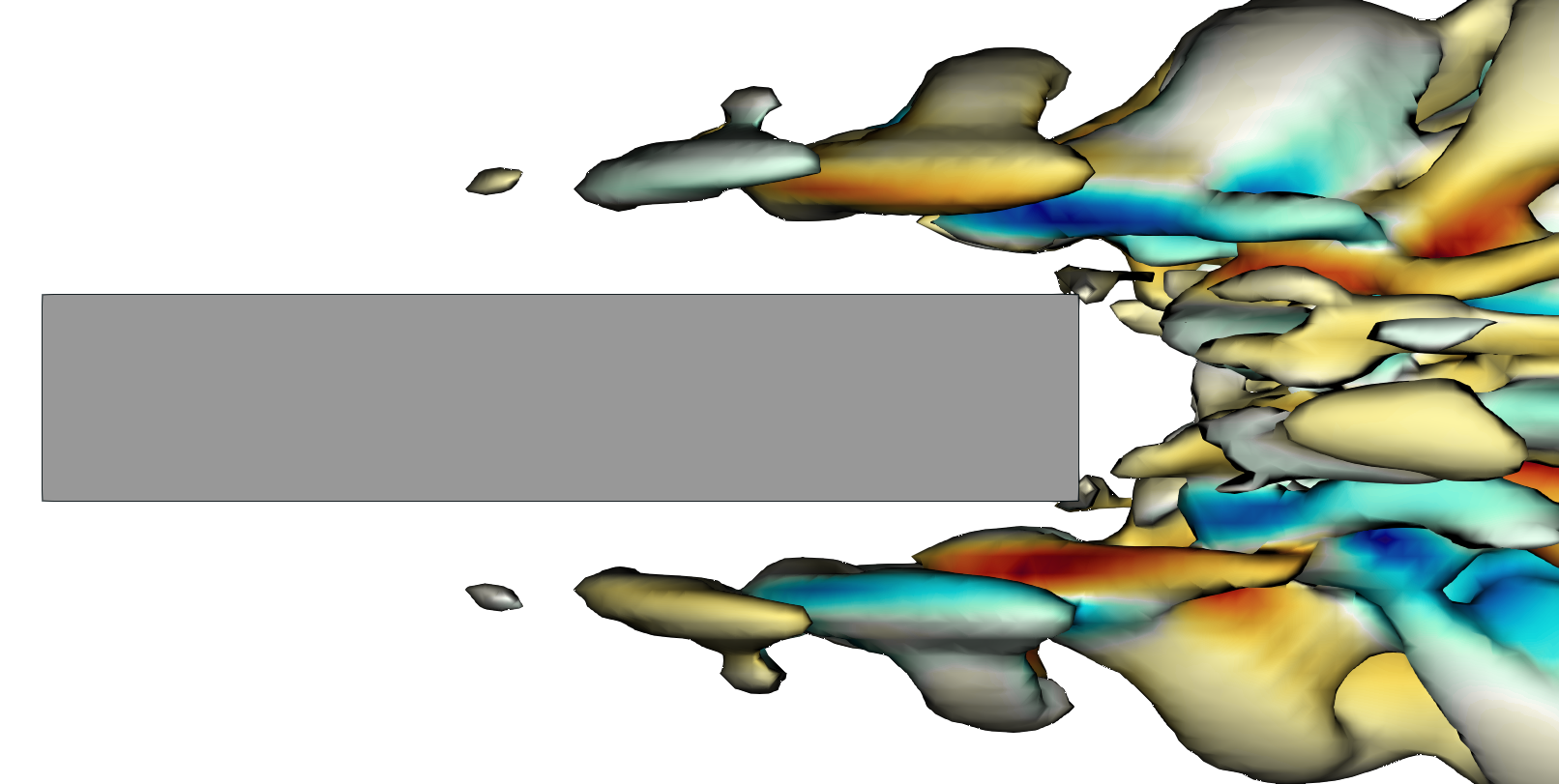}  
  \caption{POD analysis for $L=5$ and $W=5$ in the range $305 \le Re \le 385$. 
  In the top panels, circles and squares refer to the unsteady wake mode ($St \approx 0.09$) and  LE vortex shedding mode ($St \approx 0.19$), respectively. 
  When the two phenomena are distributed over several POD modes, the sum of their energy content is reported in the top right panel. 
  Bottom panels: lateral view of the POD mode associated with LE vortex shedding. Isosurfaces of $\lambda_2$ coloured by streamwise vorticity. From left to right: aperiodic regime at $Re=345$, and  periodic regimes at $Re=357$ and $Re=375$.}
  \label{fig:POD_W5_Re305_Re375}
\end{figure}    
We use POD to further examine the dependence of the flow structure on $Re$ 
(figure \ref{fig:POD_W5_Re305_Re375}). 
In agreement with the above discussion, the dominant POD modes are associated with the wake unsteadiness and with the LE vortex shedding, and exhibit a single frequency of $St \approx 0.09$ and $St \approx 0.19$ respectively. 
The POD modes associated with the wake unsteadiness mainly evolve in the wake and feature large-scale head-up and head-down HVs downstream the TE (not shown for brevity). The POD modes associated with the LE vortex shedding, instead, originate over the top and bottom sides where they feature spanwise aligned structures (see the bottom panels). 
The spatial structure of the LE vortex shedding POD modes changes with $Re$, as visible along the sides of the prism. 
In the aperiodic regime ($Re=345, 365, 385$) and in the lock-in region I ($Re=357$), the modes are antisymmetric with respect to an inversion of the $z$ axis, with $\omega_x(x,y,z) = \omega_x(x,y,-z)$. This agrees with the out-of-phase shedding of vortices from the top and bottom LE shear layers. 
In the lock-in region II ($Re = 375$), instead, the modes exhibit a top/down symmetry with $\omega_x(x,y,z) = -\omega_x(x,y,-z)$, which is consistent with in-phase LE vortex shedding. 
Also, the POD confirms that in the aperiodic $aS_yS_z$ regime the relative intensity of the LE vortex shedding  increases with $Re$ (see the top right panel). 
However, the wake unsteadiness dominates in the periodic regimes.

\subsubsection{Large $Re$: Aperiodic regime}

\begin{figure}
\centering
\includegraphics[width=0.49\textwidth]{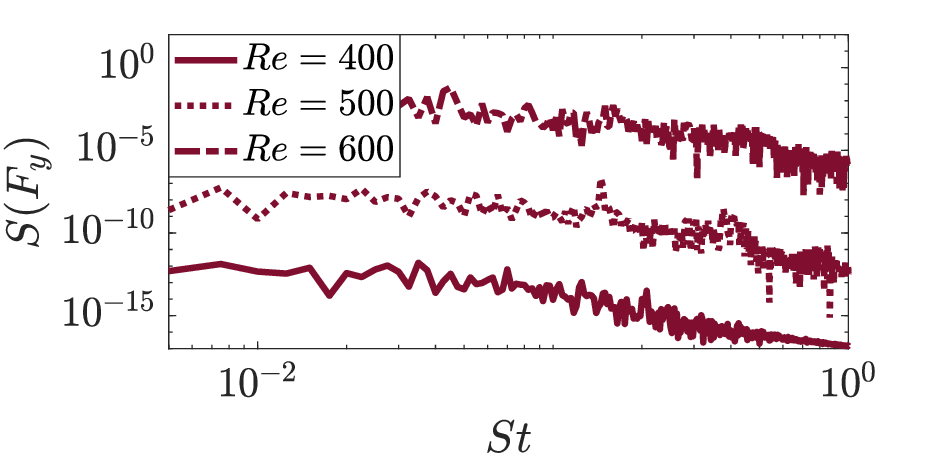}
\includegraphics[width=0.49\textwidth]{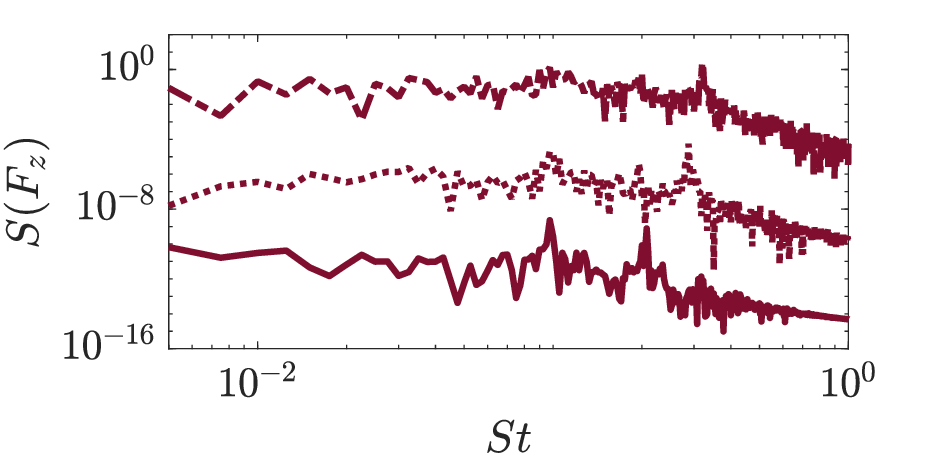}
\includegraphics[trim={0 0 0 0},clip,width=0.49\textwidth]{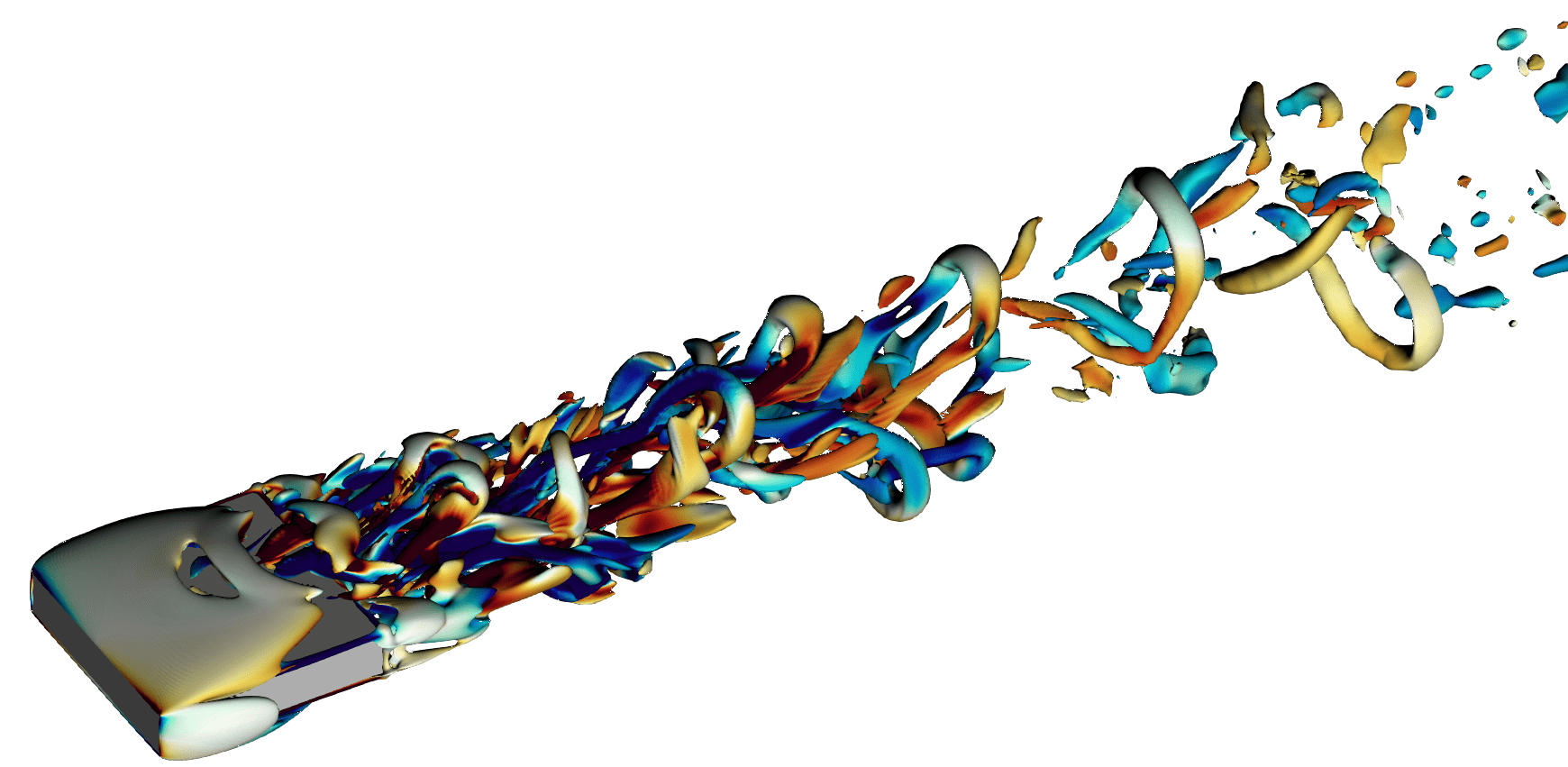}
\includegraphics[width=0.24\textwidth]{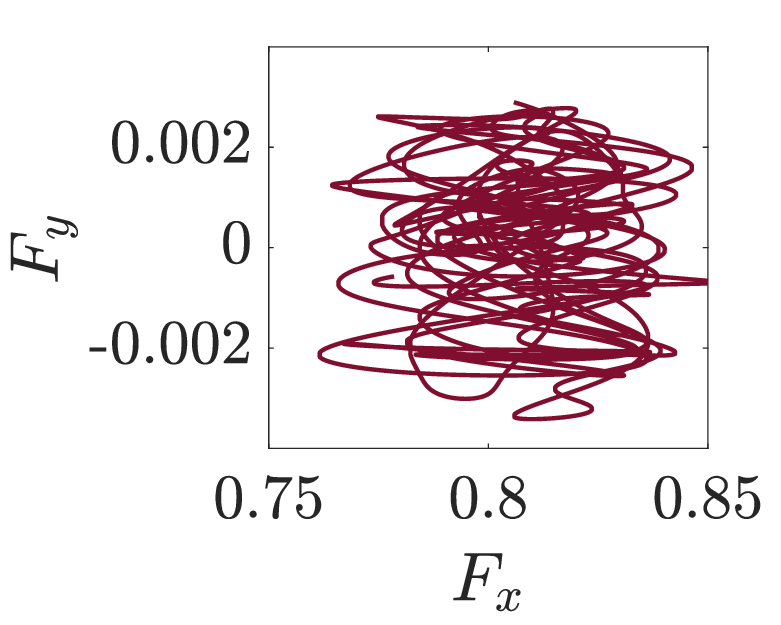}
\includegraphics[width=0.24\textwidth]{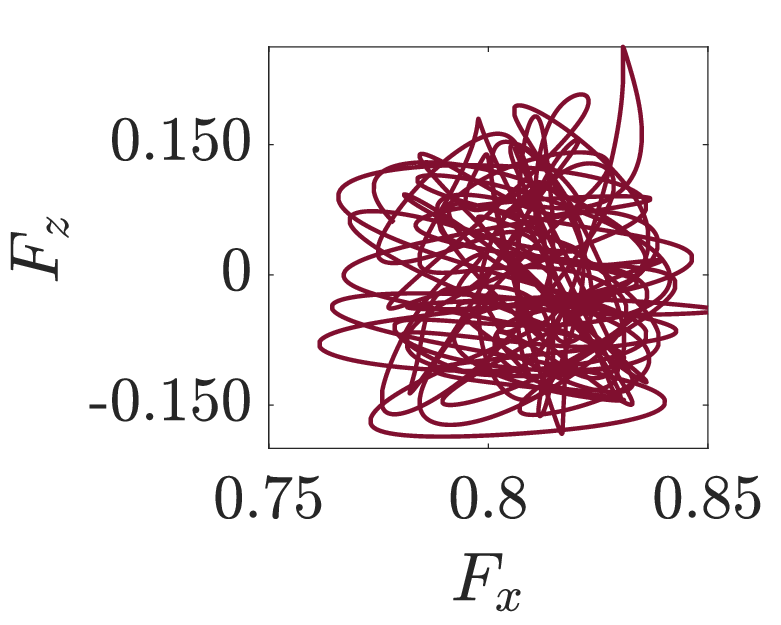}
\includegraphics[trim={0 0 0 0},clip,width=0.49\textwidth]{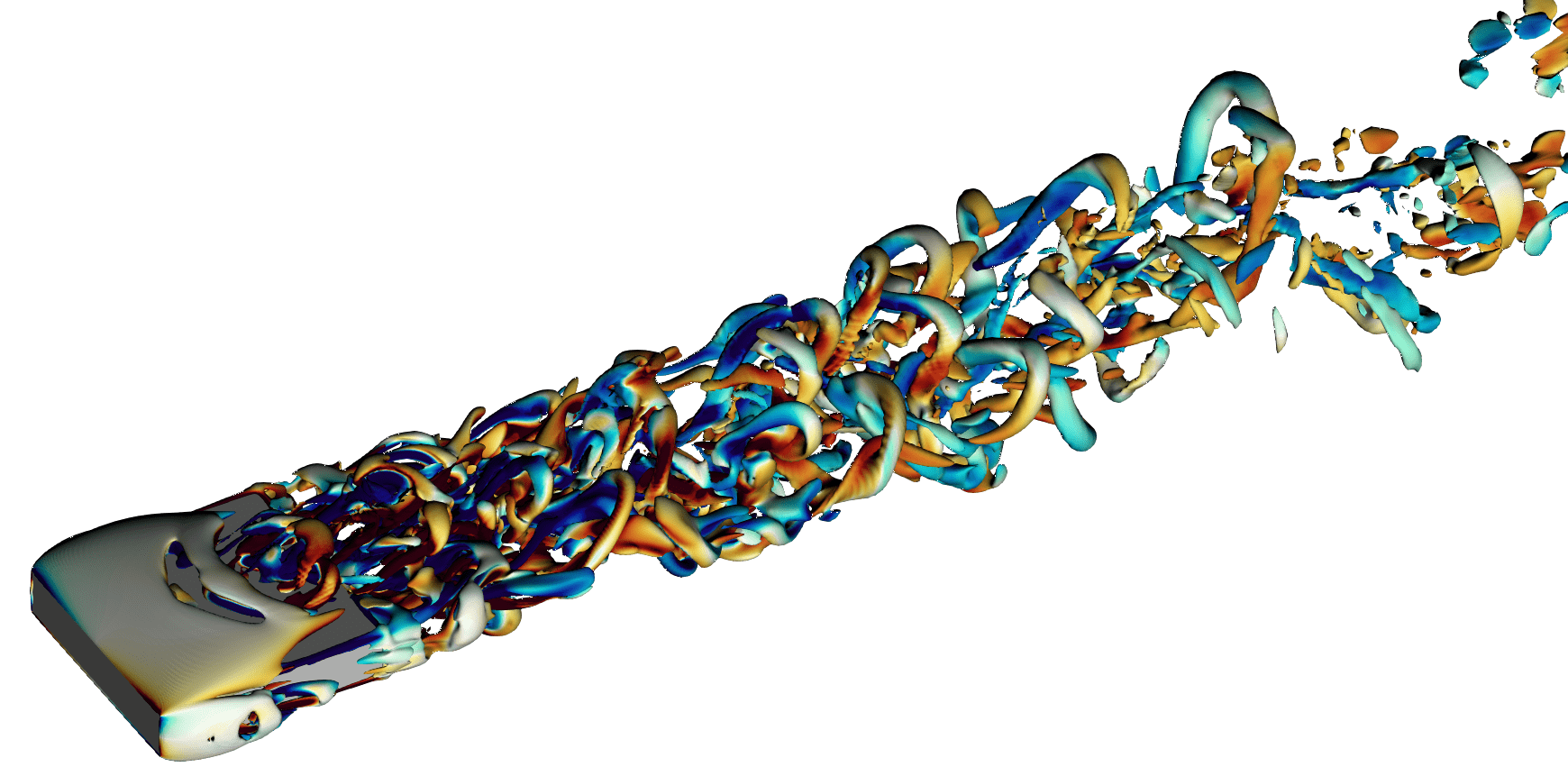}
\includegraphics[width=0.24\textwidth]{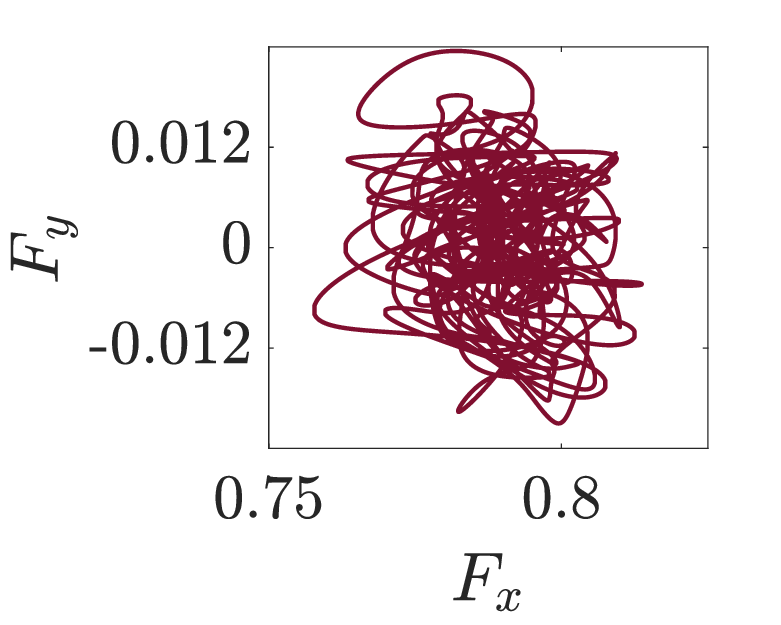}
\includegraphics[width=0.24\textwidth]{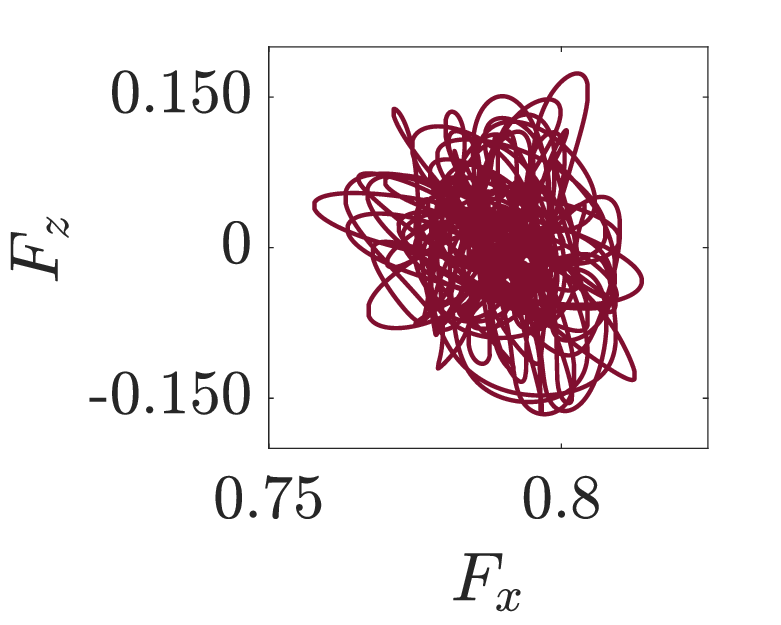}
\caption{
Unsteady regimes for $L=5$ and $W=5$ at larger Reynolds number $400 \le Re \le 600$.
Top panels: frequency spectra of the aerodynamic forces $F_y$ (left) and $F_z$ (right). 
Bottom panels: structure of the flow at $Re=400$ (centre) and $Re=500$ (bottom). 
Left: isosurfaces of $\lambda_2=-0.05$ coloured by streamwise vorticity. 
Right: force diagrams $F_y-F_x$ and $F_z-F_x$.}
\label{fig:Ary5-lambda2-Re400-600}
\end{figure}

For $Re \ge 400$ the synchronisation between the LE and TE vortex shedding is lost, and the flow becomes progressively more chaotic. 
Like for lower $Re$, the frequency of the wake oscillation 
remains approximately constant   ($St \approx 0.095-0.1$ for $Re\in[400,600]$), while that of the LE vortex shedding 
keeps increasing ($St \approx 0.21-0.32$). 
%
The increase in $Re$ is accompanied by a slight increase in the oscillation amplitude in the $y$ direction, as visualised in figure \ref{fig:AR5_forces}. 
For $Re \ge 500$, a third peak 
appears in $S(F_y)$ at $St \approx 0.125$, 
and a shedding of HVs occurs also along the lateral sides of the prism (figure \ref{fig:Ary5-lambda2-Re400-600}).

\section{Conclusion}
\label{sec:conclusions}

\begin{figure}
\centering
\includegraphics[trim={20 0 110 0},clip,width=1\textwidth]{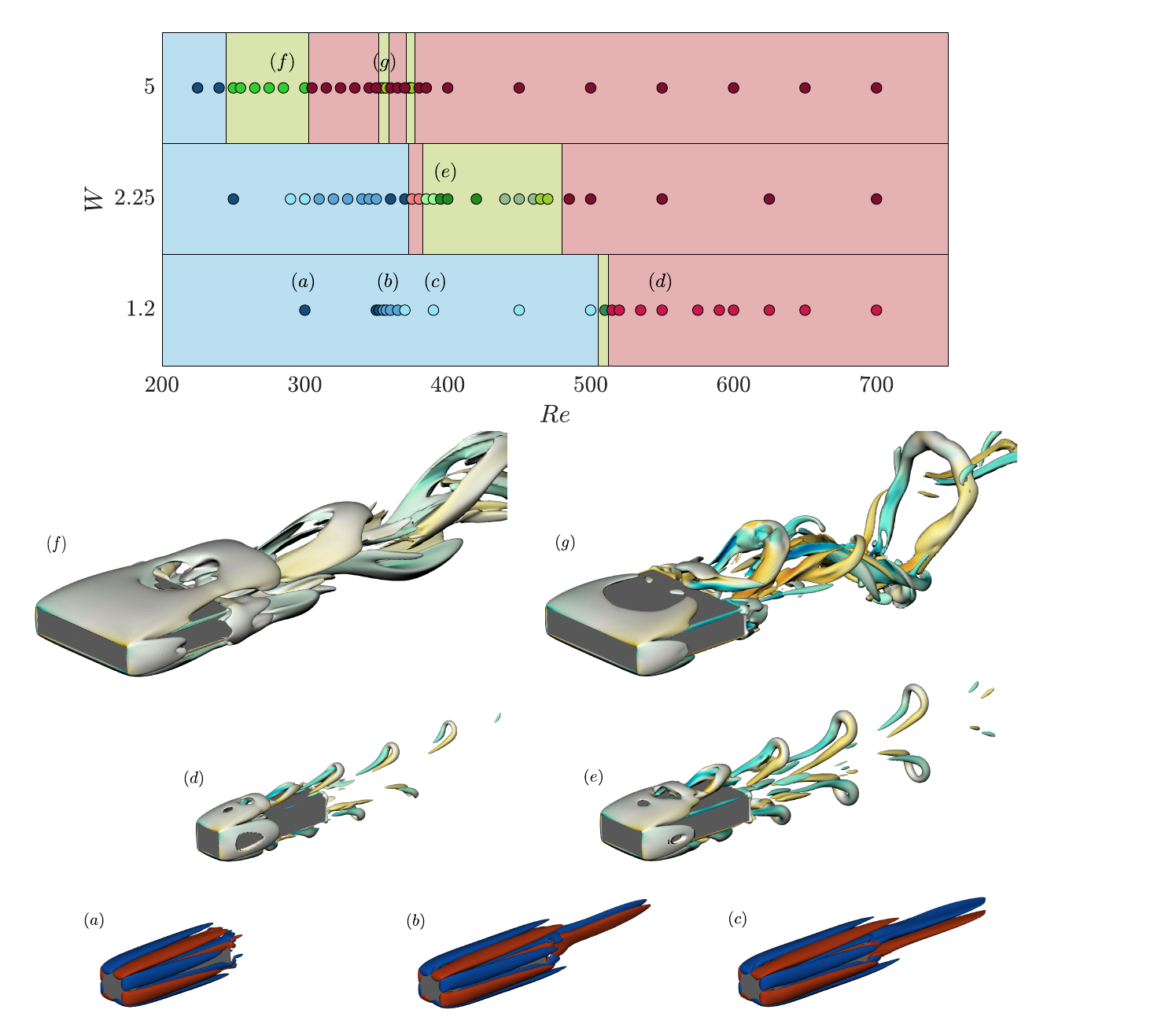}
\caption{Regimes observed in the flow past rectangular prisms with $L=5$ for Reynolds numbers up to $Re=700$. 
Blue, green and red shaded areas refer to steady, periodic and aperiodic states, respectively. Different tones refer to different regimes, accordingly to figures \ref{fig:3D_Ary1_forces}, \ref{fig:AR225_forces} and \ref{fig:AR5_forces}. The solid lines separating the regimes are a guide to the eye. Vertical axis not to scale. $(a)$-$(c)$~Steady flows are represented with $\omega_x = \pm 0.1$ isosurfaces, and $(d)$-$(g)$~unsteady flows  with snapshots of $\lambda_2$ isosurfaces coloured by streamwise vorticity.}
\label{fig:recap}
\end{figure}

In this study we have investigated the sequence of bifurcations of the laminar flow past 3D rectangular prisms 
for length-to-height and width-to-height ratios $1 \le L \le 5$ and $1.2 \le W \le 5$, and  Reynolds number up to $Re \approx 700$.

LSA shows that the nature of the primary bifurcation changes with the geometry. 
The flow past wide prisms with large $W/L$ experiences first a Hopf bifurcation and  becomes unstable to oscillatory perturbations that break the top/bottom planar symmetry, leading to a periodic vortex shedding across the smaller dimension of the body. 
For smaller $W/L$, the primary instability consists of a pitchfork bifurcation: 
for intermediate (small) $W/L$, the flow becomes unstable to stationary perturbations that break the top/bottom (left/right) planar symmetry and 
lead to a static vertical (horizontal) deflection of the wake. 
In all cases, the critical $Re$ of the primary instability increases with $L$, and depends non-monotonically on $W$.

A WNL analysis has been performed for $(L,W)=(5,1.2)$ and $(L,W)=(5,2.25)$ in the vicinity of the critical $Re$ of the static modes.
Two coupled amplitude equations are obtained, revealing the sequence of bifurcations close to the bifurcation points. 
For $W=1.2$, third-order amplitude equations yield a supercritical bifurcation scenario similar to that of Ahmed bodies \cite{zampogna_boujo_2023}, with an exchange of stability between the vertically and horizontally deflected states at $Re \approx 353$. 
For $W=2.25$, fifth-order amplitude equations yield a subcritical bifurcation for the horizontally deflected state, with a  narrow interval of bistability where the vertically  and horizontally deflected states coexist.
Fully nonlinear DNS confirm the bifurcation sequence predicted by the LSA and the WNL analysis, including the bistability-induced hysteresis for $W=2.25$ and the bifurcated states. 

At larger $Re$, nonlinear DNS revealed a rich sequence of bifurcations (figure \ref{fig:recap}), that has been investigated by means of frequency spectra, force diagrams and POD.
Overall, the flow dynamics is driven by 
five different modes: 
(i)-(ii)~static deflections of the wake in the vertical and horizontal directions, 
(iii)~vortex shedding of hairpin vortices from the LE shear layers, and
(iv)-(v)~unsteady flapping of the wake in the  vertical and horizontal directions. 
Their nonlinear interaction changes with $W$ and $Re$ giving rise to several regimes, ranging from steady and periodic regimes at small $Re$ to aperiodic and chaotic regimes at larger $Re$. For intermediate $W$ we have observed, for the first time, a periodic LE shedding of hairpin vortices that instantaneously preserves all spatial symmetries and generates neither lift nor side forces. Interestingly, in some portions of the parameter space the different modes synchronise, and give origin to periodic regimes also at relatively large $Re$.

The mechanism sustaining the shedding of the hairpin vortices from the LE shear layers has been investigated. Unlike in the flow past two-dimensional rectangular cylinders, this LE vortex shedding does not require the presence of a sharp TE, and does not necessarily lock with the TE vortex shedding. 
Indeed, for $W=2.25$ the same vortex shedding has been found for a three-dimensional rectangular flat plate without TE ($L = \infty$). By rounding the LE corners, we have also shown that the LE vortex shedding  is the result of a feedback mechanism embedded within the recirculating regions that form over the lateral sides of the prisms.

Having characterised the flow stability and dynamics 
over a wide range of body widths and lengths, the present study will serve as a stepping stone to further investigation in more complex settings.
For instance, the laminar wakes of  spheres and 2D cylinders translating close to a solid wall have been studied extensively \citep{thompson-etal-2021}. 
Ground proximity is by definition an essential feature for ground vehicles such as cars and trains, and although many numerical simulations have been carried out, 
the stability of these flows  remains largely unexplored in both the laminar and turbulent regimes,  except for a few studies \citep[e.g. simplified train geometry with 
$(L,W)=(6.9,0.83)H$ 
and ground clearance 0.15$H$ in][]{li2024linear}.
Another topic of interest is that of body attitude (yaw and pitch): 
in the laminar regime a small misalignment may significantly alter the onset and sequence of bifurcations;  in the turbulent regime ground vehicles may be subject to crosswind or varying front/rear load distribution.
Finally, the effect of stochastic  conditions (e.g. disturbances in the incoming flow) on the laminar bifurcations and wake dynamics could be studied, especially near codimension-two bifurcations  where two stationary modes become unstable simultaneously 
(e.g. $(L,W)=(3,1.2)$ at $Re \approx 330$, $(5,1.2)$ at $Re \approx 350$, and $(5,2.25)$ at $Re \approx 300$),
which may result in  multistability, with the wake switching randomly between two or more static deflected states. A treatment in the spirit of \cite{ducimetiere2024noiseinduced} would yield a rigorously derived reduced-order model in the form of coupled stochastic amplitude equations, one for each symmetry-breaking mode, which could then be used to predict flow statistics.

\section*{Acknowledgments}
A.C. acknowledges the computer time provided by the Scientific Computing \& Data Analysis section of the Core Facilities at OIST.

\section*{Funding} 
This research received no specific grant from any funding agency, commercial or not-for-profit sectors.

\section*{Declaration of Interests} 
The authors report no conflict of interest.

\appendix
\section{Convergence study} 
\label{sec:app-conv}

In this appendix, we report the sensitivity of the results on the grid resolution, for the LSA and the 3D nonlinear DNS.

For the  LSA, tables \ref{tab:mesh_convergence}-\ref{tab:domain_convergence} show how the critical Reynolds numbers of the first two bifurcations vary with the mesh size and domain size.
The numerical domain is $\{ x,y,z \, | \,  x_{min} \leq x \leq x_{max}; \, 0 \leq y,z \leq y_{max}=z_{max} \}$.
The mesh density is $n_1$ on the prism surface, $n_2$ on
the boundaries of the sub-domain $\{ x,y,z \, | \,  -5 \leq x \leq 15; \, 0 \leq y,z \leq 2 \}$, and $n_3=1$ on the outermost boundaries. 
Given the weak influence of the mesh size and domain size on $Re_{c,1}$ and $Re_{c,2}$, we choose mesh $M3$ and domain $D1$ throughout the linear and weakly nonlinear analyses, similar to \cite{zampogna_boujo_2023}.

\begin{table}
\centering
\begin{tabular}{l l ccccccccc} 
& Mesh                         & M1     & M2     & M3     & M4     & M5    & M6    & M7\\
& $(n_1,n_2)$                  & (30,10)& (40,10)& (60,10)& (60,12)& (60,15)& (80,10)& (80,12)   \\
\\
$W=1.2$, $L=1/6$ & $N_{elmts}$  & 0.76   & 0.83   & 0.97   & 1.46   & 2.49 & - & -\\ 
& $Re_{c,1} \, (S_y A_z)$      & 109    & 109    & 108    & 108    & 108  & - & -\\ 
& $Re_{c,2} \, (A_y S_z)$      & 113    & 113    & 113    & 113    & 113  & - & -\\ 
\\
$W=1.2$, $L=5$ & $N_{elmts}$  & 0.92   & 1.14   & 1.63   & 2.23   &  -   & 2.27  & 3.04\\ 
& $Re_{c,1} \, (S_y A_z)$     &  357   & 354    & 352    & 352    &  -   & 352   & 352 \\ 
& $Re_{c,2} \, (A_y S_z)$     &  354   & 354    & 353    & 353    &  -   & 353   & 353 \\ 
\\
$W=5$, $L=1/6$ & $N_{elmts}$  & 0.76   & 0.82   & 0.95   & 1.43   & 2.46 & - & -\\ 
& $Re_{c,1} \, (S_y A_z)$     & 61     & 60     & 60     & 60     & 60   & - & -\\ 
& $Re_{c,2} \, (S_y S_z)$     & 80     & 79     & 79     & 79     & 79   & - & -\\ 
\\
$W=5$, $L=5$ & $N_{elmts}$ & 0.78   & 0.86   & 1.04   & 1.55   & 2.61   & 1.26   & 1.83 \\ 
& $Re_{c,1} \, (S_y A_z)$     & 233    & 234    & 235    & 235    & 235    & 235    & 236  \\
& $Re_{c,2} \, (S_y S_z)$     & 259    & 259    & 263    & 264    & 264    & 263    & 265  \\ 
\end{tabular} 
\caption{Influence of the mesh size on the first two critical Reynolds numbers for four different prism geometries.
Domain D1: $(x_{min},x_{max})=(-10,20)$, $y_{max}=z_{max}=10$.
$N_{elmts}$ in millions.}
\label{tab:mesh_convergence}
\end{table}

\begin{table}
\centering
\begin{tabular}{l l cccccccccc} 
 Domain                         && D1      & D2      & D3        &&D1      & D2      & D3 \\
 $(x_{min},x_{max})$            &&(-10,20) &(-15,30) &(-20,40)   &&(-10,20) &(-15,30) &(-20,40)\\
 $y_{max}$, $z_{max}$           &&  10     &  15     &  20       &&10     &  15     &  20 \\
\\
&& \multicolumn{3}{c}{$L=1/6$}  && \multicolumn{3}{c}{$L=5$}
\\
$W=1.2$ & $N_{elmts}$                  &  0.97   & 1.16    & 1.49 &         & 1.63    & 1.81    & 2.14\\ 
& $Re_{c,1} \, (S_y A_z)$              &   108   & 109     & 109  &         & 352     & 354     & 354 \\ 
& $Re_{c,2} \, (A_y S_z)$              &   113   & 113     & 113  &         & 353     & 354     & 354 \\ 
\\
$W=5$ & $N_{elmts}$                    & 0.95    & 1.14  & 1.46 &         & 1.04    & 2.06    & 2.06 \\ 
& $Re_{c,1} \, (S_y A_z)$              & 60      & 61 	 & 61   &         & 235     & 236     & 236\\ 
& $Re_{c,2} \, (S_y S_z)$              & 79      & 79 	 & 80   &         & 263     & 262     & 263 \\ 
\end{tabular} 
\caption{Influence of the domain size on the first two critical Reynolds numbers for four different prism geometries.
Mesh M3: $(n_1,n_2)=(60,10)$.
$N_{elmts}$ in millions.}
\label{tab:domain_convergence}
\end{table}

For the 3D nonlinear DNS, sensitivity to the grid resolution and to time step has been investigated  with six additional simulations, two for each considered geometry. 
The number of points in the streamwise and vertical directions has been decreased from $(N_x,N_z)=(1072,590)$ on the standard grids to $(N_x,N_z)=(860,400)$ for all three geometries; 
in the spanwise direction,  the number of points has been decreased from $N_y=666,720$ and $804$ to $N_y=460,500$ and $560$, for $W=1.2,2.25$ and $5$ respectively. 
On the coarser grid, two different time steps have been used, $\Delta t \approx 0.0066$ and $0.0033$, leading to an average 
CFL number of approximately $1$ and $0.5$. 
For each geometry we have considered a single Reynolds number: 
$Re = 535$ for $W=1.2$ ($aA_yS_z$ regime), 
$Re=450$ for $W=2.25$ ($pS_yS_zlb$) and 
$Re=385$ for $W=5$ ($aS_yS_z$). 
In all cases, we find an excellent qualitative and quantitative agreement on the mean and fluctuating  forces  between the standard and coarse grids, and between the two $\Delta t$ as well (table \ref{tab:DNS_conv}).
\begin{table}
\centering
  \setlength{\tabcolsep}{6pt}
\begin{tabular}{lccccccccccc} 
\vspace{0.2cm}
$W$   & $Re$  & $N_x$ & $N_y$ & $N_z$ & $\Delta t$ & $F_x$ & $F_y$ & $F_z$ & $F'_{x,rms}$ & $ F'_{y,rms}$ & $F'_{z,rms}$ \\
$1.2$ & $535$ & $860$ & $460$ & $400$ & $0.0066$   &$0.7833$ & $0.0202$ & $0$ & $0.0011$ & $0.0027$ & $0.0144$ \\
$1.2$ & $535$ & $860$ & $460$ & $400$ & $0.0033$   &$0.7834$ & $0.0203$ & $0$ & $0.0010$ & $0.0026$ & $0.0144$ \\
\vspace{0.2cm}
$1.2$ & $535$ & $1072$& $666$ & $590$ & $0.0066$   &$0.7831$ & $0.0201$ & $0$ & $0.0010$ & $0.0028$ & $0.0143$ \\
$2.25$& $450$ & $860$ & $500$ & $400$ & $0.0066$   &$0.7660$ & $0$      & $0$ & $0.0021$ & $0$      & $0.0412$  \\
$2.25$& $450$ & $860$ & $500$ & $400$ & $0.0033$   &$0.7660$ & $0$      & $0$ & $0.0021$ & $0$      & $0.0412$  \\
\vspace{0.2cm}
$2.25$& $450$ & $1072$& $720$ & $590$ & $0.0066$   & $0.7658$ & $0$     & $0$ & $0.0020$ & $0$     & $0.0410$  \\
$5$   & $385$ & $860$ & $560$ & $400$ & $0.0066$   & $0.8116$ & $0$     & $0$ & $0.0117$ & $0.0015$ & $0.0886$ \\
$5$   & $385$ & $860$ & $560$ & $400$ & $0.0033$   & $0.8118$ & $0$     & $0$ & $0.0118$ & $0.0016$ & $0.0890$ \\
$5$   & $385$& $1072$ & $804$ & $590$ & $0.0066$   & $0.8131$ & $0$     & $0$ & $0.0116$ & $0.0011$ & $0.0760$ \\
\end{tabular}
\caption{Convergence study for the three-dimensional non linear simulations. Influence of the grid resolution on the aerodynamic forces for $(L,W)=(5,1.2)$ and $Re=535$ ($aA_yS_z$ regime), $(L,W)=(5,2.25)$ and $Re=450$ ($pS_yS_zlb$ regime), and $(L,W)=(5,5)$ and $Re=385$ ($aS_yS_z$ regime).}
\label{tab:DNS_conv}
\end{table}
%
We observe an excellent agreement  in terms of frequencies too, with differences smaller than 0.7\% between the coarse  and standard grids (not shown).

\section{Weakly nonlinear analysis}
\label{sec:app-WNL}

We detail here the derivation of the systems of third- and fifth-order amplitude equations (\ref{eq:A})-(\ref{eq:B}) and (\ref{AE_dAdt_2})-(\ref{AE_dBdt_2}) in the vicinity of a codimension-two bifurcation for two stationary modes $A$ and $B$.
The procedure is similar to that in \cite{zampogna_boujo_2023}.

\subsection{Derivation of the amplitude equations}
\label{sec:app-WNL-deriv}

When the two modes of interest do not bifurcate exactly at the same Reynolds number, we introduce a reference critical Reynolds number $Re_c$, typically chosen between the critical Reynolds numbers of the two modes.
Departure from criticality is measured as
\begin{equation}
    Re_c^{-1}-Re^{-1}=\epsilon^2 \tilde\alpha,
\end{equation}
which defines the small parameter  $0<\epsilon\ll1$ and the order-one parameter $\tilde\alpha=O(1)$.
We also define a shift operator $\mathcal{S}$ such that, unlike the original linearised NS operator ${\mathcal{A}}$, the shifted linearised NS operator $\widetilde{\mathcal{A}} = \mathcal{A}   + \epsilon^2 \mathcal{S}$ is singular exactly at $Re=Re_c$ for both modes.
In other words, denoting $\mathcal{B} \partial_t \bm{q} + {\mathcal{A}} \bm{q} = \bm{0}$ the linearised NS equations
and $\hat{\bm{q}} = \{\hat{\bm{u}},\hat{p}\}$,
one has
$\widetilde{\mathcal{A}} \hat{\bm{q}}_1^A = \widetilde{\mathcal{A}} \hat{\bm{q}}_1^B = \bm{0}$,
while
$\mathcal{A} \hat{\bm{q}}_1^A = -\sigma_A \mathcal{B} \hat{\bm{q}}_1^A$ and 
$\mathcal{A} \hat{\bm{q}}_1^B = -\sigma_B \mathcal{B} \hat{\bm{q}}_1^B$.

We use the method of multiple scales. 
We introduce slow time scales $T_1 = \epsilon^2 t$, $\quad T_2 = \epsilon^4 t, \ldots$
and inject the expansion 
\begin{equation}
  \bm{q}(\bm x,t,T_1,T_2\ldots) = \bm{q}_0 + \epsilon \bm{q}_1 + \epsilon^2 \bm{q}_2 + \epsilon^3 \bm{q}_3 + \ldots  
\end{equation}
in the NS equations at $Re=Re_c$, where the time derivative now reads $\partial_t + \epsilon^2 \partial_{T_1} + \epsilon^4 \partial_{T_2} + \ldots$. 
Collecting like-order terms yields the following series of problems.

\subsubsection{Zeroth and first orders}

At  order $\epsilon^0$ we find the nonlinear NS equations, and the zeroth-order field $\bm{q}_0(\bm{x})$ is the steady $S_yS_z$ base flow at $Re=Re_c$.
%
%
At  order $\epsilon^1$ we obtain the linearised and shifted NS equations
\begin{equation}
\partial_t \bm{q}_1 +  \widetilde{\mathcal{A}} \bm{q}_1 = 0.
\end{equation}
Since $\widetilde{\mathcal{A}}$ is singular at $Re=Re_c$, the first-order field is a superposition of the $S_yA_z$ and $A_yS_z$ eigenmodes:
\begin{equation}
\bm{q}_1(\bm{x},T_1,T_2,\ldots) = A \hat{\bm{q}}_1^A + B \hat{\bm{q}}_1^B,
\end{equation}
where $A(T_1,T_2,\ldots)$ and $B(T_1,T_2,\ldots)$ are real-valued slowly-varying amplitudes to be determined.

\subsubsection{Second order}

At order $\epsilon^2$ the 
field $\bm{q}_2$ is a solution of the linearised and shifted NS equations 
\begin{align}
\mathcal{B} \partial_t \bm{q}_2 +  \widetilde{\mathcal{A}} \bm{q}_2 = \{\bm{F}_2,0\},
\label{eq:WNL_eps2_eq}
\end{align}
where
\begin{align}
\bm F_2 = & -\tilde\alpha \nabla^2 \bm u_0 - (\bm u_1 \cdot\nabla) \bm u_1
\nonumber\\
= & -\tilde\alpha \nabla^2 \bm u_0
- A^2 \hat{\bm u}_1^A \cdot\nabla \hat{\bm u}_1^A
- AB \, \mathcal{C}(\hat{\bm u}_1^A, \hat{\bm u}_1^B)
- B^2 \hat{\bm u}_1^B \cdot\nabla \hat{\bm u}_1^B,
\end{align}
and introducing the notation
$\mathcal{C}( \bm a , \bm b)  = (\bm a \cdot\nabla) \bm b + (\bm b \cdot\nabla) \bm a.$ 
Symmetry considerations show that (\ref{eq:WNL_eps2_eq}) can be inverted:
all the terms in $\bm F_2$ are either $S_y S_z$ or $A_y A_z$ and  cannot resonate since $\widetilde{\mathcal{A}}$ is singular to $S_y A_z$ and $A_y S_z$ modes.
Therefore, 
\begin{align}
\bm q_2(\bm x, T_1, T_2, \ldots) = \tilde\alpha  \bm q_2^\alpha
+ A^2 \bm q_2^{A^2}
+ B^2 \bm q_2^{B^2}
+ AB  \bm q_2^{AB},
\end{align}
where the $\bm q_2^*$ fields are solution of:
%
\begin{align}
& \widetilde{\mathcal{A}} \bm q_2^\alpha = - \{\nabla^2 \bm u_0, 0\},
& \widetilde{\mathcal{A}} \bm q_2^{A^2} = - \{ (\hat{\bm u}_1^A \cdot\nabla) \hat{\bm u}_1^A, 0\},
\nonumber\\
& \widetilde{\mathcal{A}} \bm q_2^{B^2} = - \{ (\hat{\bm u}_1^B \cdot\nabla) \hat{\bm u}_1^B, 0\},
& \widetilde{\mathcal{A}} \bm q_2^{AB} = - \{ \mathcal{C}(\hat{\bm u}_1^A, \hat{\bm u}_1^B), 0\}.
\end{align}

\subsubsection{Third order}

At order $\epsilon^3$ the 
field $\bm{q}_3$ is a solution of the linearised and shifted NS equations 
\begin{align}
\mathcal{B} \partial_t \bm{q}_3 +  \widetilde{\mathcal{A}} \bm{q}_3 = \{\bm{F}_3,0\},
\label{eq:WNL_eps3_eq}
\end{align}
where
\begin{align}
\bm F_3 = 
- & (\partial_{T_1}A)  {\hat{\bm u}_1^A} 
-   (\partial_{T_1}B)  {\hat{\bm u}_1^B}
+  A \left(
\tilde\sigma_A  {\hat{\bm u}_1^A} 
-\tilde{\alpha} \mathcal{C}( {\hat{\bm u}_1^A}  , {\bm u_2^\alpha}) 
-\tilde{\alpha} \nabla^2 {\hat{\bm u}_1^A}
\right)
\nonumber\\
+ & B \left(
\tilde\sigma_B  {\hat{\bm u}_1^B} 
-\tilde{\alpha } \mathcal{C}( {\hat{\bm u}_1^B} ,  {\bm u_2^\alpha}) 
-\tilde{\alpha } \nabla^2 {\hat{\bm u}_1^B}
\right)
-  A^3 
 \mathcal{C}({\hat{\bm u}_1^A} ,  {\bm u_2^{A^2}}) 
- B^3 
 \mathcal{C}({\hat{\bm u}_1^B} ,  {\bm u_2^{B^2}}) 
\nonumber\\
- & A B^2 \left(
 \mathcal{C}( {\hat{\bm u}_1^A}  , {\bm u_2^{B^2}})
+\mathcal{C}( {\hat{\bm u}_1^B}  , {\bm u_2^{AB}})  
\right)
-  A^2 B \left(
 \mathcal{C}({\hat{\bm u}_1^B} ,  {\bm u_2^{A^2}}) 
+\mathcal{C}({\hat{\bm u}_1^A} ,  {\bm u_2^{AB}}) 
\right),
\end{align}
and where $\tilde\sigma_A=\sigma_A/\epsilon^2$ and $\tilde\sigma_B=\sigma_B/\epsilon^2$.
All the terms in $\bm F_3$ are resonant with either  $\hat{\bm u}_1^A$ or  $\hat{\bm u}_1^B$.
To avoid secular terms we impose a compatibility condition, i.e. we require  the respective inner products with either $\hat{\bm u}_1^{A\dag}$ or $\hat{\bm u}_1^{B\dag}$ be zero.
With the normalisation $\langle \hat{\bm u}_1^{A\dag}, \hat{\bm u}_1^A \rangle = \langle \hat{\bm u}_1^{B\dag}, \hat{\bm u}_1^B \rangle = 1$, we obtain
\begin{align}
\partial_{T_1}A
= \tilde\lambda_A A - \tilde\chi_A A^3 - \tilde\eta_A A B^2,
\label{AE_dAdT1}
\\
\partial_{T_1}B
= \tilde\lambda_B B - \tilde\chi_B B^3 - \tilde\eta_B A^2 B,
\label{AE_dBdT1}
\end{align}
with the coefficients
\begin{align}
\tilde\lambda_A &=  \tilde\sigma_A   
 + \tilde{\alpha} \langle \hat{\bm u}_1^{A\dag}, 
- \mathcal{C}( {\hat{\bm u}_1^A}  , {\bm u_2^\alpha}) 
- \nabla^2 {\hat{\bm u}_1^A}
\rangle,
\nonumber\\
 \tilde\chi_A &=  \langle \hat{\bm u}_1^{A\dag},
- \mathcal{C}({\hat{\bm u}_1^A} ,  {\bm u_2^{A^2}}) 
\rangle,
\nonumber\\
 \tilde\eta_A &=   \langle \hat{\bm u}_1^{A\dag},
-\mathcal{C}( {\hat{\bm u}_1^A}  , {\bm u_2^{B^2}}) 
-\mathcal{C}( {\hat{\bm u}_1^B}  , {\bm u_2^{AB}}) 
\rangle,
\end{align}
and similar expressions for $\tilde\lambda_B$, $\tilde\chi_B$ and $\tilde\eta_B$  upon exchange of indices $A$ and $B$.

Equations (\ref{AE_dAdT1})-(\ref{AE_dBdT1}) constitute the leading-order system of amplitude equations. If the derivation is stopped at this order, one can reintroduce the fast time $t=\epsilon^2 T_1$ and obtain the system (\ref{eq:A})-(\ref{eq:B}), with the coefficients $\lambda_A = \epsilon^2 \tilde\lambda_A$, $\chi_A = \epsilon^2 \tilde\chi_A$, etc.:
\begin{align}
\mathrm{d}_{t}A
= \lambda_A A - \chi_A A^3 - \eta_A A B^2,
\\
\mathrm{d}_{t}B
= \lambda_B B - \chi_B B^3 - \eta_B A^2 B.
\end{align}

To derive higher-order amplitude equations, the third-order field 
\begin{equation}
 \bm q_3(\bm x, T_1, T_2, \ldots) =
  \tilde\alpha A \bm q_3^A
+ \tilde\alpha B \bm q_3^B
+ A^3   \bm q_3^{A^3}
+ B^3   \bm q_3^{B^3}
+ A B^2 \bm q_3^{AB^2}
+ A^2 B \bm q_3^{A^2B},
\end{equation}
is needed, where the $\bm q_3^*$ fields are solution of:
\begin{align}
\widetilde{\mathcal{A}} \bm q_3^A &=  
\{ - \mathcal{C}( {\hat{\bm u}_1^A}  , {\bm u_2^\alpha}) 
-\nabla^2 {\hat{\bm u}_1^A} 
+ \langle \hat{\bm u}_1^{A\dag}, 
 \mathcal{C}( {\hat{\bm u}_1^A}  , {\bm u_2^\alpha}) 
+ \nabla^2 {\hat{\bm u}_1^A}
\rangle \hat{\bm u}_1^A , 0 \},
\nonumber\\
\widetilde{\mathcal{A}} \bm q_3^{A^3} &= \{ - \mathcal{C}({\hat{\bm u}_1^A} ,  {\bm u_2^{A^2}}) + \tilde\chi_A {\hat{\bm u}_1^A}, 0\},
\nonumber\\
\widetilde{\mathcal{A}} \bm q_3^{A^2B} &= \{  -\mathcal{C}({\hat{\bm u}_1^B} ,  {\bm u_2^{A^2}}) 
-\mathcal{C}({\hat{\bm u}_1^A} ,  {\bm u_2^{AB}})
+ \tilde\eta_B  {\hat{\bm u}_1^B}, 0\},
\end{align}
with similar expressions for 
${\bm q_3^{B}}$, ${\bm q_3^{B^3}}$ and ${\bm q_3^{A B^2}}$ 
upon exchange of $A$ and $B$.

Because the kernel of $\widetilde{\mathcal{A}}$ is not empty, the  $\bm q_3^*$  fields are defined up to an arbitrary constant along  $\hat{\bm u}_1^A$ or  $\hat{\bm u}_1^B$.
We remove this component such that 
$\langle \hat{\bm u}_1^{A\dag}, \bm u_3^A      \rangle 
=\langle \hat{\bm u}_1^{A\dag}, \bm u_3^{A^3}  \rangle
=\langle \hat{\bm u}_1^{A\dag}, \bm u_3^{AB^2} \rangle
=
\langle \hat{\bm u}_1^{B\dag}, \bm u_3^B      \rangle 
=\langle \hat{\bm u}_1^{B\dag}, \bm u_3^{B^3}  \rangle
=\langle \hat{\bm u}_1^{B\dag}, \bm u_3^{A^2B} \rangle
=0$.

\subsubsection{Fourth order}

At order $\epsilon^4$ the 
field $\bm{q}_4$ is a solution of the linearised and shifted NS equations 
\begin{align}
\mathcal{B} \partial_t \bm{q}_4 +  \widetilde{\mathcal{A}} \bm{q}_4 = \{\bm{F}_4,0\},
\label{eq:WNL_eps4_eq}
\end{align}
where 
symmetry considerations show that
$\bm F_4$ is not resonant. Therefore, one can invert (\ref{eq:WNL_eps4_eq}) and compute
\begin{align}
\bm q_4(\bm x, T_1, T_2, \ldots) = &
  \tilde\alpha^2 \bm q_4^\alpha
+  \partial_{T_1}(A^2)  {\bm q_4^{(A^2)'}} 
+  \partial_{T_1}(AB)   {\bm q_4^{(AB)'}} 
+  \partial_{T_1}(B^2)  {\bm q_4^{(B^2)'}} 
\nonumber\\
+ & \tilde\alpha A^2 \bm q_4^{A^2}
+ \tilde\alpha A B \bm q_4^{AB}
+ \tilde\alpha B^2 \bm q_4^{B^2}
+ A^4 \bm q_4^{A^4}
+ B^4 \bm q_4^{B^4}
\nonumber\\
+ & A^3 B   \bm q_4^{A^3B}
+ A^2 B^2 \bm q_4^{A^2B^2}
+ A   B^3 \bm q_4^{AB^3},
\end{align}
where the $\bm q_4^*$ fields are solution of:
\begin{align}
 \widetilde{\mathcal{A}} \bm q_4^\alpha &= \{-  ({\bm u_2^\alpha} \cdot\nabla)  {\bm u_2^\alpha} 
- \nabla^2 {\bm u_2^\alpha}, 0\},
\qquad
 \widetilde{\mathcal{A}} {\bm q_4^{(A^2)'}} = \{-  {\bm u_2^{A^2}} ,0\},
\nonumber\\
 \widetilde{\mathcal{A}} {\bm q_4^{A^2}} &=  
\{- \mathcal{C}( {\bm u_2^{A^2}} ,  {\bm u_2^\alpha}) 
- \mathcal{C}( {\hat{\bm u}_1^A} ,  {\bm u_3^{A}}) 
- \nabla^2 {\bm u_2^{A^2}} ,0\},
\qquad
\widetilde{\mathcal{A}} {\bm q_4^{(AB)'}} = \{-  {\bm u_2^{AB}}, 0\},
\nonumber\\
\widetilde{\mathcal{A}} {\bm q_4^{AB}} &= \{- \mathcal{C}( {\bm u_2^{AB}} ,  {\bm u_2^\alpha}) 
- \mathcal{C}( {\hat{\bm u}_1^B}  , {\bm u_3^{A}}) 
- \mathcal{C}( {\hat{\bm u}_1^A} ,  {\bm u_3^{B}}) 
- \nabla^2 {\bm u_2^{AB}},0\},
\nonumber\\
\widetilde{\mathcal{A}} {\bm q_4^{A^2 B^2}} &=\{ - ({\bm u_2^{AB}} \cdot\nabla ) {\bm u_2^{AB}} 
- \mathcal{C}({\bm u_2^{A^2}} ,  {\bm u_2^{B^2}}) 
- \mathcal{C}({\hat{\bm u}_1^B} ,  {\bm u_3^{A^2 B}}) 
- \mathcal{C}({\hat{\bm u}_1^A},   {\bm u_3^{AB^2}}), 0\},
\nonumber\\
\widetilde{\mathcal{A}} {\bm q_4^{A^3 B}} &= \{- \mathcal{C}({\bm u_2^{A^2}} ,  {\bm u_2^{AB}}) 
- \mathcal{C}({\hat{\bm u}_1^B} , {\bm u_3^{A^3}}) 
- \mathcal{C}({\hat{\bm u}_1^A}  , {\bm u_3^{A^2 B}}),0\},\nonumber\\
\widetilde{\mathcal{A}} {\bm q_4^{A^4}} &= \{- ({\bm u_2^{A^2}} \cdot\nabla)  {\bm u_2^{A^2}} 
- \mathcal{C}({\hat{\bm u}_1^A} ,  {\bm u_3^{A^3}}) ,0\}.
\end{align}
with similar expressions for 
${\bm q_4^{(B^2)'}}$,
${\bm q_4^{B^2}}$,
${\bm q_4^{A B^3}}$,
${\bm q_4^{B^4}}$ 
upon exchange of $A$ and $B$.

\subsubsection{Fifth order}

At order $\epsilon^5$ the 
field $\bm{q}_5$ is a solution of the linearised and shifted NS equations 
\begin{align}
\mathcal{B} \partial_t \bm{q}_5 +  \widetilde{\mathcal{A}} \bm{q}_5 = \{\bm{F}_5,0\},
\label{eq:WNL_eps5_eq}
\end{align}
where all the terms in $\bm F_5 $ resonate with either $\hat{\bm u}_1^A$ or $\hat{\bm u}_1^B$.
Imposing a compatibility condition yields
%
%
\begin{align}
\partial_{T_2}A
=
& \langle \hat{\bm u}_1^{A\dag}, 
  A       \bm a 
+ A^3     \bm b 
+ A B^2   \bm c 
+ A B^4   \bm d 
+ A^3 B^2 \bm e 
+ A^5     \bm f \rangle,
\end{align}
where
\begin{align}
\bm a =& 
\tilde{\alpha }^2 \left[
 -  \mathcal{C}( {\bm u_2^{\alpha}} ,  {\bm u_3^{A}}) 
 -  \mathcal{C}( {\hat{\bm u}_1^A} ,  {\bm u_4^{\alpha}})
 - \nabla^2 {\bm u_3^{A}} 
\right],
\nonumber\\
\bm b =& 
\tilde{\alpha} \left[
 -  \mathcal{C}( {\bm u_2^{A^2}}  , {\bm u_3^{A}}) 
 -  \mathcal{C}( {\bm u_2^{\alpha}} ,  {\bm u_3^{A^3}}) 
 -  \mathcal{C}( {\hat{\bm u}_1^A} ,  {\bm u_4^{A^2}}) 
 -   \nabla^2{\bm u_3^{A^3}} 
 \right] 
 - 2\tilde\lambda_A \mathcal{C}( {\hat{\bm u}_1^A}, {\bm u_4^{(A^2)'}} ) ,
\nonumber\\
\bm c =& 
\tilde{\alpha}  \left[
 -\mathcal{C}( {\bm u_2^{B^2}} ,  {\bm u_3^{A}}) 
 -\mathcal{C}( {\bm u_2^{\alpha}} ,  {\bm u_3^{A B^2}}) 
 -\mathcal{C}( {\bm u_2^{AB}} ,  {\bm u_3^{B}})
 -\mathcal{C}( {\hat{\bm u}_1^B} , {\bm u_4^{AB}}) 
 -\mathcal{C}( {\hat{\bm u}_1^A} ,  {\bm u_4^{B^2}}) 
\right.
\nonumber\\
& \quad \left.
 -\nabla^2{\bm u_3^{A B^2}} 
\right] 
- 2\tilde\lambda_B \mathcal{C}( {\hat{\bm u}_1^A},  {\bm u_4^{(B^2)'}}  )
- (\tilde\lambda_A + \tilde\lambda_B)\mathcal{C}( {\hat{\bm u}_1^B}, {\bm u_4^{(AB)'}} ),
\nonumber\\
\bm d =& 
 - \mathcal{C}({\bm u_2^{B^2}} ,{\bm u_3^{A B^2}}) 
 - \mathcal{C}({\bm u_2^{AB}} ,  {\bm u_3^{B^3}}) 
 - \mathcal{C}({\hat{\bm u}_1^B} ,  {\bm u_4^{A B^3}}) 
 - \mathcal{C}({\hat{\bm u}_1^A} ,  {\bm u_4^{B^4}})  
\nonumber\\
& \quad 
 +   2\tilde\chi_B \mathcal{C}( {\hat{\bm u}_1^A}, {\bm u_4^{(B^2)'}} ) 
+ (\tilde\eta_A + \tilde\chi_B) \mathcal{C}( {\hat{\bm u}_1^B},  {\bm u_4^{(AB)'}} ) , 
\nonumber\\
\bm e =&  
- \mathcal{C}({\bm u_2^{B^2}}  , {\bm u_3^{A^3}}) 
 - \mathcal{C}({\bm u_2^{AB}}  , {\bm u_3^{A^2 B}}) 
 - \mathcal{C}({\bm u_2^{A^2}} ,  {\bm u_3^{A B^2}}) 
 - \mathcal{C}({\hat{\bm u}_1^B} ,  {\bm u_4^{A^3 B}}) 
 - \mathcal{C}({\hat{\bm u}_1^A}, {\bm u_4^{A^2 B^2}}) 
\nonumber\\
& \quad 
 + 2\tilde\eta_A \mathcal{C}( {\hat{\bm u}_1^A}, {\bm u_4^{(A^2)'}} )  
      + 2\tilde\eta_B \mathcal{C}( {\hat{\bm u}_1^A},   {\bm u_4^{(B^2)'}} )
      + (\tilde\chi_A + \tilde\eta_B) \mathcal{C}( {\hat{\bm u}_1^B},  {\bm u_4^{(AB)'}} ) ,
\nonumber\\
\bm f =&  
- \mathcal{C}({\bm u_2^{A^2}} ,  {\bm u_3^{A^3}}) 
 - \mathcal{C}({\hat{\bm u}_1^A}     , {\bm u_4^{A^4}}) 
 + 2\tilde\chi_A  \mathcal{C}( {\hat{\bm u}_1^A}, {\bm u_4^{(A^2)'}} ).
\end{align}
A similar equation is obtained for $\partial_{T_2}B$ upon exchange of $A$ and $B$.

Terms proportional to $\tilde \lambda_A$, $\tilde\chi_A$, etc., come from the interaction of the eigenmodes with the fourth-order fields $ {\bm q_4^{(A^2)'}}$, ${\bm q_4^{(AB)'}}$ and ${\bm q_4^{(B^2)'}}$, and are obtained after developing $ \partial_{T_1}(A^2)$, $\partial_{T_1}(AB)  
$ and $\partial_{T_1}(B^2)  $
 and using (\ref{AE_dAdT1})-(\ref{AE_dBdT1}).
 By contrast, all the forcing terms of the form $ \partial_{T_1}(A^m B^n) {\bm u_3^{*}}$ with $m+n=1$ or 3 vanish when imposing the compatibility condition because of our earlier choice $\langle \hat{\bm u}_1^{\dag}, \bm u_3^*  \rangle =0$.

Defining  coefficients $\tilde *'$, we obtain:
\begin{align}
\partial_{T_2}A
= \tilde\lambda_A' A - \tilde\chi_A' A^3 - \tilde\eta_A' A B^2
+ \tilde\kappa_A' A B^4
+ \tilde\beta_A'  A^3 B^2
+ \tilde\gamma_A' A^5,
\label{AE_dAdT2}
\\
\partial_{T_2}B
= \tilde\lambda_B' B - \tilde\chi_B' B^3 - \tilde\eta_B' A^2 B
+ \tilde\kappa_B' A^4 B
+ \tilde\beta_B'  A^2 B^3
+ \tilde\gamma_B' B^5.
\label{AE_dBdT2}
\end{align}


Finally, the total fifth-order amplitude equations are obtained by reconstructing the total derivatives $\mathrm{d}_t(A,B) = (\epsilon^2 \partial_{T_1}+ \epsilon^4 \partial_{T_2})(A,B)$, i.e. combining (\ref{AE_dAdT1})-(\ref{AE_dBdT1}) and (\ref{AE_dAdT2})-(\ref{AE_dBdT2}):
\begin{align}
\mathrm{d}_t A
=  
  \lambda_A A 
- \chi_A A^3 
- \eta_A A B^2
+ \kappa_A A B^4
+ \beta_A  A^3 B^2
+ \gamma_A A^5,
\label{AE_dAdt_22}
\\
\mathrm{d}_t B
= 
 \lambda_B B 
-\chi_B B^3 
-\eta_B A^2 B
+  \kappa_B A^4 B
+  \beta_B  A^2 B^3
+  \gamma_B B^5,
\label{AE_dBdt_22}
\end{align}
where now $\lambda_A = \epsilon^2\tilde\lambda_A + \epsilon^4\tilde\lambda_A'$, $\chi_A = \epsilon^2\tilde\chi_A + \epsilon^4\tilde\chi_A'$, etc. 
Therefore, compared to the third-order system, the fifth-order system has not only higher-order terms but also corrected coefficients.

\subsection{Equilibrium solutions }

\subsubsection{Third-order system}

The third-order system (\ref{eq:A})-(\ref{eq:B}) has four possible equilibrium solutions:
\begin{itemize}
\item 
~Symmetric state: $ A=B=0$;

\item 
~Pure vertical symmetry breaking: $A^2 = \lambda_A/\chi_A$, $B=0$;

\item 
~Pure horizontal symmetry breaking: $A=0$, $B^2 = \lambda_B/\chi_B$;

\item 
~Mixed state (double symmetry breaking):
$A^2 =  (\chi_B \lambda_A - \eta_A \lambda_B)/(\chi_A \chi_B -\eta_A\eta_B)$,   
$B^2 = (\chi_A \lambda_B - \eta_B \lambda_A)/(\chi_A \chi_B -\eta_A\eta_B).$
\end{itemize}

\subsubsection{Fifth-order system}

The fifth-order system (\ref{AE_dAdt_2})-(\ref{AE_dBdt_2}) has four types of equilibrium states, with multiple possible amplitudes:
\begin{itemize}
\item
~Symmetric state: $A=B=0$;

\item
~Pure vertical symmetry breaking: 
$A^2 = \pm \sqrt{ (\chi_A \pm  \sqrt{\Delta})/(2\gamma_A) }$,
$B=0,$
where $\Delta =  \chi_A^2 - 4\lambda_A\gamma_A$.
Real-valued solutions exist when $\Delta>0$ and  $(\chi_A \pm  \sqrt{\Delta})/\gamma_A>0$. Focusing on positive values, this yields up to two different amplitudes $A_1$, $A_2$;

\item
~Pure horizontal symmetry breaking: 
$A=0$,
$B^2 = \pm \sqrt{ (\chi_B \pm  \sqrt{\Delta})/(2\gamma_B) },$
where $\Delta =  \chi_B^2 - 4\lambda_B\gamma_B$.
Real-valued solutions exist when $\Delta>0$ and  $(\chi_B \pm  \sqrt{\Delta})/\gamma_B>0$. 
Again, there are up to two different positive amplitudes $B_1$, $B_2$;

\item
~Mixed states (double symmetry breaking)  $A\neq0, B\neq0$. These states correspond to the intersections of conic sections, since  $u=A^2$ and $v=B^2$ satisfy the  bivariate quadratic equations
\begin{align}
0 &=  \lambda_A  - \chi_A u - \eta_A  v + \kappa_A  v^2 + \beta_A  u v + \gamma_A u^2,
\\
0 &=  \lambda_B  -\chi_B v  -\eta_B u  +  \kappa_B u^2  +  \beta_B  u v +  \gamma_B v^2.
\end{align}
An analytic treatment of these equations can be found for example in chapter 11 of 
\cite{RichterGebert}.
There are in general zero, two or four real-valued intersections.
Here, we compute these intersections numerically with the algorithm of 
\cite{Nakatsukasa2015}.
\end{itemize}

\section{Proper orthogonal decomposition}
\label{sec:POD}

\subsection{Numerical implementation}

To perform the POD analysis of \S\ref{sec:simulations}, we have used the method of snapshots \citep{sirovich-1987} on a portion of the domain ($-3.7 \le x \le 17.7$,  $|y|,|z| \le 3.2$ for $W=1.2,2.25$ and $-3.7 \le x \le 17.7$, $|y| \le 4.2$, $|z| \le 6.2$ for $W=5$) and, to avoid the need of weight matrices for scalar products, interpolated on a uniform cubic grid (with $N_x=410$ and $N_y=N_z=120$ for $W=1.2,2.25$ and $N_x=318$, $N_y=100$ and $N_z=150$ for $W=5$).
This yields approximately $5.9 \times 10^6$ (for $W=1.2,2.25$) and $4.8 \times 10^6$ (for $W=5$) grids points and, with three velocity components at each point, $N_{dof} \approx 17.7 \times 10^6$ and $N_{dof} \approx 14.4 \times 10^6$ degrees of freedom. 
The number of snapshots and their time separation has been chosen carefully:
the sampling time  $\delta t_s \simeq 1/(20 f_l)$ ensures that the largest frequency $f_l$ of interest detected in the spectra is captured;
similarly, the total time $T \simeq 8/f_s$, ensures that the smallest frequency $f_s$ is captured. 
For example, 
for $W=2.25$ and $Re=380$, this yields $\delta t_s = 0.195$ and $T= 600 \delta t_s$ (i.e. $m=600$ snapshots).

With the snapshots collected in the $N_{dof}\times m$  matrix $S$, the $m\times m$ covariance matrix $S^T S$ is formed and the eigenvalue problem  $S^T S Z = Z \Lambda$ is solved for the $r$ largest eigenvalues $\lambda_j$ (components of the diagonal matrix $\Lambda$) and corresponding eigenvectors $\bm{z}_j$ (columns of $Z$). The positive square roots of the eigenvalues of $S^TS$ are the singular values of $S$, 
and the columns of the matrix $\Phi = SZ$ are the POD modes.
%
%
Frequencies associated with a given POD mode are detected by projecting the snapshots on that mode and performing a frequency analysis of the so obtained time-dependent coefficient.

\subsection{Effect of $Re$ on the LE vortex shedding POD mode for $W=2.25$}
\label{app:POD_LE_VS}

\begin{figure}
    \centering
    \begin{tikzpicture}

\definecolor{clr1}{RGB}{18 78 128}
\definecolor{clr2}{RGB}{89 165 216}
\definecolor{clr3}{RGB}{145 229 246}
\definecolor{clr4}{RGB}{34 139 34}
\definecolor{clr5}{RGB}{255 77 109}
\definecolor{clr6}{RGB}{201 24 74}
\definecolor{clr7}{RGB}{128 15 47}
\definecolor{clr8}{RGB}{174 32 18}
\definecolor{clr9}{RGB}{155 34 38}
\definecolor{clr10}{RGB}{46 139 87}
\definecolor{clr11}{RGB}{240 128 128}
\definecolor{clr12}{RGB}{152 251 152}

\begin{axis}[%
width=0.35\textwidth,
height=0.15\textwidth,
scale only axis,
ymode = log,
xmin=1,
xmax=8,
xtick={1,2,3,4,5,6,7,8},
xlabel={$mode$},
ylabel={$\lambda/\sum \lambda$},
ylabel style={at={(0.05,0.5)}},
axis background/.style={fill=white},
legend columns=3,transpose legend,
legend style={at={(0.99,0.25)}, anchor=east, legend cell align=left, align=left, fill=none, draw=none}
]

\addplot [color=black,solid,draw=none,mark=*,mark options={scale=1.4,black,fill=clr11}]
  table[row sep=crcr]{%
  1.0000 0.4393 \\
  2.0000 0.4386 \\
  3.0000 0.0496 \\
  4.0000 0.0495 \\
  5.0000 0.0087 \\
  6.0000 0.0087 \\
  7.0000 0.0020 \\
  8.0000 0.0020 \\
};

\end{axis}

\end{tikzpicture}%
    \begin{tikzpicture}

\definecolor{clr1}{RGB}{18 78 128}
\definecolor{clr2}{RGB}{89 165 216}
\definecolor{clr3}{RGB}{145 229 246}
\definecolor{clr4}{RGB}{34 139 34}
\definecolor{clr5}{RGB}{255 77 109}
\definecolor{clr6}{RGB}{201 24 74}
\definecolor{clr7}{RGB}{128 15 47}
\definecolor{clr8}{RGB}{174 32 18}
\definecolor{clr9}{RGB}{155 34 38}
\definecolor{clr10}{RGB}{46 139 87}
\definecolor{clr11}{RGB}{240 128 128}
\definecolor{clr12}{RGB}{152 251 152}

\begin{axis}[%
width=0.35\textwidth,
height=0.15\textwidth,
scale only axis,
ymode = log,
xmin=1,
xmax=8,
xtick={1,2,3,4,5,6,7,8},
xlabel={$mode$},
ylabel={$\lambda/\sum \lambda$},
ylabel style={at={(0.05,0.5)}},
axis background/.style={fill=white},
legend columns=3,transpose legend,
legend style={at={(0.99,0.25)}, anchor=east, legend cell align=left, align=left, fill=none, draw=none}
]

\addplot [color=black,solid,draw=none,mark=*,mark options={scale=1.4,black,fill=clr11}]
  table[row sep=crcr]{%
  1.0000 0.4411 \\
  2.0000 0.4334 \\
  3.0000 0.0509 \\
  4.0000 0.0506 \\
  5.0000 0.0094 \\
  6.0000 0.0093 \\
  7.0000 0.0020 \\
  8.0000 0.0020 \\
};

\end{axis}

\end{tikzpicture}%
    \begin{tikzpicture}

\definecolor{clr1}{RGB}{18 78 128}
\definecolor{clr2}{RGB}{89 165 216}
\definecolor{clr3}{RGB}{145 229 246}
\definecolor{clr4}{RGB}{34 139 34}
\definecolor{clr5}{RGB}{255 77 109}
\definecolor{clr6}{RGB}{201 24 74}
\definecolor{clr7}{RGB}{128 15 47}
\definecolor{clr8}{RGB}{174 32 18}
\definecolor{clr9}{RGB}{155 34 38}
\definecolor{clr10}{RGB}{46 139 87}
\definecolor{clr11}{RGB}{240 128 128}
\definecolor{clr12}{RGB}{152 251 152}

\begin{axis}[%
width=0.35\textwidth,
height=0.15\textwidth,
scale only axis,
xmin=1,
xmax=8,
ymin=0.0,
ymax=1.07,
xtick={1,2,3,4,5,6,7,8},
xlabel={$mode$},
ylabel={$St$},
ylabel style={at={(0.05,0.5)}},
axis background/.style={fill=white},
legend columns=3,transpose legend,
legend style={at={(0.99,0.25)}, anchor=east, legend cell align=left, align=left, fill=none, draw=none}
]

\addplot[red, dashed, domain=1:8] {0.259};

\addplot [color=black,solid,draw=none,mark=*,mark options={scale=1.4,black,fill=clr11}]
  table[row sep=crcr]{%
  1.0000 0.259 \\
  2.0000 0.259 \\
  3.0000 0.514 \\
  4.0000 0.514\\
  5.0000 0.7692 \\
  6.0000 0.7692 \\
  7.0000 1.024 \\
  8.0000 1.024 \\
};

\end{axis}

\end{tikzpicture}%
    \begin{tikzpicture}

\definecolor{clr1}{RGB}{18 78 128}
\definecolor{clr2}{RGB}{89 165 216}
\definecolor{clr3}{RGB}{145 229 246}
\definecolor{clr4}{RGB}{34 139 34}
\definecolor{clr5}{RGB}{255 77 109}
\definecolor{clr6}{RGB}{201 24 74}
\definecolor{clr7}{RGB}{128 15 47}
\definecolor{clr8}{RGB}{174 32 18}
\definecolor{clr9}{RGB}{155 34 38}
\definecolor{clr10}{RGB}{46 139 87}
\definecolor{clr11}{RGB}{240 128 128}
\definecolor{clr12}{RGB}{152 251 152}

\begin{axis}[%
width=0.35\textwidth,
height=0.15\textwidth,
scale only axis,
xmin=1,
xmax=8,
ymin=0.0,
ymax=1.07,
xtick={1,2,3,4,5,6,7,8},
xlabel={$mode$},
ylabel={$St$},
ylabel style={at={(0.05,0.5)}},
axis background/.style={fill=white},
legend columns=3,transpose legend,
legend style={at={(0.99,0.25)}, anchor=east, legend cell align=left, align=left, fill=none, draw=none}
]

\addplot[red, dashed, domain=1:8] {0.2645};

\addplot [color=black,solid,draw=none,mark=*,mark options={scale=1.4,black,fill=clr11}]
  table[row sep=crcr]{%
  1.0000 0.2645 \\
  2.0000 0.2645 \\
  3.0000 0.5289 \\
  4.0000 0.5289\\
  5.0000 0.7987 \\
  6.0000 0.7987 \\
  7.0000 1.0632 \\
  8.0000 1.0632\\
};

\end{axis}

\end{tikzpicture}%
    \includegraphics[width=0.49\textwidth]{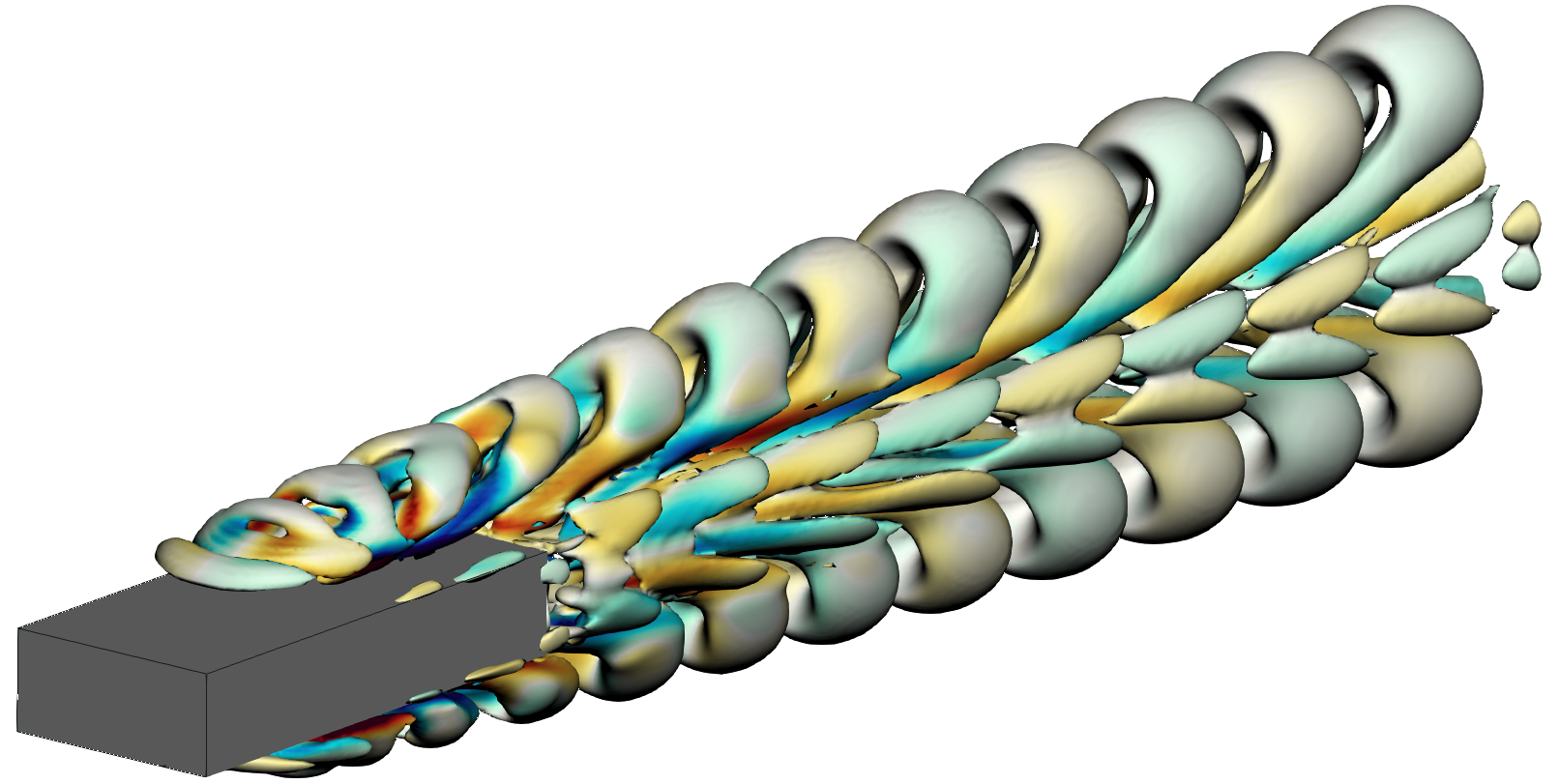}
    \includegraphics[width=0.49\textwidth]{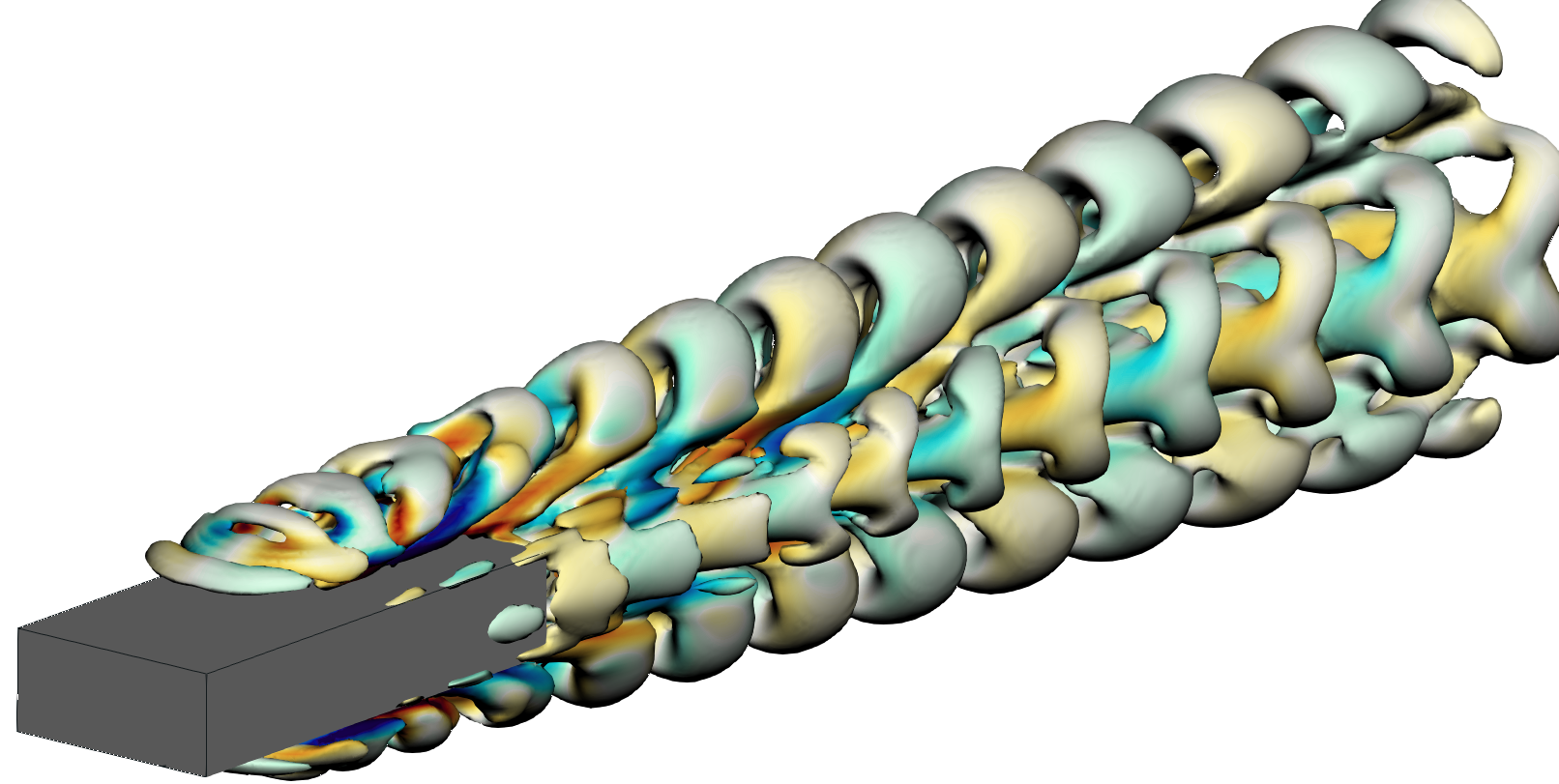}
    \caption{POD analysis for $L=5$ and $W=2.25$ at $Re=420$ (left) and $Re=450$ (right). In both cases the flow is periodic and has a single dominant mode, of frequency  $St \approx 0.255$ and $St \approx 0.265$, respectively (see figure \ref{fig:W225_FlowStr_375_470}). 
    Top panels:  energy fractions of the first 8 modes and corresponding frequencies. 
    Bottom panels: structure of POD mode 1, to be compared with 
    figure \ref{fig:W225_POD}. Isosurfaces of $\lambda_2$ coloured by streamwise vorticity $\omega_x$ (blue-to-red colormap ranges from negative to positive values). 
    For $Re=420$ the structure of the mode is the same as for $Re=380$, and shows the same symmetries. For $Re=450$, instead, the symmetries of the mode change.}
    \label{fig:W225_POD_Re420_Re450}
\end{figure}

Figure \ref{fig:W225_POD_Re420_Re450} considers $L=5$ and $W=2.25$ and shows the effect of $Re$ on the spatial structure of the POD mode associated with the LE vortex shedding.
The structure of that mode does not change when $Re$ is increased from $380$ to $420$. 
The flow is periodic and the unsteadiness is driven by the in-phase shedding of hairpin vortices from the top/bottom LE shear layers (figure \ref{fig:W225_FlowStr_375_470}). 
When $Re$ is further increased, however, the symmetries of the mode change, in agreement with a shedding of hairpin vortices in phase opposition from the top /bottom LE shear layers:
for $380 \le Re \le 420$, the POD mode is symmetric with respect to $z=0$, with $\omega_x(x,y,z) = - \omega_x(x,y,-z)$. 
whereas for $440 \le Re \le 460$, it is antisymmetric, with $\omega_x(x,y,z) = \omega_x(x,y,-z)$. 

\bibliographystyle{jfm}
\bibliography{Stability}

\end{document}